\documentclass[a4paper,12pt,english,greek]{report}
\usepackage{cite}
\usepackage{amssymb}
\usepackage{amsmath}
\usepackage{amsfonts}
\usepackage{amsthm}
\usepackage{mathtools}
\usepackage[utf8]{inputenc}
\usepackage{color}
\usepackage{graphicx}
\usepackage[dvipsnames]{xcolor}
\usepackage[active]{srcltx}
\usepackage{here}
\usepackage{cite}
\usepackage{soul}
\usepackage[affil-it]{authblk}

\input paperdef

\usepackage{hyperref}
\hypersetup{
    colorlinks=true, 
    linktoc=all,     
    citecolor=BrickRed,
    filecolor=black,
    linkcolor=blue,
    urlcolor=cyan
}

\usepackage{kerkis}

\usepackage[T1]{fontenc}
\usepackage{kerkis}
\usepackage[english,greek]{babel}
\usepackage{amsfonts,bm,amssymb,array,cite}
\usepackage[mathscr]{euscript}
\usepackage[all,color]{xy}
\usepackage{relsize}
\usepackage{tikz}
\usepackage{url}
\usepackage{amsmath,amsthm,bm}
\usepackage{cancel}
\usepackage{accents}
\usepackage[]{graphicx}
\usepackage{mwe} 
\usepackage{caption}

\usepackage{empheq}

\begin{document}
\begin{titlepage}
\!\!\!\!\!\!\!\!\!\!\!\!\!\!\!
\selectlanguage{greek}
\begin{minipage}{0.22\textwidth}
\begin{flushleft}
\includegraphics[width=0.82\textwidth]{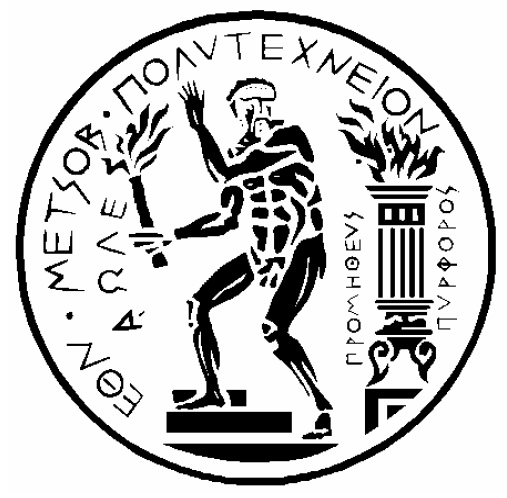}\quad
\end{flushleft}
\end{minipage}\!\!\!\!\!\!\!\!\!\!
\begin{minipage}{0.8 \textwidth}
\begin{center}
\textsc{\LARGE ΕΘΝΙΚΟ ΜΕΤΣΟΒΙΟ ΠΟΛΥΤΕΧΝΕΙΟ}\\[0.2cm]
\textsc{\Large ΣΧΟΛΗ ΕΦΑΡΜΟΣΜΕΝΩΝ ΜΑΘΗΜΑΤΙΚΩΝ   \\[0.1cm] ΚΑΙ ΦΥΣΙΚΩΝ ΕΠΙΣΤΗΜΩΝ}\\[0.2cm]
\end{center}
\end{minipage}
\\[3cm]
\begin{center}
\textsc{\textbf{\LARGE \eng Phenomenological Consequences of Supersymmetric Theories from Dimensional Reduction and Reduction of Couplings}}\\[3cm]
{\Large  \eng PhD Thesis of \\[0.1cm]
\Large GREGORY PATELLIS\\[0.5cm]
\large Department of Physics \\[0.1cm]
National Technical University of Athens}
\\[2.5cm]
\end{center}
\begin{flushleft}
{\Large \eng
SUPERVISOR:\\[0.1cm] George Zoupanos\\[0.1cm]
Professor Emeritus NTUA \\}
\end{flushleft}
\begin{center}
\vfill
\large  \eng December 2020
\end{center}
\end{titlepage}
\newpage
\thispagestyle{empty}
\mbox{}
\newpage

\begin{titlepage}
\!\!\!\!\!\!\!\!\!\!\!\!\!\!\!
\selectlanguage{greek}
\begin{minipage}{0.22\textwidth}
\begin{flushleft}
\includegraphics[width=0.82\textwidth]{./prometheus}\quad
\end{flushleft}
\end{minipage}\!\!\!\!\!\!\!\!\!\!
\begin{minipage}{0.8 \textwidth}
\begin{center}
\textsc{\LARGE ΕΘΝΙΚΟ ΜΕΤΣΟΒΙΟ ΠΟΛΥΤΕΧΝΕΙΟ}\\[0.2cm]
\textsc{\Large ΣΧΟΛΗ ΕΦΑΡΜΟΣΜΕΝΩΝ ΜΑΘΗΜΑΤΙΚΩΝ   \\[0.1cm] 
ΚΑΙ ΦΥΣΙΚΩΝ ΕΠΙΣΤΗΜΩΝ}\\[0.2cm]
\end{center}
\end{minipage}
\\[2.2cm]
\begin{center}
\textsc{\textbf{\LARGE \eng Phenomenological Consequences of Supersymmetric Theories from Dimensional Reduction and Reduction of Couplings}}\\[2cm]
{\Large  \eng PhD Thesis of \\[0.1cm]
\Large GREGORY PATELLIS\\[0.1cm]
\large Department of Physics \\[0.1cm]
National Technical University of Athens}
\\[2cm]
\end{center}

\begin{minipage}{1.1\textwidth}
\begin{minipage}{0.5\textwidth} 
{\large \eng 3-MEMBER ADVISORY COMMITTEE:}
\begin{flushleft}
\eng
1.............George Koutsoumbas, Prof. NTUA \\[0.2cm]
2.............Nicholas Tracas, Prof. NTUA \\[0.2cm]
3.............George Zoupanos, Prof. Emer. NTUA \\[0.2cm]
\phantom{3.............}(Supervisor) \\[2.5cm]
\phantom{l}
\end{flushleft}
\end{minipage}
\begin{minipage}{0.5\textwidth}
{\large \eng \vspace{-0.3cm} 7-MEMBER EXAMINING COMMITTEE:}
\begin{flushleft}
\eng
1............Konstantinos Anagnostopoulos,\\ \phantom{4............}Assoc. Prof. NTUA \\[0.2cm]
2............Konstantinos Farakos, Prof. NTUA \\[0.2cm]
3............Nikos Irges, Assoc. Prof. NTUA \\[0.2cm]
4............George Koutsoumbas, Prof. NTUA  \\[0.2cm]
5............George Leontaris, Professor UOI \\[0.2cm]
6............Myriam Mondrag\'on, Prof. UNAM \\[0.2cm]
7............Nicholas Tracas, Prof. NTUA \\[0.2cm]
\end{flushleft}
\end{minipage}
\end{minipage}

\begin{center}
\large \eng  Athens, December 2020
\end{center}
\end{titlepage}

\newpage
\thispagestyle{empty}
\mbox{}
\newpage 

\pagenumbering{arabic}
\setcounter{page}{1}

\selectlanguage{english}

\phantom{a}
\vspace{4cm}

\begin{flushright}
\textit{“Man can will nothing unless he has first understood that he must\\ count on no one but himself; that he is alone, abandoned on earth in the\\ midst of his infinite responsibilities, without help, with no other aim than the one\\ he sets himself, with no other destiny than the one he forges for himself on this earth.”}\\
JP. S.
\end{flushright}

\newpage
\mbox{}
\newpage 

\subsection*{\hel Ευχαριστίες}
\hel
Καταρχάς, θα ήθελα να ευχαριστήσω από καρδιάς τον επιβλέποντά μου, Ομ. Καθηγητή Γιώργο Ζουπάνο, για την καθοδήγησή του ολόκληρη την τελευταία δεκαετία. Στο επιστημονικό κομμάτι, μεταξύ πολλών άλλων, για την -άμεση ή έμμεση- προτροπή στη σφαιρική κατανόηση αντί της εμμονής στο υπέρ του δέοντος συγκεκριμένο. Κυρίως όμως για την σπάνια ηθική του ακεραιότητα σε κάθε τομέα, επιστημονικό και κοινωνικό.
Τον ευχαριστώ επίσης για την δυνατότητα συμμετοχής -αλλά και συνεισφοράς- που μου έδωσε στο ετήσιο διεθνές συνέδριο της Κέρκυρας, αλλά και για την ενθάρρυνση για επισκέψεις σε διακεκριμένα ιδρύματα του εξωτερικού, στα οποία βρήκα συνεργάτες διεθνούς δραστηριότητας και αναγνώρισης.

Θα ήθελα να ευχαριστήσω τους δύο συνεργάτες μου, την Καθηγήτρια \eng Myriam Mondrag\'on \hel και τον Καθηγητή \eng Sven Heinemeyer, \hel για την πολύτιμη βοήθεια και τις συμβουλές που μου προσέφεραν καθόλη τη διάρκεια της συνεργασίας μας.

Επίσης, ευχαριστώ τους συνεπιβλέποντές μου, τον Καθηγητή Νικόλα Τράκα και τον Καθηγητή Γιώργο Κουτσούμπα, καθώς ήταν πάντα πρόθυμοι να απαντήσουν σε κάθε μου ερώτηση, αλλά και τον στενό μου φίλο και συνεργάτη, Γιώργο Μανωλάκο, διότι, πέρα από τη βοήθειά του σε επιστημονικό επίπεδο, άντεξε όλα μου τα παράπονα τον τελευταίο καιρό.

Ακόμα, ευχαριστώ όλους τους καλούς φίλους που απέκτησα αυτά τα χρόνια, πανεπιστημιακούς και χορευτικούς, καθώς η παρέα και οι συζητήσεις μαζί τους διαμόρφωσαν σε μεγάλο βαθμό το πλαίσιο μέσα στο οποίο η προσπάθειά μου αυτή αποκτά νόημα.

Τέλος, δε γίνεται να μην αφιερώσω την παρούσα διατριβή στη μητέρα μου και τη γιαγιά μου, χωρίς την ακούραστη στήριξη των οποίων δε θα ήταν δυνατό να ολοκληρωθεί.

\eng

\newpage 
\mbox{} 
\newpage

\section*{\hel Περίληψη}

\hel
Η ιδέα της «ελάττωσης παραμέτρων» βασίζεται στην εύρεση σχέσεων αναλλοίωτων κάτω από την ομάδα επανακανονικοποίησης μεταξύ παραμέτρων μιας επανακανονικοποιήσιμης θεωρίας, οι οποίες ισχύουν σε όλες τις τάξεις θεωρίας διαταραχών. Αυτή η μέθοδος μπορεί να εφαρμοστεί σε $N=1$ υπερσυμμετρικές θεωρίες μεγάλης ενοποίησης, καθιστώντας τις πεπερασμένες σε κάθε επίπεδο βρόχων. Στο πρώτο μέρος της παρούσας διατριβής, μετά από σύντομη αναδρομή στις βασικές αρχές της ελάττωσης παραμέτρων και της περατότητας, εξετάζονται τέσσερα μοντέλα που παρουσιάζουν μεγάλο φαινομενολογικό ενδιαφέρον: μια ελαχιστοποιημένη εκδοχή του $N=1$ υπερσυμμετρικού μοντέλου $SU(5)$, το πεπερασμένο $N=1$ υπερσυμμετρικό μοντέλο $SU(5)$, το πεπερασμένο (σε επίπεδο δύο βρόχων) $N=1$ υπερσυμμετρικό μοντέλο $SU(3)\otimes SU(3)\otimes SU(3)$ και μια ελαχιστοποιημένη εκδοχή του Ελάχιστου Υπερσυμμετρικού Καθιερωμένου Προτύπου (ΕΥΚΠ). Η επανεξέταση αυτών των μοντέλων παρουσιάζει μια βελτιωμένη τιμή της μάζας του ελαφρού σωματιδίου \eng Higgs, \hel όπου για την ανάλυση χρησιμοποιήθηκε η καινούρια εκδοχή του προγράμματος \eng {\tt FeynHiggs}. \hel Το ελαφρύτερο υπερσυμμετρικό σωματίδιο, εφόσον είναι \eng neutralino, \hel μπορεί να θεωρηθεί ως υποψήφιο σωματίδιο σκοτεινής ύλης, υπόθεση που εξετάζεται με το πρόγραμμα \eng \MO, \hel αν και κανένα μοντέλο δεν δίνει ικανοποιητικά αποτελέσματα σε αυτήν την κατεύθυνση. Στα τρία ενοποιημένα μοντέλα παρατηρούνται σχετικά βαριά υπερσυμμετρικά φάσματα (τα οποία παρήχθησαν με τα προγράμματα \eng {\tt FeynHiggs} \hel και  \eng {\tt SPheno}) \hel που ξεκινούν πάνω από το $1~TeV$, οπότε είναι συνεπή με την μη παρατήρησή τους από τον Μεγάλο Επιταχυντή Αδρονίων \eng (LHC), \hel ενώ το ελαχιστοποιημένο ΕΥΚΠ παρουσιάζει τη μάζα του ψευδοβαθμωτού μποζονίου \eng Higgs, \hel $M_A$, σε περιοχή που αποκλείεται από πρόσφατα αποτελέσματα του πειράματος \eng ATLAS. \hel Η δυνατότητα ανακάλυψης καθενός από τα τρία μοντέλα ενοποίησης στον \eng HL-LHC \hel είναι μηδαμινή, όμως ο \eng FCC-hh \hel πιστεύεται ότι θα είναι σε θέση να παρατηρήσει μεγάλο μέρος των φασμάτων των τριών μοντέλων.

Στο δεύτερο μέρος της διατριβής, μετά από μία αναδρομή στις βασικές έννοιες της Διαστατικής Ελάττωσης σε Χώρους Πηλίκου, παρουσιάζεται μια επέκταση του Καθιερωμένου Προτύπου, η οποία προκύπτει από τη διαστατική ελάττωση μιας $N=1$,~$10D$ $E_8$ θεωρίας σε ένα χώρο της μορφής $M_4 \times B_0/ \mathbf{Z}_3 $, όπου $B_0$ είναι η \eng nearly-K\"ahler \hel πολλαπλότητα $SU(3)/U(1) \times U(1)$ και $\mathbf{Z}_3$ είναι μια \eng "freely-acting" \hel διακριτή συμμετρία του $B_0$. Η εφαρμογή του μηχανισμού σπασίματος \eng Wilson flux \hel καταλήγει σε μια $N=1$ θεωρία βαθμίδας $SU(3)^3$, με δύο επιπλέον καθολικές συμμετρίες $U(1)$. Κάτω από την κλίμακα ενοποίησης έχουμε ενα μοντέλο με δύο διπλέτες \eng Higgs \hel σε μια \eng split-like \hel υπερσυμμετρική εκδοχή του Καθιερωμένου Προτύπου η οποία είναι φαινομενολογικά συνεπής, εφόσον παράγει μάζες για το μποζόνιο \eng Higgs \hel και τα κουάρκς της τρίτης οικογένειας σε συμφωνία με τις πρόσφατες πειραματικές μετρήσεις, ενώ η τιμή της μάζας του ελαφρύτερου υπερσυμμετρικού σωματιδίου προβλέπεται στα $\sim 1500~GeV$.
\eng

\newpage 
\mbox{} 
\newpage

\section*{Abstract}
The 'reduction of couplings' idea~consists~in the search for
renormalization~group invariant~(RGI) relations between seemingly independent parameters of a~renormalizable~theory that hold to all orders of perturbation~theory. This~concept can be applied to $N=1$ supersymmetric~Grand Unified~Theories (GUTs) and make them finite at all loops.~In the~first part of this thesis, after a review of the~reduction of~couplings method and the properties of the resulting~finiteness~in supersymmetric theories, four phenomenologically~interesting models~are analysed: a minimal version of the $N=1$~supersymmetric~$SU(5)$, a finite $N=1$ supersymmmetric~$SU(5)$, a two-loop~finite $N=1$ supersymmetric~$SU(3)\otimes SU(3)\otimes SU(3)$ model~and a reduced version~of the~Minimal Supersymmetric Standard Model (MSSM).~A relevant update~in the phenomenological evaluation has been~the improved~light Higgs-boson mass prediction as provided~by the latest~version of {\tt FeynHiggs}. The lightest supersymmetric~particle (LSP),~which is a neutralino, is considered as a cold~dark matter (CDM) candidate and put to test using the latest~\MO~code, although no model supports an acceptable value of~the CDM relic density. The first three models predict relatively~heavy supersymmetric spectra (produced using~the {\tt FeynHiggs} and {\tt SPheno} codes) that start just above~the TeV~scale, consistent with the non-observation LHC~results, while the reduced MSSM results in a pseudoscalar Higgs~boson mass $M_A$ that is ruled out by recent results~of the~ATLAS experiment. The prospects of discovery of the three~models at the HL-LHC vary from very dim to none, depending on~the model. The FCC-hh, however, will in principle be able to~test large parts of the predicted spectrum of each~unified model. 

In the second~part of the~thesis, after a review of the Coset Space Dimensional Reduction~(CSDR) scheme, an extension of the Standard Model is~presented~that results from the dimensional reduction of~the $N=1$,~$10D$ $E_8$ group over a $M_4 \times B_0/ \mathbf{Z}_3 $~space,~where $B_0$ is the nearly-K\"ahler manifold $SU(3)/U(1) \times U(1)$ and $\mathbf{Z}_3$ is~a freely acting~discrete group~on $B_0$. Using the Wilson flux breaking mechanism~we are~left in four dimensions with an $N=1$ $SU(3)^3$ gauge~theory~plus two global $U(1)$s. Below the unification scale~we have~a two Higgs doublet model in a split-like supersymmetric~version of~the Standard Model which is phenomenologically~consistent,~since it produces masses of a light Higgs,~the top and the~bottom particles within the experimental~limits and~predicts the LSP $\sim 1500~GeV$.

\newpage 
\mbox{} 
\newpage


{
  \hypersetup{linkcolor=black}
  \tableofcontents
}

\newpage

\chapter{Introduction}\label{intro}


An essential direction of the theoretical efforts of the last decades in Elementary~Particle Physics (EPP) is to understand the  free parameters of the Standard Model (SM)~in terms of few
fundamental~ones, i.e. to achieve~{\it reductions of couplings}\cite{book}. However, despite the many successes of the SM in describing elementary particles and their interactions, there is little to no progress when it comes to the plethora~of free parameters. The pathology of the large number of free parameters is deeply correlated to the {\it infinities} that emerge at the quantum level. Renormalization removes the infinities, but only at the cost of introducing counterterms that add to the number of~free parameters.

Although the SM is a successful theory, the widespread belief suggests that it will be proven to be the low energy  limit of a fundamental theory. The search for physics beyond the Standard Model (BSM)~expands in various directions.~One of the most efficient ways to reduce the number of~free parameters of a theory (and thus render it more~predictive) is the introduction of~a symmetry. A celebrated application of such an idea are~the Grand Unified Theories (GUTs)~\cite{Pati:1973rp,Georgi:1974sy,Georgi:1974yf,Fritzsch:1974nn,Gursey:1975ki,Achiman:1978vg}. Decades ago, an early version of $SU(5)$ reduced the gauge~couplings of the SM (featuring an -approximate-~gauge unification), predicting~one of them.~It was the addition of an $N=1$ (softly broken) supersymmetry~(SUSY)
\cite{Amaldi:1991cn,Dimopoulos:1981zb,Sakai:1981gr} that~made the prediction viable. In the framework~of GUTs, the Yukawa couplings can also be related among~themselves. Again, $SU(5)$ demonstrated~this by predicting the~ratio of the tau to bottom mass \cite{Buras:1977yy} for the~SM.
Unfortunately,~the introduction of additional gauge symmetry does not seem~to help, since new~complications arise due to new degrees of freedom, i.e. the channels of breaking the symmetry.

An alternative way to look for~relations among seemingly unrelated~parameters is the reduction of couplings method
\cite{Zimmermann:1984sx,Oehme:1984yy,Oehme:1985jy} (see~also refs \cite{Ma:1977hf,Chang:1974bv,Nandi:1978fw}).
This technique reduces the number of parameters of a
theory by relating~either all or a number~of couplings to a single~parameter, the ``primary coupling''.  This method can~identify hidden symmetries~in a system, but it is~also possible to have~reduction of couplings~in systems with no apparent~symmetry. It is necessary,~though, to make two assumptions:~both the
original and the reduced theory have to be~renormalizable and the relations among~parameters should be Renormalization Group Invariant (RGI).

A natural extension of the idea of Grand~Unification is to achieve gauge-Yukawa Unification~(GYU), that is to~relate the gauge sector to the Yukawa sector. This is a~feature of theories in which~\textit{reduction of couplings} has been~applied. The~original suggestion for~the reduction of couplings in GUTs leads to the search for~RGI relations that hold below the Planck scale, which~are in turn preserved down to the~unification scale. Impressively,~this observation guarantees~validity of such relations to all-orders in perturbation theory.~This is done by studying
their uniqueness~at one-loop level. Even more remarkably,
one can find~such RGI relations that result in all-loop finiteness. The above principles~have
only been applied in $N=1$~supersymmetric GUTs for reasons that will be transparent~in the following chapters. The application of the GYU~programme in the dimensionless couplings of~such theories has been very successful, including~the prediction of the mass of the top quark in the minimal~\cite{Kubo:1994bj} (see also Chapter \ref{roc_pheno} for the latest~update) and~in the finite $N = 1$~$SU(5)$ \cite{Kapetanakis:1992vx,Mondragon:1993tw} before its~experimental discovery~\cite{Lancaster:2011wr}.

In order to successfully~apply the above-mentioned programme, SUSY appears to be~essential. However, one has~to understand its breaking as well, since it has the~ambition to
supply the SM with predictions for~several of its free parameters. Indeed, the search for~RGI relations has been extended to the soft supersymmetry~breaking sector (SSB) of these~theories, which involves~parameters of dimension one and two. In addition,~there was important progress concerning
the renormalization~properties of the SSB parameters, based on the powerful~supergraph method~for studying supersymmetric theories, and it~was applied to the softly broken ones by~using the``spurion'' external space-time independent~superfields.
According to this method a~softly broken supersymmetric gauge theory is considered as a~supersymmetric one in which the various parameters,~such as couplings and masses, have
been promoted to~external superfields. Then, relations among the soft term~renormalization~and that of an unbroken supersymmetric theory have~been derived. In particular the
$\beta$-functions of the parameters of~the softly broken theory are expressed in terms~of partial differential operators involving the dimensionless~parameters of the unbroken
theory. The key point in~solving the set of coupled differential equations so as to be~able to express all parameters in a RGI way,~was to transform the partial differential~operators involved to~total derivative operators. It is indeed possible to do this~by~choosing a suitable RGI surface.

The application of~\textit{reduction of couplings} on $N=1$ SUSY theories has~led to many interesting phenomenological developments, too. In past work, an~appealing ``universal'' set of soft scalar masses was assumed~in the SSB sector of~supersymmetric theories, given that apart from economy and simplicity~(1) they are~part of the constraints that preserve finiteness up to~two loops, (2) they appear~in the attractive dilaton dominated~supersymmetry breaking superstring~scenarios.
However, further studies have~exhibited a number of problems, all due to the restrictive~nature of the ``universality'' assumption for the soft~scalar masses. Subsequently,~this constraint was replaced by~a more ``relaxed'' all-loop RGI soft scalar squared masses sum rule that keeps the most~attractive features of the universal case and overcomes~the unpleasant~phenomenological consequences. This arsenal~of tools and results opened the way~for the study of full finite models with few~free parameters, with~emphasis given on~predictions for the SUSY spectrum~and the light Higgs boson mass.

The Higgs mass prediction,~which coincided with the LHC results (ATLAS \cite{Aad:2012tfa,ATLAS:2013mma}~and CMS \cite{Chatrchyan:2012ufa,Chatrchyan:2013lba}) - combined with a predicted~relatively heavy spectrum of SUSY particles - was a success of the all-loop finite $N = 1$ SUSY $SU(5)$ model \cite{Heinemeyer:2007tz}, while another finite theory,~namely the $N=1$ (two-loop) finite~$SU(3)\otimes SU(3)\otimes SU(3)$ \cite{Ma:2004mi} (see also \cite{Mann:2017wzh,Wang:2018yer} for alternative versions of the model), also confirmed the~LHC results. Additionally, the~above programme was also~applied in the MSSM \cite{Mondragon:2013aea,Heinemeyer:2017gsv}, with~successful results concerning the top, bottom and Higgs masses,~also featuring a relatively heavy SUSY spectrum (a review of past analyses can be found in \cite{Heinemeyer:2019vbc}). Furthermore, it is~a well~known fact~that the lightest~neutralino, being the Lightest~SUSY Particle (LSP),~is an~excellent candidate for~Cold Dark~Matter (CDM)~\cite{EHNOS}. \\


On the other hand, apart from the efforts to reduce the arbitrariness in our description of Nature, the unification of all fundamental interactions has been - for more than half a century- the theoretical~physicists' desideratum. The community sets this direction in high priority, and as a result several interesting approaches have been developed over the past decades. Out of all of them, the ones that support and/or employ the existence of extra dimensions are perhaps the most prominent. The extra-dimensional scenarios are encouraged by a~very consistent~framework, the superstring theories \cite{strings}, with the~most promising being the~heterotic string \cite{gross-harvey} (defined~ in ten~dimensions), due to its potential~experimental~testability. 
Specifically, the~phenomenological~aspect of the heterotic string is designated in~the resulting GUTs, containing~the SM gauge group, which are obtained after compactifying~the 10D spacetime and performing a~dimensional reduction of the $E_8\times E_8$ initial~gauge theory. In~addition,~a~few years before the formulation~of superstring~theories, an~alternative framework of dimensional reduction~of higher-dimensional~gauge theories emerged. This~important undertaking, which~shared common aims with one of the many superstring~theories, was~initially explored by Forgacs-Manton~(F-M) and Scherk-Schwartz~(S-S) studying~the Coset Space Dimensional Reduction (CSDR) \cite{Forgacs:1979zs,Kapetanakis:1992hf, Kubyshin:1989vd} and the~group manifold reduction \cite{Scherk:1979zr},~respectively.

In the higher-dimensional theory, the gauge~fields unify the gauge~and scalar sector, while, after the~dimensional reduction,~the resulting 4D theory features a particle~spectrum of the surviving~components. This holds for both~approaches (F-M and S-S). Moreover, in the CSDR~programme, inclusion~of fermions in~the higher-dimensional theory~gives rise to~Yukawa interactions~in the resulting 4D theory. Furthermore,~for a specific choice of initial~dimensions, one can unify~the fields even more if~the  higher-dimensional gauge~theory is  $N=1$ supersymmetric, in the way  that both gauge and fermionic fields are accommodated in the same vector supermultiplet. It is important to note that the CSDR~scheme allows the~possibility of obtaining chiral 4D  theories~\cite{Manton:1981es,Chapline:1982wy}.

In the CSDR context, some notions inspired from the heterotic~string  were~incorporated, specifically~the dimensionality of the spacetime~and the gauge group~of the higher-dimensional~theory. Therefore, taking into consideration that superstring~theories are consistent~in ten dimensions, the extra~dimensions have to be~distinguished from the four observable~ones with a~compactification of the~metric and then the~resulting 4D theory is determined. Furthermore, a suitable~choice of compactification~manifold can~lead to 4D, $N=1$ supersymmetric~theories, aiming~for realistic~GUTs.

When dimensionally reducing an $N=1$ supersymmetric gauge theory, a~very~important and desired~property is the amount of supersymmetry of the~initial theory to be~preserved in the 4D one. Some very good candidates that preserve it are the compact~internal Calabi--Yau (CY) manifolds~\cite{Candelas:1985en}. However, the moduli~stabilization problem~that emerges led to the~study of flux compactification~in~a wider class of internal~spaces, namely manifolds with an SU(3)-structure. In these~manifolds a background~non-vanishing, globally defined spinor is considered. This spinor is covariantly~constant with respect to a connection including~torsion, while, in~the CY~case, this holds for~a Levi--Civita~connection. Here, the nearly-K\"ahler manifolds class~of the SU(3)-structure~manifolds is considered~\cite{LopesCardoso:2002vpf,Strominger:1986uh,Lust:1986ix,Castellani:1986rg} , \cite{Becker:2003yv,Becker:2003sh,Gurrieri:2004dt}, \cite{Benmachiche:2008ma,Micu:2004tz,Frey:2005zz,Manousselis:2005xa} , \cite{Chatzistavrakidis:2008ii,Chatzistavrakidis:2009mh,Dolan:2009nz} , \cite{Lechtenfeld:2010dr,Popov:2010rf,Klaput:2011mz,Chatzistavrakidis:2012qb,Gray:2012md,Klaput:2012vv,Irges:2011de,Irges:2012ze, Butruille}. Specifically,~the class of 6D homogeneous nearly-K\"ahler manifolds includes the non-symmetric~coset spaces $G_2/SU(3)$,~$Sp(4)/SU(2)\times U(1)_{non-max}$,~$ SU(3)/U(1)\times U(1)$ and~the group~manifold $SU(2)\times SU(2)$  \cite{Butruille} (see~also \cite{LopesCardoso:2002vpf,Strominger:1986uh,Lust:1986ix,Castellani:1986rg, Becker:2003yv,Becker:2003sh,Gurrieri:2004dt,Benmachiche:2008ma,Micu:2004tz,Frey:2005zz,Manousselis:2005xa,Chatzistavrakidis:2008ii,Chatzistavrakidis:2009mh,Dolan:2009nz,Lechtenfeld:2010dr,Popov:2010rf,Klaput:2011mz,Chatzistavrakidis:2012qb,Gray:2012md,Klaput:2012vv,Irges:2011de,Irges:2012ze}). It is~worth mentioning~that, contrary~to the CY case,~the dimensional reduction of a 10D $N=1$ supersymmetric~gauge theory~over a non-symmetric coset~space, leads to 4D theories which~include supersymmetry breaking~terms \cite{Manousselis:2001xb,Manousselis:2000aj,Manousselis:2004xd}.

The above-mentioned framework has a very interesting application in the~dimensional reduction of an $N-1$, 10D $E_8$ over the compactification space $SU(3)/ U(1) \times U(1) \times \mathbf{Z}_3$, where~the latter~is the non-symmetric coset~space $SU(3)/U(1)\times U(1)$ equipped with the freely-acting discrete symmetry~$Z_3$. This extra symmetry is needed in order to employ the Wilson flux breaking~mechanism (\cite{Hosotani:1983xw,Zoupanos:1987wj,Kozimirov:1989kn}) to further reduce the gauge symmetry of the $4D$ GUT.  Specifically, the $E_6$ GUT (along with two $U(1)$ global~symmetries) is broken to the  trinification group, $SU(3)^3$ \cite{Kapetanakis:1992hf,Manousselis:2001xb,Chatzistavrakidis:2008ii,Irges:2011de} (see also \cite{Lust:1985be}). The potential of the~resulting 4D theory contains terms that~can be identified as $F$-, $D$- and soft~breaking terms, which means that the~resulting theory is a (broken) $N=1$ supersymmetric theory. \\


\noindent The present thesis in organised as follows:\\
Starting from \refpa{part1}, in \refcha{RoC} there is a brief review of the reduction of couplings principle and method. In \refcha{finiteness} the notion of finiteness is explained and the way to obtain finite theories with dimensionless and dimensionful couplings is reviewed. \refcha{roc_models} reviews four phenomenologically promising models, namely the Minimal $N=1$ $SU(5)$, the all-loop Finite $N=1$ $SU(5)$, the two-loop Finite $SU(3)\otimes SU(3)\otimes SU(3)$ and the Reduced MSSM. Their analysis follows in \refcha{roc_pheno} (see the original publications \cite{Heinemeyer:2018zpw,Heinemeyer:2020ftk,Heinemeyer:2020nzi}). Our analyses predict the top, bottom and light Higgs masses within experimental limits (with the exception of the bottom mass in the Minimal $SU(5)$ model), produce the full supersymmetric spectrum and the SUSY breaking scale, the CDM relic density for the LSP and the cross sections and branching rations necessary to determine the discovery potential of each model in (near) future colliders. While our analysis is based on the Higgs sector results obtained with the new version of {\tt FeynHiggs 2.16.0}  code \cite{Degrassi:2002fi,BHHW,FeynHiggs,Bahl:2019hmm}, other software was used as well. In particular, {\tt SPheno 4.0.4}  \cite{Porod:2003um,Porod:2011nf} was used for the calculation of branching ratios, {\tt MadGraph} \cite{Alwall:2014hca} for the calculation of the cross sections and the {\tt MicrOMEGAs 5.0} \cite{Belanger:2001fz,Belanger:2004yn,Barducci:2016pcb} code was used for the calculation of the CDM relic density. As will be~discussed in \refcha{roc_pheno}, none of the models~satisfy the experimental bounds of~the relic density.

Proceeding to \refpa{part2}, \refcha{csdr} reviews the CSDR scheme and the reduction of a $D$-dimensional Yang-Mills-Dirac theory accordingly. In \refcha{e8_coset} the application of the CSDR programme in the case of a $N=1$, $10D$ $E_8$ group reduced over the non-symmetric coset $SU(3)/U(1)\times U(1)$ in demonstrated. The necessary Wilson flux breaking using a $Z_3$ discrete symmetry is demonstrated in \refcha{wilson} and the $SU(3)^3$ GUT is obtained.
The radii of the coset are chosen to be small in a specific configuration, while the compactification~and unification scales coincide, leading~to a split-like supersymmetry scenario in which~some supersymmetric particles are superheavy and others~obtain mass in the $TeV$ region, as described in \refcha{selection-breaking}. After the employment of~the spontaneous symmetry breaking of the GUT, the model~can be viewed as a two Higgs doublet~model (2HDM).
The detailed phenomenological analysis of the model is presented in \refcha{pheno-su33} (see \cite{Manolakos:2020cco} for the original work). Finally, some closing remarks can be found in \ref{conclusions}.

\newpage
\mbox{} 
\newpage

\part{Theories with Reduced Couplings - Finite Theories}\label{part1}

\newpage
\mbox{} 
\newpage

\chapter{Reduction of Couplings}\label{RoC}

The basic idea of \textit{reduction of couplings} should be the first to be reviewed, as first introduced in \cite{Zimmermann:1984sx} and consequently evolved and expanded over the next two decades. The aim is to express the parameters~of a theory that~are considered free in~terms of one basic~parameter called~primary coupling. The basic idea is~to search for RGI relations~among couplings and~use them to reduce~the seemingly independent~parameters.
It is first applied to parameters without mass dimension, then extended to parameters of dimension one or two, i.e. the couplings and masses of the soft breaking sector of an $N=1$ supersymmetric theory.

\section{Reduction of Dimensionless~Parameters}\label{roc_0}

Any RGI~relation (i.e. which does not depend on the  renormalization~scale $\mu$ explicitly) among couplings $g_1,...,g_A$ of a given~renormalizable theory
 can be~expressed in the  implicit~form $\Phi (g_1,\cdots,g_A) ~=~\mbox{const.}$,~which has to~satisfy the ~partial differential equation~(PDE)
\beq
\mu\,\frac{d \Phi}{d \mu} = {\vec \nabla}\Phi\cdot {\vec \beta} ~=~
\sum_{a=1}^{A}
\,\beta_{a}\,\frac{\partial \Phi}{\partial g_{a}}~=~0~,
\eeq
where $\beta_a$ is the~$\beta$-function of $g_a$.
Solving this PDE is equivalent~to solving a set of ordinary~differential equations, the so-called~reduction equations~(REs) \cite{Zimmermann:1984sx,Oehme:1984yy,Oehme:1985jy},
\beq
\beta_{g} \,\frac{d g_{a}}{d g} =\beta_{a}~,~a=1,\cdots,A~,
\label{redeq}
\eeq
where $g$ and  $\beta_{g}$ are~the  primary
coupling and its~$\beta$-function, respectively, 
and the~counting on $a$ does not~include $g$.
Since maximally~($A-1$)~independent
RGI ``constraints''~in the  $A$-dimensional~space of~couplings
can be imposed by~the  $\Phi_a$'s, one~could in~principle
express all the~couplings~in terms of
a single coupling $g$.~However, a closer~look to the  set of \refeqs{redeq} reveals~that their~general solutions~contain as many integration constants~as the  number of~equations themselves. Thus, using such~integration constants~we have just~traded an integration~constant for each ordinary renormalized coupling,~and consequently, these~general solutions cannot be considered~as reduced ones.~The  crucial requirement in the~search for RGE relations~is to demand~power series solutions to the~REs,
\beq
g_{a} = \sum_{n}\rho_{a}^{(n)}\,g^{2n+1}~,
\label{powerser}
\eeq
which~preserve perturbative~renormalizability.
Such an ansatz fixes the~corresponding integration constant in each of~the REs and~picks up a special~solution out of the~general one.~Remarkably, the  uniqueness~of such power series solutions can be~decided already at the~one-loop level
\cite{Zimmermann:1984sx,Oehme:1984yy,Oehme:1985jy}.  To illustrate~this, we may~assume   $\beta$-functions of the~form
\beq
\begin{split}
\beta_{a} &=\frac{1}{16 \pi^2}\left[ \sum_{b,c,d\neq
  g}\beta^{(1)\,bcd}_{a}g_b g_c g_d+
\sum_{b\neq g}\beta^{(1)\,b}_{a}g_b g^2\right]+\cdots~,\\
\beta_{g} &=\frac{1}{16 \pi^2}\beta^{(1)}_{g}g^3+ \cdots~.
\end{split}
\eeq
Here
$\cdots$ stands~for higher-order terms, and~$\beta^{(1)\,bcd}_{a}$'s
are~symmetric in $ b,c,d$.  We then~assume that the  $\rho_{a}^{(n)}$'s~with $n\leq r$ have~been uniquely determined. To obtain~$\rho_{a}^{(r+1)}$'s, we insert~the  power series (\ref{powerser}) into~the REs (\ref{redeq}) and  collect terms of ${\cal O}(g^{2r+3})$. Thus, we~find
\beq
\sum_{d\neq g}M(r)_{a}^{d}\,\rho_{d}^{(r+1)} = \mbox{lower
  order~quantities}~,\non
\eeq
where the  right-hand side is known~by assumption and
\begin{align}
M(r)_{a}^{d} &=3\sum_{b,c\neq
  g}\,\beta^{(1)\,bcd}_{a}\,\rho_{b}^{(1)}\,
\rho_{c}^{(1)}+\beta^{(1)\,d}_{a}
-(2r+1)\,\beta^{(1)}_{g}\,\delta_{a}^{d}~,\label{M}\\
0 &=\sum_{b,c,d\neq g}\,\beta^{(1)\,bcd}_{a}\,
\rho_{b}^{(1)}\,\rho_{c}^{(1)}\,\rho_{d}^{(1)} +\sum_{d\neq
  g}\beta^{(1)\,d}_{a}\,\rho_{d}^{(1)}
-\beta^{(1)}_{g}\,\rho_{a}^{(1)}~.
\end{align}
Therefore,~the  $\rho_{a}^{(n)}$'s for all $n > 1$ for a
given set~of $\rho_{a}^{(1)}$'s can be uniquely~determined if $\det M(n)_{a}^{d} \neq 0$ for all $n \geq 0$. This is checked~in all models that reduction~of couplings is applied.

The  various~couplings in supersymmetric theories~have the  same asymptotic~behaviour.  Therefore, searching for~a power series solution of the~form (\ref{powerser}) to the  REs (\ref{redeq}) is~justified.

The possibility~of coupling unification described in this section~is without any doubt very attractive because~the  ``completely reduced'' theory contains~only one independent coupling, but it can be~unrealistic. Therefore, one often would like~to impose fewer RGI~constraints, and  this is~the idea of partial~reduction \cite{Kubo:1985up,Kubo:1988zu}.\\

All the above~give rise to hints towards~an underlying connection among~the~requirement~of~reduction~of couplings~and
SUSY.  
As an~example, one can consider a $SU(N)$ gauge theory~with $\phi^{i}({\bf N})$ and $\hat{\phi}_{i}(\overline{\bf N})$ complex scalars,~$\psi^{i}({\bf N})$ and $\hat{\psi}_{i}(\overline{\bf N})$ left-handed~Weyl spinors
and $\lambda^a~(a=1,\dots,N^2-1)$~right-handed Weyl spinors~in the adjoint~representation~of $SU(N)$.

\noindent The Lagrangian~(kinetic terms are omitted)~includes
\beq
{\cal L} \supset 
i \sqrt{2} \{~g_Y\overline{\psi}\lambda^a T^a \phi
-\hat{g}_Y\overline{\hat{\psi}}\lambda^a T^a \hat{\phi}
+\mbox{h.c.}~\}-V(\phi,\overline{\phi}),
\eeq
where
\beq
V(\phi,\overline{\phi}) =
\frac{1}{4}\lambda_1(\phi^i \phi^{*}_{i})^2+
\frac{1}{4}\lambda_2(\hat{\phi}_i \hat{\phi}^{*~i})^2
+\lambda_3(\phi^i \phi^{*}_{i})(\hat{\phi}_j \hat{\phi}^{*~j})+
\lambda_4(\phi^i \phi^{*}_{j})
(\hat{\phi}_i \hat{\phi}^{*~j}),
\eeq
This is the~most~general renormalizable form~in~four dimensions. In search~of a solution of the~form of \refeq{powerser} for~the REs, among other~solutions, one finds in lowest~order:
\beq
\begin{split}
g_{Y}&=\hat{g}_{Y}=g~,\\
\lambda_{1}&=\lambda_{2}=\frac{N-1}{N}g^2~,\\
\lambda_{3}&=\frac{1}{2N}g^2~,~
\lambda_{4}=-\frac{1}{2}g^2~,
\end{split}
\eeq
which~corresponds~to a $N=1$ SUSY gauge~theory. While~these remarks do not provide~an answer about the relation~of reduction of couplings~and SUSY, they certainly~point to further study~in that direction.

\section{Reduction~of Couplings in \texorpdfstring{$N=1$}{Lg} SUSY Gauge~Theories - Partial Reduction}\label{roc_susy}

Consider a~chiral,~anomaly-free, $N=1$ globally~supersymmetric
gauge~theory that is~based on a group G and has~gauge coupling $g$.~The~superpotential of the theory~is:
\bea
W&=& \frac{1}{2}\,m_{ij} \,\phi_{i}\,\phi_{j}+
\frac{1}{6}\,C_{ijk} \,\phi_{i}\,\phi_{j}\,\phi_{k}~,
\label{supot0}
\eea
where $m_{ij}$ and $C_{ijk}$ are gauge~invariant tensors and~the chiral superfield $\phi_{i}$ belongs to the irreducible representation~$R_{i}$ of the gauge group. The~renormalization~constants associated with the~superpotential,~for preserved SUSY,~are:
\begin{align}
\phi_{i}^{0}&=\left(Z^{j}_{i}\right)^{(1/2)}\,\phi_{j}~,~\\
m_{ij}^{0}&=Z^{i'j'}_{ij}\,m_{i'j'}~,~\\
C_{ijk}^{0}&=Z^{i'j'k'}_{ijk}\,C_{i'j'k'}~.
\end{align}

\noindent By virtue of~the $N=1$ non-renormalization theorem \cite{Wess:1973kz,Iliopoulos:1974zv,Ferrara:1974fv,Fujikawa:1974ay} there~are no~mass and~cubic~interaction~term infinities. Therefore:
\begin{equation}
\begin{split}
Z_{ij}^{i'j'}\left(Z^{i''}_{i'}\right)^{(1/2)}\left(Z^{j''}_{j'}\right)^{(1/2)}
&=\delta_{(i}^{i''}
\,\delta_{j)}^{j''}~,\\
Z_{ijk}^{i'j'k'}\left(Z^{i''}_{i'}\right)^{(1/2)}\left(Z^{j''}_{j'}\right)^{(1/2)}
\left(Z^{k''}_{k'}\right)^{(1/2)}&=\delta_{(i}^{i''}
\,\delta_{j}^{j''}\delta_{k)}^{k''}~.
\end{split}
\end{equation}
The only~surviving~infinities are
the wave~function~renormalization~constants $Z^{j}_{i}$, so just~one~infinity
per field. The  $\beta$-function of the gauge~coupling $g$ at the one-loop level~is given~by
\cite{Parkes:1984dh,West:1984dg,Jones:1985ay,Jones:1984cx,Parkes:1985hh}
\beq
\beta^{(1)}_{g}=\frac{d g}{d t} =
\frac{g^3}{16\pi^2}\left[\,\sum_{i}\,T(R_{i})-3\,C_{2}(G)\right]~,
\label{betag}
\eeq
where~$C_{2}(G)$ is the~quadratic~Casimir operator of the adjoint~representation~of the gauge~group $G$ and $\textrm{Tr}[T^aT^b]=T(R)\delta^{ab}$,~where  $T^a$ are the group~generators in the appropriate~representation.
The $\beta$-functions of $C_{ijk}$ are related~to the
anomalous dimension~matrices $\gamma_{ij}$ of the matter~fields~as:
\beq
\beta_{ijk} =
 \frac{d C_{ijk}}{d t}~=~C_{ijl}\,\gamma^{l}_{k}+
 C_{ikl}\,\gamma^{l}_{j}+
 C_{jkl}\,\gamma^{l}_{i}~.
\label{betay}
\eeq
The one-loop~$\gamma^i_j$ is~given by \cite{Parkes:1984dh}:
\beq
\gamma^{(1)}{}_{j}^{i}=\frac{1}{32\pi^2}\,[\,
C^{ikl}\,C_{jkl}-2\,g^2\,C_{2}(R_{i})\delta^i_j\,],
\label{gamay}
\eeq
where~$C^{ijk}=C_{ijk}^{*}$. We take $C_{ijk}$ to be real so that~$C_{ijk}^2$  are always~positive. The squares of the couplings~are convenient to work with, and the $C_{ijk}$ can be covered~by~a single~index $i~(i=1,\cdots,n)$:
\beq
\alpha = \frac{g^2}{4\pi}~,~
\alpha_{i} ~=~ \frac{g_i^2}{4\pi}~.
\label{alfas}
\eeq
Then the~evolution of $\alpha$'s in perturbation theory will take~the~form
\beq
\begin{split}
\frac{d\alpha}{d t}&=\beta~=~ -\beta^{(1)}\alpha^2+\cdots~,\\
\frac{d\alpha_{i}}{d t}&=\beta_{i}~=~ -\beta^{(1)}_{i}\,\alpha_{i}\,
\alpha+\sum_{j,k}\,\beta^{(1)}_{i,jk}\,\alpha_{j}\,
\alpha_{k}+\cdots~,
\label{eveq}
\end{split}
\eeq
Here, $\cdots$ denotes~higher-order contributions~and
$ \beta^{(1)}_{i,jk}=\beta^{(1)}_{i,kj} $.
For the evolution~equations~(\ref{eveq}) we investigate the
asymptotic  properties.~First, we~define
\cite{Zimmermann:1984sx,Oehme:1985jy,Oehme:1984iz,Cheng:1973nv,Chang:1974bv}
\beq
\tilde{\alpha}_{i} \equiv \frac{\alpha_{i}}{\alpha}~,~i=1,\cdots,n~,
\label{alfat}
\eeq
and~derive~from~\refeq{eveq}
\beq
\begin{split}
\alpha \frac{d \tilde{\alpha}_{i}}{d\alpha} &=
-\tilde{\alpha}_{i}+\frac{\beta_{i}}{\beta}= \left(-1+\frac{\beta^{(1)}_{i}}{\beta^{(1)}}\,\right) \tilde{\alpha}_{i}\\
&
-\sum_{j,k}\,\frac{\beta^{(1)}_{i,jk}}{\beta^{(1)}}
\,\tilde{\alpha}_{j}\, \tilde{\alpha}_{k}+\sum_{r=2}\,
\left(\frac{\alpha}{\pi}\right)^{r-1}\,\tilde{\beta}^{(r)}_{i}(\tilde{\alpha})~,
\label{RE}
\end{split}
\eeq
where~$\tilde{\beta}^{(r)}_{i}(\tilde{\alpha})~(r=2,\cdots)$
are~power~series of $\tilde{\alpha}$'s and~can be~computed
from~the $r^{th}$-loop $\beta$-functions.
We then~search for~fixed points $\rho_{i}$ of \refeq{alfat} at $ \alpha = 0$. We have~to solve the~equation
\beq
\left(-1+\frac{\beta ^{(1)}_{i}}{\beta ^{(1)}}\right) \rho_{i}
-\sum_{j,k}\frac{\beta ^{(1)}_{i,jk}}{\beta ^{(1)}}
\,\rho_{j}\, \rho_{k}=0~,
\label{fixpt}
\eeq
assuming~fixed~points of the~form
\beq
\rho_{i}=0~\mbox{for}~ i=1,\cdots,n'~;~
\rho_{i} ~>0 ~\mbox{for}~i=n'+1,\cdots,n~.
\eeq
Next, we~treat $\tilde{\alpha}_{i}$ with $i \leq n'$
as small~perturbations  to the~undisturbed~system (defined by~setting~$\tilde{\alpha}_{i}$  with $i \leq n'$ equal~to zero).~It is possible~to verify the
existence~of the unique~power~series solution of the reduction~equations~(\ref{RE}) to all orders already at one-loop level~\cite{Zimmermann:1984sx,Oehme:1984yy,Oehme:1985jy,Oehme:1984iz}:
\beq
\tilde{\alpha}_{i}=\rho_{i}+\sum_{r=2}\rho^{(r)}_{i}\,
\alpha^{r-1}~,~i=n'+1,\cdots,n~.
\label{usol}
\eeq
 These~are RGI~relations among~parameters, and preserve formally~perturbative~renormalizability.
So, in~the~undisturbed system~there is only one~independent
parameter,~the primary coupling $\alpha$.

The non-vanishing $\tilde{\alpha}_{i}$~with $i \leq n'$ cause small perturbations that enter in a~way that the reduced~couplings~($\tilde{\alpha}_{i}$  with $i > n'$) become functions both of~$\alpha$ and $\tilde{\alpha}_{i}$  with $i \leq n'$.~Investigating such systems with partial reduction is very convenient to~work with the following PDEs:
\beq
\begin{split}
\left\{ \tilde{\beta}\,\frac{\partial}{\partial\alpha}
+\sum_{a=1}^{n'}\,
\tilde{\beta_{a}}\,\frac{\partial}{\partial\tilde{\alpha}_{a}}\right\}~
\tilde{\alpha}_{i}(\alpha,\tilde{\alpha})
&=\tilde{\beta}_{i}(\alpha,\tilde{\alpha})~,\\
\tilde{\beta}_{i(a)}~=~\frac{\beta_{i(a)}}{\alpha^2}
-\frac{\beta}{\alpha^{2}}~\tilde{\alpha}_{i(a)}
&,\qquad
\tilde{\beta}~\equiv~\frac{\beta}{\alpha}~.
\end{split}
\eeq
These equations~are~equivalent~to the REs (\ref{RE}), where, in order to~avoid~any confusion, we let~$a,b$ run from $1$ to $n'$ and $i,j$ from $n'+1$ to $n$. Then,~we search for solutions of the~form
\beq
\tilde{\alpha}_{i}=\rho_{i}+
\sum_{r=2}\,\left(\frac{\alpha}{\pi}\right)^{r-1}\,f^{(r)}_{i}
(\tilde{\alpha}_{a})~,~i=n'+1,\cdots,n~,
\label{algeq}
\eeq
where~$ f^{(r)}_{i}(\tilde{\alpha}_{a})$ are power~series of $\tilde{\alpha}_{a}$.~The requirement that~in the~limit of~vanishing~perturbations~we obtain~the undisturbed~solutions (\ref{usol})~\cite{Kubo:1988zu,Zimmermann:1993ei} suggests this type of~solutions. Once more, one can obtain  the conditions for~uniqueness of $ f^{(r)}_{i}$ in~terms of~the lowest~order~coefficients.

\section{Reduction of Dimension-1 and -2 Parameters}\label{roc_dim_1-2}

The extension of the reduction of couplings method to massive parameters is~not straightforward, since the technique~was originally aimed at massless~theories~on~the basis~of
the~Callan-Symanzik~equation~\cite{Zimmermann:1984sx,Oehme:1984yy}. Many requirements ~ave to be met, such  as the normalization conditions ~mposed on irreducible Green's functions~\cite{Piguet:1989pc}, etc. ~ignificant progress has been made towards this goal, ~tarting from~\cite{Kubo:1996js}, where, as an assumption, a~mass-independent renormalization scheme~renders all RG~functions only trivially dependent on dimensional~parameters.~Mass parameters can~then~be introduced~similarly to couplings. 

This  was~justified~later~\cite{Breitenlohner:2001pp,Zimmermann:2001pq}, where it~was demonstrated that, apart from dimensionless parameters,~pole~masses and gauge couplings, the model can also include~couplings carrying a dimension and masses.~To simplify the~analysis, we follow~\citere{Kubo:1996js} and~use a~mass-independent renormalization~scheme~as well.

Consider a renormalizable theory~that contains $(N + 1)$
dimension-0~couplings, $\left(\hat g_0,\hat g_1, ...,\hat g_N\right)$,~$L$ parameters~with~mass dimension-1, $\left(\hat h_1,...,\hat h_L\right)$,~and  $M$ parameters~with mass dimension-2,~$\left(\hat m_1^2,...,\hat m_M^2\right)$.
The~renormalized~irreducible vertex~function $\Gamma$ satisfies~the RGE
\beq
\label{RGE_OR_1}
\mathcal{D}\Gamma\left[\Phi's;\hat g_0,\hat g_1, ...,\hat g_N;\hat h_1,...,\hat h_L;\hat m_1^2,...,\hat m_M^2;\mu\right]=0~,
\eeq
with
\beq
\label{RGE_OR_2}
\mathcal{D}=\mu\frac{\partial}{\partial \mu}+
\sum_{i=0}^N \beta_i\frac{\partial}{\partial \hat g_i}+
\sum_{a=1}^L \gamma_a^h\frac{\partial}{\partial \hat h_a}+
\sum_{\alpha=1}^M \gamma_\alpha^{m^2}\frac{\partial}{\partial \hat m_\alpha ^2}+
\sum_J \Phi_I\gamma^{\phi I}_{\,\,\,\, J}\,\frac{\delta}{\delta\Phi_J}~,
\eeq
where~ $\beta_i$ are the $\beta$-functions of the dimensionless couplings~$g_i$ and $\Phi_I$  are
the~matter~fields. The mass,~trilinear~coupling and wave~function~anomalous~dimensions,~respectively, are denoted by~$\ga_\alpha^{m^2}$, $\ga_a^h$ and $\ga^{\phi I}_{\,\,\,\, J}$~ and $\mu$ denotes the energy scale.~For a~mass-independent renormalization scheme, the $\gamma$'s are given~by
\beq
\label{gammas}
\begin{split}
\gamma^h_a&=\sum_{b=1}^L\gamma_a^{h,b}(g_0,g_1,...,g_N)\hat h_b,\\
\gamma_\alpha^{m^2}&=\sum_{\beta=1}^M \gamma_\alpha^{m^2,\beta}(g_0,g_1,...,g_N)\hat m_\beta^2+
\sum_{a,b=1}^L \gamma_\alpha^{m^2,ab}(g_0,g_1,...,g_N)\hat h_a\hat h_b~.
\end{split}
\eeq
The $\gamma_a^{h,b}$,~$\gamma_\alpha^{m^2,\beta}$ and $\gamma_\alpha^{m^2,ab}$~are power~series of the~(dimensionless)~$g$'s.

\vspace{0.35cm}

\noindent We search~for~a reduced~theory~where
\[
g\equiv g_0,\qquad h_a\equiv \hat h_a\quad \textrm{for $1\leq a\leq P$},\qquad
m^2_\alpha\equiv\hat m^2_\alpha\quad \textrm{for $1\leq \alpha\leq Q$}
\]
are~independent parameters. The reduction~of the rest of the parameters,~namely
\beq
\label{reduction}
\begin{split}
\hat g_i &= \hat g_i(g), \qquad (i=1,...,N),\\
\hat h_a &= \sum_{b=1}^P f_a^b(g)h_b, \qquad (a=P+1,...,L),\\
\hat m^2_\alpha &= \sum_{\beta=1}^Q e^\beta_\alpha(g)m^2_\beta + \sum_{a,b=1}^P k^{ab}_\alpha(g)h_ah_b,
\qquad (\alpha=Q+1,...,M)
\end{split}
\eeq
is consistent~with~the RGEs~(\ref{RGE_OR_1},\ref{RGE_OR_2}). The~following~relations~should~be satisfied
\beq
\label{relation}
\begin{split}
\beta_g\,\frac{\partial\hat g_i}{\partial g} &= \beta_i,\qquad (i=1,...,N),\\
\beta_g\,\frac{\partial \hat h_a}{\partial g}+\sum_{b=1}^P \gamma^h_b\,\frac{\partial\hat h_a}{\partial h_b} &= \gamma^h_a,\qquad (a=P+1,...,L),\\
\beta_g\,\frac{\partial\hat m^2_\alpha}{\partial g} +\sum_{a=1}^P \gamma_a^h\,\frac{\partial\hat m^2_\alpha}{\partial h_a} +  \sum_{\beta=1}^Q \gamma_\beta ^{m^2}\,\frac{\partial\hat m_\alpha^2}{\partial m_\beta^2} &= \gamma_\alpha^{m^2}, \qquad (\alpha=Q+1,...,M).
\end{split}
\eeq
Using~Eqs. (\ref{gammas}) and~(\ref{reduction}),~they reduce to
\beq
\label{relation_2}
\begin{split}
&\beta_g\,\frac{df^b_a}{dg}+ \sum_{c=1}^P f^c_a\left[\gamma^{h,b}_c + \sum_{d=P+1}^L \gamma^{h,d}_c f^b_d\right] -\gamma^{h,b}_a - \sum_{d=P+1}^L \gamma^{h,d}_a f^b_d=0,\\
&\hspace{8.6cm} (a=P+1,...,L;\, b=1,...,P),\\
&\beta_g\,\frac{de^\beta_\alpha}{dg} + \sum_{\gamma=1}^Q e^\gamma_\alpha\left[\gamma_\gamma^{m^2,\beta} +
\sum _{\delta=Q+1}^M\gamma_\gamma^{m^2,\delta} e^\beta_\delta\right]-\gamma_\alpha^{m^2,\beta} -
\sum_{\delta=Q+1}^M \gamma_\alpha^{m^2,d}e^\beta_\delta =0,\\
&\hspace{8.3cm} (\alpha=Q+1,...,M ;\, \beta=1,...,Q),\\
&\beta_g\,\frac{dk_\alpha^{ab}}{dg}
+ 2\sum_{c=1}^P \left(\gamma_c^{h,a} + \sum_{d=P+1}^L \gamma_c^{h,d} f_d^a\right)k_\alpha^{cb}
+\sum_{\beta=1}^Q e^\beta_\alpha\left[\gamma_\beta^{m^2,ab} + \sum_{c,d=P+1}^L \gamma_\beta^{m^2,cd}f^a_cf^b_d \right.\\
&\left. +2\sum_{c=P+1}^L \gamma_\beta^{m^2,cb}f^a_c + \sum_{\delta=Q+1}^M \gamma_\beta^{m^2,d} k_\delta^{ab}\right]- \left[\gamma_\alpha^{m^2,ab}+\sum_{c,d=P+1}^L \gamma_\alpha^{m^2,cd}f^a_c f^b_d\right.\\
&\left. +2 \sum_{c=P+1}^L \gamma_\alpha^{m^2,cb}f^a_c + \sum_{\delta=Q+1}^M \gamma_\alpha^{m^2,\delta}k_\delta^{ab}\right]=0,\\
&\hspace{8cm} (\alpha=Q+1,...,M;\, a,b=1,...,P)~.
\end{split}
\eeq
The above~relations ensure~that the~irreducible vertex function~of~the~reduced theory
\beq
\label{Green}
\begin{split}
\Gamma_R&\left[\Phi\textrm{'s};g;h_1,...,h_P; m_1^2,...,m_Q^2;\mu\right]\equiv\\
&\Gamma \left[  \Phi\textrm{'s};g,\hat g_1(g)...,\hat g_N(g);
h_1,...,h_P,\hat h_{P+1}(g,h),...,\hat h_L(g,h);\right.\\
& \left.  \qquad\qquad\qquad m_1^2,...,m^2_Q,\hat m^2_{Q+1}(g,h,m^2),...,\hat m^2_M(g,h,m^2);\mu\right]
\end{split}
\eeq
has~the~same~renormalization group~flow as~the~original~one.

Assuming~a perturbatively~renormalizable reduced~theory, the functions~$\hat g_i$, $f^b_a$, $e^\beta_\alpha$ and $k_\alpha^{ab}$ are~expressed~as power~series in~the primary coupling:
\beq
\label{pert}
\begin{split}
\hat g_i & = g\sum_{n=0}^\infty \rho_i^{(n)} g^n,\qquad
f_a^b  =  g \sum_{n=0}^\infty \eta_a^{b(n)} g^n,\\
e^\beta_\alpha & = \sum_{n=0}^\infty \xi^{\beta(n)}_\alpha g^n,\qquad
k_\alpha^{ab}=\sum_{n=0}^\infty \chi_\alpha^{ab(n)} g^n.
\end{split}
\eeq
These~expansion~coefficients are found by~inserting~the above power series~into Eqs. (\ref{relation}), (\ref{relation_2}) and~requiring~the equations to~be satisfied~at each~order of~$g$. It is~not trivial to have a unique power series solution;~it depends both on the theory and  the choice of independent~couplings.

If~there are no independent~dimension-1 parameters ($\hat h$), their~reduction becomes
\[
\hat h_a = \sum_{b=1}^L f_a^b(g)M,
\]
where~$M$ is a dimension-1 parameter (i.e. a gaugino~mass, corresponding~to the independent gauge coupling). If there
are no independent~dimension-2~parameters ($\hat m^2$), their reduction~takes the~form
\[
\hat m^2_a=\sum_{b=1}^M e_a^b(g) M^2.
\]

\section{Reduction of Couplings of Soft Breaking Terms in \texorpdfstring{$N=1$}~ SUSY Theories}\label{roc_soft}

The~reduction of dimensionless couplings was~extended \cite{Kubo:1996js,Jack:1995gm} to~the SSB~dimensionful parameters of $N=1$~supersymmetric~theories. It was also found \cite{Kawamura:1997cw,Kobayashi:1997qx} that~soft scalar masses~satisfy~a universal~sum rule.\\
Consider~the superpotential (\ref{supot0})
\beq
W= \frac{1}{2}\,\mu^{ij} \,\Phi_{i}\,\Phi_{j}+
\frac{1}{6}\,C^{ijk} \,\Phi_{i}\,\Phi_{j}\,\Phi_{k}~,
\label{supot-prime}
\eeq
and the SSB~Lagrangian 
\beq
-{\cal L}_{\rm SSB} =
\frac{1}{6} \,h^{ijk}\,\phi_i \phi_j \phi_k
+
\frac{1}{2} \,b^{ij}\,\phi_i \phi_j
+
\frac{1}{2} \,(m^2)^{j}_{i}\,\phi^{*\,i} \phi_j+
\frac{1}{2} \,M\,\lambda_i \lambda_i+\mbox{h.c.}
\label{supot_l}
\eeq
The $\phi_i$'s are~the scalar~parts~of chiral~superfields $\Phi_i$, $\lambda$ are gauginos~and $M$ the unified gaugino mass.

The~one-loop gauge $\beta$-function (\ref{betag}) is given~by
\cite{Parkes:1984dh,West:1984dg,Jones:1985ay,Jones:1984cx,Parkes:1985hh}
\bea
\beta^{(1)}_{g}=\frac{d g}{d t} =
  \frac{g^3}{16\pi^2}\,\left[\,\sum_{i}\,T(R_{i})-3\,C_{2}(G)\,\right]~,
\eea
whereas~the one-loop $C_{ijk}$'s $\beta$-function~(\ref{betay}) is~given by
\beq
\beta_C^{ijk} =
  \frac{d C_{ijk}}{d t}~=~C_{ijl}\,\gamma^{l}_{k}+
  C_{ikl}\,\gamma^{l}_{j}+
  C_{jkl}\,\gamma^{l}_{i}~,
\eeq
and~the (one-loop)~anomalous dimension~$\gamma^{(1)}\,^i_j$ of a chiral~superfield (\ref{gamay})~is
\beq
\gamma^{(1)}\,^i_j=\frac{1}{32\pi^2}\,\left[\,
C^{ikl}\,C_{jkl}-2\,g^2\,C_{2}(R_{i})\delta^i_j\,\right]~.
\label{allolabel}
\eeq
Then the~$N = 1$ non-renormalization theorem~\cite{Wess:1973kz,Iliopoulos:1974zv,Fujikawa:1974ay} guarantees~that the $\beta$-functions of $C_{ijk}$ are expressed~in terms of the anomalous dimensions.\\
We make~the assumption that the REs admit~power~series solutions:
\beq
C^{ijk} = g\,\sum_{n=0}\,\rho^{ijk}_{(n)} g^{2n}~.
\label{Yg-prime}
\eeq
Since~we want to obtain higher-loop results instead~of~knowledge of explicit~$\beta$-functions, we~require relations among~$\beta$-functions. The spurion~technique
\cite{Fujikawa:1974ay,Delbourgo:1974jg,Salam:1974pp,Grisaru:1979wc,Girardello:1981wz} gives~all-loop relations among SSB~$\beta$-functions
\cite{Yamada:1994id,Kazakov:1997nf,Jack:1997pa,Hisano:1997ua,Jack:1997eh,Avdeev:1997vx,Kazakov:1998uj,Karch:1998qa}:
\begin{align}
\beta_M &= 2{\cal O}\left(\frac{\beta_g}{g}\right)~,
\label{betaM}\\
\beta_h^{ijk}&=\gamma^i_l h^{ljk}+\gamma^j_l h^{ilk}
+\gamma^k_l h^{ijl}\non\\
&\,-2\left(\gamma_1\right)^i_l C^{ljk}
-2\left(\gamma_1\right)^j_l C^{ilk}-2\left(\gamma_1\right)^k_l C^{ijl}~,\label{betah}\\
(\beta_{m^2})^i_j &=\left[ \Delta
+ X \frac{\partial}{\partial g}\right]\gamma^i_j~,
\label{betam2}
\end{align}
where
\begin{align}
{\cal O} &=\left(Mg^2\frac{\partial}{\partial g^2}
-h^{lmn}\frac{\partial}{\partial C^{lmn}}\right)~,
\label{diffo}\\
\Delta &= 2{\cal O}{\cal O}^* +2|M|^2 g^2\frac{\partial}
{\partial g^2} +\tilde{C}_{lmn}
\frac{\partial}{\partial C_{lmn}} +
\tilde{C}^{lmn}\frac{\partial}{\partial C^{lmn}}~,\\
(\gamma_1)^i_j&={\cal O}\gamma^i_j,\\
\tilde{C}^{ijk}&=
(m^2)^i_l C^{ljk}+(m^2)^j_l C^{ilk}+(m^2)^k_l C^{ijl}~.
\label{tildeC}
\end{align}

\noindent Assuming~(following \cite{Jack:1997pa}) that~the relation among~couplings
\beq
h^{ijk} = -M (C^{ijk})'
\equiv -M \frac{d C^{ijk}(g)}{d \ln g}~,
\label{h2}
\eeq
is RGI~and  the~use~of~the all-loop~gauge $\beta$-function of~\cite{Novikov:1983ee,Novikov:1985rd,Shifman:1996iy}
\beq
\beta_g^{\rm NSVZ} =
\frac{g^3}{16\pi^2}
\left[ \frac{\sum_l T(R_l)(1-\gamma_l /2)
-3 C_2(G)}{ 1-g^2C_2(G)/8\pi^2}\right]~,
\label{bnsvz}
\eeq
we~are led to an~all-loop RGI sum rule~\cite{Kobayashi:1998jq} (assuming $(m^2)^i_j=m^2_j\delta^i_j$),
\begin{equation}
\begin{split}
m^2_i+m^2_j+m^2_k &=
|M|^2 \left\{~
\frac{1}{1-g^2 C_2(G)/(8\pi^2)}\frac{d \ln C^{ijk}}{d \ln g}
+\frac{1}{2}\frac{d^2 \ln C^{ijk}}{d (\ln g)^2}~\right\}\\
& \qquad\qquad +\sum_l
\frac{m^2_l T(R_l)}{C_2(G)-8\pi^2/g^2}
\frac{d \ln C^{ijk}}{d \ln g}~.
\label{sum2}
\end{split}
\end{equation}
It is~worth noting that~the all-loop result of \refeq{sum2} coincides~with~the~superstring result~for~the~finite case~in a certain~class of orbifold~models~\cite{Ibanez:1992hc,Brignole:1995fb,Kobayashi:1997qx}~if $\frac{d \ln C^{ijk}}{d \ln g}=1$~\cite{Mondragon:1993tw}.

As~mentioned above, the~all-loop~results~on the~SSB~$\beta$-functions, Eqs.(\ref{betaM})-(\ref{tildeC}),
lead~to all-loop~RGI~relations.~We assume:\\
(a) the~existence~of~an RGI~surface~on~which $C = C(g)$, or~equivalently~that~the expression
\beq
\label{Cbeta}
\frac{dC^{ijk}}{dg} = \frac{\beta^{ijk}_C}{\beta_g}
\eeq
holds~(i.e. reduction~of couplings is~possible)\\
(b)~the existence of a~RGI~surface on~which
\beq
\label{h2NEW}
h^{ijk} = - M \frac{dC(g)^{ijk}}{d\ln g}
\eeq
holds~to all orders.\\
Then it~can be proven~\cite{Jack:1999aj,Kobayashi:1998iaa} that the~relations that~follow are all-loop~RGI (note that in
both~assumptions we~do not rely on specific~solutions of these equations)
\begin{align}
M &= M_0~\frac{\beta_g}{g} ,  \label{M-M0} \\
h^{ijk}&=-M_0~\beta_C^{ijk},  \label{hbeta}  \\
b^{ij}&=-M_0~\beta_{\mu}^{ij},\label{bij}\\
(m^2)^i_j&= \frac{1}{2}~|M_0|^2~\mu\frac{d\gamma^i{}_j}{d\mu},
\label{scalmass}
\end{align}
where~$M_0$ is~an~arbitrary reference~mass~scale~to be~specified shortly.~Assuming
\beq
C_a\frac{\partial}{\partial C_a}
= C_a^*\frac{\partial}{\partial C_a^*} \label{dc/dc}
\eeq
for~an RGI~surface $F(g,C^{ijk},C^{*ijk})$ we~are led to
\begin{equation}
\label{F}
\frac{d}{dg} = \left(\frac{\partial}{\partial g} + 2\frac{\partial}{\partial C}\,\frac{dC}{dg}\right)
= \left(\frac{\partial}{\partial g} + 2 \frac{\beta_C}{\beta_g}
\frac{\partial}{\partial C} \right)\, ,
\end{equation}
where~\refeq{Cbeta} was used.~Let us now consider~the~partial~differential~operator ${\cal O}$ in
\refeq{diffo}~which (assuming~\refeq{h2}), becomes
\beq
{\cal O} = \frac{1}{2}M\frac{d}{d\ln g}\, 
\eeq
and $\beta_M$,~given~in \refeq{betaM}, becomes
\beq
\beta_M = M\frac{d}{d\ln g} \big( \frac{\beta_g}{g}\big) ~, \label{betaM2}
\eeq
which~by~integration~provides us \cite{Karch:1998qa,Jack:1999aj} with~the
generalized,~i.e. including Yukawa couplings,~all-loop RGI Hisano - Shifman relation \cite{Hisano:1997ua}
\beq
 M = \frac{\beta_g}{g} M_0~. \nonumber
\eeq
$M_0$ is the~integration~constant~and can~be associated~to the unified gaugino mass $M$~(of an assumed covering GUT), or to~the gravitino~mass~$m_{3/2}$ in~a supergravity~framework. Therefore, \refeq{M-M0} becomes the~all-loop RGI \refeq{M-M0}.  $\beta_M$, using~Eqs.(\ref{betaM2}) and (\ref{M-M0}) can be written as~follows:
\beq \beta_M =
M_0\frac{d}{dt} (\beta _g/g)~.
\eeq
Similarly
\beq (\gamma_1)^i_j =
{\cal O} \gamma^i_j = \frac{1}{2}~M_0~\frac{d
  \gamma^i_j}{dt}~. \label{gammaO}
\eeq
Next,~from Eq.(\ref{h2}) and Eq.(\ref{M-M0}) we get
\beq
 h^{ijk} = - M_0 ~\beta_C^{ijk}~,  \label{hm32}
\eeq
while~$\beta^{ijk}_h$, using Eq.(\ref{gammaO}), becomes \cite{Jack:1999aj}
\beq
  \beta_h^{ijk} = - M_0~\frac{d}{dt} \beta_C^{ijk},
\eeq
which~shows that \refeq{hm32} is  RGI to all loops.~\refeq{bij} can similarly be~shown to~be all-loop~RGI as well.

Finally, it~is important to note~that,~under the assumptions (a)~and (b), the sum rule~of \refeq{sum2} has been~proven
\cite{Kobayashi:1998jq} to~be RGI to all loops, which~(using \refeq{M-M0}) generalizes \refeq{scalmass} for~application in cases with non-universal soft~scalar masses,~a necessary ingredient in~the models that will be~examined in the next Sections. Another~important point to note is the~use of \refeq{M-M0}, which,~in the case of product~gauge groups (as in the MSSM),~takes the form 
\beq
M_i=\frac{\beta_{g_i}}{g_i}M_0~,
\eeq
where $i=1,2,3$~denotes each gauge group, and~will be used in the Reduced~MSSM case.

\newpage
\mbox{} 
\newpage

\chapter{Finiteness}\label{finiteness}

The principle of~finiteness requires perhaps~some more motivation to be~considered and generally accepted~these days than when~it was first envisaged, since in~recent years we have a more relaxed~attitude towards divergencies. Most~theorists believe that~the divergencies are signals~of the existence of a higher scale,~where new degrees of freedom~are excited. Even accepting~this dogma, we are naturally led to the~conclusion that beyond~the unification scale, i.e. when all~interactions have been taken~into account in a unified scheme,~the theory should be completely~finite. In fact, this is one~of the main motivations and~aims of string, non-commutative~geometry, and quantum group~theories, which include~also gravity in the unification~of the interactions. The work on~reduction of couplings and finiteness~reviewed in this chapter is~restricted  to the unification of~the known gauge interactions.

\section{The idea~behind finiteness}

Finiteness is based on~the fact that it is possible to find renormalization group~invariant (RGI) relations among couplings that keep~finiteness in perturbation theory, even~to all orders. Accepting~finiteness as a guiding principle~in constructing realistic~theories of EPP, the first~thing that comes to mind~is to look for an $N=4$~supersymmetric unified
gauge theory, since~any ultraviolet (UV) divergencies~are absent in these~theories. However nobody~has managed so far to produce~realistic models in the~framework of $N=4$~SUSY. In the best~of cases one could try to~do a drastic truncation of~the theory like the~orbifold projection of~refs. \cite{Kachru:1998ys,Chatzistavrakidis:2010xi},~but this is already a different~theory than the original~one. The next possibility~is to consider an~$N=2$ supersymmetric gauge theory,~whose $\beta$ function receives~corrections only at one loop.~Then it is~not hard to select a~spectrum to make~the theory~all-loop finite. However~a serious obstacle in these~theories is their mirror~spectrum, which in the~absence of a mechanism to make it~heavy, does not permit the~construction of realistic models.~Therefore, one is naturally led to~consider $N=1$ supersymmetric gauge~theories, which can be chiral~and in principle realistic.

It should be noted~that in the approach followed~here (UV)~finiteness means the vanishing of~all the $\beta$ functions, i.e. the non-renormalization~of the coupling constants, in contrast to a complete~(UV) finiteness where even
field~amplitude renormalization is~absent.
Before the work of~several members of our group,~the studies on $N=1$ finite~theories were following two~directions: (i) construction of finite~theories up to two loops examining various~possibilities to make them phenomenologically~viable, (ii) construction of all-loop~finite models without particular emphasis~on the phenomenological consequences. The success~of their work was the~construction of the first realistic~all-loop finite model, based on the theorem~presented in the \refse{finite-dimless} below, realising in this~way an old theoretical dream of field~theorists.

\section{Finiteness in N=1 Supersymmetric~Gauge Theories}
\label{finite-dimless}

Let us,~once more, consider a chiral,~anomaly free, $N=1$ globally supersymmetric~gauge theory based on a group G with gauge coupling~constant $g$.
The superpotential~of the theory is given by (see \refeq{supot0})
\beq
W= \frac{1}{2}\,m_{ij} \,\phi_{i}\,\phi_{j}+
\frac{1}{6}\,C_{ijk} \,\phi_{i}\,\phi_{j}\,\phi_{k}~.
\label{supot}
\eeq
The $N=1$~non-renormalization theorem, ensuring the absence~of mass and cubic-interaction-term~infinities, leads to~wave-function infinities.~The one-loop $\beta$-function is~given by (see \refeq{betag})
\beq
\beta^{(1)}_{g}=\frac{d g}{d t} =
\frac{g^3}{16\pi^2}\left[\,\sum_{i}\,T(R_{i})-3\,C_{2}(G)\right]~,
\label{betagP}
\eeq
the~$\beta$ function of~$C_{ijk}$ by (see Eq. (\ref{betay}))
\beq
\beta_{ijk} =
 \frac{d C_{ijk}}{d t}~=~C_{ijl}\,\gamma^{l}_{k}+
 C_{ikl}\,\gamma^{l}_{j}+
 C_{jkl}\,\gamma^{l}_{i}~
\label{betayP}
\eeq
and~the one-loop wave function anomalous dimensions~by (see \refeq{gamay})
\beq
\gamma^{(1)}{}_{j}^{i}=\frac{1}{32\pi^2}\,[\,
C^{ikl}\,C_{jkl}-2\,g^2\,C_{2}(R)\delta_{j}^i\,]~. \label{gamayP}
\eeq
As one can~see from Eqs. (\ref{betagP}) and~(\ref{gamayP}),
all the  one-loop $\beta$-functions of the theory vanish if
$\beta_g^{(1)}$~and  $\gamma^{(1)}{}_{j}^{i}$ vanish, i.e.
\begin{align}
\sum _i T(R_{i})& = 3 C_2(G) \,,
\label{1st}     \\
 C^{ikl} C_{jkl} &= 2\delta ^i_j g^2  C_2(R_i)\,,
\label{2nd}
\end{align}
The conditions~for finiteness for $N=1$ field theories with $SU(N)$ gauge~symmetry are discussed in \cite{Rajpoot:1984zq}, and  the~analysis of the  anomaly-free and  no-charge renormalization~requirements for these theories can be found in \cite{Rajpoot:1985aq}.~A very interesting result is that the  conditions (\ref{1st},\ref{2nd})~are necessary and  sufficient for finiteness at the~two-loop level
\cite{Parkes:1984dh,West:1984dg,Jones:1985ay,Jones:1984cx,Parkes:1985hh}.

In case SUSY~is broken by soft terms, the~requirement of
finiteness in the~one-loop soft breaking terms imposes~further
constraints among~themselves \cite{Jones:1984cu}.  In addition, the~same set~of conditions that are~sufficient for one-loop finiteness of the~soft~breaking terms render the~soft sector of the theory two-loop~finite \cite{Jack:1994kd}.

The one- and  two-loop~finiteness conditions~of Eqs. (\ref{1st},\ref{2nd}) restrict~considerably the  possible choices of the  irreducible~representations~(irreps)
$R_i$ for a given~group $G$ as well as the  Yukawa couplings in the~superpotential (\ref{supot}).~Note in particular~that the  finiteness conditions cannot be~applied to the  minimal supersymmetric~standard model (MSSM), since the~presence
of a $U(1)$ gauge~group is incompatible with the~condition
(\ref{1st}), due to $C_2[U(1)]=0$.~This naturally leads to the
expectation that finiteness should~be attained at the  grand unified~level only, the MSSM being just the~corresponding, low-energy,~effective theory.

Another important~consequence of one- and two-loop finiteness is that~SUSY (most probably) can only be broken~due to the  soft~breaking terms.  Indeed, due to the unacceptability of gauge singlets,~F-type spontaneous symmetry~breaking \cite{ORaifeartaigh:1975nky}~terms are incompatible with finiteness, as~well as D-type \cite{Fayet:1974jb} spontaneous breaking which~requires the  existence of a $U(1)$ gauge~group.

A natural~question to ask is what happens at higher loop orders.~The answer is contained in~a theorem
\cite{Lucchesi:1987he,Lucchesi:1987ef} which~states the  necessary and~sufficient conditions to achieve finiteness~at all orders.  Before we~discuss the  theorem let us make some introductory remarks.~The finiteness conditions impose relations~between gauge and Yukawa couplings.~To require such relations which render the  couplings~mutually dependent at a given renormalization~point is trivial.  What~is not trivial is to guarantee that relations leading~to a reduction of the  couplings~hold at any renormalization point.~As we have seen
(see \refeq{Cbeta}),~the necessary~and also sufficient, condition~for this to happen is to~require that such relations are solutions to the~REs
\beq \beta _g
\frac{d C_{ijk}}{dg} = \beta _{ijk}
\label{redeq2}
\eeq
and hold at all orders.~Remarkably, the existence of~all-order power series solutions to (\ref{redeq2})~can be decided at
one-loop level, as already~mentioned.

Let us now turn to the~all-order finiteness theorem
\cite{Lucchesi:1987he,Lucchesi:1987ef}, which~states under which conditions an $N=1$~supersymmetric gauge theory can become finite to all~orders in perturbation theory, that is attain physical~scale invariance.  It is based on (a) the  structure of the~supercurrent in $N=1$ supersymmetric~gauge theory \cite{Ferrara:1974pz,Piguet:1981mu,Piguet:1981mw}, and  on (b) the~non-renormalization properties of~$N=1$ chiral anomalies \cite{Lucchesi:1987he,Lucchesi:1987ef,Piguet:1986td,Piguet:1986pk,Ensign:1987wy}.~Details of the  proof can be found in~refs. \cite{Lucchesi:1987he,Lucchesi:1987ef} and  further discussion in~\citeres{Piguet:1986td,Piguet:1986pk,Ensign:1987wy,Lucchesi:1996ir,Piguet:1996mx}.~Here, following mostly~\citere{Piguet:1996mx} we present a~comprehensible sketch of the  proof.

Consider~an $N=1$ supersymmetric gauge theory, with simple Lie group~$G$. The  content of this theory is~given at the  classical level by~the matter supermultiplets $S_i$, which contain a scalar field~$\phi_i$ and  a Weyl spinor $\psi_{ia}$, and  the~vector supermultiplet $V_a$, which contains a~gauge vector field $A_{\mu}^a$ and a~gaugino
Weyl spinor $\lambda^a_{\alpha}$.\\

\noindent Let us first recall certain~facts about the  theory:

\noindent (1)~A massless $N=1$ supersymmetric theory is invariant~under a $U(1)$ chiral transformation $R$ under which the~various fields transform as follows
\beq
\begin{split}
A'_{\mu}&=A_{\mu},~~\lambda '_{\alpha}=\exp({-i\theta})\lambda_{\alpha}\\
\phi '&= \exp({-i\frac{2}{3}\theta})\phi,~~\psi_{\alpha}'= \exp({-i\frac{1}
    {3}\theta})\psi_{\alpha},~\cdots
\end{split}
\eeq
The corresponding~axial Noether current $J^{\mu}_R(x)$ is
\beq
J^{\mu}_R(x)=\bar{\lambda}\gamma^{\mu}\gamma^5\lambda + \cdots
\label{noethcurr}
\eeq
is conserved classically,~while in the quantum case is violated by the~axial anomaly
\beq
\partial_{\mu} J^{\mu}_R =
r\left(\epsilon^{\mu\nu\sigma\rho}F_{\mu\nu}F_{\sigma\rho}+\cdots\right).
\label{anomaly}
\eeq
From its~known topological origin in ordinary~gauge theories
\cite{AlvarezGaume:1983cs,Bardeen:1984pm,Zumino:1983rz}, one would~expect the  axial vector current~$J^{\mu}_R$ to satisfy the Adler-Bardeen theorem and~receive corrections only at the  one-loop level.~Indeed it has been shown that the same non-renormalization~theorem holds also in supersymmetric theories~\cite{Piguet:1986td,Piguet:1986pk,Ensign:1987wy}.  Therefore
\beq
r=\hbar \beta_g^{(1)}.
\label{r}
\eeq

\noindent (2)~The massless theory we consider is scale invariant at~the classical level and, in general,~there is a scale anomaly due to~radiative corrections.  The  scale anomaly~appears in the  trace of the~energy momentum tensor $T_{\mu\nu}$, which~is traceless classically.
It has the  form
\beq
T^{\mu}_{\mu} = \beta_g F^{\mu\nu}F_{\mu\nu} +\cdots
\label{Tmm}
\eeq

\noindent (3)~Massless, $N=1$ supersymmetric gauge theories are~classically invariant under the  supersymmetric extension of the~conformal group -- the  superconformal group.  Examining the~superconformal algebra, it can be seen that the  subset of~superconformal transformations consisting of translations,~SUSY transformations, and  axial $R$ transformations~is closed under SUSY, i.e. these transformations~form a representation of SUSY.  It follows that the~conserved currents~corresponding to these transformations~make up a supermultiplet represented by an axial vector~superfield called the supercurrent~$J$,
\beq
J \equiv \left\{ J'^{\mu}_R, ~Q^{\mu}_{\alpha}, ~T^{\mu}_{\nu} , ... \right\},
\label{J}
\eeq
where~$J'^{\mu}_R$ is the  current associated to R invariance,
$Q^{\mu}_{\alpha}$~is the  one associated to SUSY invariance,
and $T^{\mu}_{\nu}$~the  one associated to translational invariance~(energy-momentum tensor).\\
The anomalies~of the  R current $J'^{\mu}_R$, the  trace
anomalies of the~SUSY current, and  the  energy-momentum tensor, form also~a second supermultiplet, called the  supertrace anomaly
\[
S =\left\{ Re~ S, ~Im~ S,~S_{\alpha}\right\}=
\left\{T^{\mu}_{\mu},~\partial _{\mu} J'^{\mu}_R,~\sigma^{\mu}_{\alpha
  \dot{\beta}} \bar{Q}^{\dot\beta}_{\mu}~+~\cdots \right\}
\]
where~$T^{\mu}_{\mu}$ is given in Eq.(\ref{Tmm}) and
\begin{align}
\partial _{\mu} J'^{\mu}_R &~=~\beta_g\epsilon^{\mu\nu\sigma\rho}
F_{\mu\nu}F_{\sigma \rho}+\cdots\\
\sigma^{\mu}_{\alpha \dot{\beta}} \bar{Q}^{\dot\beta}_{\mu}&~=~\beta_g
\lambda^{\beta}\sigma ^{\mu\nu}_{\alpha\beta}F_{\mu\nu}+\cdots
\end{align}

\noindent (4)~It is very important to note that
the Noether~current defined in (\ref{noethcurr}) is not the  same as the~current associated to R invariance that appears in the~supercurrent $J$ in (\ref{J}), but they coincide in the  tree approximation.~So starting from a unique classical Noether current~$J^{\mu}_{R(class)}$,  the  Noether
current~$J^{\mu}_R$ is defined as the  quantum extension of
$J^{\mu}_{R(class)}$~which allows for the~validity of the  non-renormalization theorem.~On the  other hand, $J'^{\mu}_R$, is defined~to belong to the  supercurrent $J$,~together with the  energy-momentum tensor.~The two requirements~cannot be fulfilled by a single~current operator at the  same time.

Although the~Noether current $J^{\mu}_R$ which obeys (\ref{anomaly})~and the current $J'^{\mu}_R$ belonging to the  supercurrent multiplet~$J$ are not the  same, there is a relation~\cite{Lucchesi:1987he,Lucchesi:1987ef} between quantities associated~with them
\beq
r=\beta_g(1+x_g)+\beta_{ijk}x^{ijk}-\gamma_Ar^A
\label{rbeta}
\eeq
where $r$ was~given in Eq.~(\ref{r}).  The  $r^A$ are the
non-renormalized~coefficients of the anomalies of the  Noether currents associated to the chiral~invariances of the  superpotential,~and --like $r$-- are strictly
one-loop quantities.~The  $\gamma_A$'s are linear
combinations of the~anomalous dimensions of the  matter fields, and~$x_g$, and  $x^{ijk}$ are radiative correction quantities.~The structure of Eq. (\ref{rbeta}) is independent of the~renormalization scheme.

One-loop~finiteness, i.e. vanishing of the  $\beta$-functions at one-loop,~implies that the  Yukawa couplings $\lambda_{ijk}$ must be functions of~the gauge coupling $g$. To find a similar~condition to all orders it is necessary and sufficient for~the Yukawa couplings to be a formal power series in~$g$, which is solution of the~REs (\ref{redeq2}).

\noindent We can now~state the theorem for all-order vanishing~$\beta$ functions \cite{Lucchesi:1987he}.
\bigskip

\noindent {\bf Theorem:}

\noindent Consider an~$N=1$ supersymmetric Yang-Mills theory, with simple gauge~group. If the  following conditions~are satisfied
\begin{enumerate}
\item There is no~gauge anomaly.
\item The gauge~$\beta$-function vanishes at one-loop
  \beq
  \beta^{(1)}_g = 0 =\sum_i T(R_{i})-3\,C_{2}(G).
  \eeq
\item There exist~solutions of the  form
  \beq
  C_{ijk}=\rho_{ijk}g,~\qquad \rho_{ijk}\in\complex
  \label{soltheo}
  \eeq
to the~conditions of vanishing one-loop matter fields anomalous~dimensions
\beq
  \gamma^{(1)}{}_{j}^{i}~=~0
  =\frac{1}{32\pi^2}~[ ~
  C^{ikl}\,C_{jkl}-2~g^2~C_{2}(R)\delta_j^i ].
\eeq
\item These solutions~are isolated and  non-degenerate when considered~as solutions of vanishing one-loop~Yukawa $\beta$ functions:
   \beq
   \beta_{ijk}=0.
   \eeq
\end{enumerate}
Then, each of the~solutions (\ref{soltheo}) can be~uniquely extended~to a formal power series in $g$, and  the  associated super Yang-Mills~models depend on the  single coupling constant $g$ with a $\beta$-function~which vanishes at all-orders.

\bigskip

It is~important to note a few things:
The~requirement of isolated and  non-degenerate
solutions guarantees the~existence of a unique formal power series solution to the reduction~equations.
The vanishing of the  gauge $\beta$ function at one-loop,~$\beta_g^{(1)}$, is equivalent to the~vanishing of the  R current anomaly (\ref{anomaly}).  The  vanishing of~the anomalous~dimensions at one-loop implies the  vanishing of the  Yukawa couplings~$\beta$~functions at that order.  It also implies the vanishing of the~chiral anomaly coefficients $r^A$. This last property is a necessary~condition for having $\beta$ functions vanishing at all orders.\footnote{There~is an alternative way to find finite~theories \cite{Ermushev:1986cu,Kazakov:1987vg,Jones:1986vp,Leigh:1995ep}.}

\bigskip

\noindent {\bf Proof:}

\noindent Insert~$\beta_{ijk}$ as given by the  REs into the relationship~(\ref{rbeta}).
Since these~chiral anomalies vanish, we get for $\beta_g$ an~homogeneous equation of the  form
\beq
0=\beta_g(1+O(\hbar)).
\label{prooftheo}
\eeq
The solution of this equation in~the sense of a formal power series in~$\hbar$ is $\beta_g=0$, order by order.  Therefore, due to the~REs (\ref{redeq2}), $\beta_{ijk}=0$ too.

Thus,~we see that finiteness and  reduction of couplings are intimately~related. Since an equation like eq.~(\ref{rbeta}) is lacking in~non-supersymmetric theories, one cannot extend the  validity of a~similar theorem in such theories.\\

A very interesting development was done in~ref~\cite{Kazakov:1997nf}.~Based on the all-loop relations among the $\beta$ functions of the soft supersymmetry breaking terms~and those of the rigid~supersymmetric theory with the help of the differential operators,~discussed in Sections \ref{roc_dim_1-2} and \ref{roc_soft}, it was shown~that certain~RGI surfaces can be chosen, so as to reach all-loop~finiteness~of the full theory. More specifically it was shown~that on certain RGI~surfaces the partial differential operators appearing in~Eqs.~(\ref{betaM},\ref{betah}) acting on the~beta- and gamma-functions of the rigid theory can be transformed to total derivatives.~Then the all-loop finiteness of the $\beta$ and $\gamma$ functions~of the rigid theory can be~transferred to the $\beta$ functions of the~soft supersymmetry breaking terms. Therefore a~totally all-loop finite $N=1$ SUSY gauge theory can be constructed,~including the soft supersymmetry breaking terms.

\newpage
\mbox{} 
\newpage

\chapter{Phenomenologically Interesting Models with Reduced Couplings}\label{roc_models}

In this chapter the basic~properties of phenomenologically~viable~SUSY~models that use the idea of reduction of couplings are reviewed, as they were developed over the previous years by members of our group. The first three use a larger gauge structure to achieve reduction of couplings (two of them also feature finiteness), while the fourth can employs it within the MSSM gauge group.

\section{The Minimal \texorpdfstring{$N=1$}~ Supersymmetric \texorpdfstring{$SU(5)$}~ }\label{sec:minimalsu5}

The first case reviewed is the partial~reduction of couplings~in~the~minimal $N = 1$ SUSY
 model~based~on~the $SU(5)$ \cite{Kubo:1994bj,Kubo:1996js}.
$\Psi^{I}({\bf 10})$~and $\Phi^{I}(\overline{\bf 5})$  accommodate~the~three~generations~of quarks~and leptons, $I$ running over~the~three~generations, an adjoint $\Sigma({\bf 24})$ breaks~$SU(5)$~down to the MSSM gauge group $SU(3)_{\rm C} \times SU(2)_{\rm L} \times U(1)_{\rm Y}$,  and~$H({\bf 5})$~and~$\overline{H}({\overline{\bf 5}})$ describe the
two~Higgs superfields of the electroweak~symmetry~breaking (ESB)~\cite{Dimopoulos:1981zb,Sakai:1981gr}.
Only~one set of $({\bf 5} + {\bf \bar{5}})$ is used to describe~the Higgs~superfields appropriate for ESB.
This~minimality~renders the present version asymptotically~free~(negative~$\beta_g$).
Its superpotential~is  \cite{Dimopoulos:1981zb,Sakai:1981gr}
\beq
\begin{split}
W =&~~ \frac{g_{t}}{4}\,
\epsilon^{\alpha\beta\gamma\delta\tau}\,
\Psi^{(3)}_{\alpha\beta}\Psi^{(3)}_{\gamma\delta}H_{\tau}+
\sqrt{2}g_b\,\Phi^{(3) \alpha}
\Psi^{(3)}_{\alpha\beta}\overline{H}^{\beta}+
\frac{g_{\lambda}}{3}\,\Sigma_{\alpha}^{\beta}
\Sigma_{\beta}^{\gamma}\Sigma_{\gamma}^{\alpha}+
g_{f}\,\overline{H}^{\alpha}\Sigma_{\alpha}^{\beta} H_{\beta}\\
&+ \frac{\mu_{\Sigma}}{2}\,
\Sigma_{\alpha}^{\gamma}\Sigma_{\gamma}^{\alpha}+
\mu_{H}\,\overline{H}^{\alpha} H_{\alpha}~.
\end{split}
\eeq
 where~$t,b$ and $f$ are~indices of the antisymmetric ${\bf 10}$ and adjoint~${\bf 24}$ tensors, $\alpha,\beta,\ldots$ are $SU(5)$~indices,~and the first two generations Yukawa couplings have been~suppressed.
The SSB~Lagrangian is
\beq
\begin{split}
-{\cal L}_{\rm soft} =&~~
m_{H_u}^{2}{\hat H}^{* \alpha}{\hat H}_{\alpha}
+m_{H_d}^{2}
\hat{\overline {H}}^{*}_{\alpha}\hat{\overline {H}}^{\alpha}
+m_{\Sigma}^{2}{\hat \Sigma}^{\dag~\alpha}_{\beta}
{\hat \Sigma}_{\alpha}^{\beta}
+\sum_{I=1,2,3}\,[\,
m_{\Phi^I}^{2}{\hat \Phi}^{* ~(I)}_{\alpha}{\hat \Phi}^{(I)\alpha}\\
& +\,m_{\Psi^I}^{2}{\hat \Psi}^{\dag~(I)\alpha\beta}
{\hat \Psi}^{(I)}_{\beta\alpha}\,]
+\{ \,
 \frac{1}{2}M\lambda \lambda+
B_H\hat{\overline {H}}^{\alpha}{\hat H}_{\alpha}
+B_{\Sigma}{\hat \Sigma}^{\alpha}_{\beta}
{\hat \Sigma}_{\alpha}^{\beta}
+h_{f}\,\hat{\overline{H}}^{\alpha}
{\hat \Sigma}_{\alpha}^{\beta} {\hat H}_{\beta}\\
& +\frac{h_{\lambda}}{3}\,{\hat \Sigma}_{\alpha}^{\beta}
{\hat \Sigma}_{\beta}^{\gamma}{\hat \Sigma}_{\gamma}^{\alpha}+
\frac{h_{t}}{4}\,
\epsilon^{\alpha\beta\gamma\delta\tau}\,
{\hat \Psi}^{(3)}_{\alpha\beta}
{\hat \Psi}^{(3)}_{\gamma\delta}{\hat H}_{\tau}+
\sqrt{2}h_{b}\,{\hat \Phi}^{(3) \alpha}
{\hat \Psi^{(3)}}_{\alpha\beta}\hat{\overline{H}}^{\beta}
+\mbox{h.c.}\, \}~,
\end{split}
\eeq
where the~hat denotes the scalar
components~of the chiral~superfields.
The $\beta$ and $\gamma$ functions and a detailed~presentation of the model can~be found in  \cite{Kubo:1994bj} and in \cite {Polonsky:1994sr,Kazakov:1995cy}.

The minimal~number~of SSB terms that do not violate perturbative~renormalizability is required in the reduced~theory.~The  perturbatively~unified SSB parameters
significantly~differ~from the~universal ones.
The gauge~coupling $g$~is assumed to be the primary~coupling. We should note that the~dimensionless sector admits reduction~solutions that~are independent of the dimensionful~sector.~Two sets of asymptotically~free (AF) solutions can achieve~a Gauge-Yukawa Unification in~this model \cite{Kubo:1994bj}:
\beq
\label{two_sol}
\begin{split}
a & : g_t=\sqrt{\frac{2533}{2605}} g + \order{g^3}~,~
g_b=\sqrt{\frac{1491}{2605}} g + \order{g^3}~,~
g_{\lambda}=0~,~
g_f=\sqrt{\frac{560}{521}} g + \order{g^3}~,\\
b & : g_t=\sqrt{\frac{89}{65}} g + \order{g^3}~,~
g_b=\sqrt{\frac{63}{65}} g + \order{g^3}~,~
g_{\lambda}=0~,~g_f=0~.
\end{split}
\eeq
The higher~order~terms~denote uniquely~computable
power~series in~$g$.
These solutions~describe~the boundaries~of an
AF RGI surface~in the parameter space, on~which $g_{\lambda}$ and $g_f$~may differ from~zero. This~fact makes possible a partial~reduction~where $g_{\lambda}$ and $g_f$ are (non-vanishing)~independent~parameters without endangering
AF. The proton-decay~safe region of that surface favours solution~$a$.~Therefore,~we choose to be exactly at~the boundary~defined by~solution $a$\footnote{
$ g_{\lambda}=0 $ is~inconsistent, but $g_{\lambda} < \sim 0.005$~is necessary in order for the proton
decay~constraint~\cite{Kubo:1995cg} to be satisfied.
A small $g_{\lambda} $~is expected not to affect the prediction of~unification of~SSB~parameters.}.

The reduction~of dimensionful couplings is performed as in \refeq{reduction}.~It is understood that $\mu_{\Sigma}$, $\mu_H$ and~$M$ cannot~be reduced in a desired~form and they are treated~as~independent~parameters.
The lowest-order~reduction~solution is found to be:
\begin{equation}\label{red_sol_1}
B_H = \frac{1029}{521}\,\mu_H M~,~
B_{\Sigma}=-\frac{3100}{521}\,\mu_{\Sigma} M~,
\end{equation}
\beq
\begin{split}
\label{red_sol}
h_t &=-g_t\,M~,~h_b =-g_b\,M~,
~h_f =-g_f\,M~,~h_{\lambda}=0~,\\
m_{H_u}^{2} &=-\frac{569}{521} M^{2}~,~
m_{H_d}^{2} =-\frac{460}{521} M^{2}~,
~m_{\Sigma}^{2} = \frac{1550}{521} M^{2}~,\\
m_{\Phi^3}^{2} & = \frac{436}{521} M^{2}~,~
m_{\Phi^{1,2}}^{2} =\frac{8}{5} M^{2}~,~
m_{\Psi^3}^{2} =\frac{545}{521} M^{2}~,~
m_{\Psi^{1,2}}^{2} =\frac{12}{5} M^{2}~.
\end{split}
\eeq
The gaugino mass~$M$ characterize~the scale of~the SUSY~breaking.~It is noted that we may include
$B_H$ and~$B_{\Sigma}$ as independent~parameters
without~changing the~one-loop~reduction~solution (\ref{red_sol}). Also note~that, although we have found specific relations among~the soft scalar masses and the unified gaugino mass,~the sum rule still holds.

\section{The Finite \texorpdfstring{$N=1$}~ Supersymmetric \texorpdfstring{$SU(5)$}~ }\label{sec:finitesu5}

Next, let us~review an $SU(5)$ gauge theory which is finite (Finite unified Theory - FUT) to all~orders, with reduction of couplings applied to the third~fermionic generation. This~FUT was selected in the past due~to agreement with experimental constraints at the time~\cite{Heinemeyer:2007tz}~and predicted the  light Higgs mass between 121-126 GeV~almost~five~years prior to the~discovery.\footnote{Improved~Higgs mass calculations would yield a different~interval, still compatible with current experimental data~(see \refcha{roc_pheno}).}
The particle~content~consists of three ($\overline{\bf 5} + \bf{10}$) supermultiplets,~a pair for each generation of~quarks and~leptons, four~($\overline{\bf 5} + {\bf 5}$) and~one ${\bf 24}$~considered as Higgs~supermultiplets.
When the finite~GUT~group is broken, the~theory is no
longer~finite, and~we are left~with the~MSSM
\cite{Kapetanakis:1992vx,Kubo:1994bj,Kubo:1994xa,Kubo:1995hm,Kubo:1997fi,Mondragon:1993tw}.

A predictive~all-order finite GYU~$SU(5)$ model should~also
have~the~following~properties:
\begin{enumerate}
\item
One-loop~anomalous~dimensions are~diagonal,
i.e.,~$\gamma_{i}^{(1)\,j} \propto \delta^{j}_{i} $.
\item The~fermions in the irreps
 $\overline{\bf 5}_{i},{\bf 10}_i~(i=1,2,3)$ do
 not~couple~to the~adjoint ${\bf 24}$.
\item The~two Higgs doublets of the MSSM are mostly made~out of a~pair~of Higgs~quintet and anti-quintet,~which~couple to the~third~generation.
\end{enumerate}
Reduction~of couplings enhances the symmetry, and the superpotential~is then given by \cite{Kobayashi:1997qx,Mondragon:2009zz}:
\begin{align}
W &= \sum_{i=1}^{3}\,[~\frac{1}{2}g_{i}^{u}
\,{\bf 10}_i{\bf 10}_i H_{i}+
g_{i}^{d}\,{\bf 10}_i \overline{\bf 5}_{i}\,
\overline{H}_{i}~] +
g_{23}^{u}\,{\bf 10}_2{\bf 10}_3 H_{4} \\
 &+g_{23}^{d}\,{\bf 10}_2 \overline{\bf 5}_{3}\,
\overline{H}_{4}+
g_{32}^{d}\,{\bf 10}_3 \overline{\bf 5}_{2}\,
\overline{H}_{4}+
g_{2}^{f}\,H_{2}\,
{\bf 24}\,\overline{H}_{2}+ g_{3}^{f}\,H_{3}\,
{\bf 24}\,\overline{H}_{3}+
\frac{g^{\lambda}}{3}\,({\bf 24})^3~.\nonumber
\label{w-futb}
\end{align}
A more detailed~description of the model and its properties can be found in~\cite{Kapetanakis:1992vx,Kubo:1994bj,Mondragon:1993tw}. The~non-degenerate and isolated~solutions to
$\gamma^{(1)}_{i}=0$~give:
\bea
&& (g_{1}^{u})^2
=\frac{8}{5}~ g^2~, ~(g_{1}^{d})^2
=\frac{6}{5}~g^2~,~
(g_{2}^{u})^2=(g_{3}^{u})^2=\frac{4}{5}~g^2~,\label{zoup-SOL52}\\
&& (g_{2}^{d})^2 = (g_{3}^{d})^2=\frac{3}{5}~g^2~,~
(g_{23}^{u})^2 =\frac{4}{5}~g^2~,~
(g_{23}^{d})^2=(g_{32}^{d})^2=\frac{3}{5}~g^2~,
\nonumber\\
&& (g^{\lambda})^2 =\frac{15}{7}g^2~,~ (g_{2}^{f})^2
=(g_{3}^{f})^2=\frac{1}{2}~g^2~,~ (g_{1}^{f})^2=0~,~
(g_{4}^{f})^2=0~.\nonumber
\eea
Furthermore, we~have the $h=-MC$ relation, while from the sum~rule (see \refse{roc_soft}) we obtain:
\beq
m^{2}_{H_u}+
2 m^{2}_{{\bf 10}} =M^2~,~
m^{2}_{H_d}-2m^{2}_{{\bf 10}}=-\frac{M^2}{3}~,~
m^{2}_{\overline{{\bf 5}}}+
3m^{2}_{{\bf 10}}=\frac{4M^2}{3}~.
\label{sumrB}
\eeq
This shows that~we have~only two free~parameters
$m_{{\bf 10}}$~and $M$ for the dimensionful~sector.

The GUT symmetry~breaks to the MSSM, where we want only two Higgs~doublets.~This is achieved with the introduction of appropriate~mass~terms~that allow a~rotation in the Higgs~sector~\cite{Leon:1985jm,Kapetanakis:1992vx,Mondragon:1993tw,Hamidi:1984gd,Jones:1984qd},~that permits only~one pair of Higgs~doublets~(which couple mostly to~the third~family) to remain light~and~acquire~vacuum expectation~values.
The usual fine~tuning to achieve~doublet-triplet splitting helps the model to~avoid~fast proton~decay (but this mechanism has differences~compared to the one used in the minimal $SU(5)$ because~of the extended~Higgs sector of the finite case).

Thus, below~the GUT scale we have the MSSM with the first two generations~unrestricted, while the third is~given by the finiteness conditions.

\section{Finite \texorpdfstring{$SU(N)^3$}~ Unification}\label{sec:su33}

One can~consider the construction of FUTs
that have~a product gauge group. Let us consider an $N=1$ theory with~a $SU(N)_1 \times SU(N)_2 \times \cdots \times SU(N)_k$~and $n_f$ copies (number of~families)
of the~supermultiplets $(N,N^*,1,\dots,1) + (1,N,N^*,\dots,1) + \cdots + (N^*,1,1,\dots,N)$.  Then, the one-loop $\beta$ function~coefficient of  the RGE of each $SU(N)$~gauge coupling~is
\begin{equation}
b = \left( -\frac{11}{3} + \frac{2}{3} \right) N + n_f \left( \frac{2}{3}
 + \frac{1}{3} \right) \left( \frac{1}{2} \right) 2 N = -3 N + n_f
N\,.
\label{3gen}
\end{equation}
The necessary~condition~for finiteness is~$b=0$, which occurs only for the choice~$n_f = 3$.  Thus, it~is natural to consider three~families of~quarks and leptons.

From~a phenomenological point of view, the~choice is the $SU(3)_C \times SU(3)_L \times~SU(3)_R$ model, which is 
discussed in~detail in \citere{Ma:2004mi}. The discussion of the general~well-known~example can be found in~\cite{Derujula:1984gu,Lazarides:1993sn,Lazarides:1993uw,Ma:1986we}.~ The quarks and the leptons of the model transform~as follows:
\begin{equation}
  Q = \begin{pmatrix} -d^1_L & -d^2_L & -d^3_L \\
 u^1_L & u^2_L & u^3_L \\
 D^1_L & D^2_L & D^3_L \end{pmatrix}
\sim (3,3^*,1), ~~~
    q^c = \begin{pmatrix}  d^{c1}_R & u^{c1}_R & D^{c1}_R \\
 d^{c2}_R & u^{c2}_R & D^{c2}_R \\
 d^{c3}_R & u^{c3}_R & D^{c3}_R
\end{pmatrix}
    \sim (3^*,1,3),
\label{2quarks}
\end{equation}
\begin{equation}
L = \begin{pmatrix}  H_d^0 & H_u^+ & \nu_L \\
 H_d^- & H_u^0 & e_L \\
 \nu^c_R & e^c_R & S
\end{pmatrix}
\sim (1,3,3^*),
\label{3leptons}
\end{equation}
where $D$ are~down-type quarks that acquire masses close to $M_{GUT}$. We~have to impose a cyclic  $Z_3$ symmetry in order to have equal~gauge couplings at the GUT scale, i.e.
\begin{equation}
Q \to L \to q^c \to Q,
\label{15}
\end{equation}
where $Q$ and $q^c$~are given in \refeq{2quarks} and $L$ in
\refeq{3leptons}.~Then the vanishing of the one-loop gauge $\beta$ function,~which is the first finiteness condition (\ref{1st}),~is~satisfied.~This leads us to the second condition,~namely the vanishing of the~anomalous dimensions of all superfields \refeq{2nd}.~Let us write down the superpotential first.~For one family we have just two trilinear invariants~that can~be~used in the~superpotential as follows:
\begin{equation}
f ~Tr (L q^c Q) + \frac{1}{6} f' ~\epsilon_{ijk} \epsilon_{abc}
(L_{ia} L_{jb} L_{kc} + q^c_{ia} q^c_{jb} q^c_{kc} +
Q_{ia} Q_{jb} Q_{kc}),
\label{16}
\end{equation}
where $f$ and $f'$~are~the Yukawa~couplings associated~to each~invariant.~The quark and~leptons obtain~masses when~the scalar~parts~of~the~superfields $(H_d^0,H_u^0)$ obtain~vacuum expectation~values (vevs),
\begin{equation}
m_d = f \langle H_d^0 \rangle, ~~ m_u = f \langle
H_u^0 \rangle, ~~
m_e = f' \langle H_d^0 \rangle, ~~ m_\nu = f' \langle H_u^0 \rangle.
\label{18}
\end{equation}
For three~families,~the most general superpotential has 11 $f$
couplings and 10~$f'$ couplings. Since anomalous dimensions of each superfield~vanish, 9~conditions are imposed on these couplings:
\begin{equation}
\sum_{j,k} f_{ijk} (f_{ljk})^* + \frac{2}{3} \sum_{j,k} f'_{ijk}
(f'_{ljk})^* = \frac{16}{9} g^2 \delta_{il}\,,
\label{19}
\end{equation}
where
\begin{eqnarray}
&& f_{ijk} = f_{jki} = f_{kij}, \label{20}\\
&& f'_{ijk} = f'_{jki} = f'_{kij} = f'_{ikj} = f'_{kji} = f'_{jik}.
\label{21}
\end{eqnarray}
Quarks~and~leptons~receive  masses~when  the~scalar part~of the~superfields~$(H_d^0)_{1,2,3}$ and $(H_u^0)_{1,2,3}$ obtain~vevs:
\begin{eqnarray}
&& ({\cal M}_d)_{ij} = \sum_k f_{kij} \langle  (H_d^0)_k \rangle, ~~~
   ({\cal M}_u)_{ij} = \sum_k f_{kij} \langle  (H_u^0)_k \rangle, \label{22} \\
&& ({\cal M}_e)_{ij} = \sum_k f'_{kij} \langle (H_d^0)_k \rangle, ~~~
   ({\cal M}_\nu)_{ij} = \sum_k f'_{kij} \langle (H_u^0)_k \rangle.
\label{23}
\end{eqnarray}
When the FUT~breaks at $M_{\rm GUT}$, we are left with the MSSM\footnote{\cite{Irges:2011de,Irges:2012ze} and refs therein~discuss in detail the spontaneous breaking of $SU(3)^3$.},~where~both Higgs~doublets couple maximally to~the third~generation.~ These doublets are the linear~combinations $H_u^0 = \sum_i a_i (H_u^0)_i$ and
$H_d^0 = \sum_i b_i (H_u^0)_i$.~For the choice of the particular~combinations we can use the appropriate masses in the~superpotential~\cite{Leon:1985jm}, since~they are~not
constrained~by the~finiteness~conditions.~The FUT breaking leaves remnants~in the form of the boundary~conditions on~the
gauge~and~Yukawa~couplings, i.e. \refeq{19}, the~$h=-Mf$
relation~and~the soft~scalar mass~sum rule~at $M_{\rm GUT}$. The latter~takes the  following form in this model:
\begin{eqnarray}
m^2_{H_u} + m^2_{\tilde t^c} + m^2_{\tilde q} = M^2 =
m^2_{H_d} + m^2_{\tilde b^c} + m^2_{\tilde q}~.
\end{eqnarray}
If the solution~of \refeq{19} is both \textit{unique} and \textit{isolated},~the model is finite at all orders of perturbation theory. This leads $f'$~to vanish~and we are left with the relations
\begin{equation}
f^2 = f^2_{111} = f^2_{222} = f^2_{333} = \frac{16}{9} g^2\,.
\label{isosol}
\end{equation}
Since all $f'$~parameters are zero in one-loop level, the lepton masses are zero.~They cannot appear
radiatively~(as one would expect) due to the finiteness
conditions,~and remain as a~problem for further study.

If the solution~is just unique (but not isolated, i.e. parametric) we~can keep non-vanishing $f'$ 
and achieve~two-loop finiteness, in which case lepton masses are not fixed~to zero. Then we have a slightly different set of conditions~that restrict the Yukawa couplings:
\begin{eqnarray}
f^2 = r \left(\frac{16}{9}\right) g^2\,,\quad
f'^2 = (1-r) \left(\frac{8}{3}\right) g^2\,,
\label{fprime}
\end{eqnarray}
where  $r$ is~free and parametrizes the~different
solutions~to~the~finiteness conditions. It is~important to note that we~use the sum rule as boundary condition~to the soft scalars.

\section{Reduction of Couplings in the MSSM}\label{sec:mssm}

Finally, a version of~the MSSM with reduced couplings is reviewed. All work is~carried out in the framework~of the~MSSM, but with~the assumption of a covering~GUT. The original partial~reduction in this model was done and analysed in \cite{Mondragon:2013aea,Mondragon:2017hki} and is once more restricted to the~third fermionic generation. The superpotential~in given by
\beq
\label{supot2}
W = Y_tH_2Qt^c+Y_bH_1Qb^c+Y_\tau H_1L\tau^c+ \mu H_1H_2\, ,
\eeq
where $Q,L,t,b,\tau, H_1,H_2$ are the usual superfields of MSSM,
while the SSB Lagrangian is given by
\beq
\label{SSB_L}
\begin{split}
-\mathcal{L}_{\rm SSB} &= \sum_\phi m^2_\phi\hat{\phi^*}\hat{\phi}+
\left[m^2_3\hat{H_1}\hat{H_2}+\sum_{i=1}^3 \frac 12 M_i\lambda_i\lambda_i +\textrm{h.c}\right]\\
&+\left[h_t\hat{H_2}\hat{Q}\hat{t^c}+h_b\hat{H_1}\hat{Q}\hat{b^c}+h_\tau \hat{H_1}\hat{L}\hat{\tau^c}+\textrm{h.c.}\right] ,
\end{split}
\eeq
where $\hat{\phi}$~represents the scalar component of all superfields,~$\lambda$ refers to the gaugino fields
while all~hatted fields refer to the scalar components of the corresponding~superfield.
The Yukawa~$Y_{t,b,\tau}$ and the trilinear $h_{t,b,\tau}$ couplings~refer to the third family.

Let us start~with the dimensionless couplings, i.e.~gauge and Yukawa. As a~first step we consider~only the strong coupling and the top and bottom~Yukawa couplings, while the other two gauge couplings and the tau~Yukawa will be treated as corrections.~Following the above line, we reduce the Yukawa couplings in favour of~the strong coupling~$\al_3$
\[
\frac{Y^2_i}{4\pi}\equiv \al_i=G_i^2\al_3,\qquad i=t,b,
\]
and using the RGE for the Yukawa, we get
\[
G_i^2=\frac 13 ,\qquad i=t,b.
\]
This system of the top~and bottom Yukawa couplings reduced with the strong one~is dictated by (i) the different running behaviour of the~$SU(2)$ and $U(1)$ coupling compared to the strong one~\cite{Kubo:1985up}
and (ii)~the incompatibility of applying the above reduction
for the tau~Yukawa since the corresponding $G^2$ turns negative~\cite{{Mondragon:2013aea}}. Adding now the two~other gauge couplings~and the tau Yukawa in the RGE as~corrections, we obtain
\beq
\label{Gt2_Gb2}
G_t^2=\frac 13+\frac{71}{525}\rho_1+\frac 37 \rho_2 +\frac 1{35}\rho_\tau,\qquad
G_b^2=\frac 13+\frac{29}{525}\rho_1+\frac 37 \rho_2 -\frac 6{35}\rho_\tau
\eeq
where
\beq
\label{r1_r2_rtau}
\rho_{1,2}=\frac{g_{1,2}^2}{g_3^2}=\frac{\al_{1,2}}{\al_3},\qquad
\rho_\tau=\frac{g_\tau^2}{g_3^2}=\frac{\displaystyle{\frac{Y^2_\tau}{4\pi}}}{\al_3}
\eeq
Note that the~corrections in \refeq{Gt2_Gb2} are taken at the GUT scale and under~the assumption that
\[
\frac{d}{dg_3}\left(\frac {Y_{t,b}^2}{g_3^2}\right)=0.
\]

A comment on the assumption above, which led to the \refeq{Gt2_Gb2}, is in order.
In practice, we~assume that even including the corrections from the rest~of the gauge as well as the tau Yukawa couplings,~at the GUT scale the ratio of the top and bottom couplings~$\al_{t,b}$ over the strong coupling are still constant,~i.e. their scale dependence is negligible.
Or, rephrasing~it, our assumption can be understood as a requirement that~in the ultraviolet (close to the GUT scale)
the ratios of~the top and bottom Yukawa couplings over the strong~coupling become least sensitive against the change
of the~renormalization scale. This requirement sets the boundary~condition at the GUT scale, given in \refeq{Gt2_Gb2}.
Alternatively,~one could follow the systematic method to include the~corrections to a non-trivially reduced system developed~in \cite{Kubo:1988zu}, but considering two reduced systems:~the first one consisting of the ``top, bottom'' couplings~and the second of the~``strong, bottom'' ones.

\smallskip

\noindent At two-loop level the~corrections are assumed to be of the form
\[
\al_i=G_i^2\al_3+J_i^2 \al_3^2,\qquad i=t,b.
\]
Then, the coefficients $J_i$ are given by
\[
J_i^2=\frac 1{4\pi}\,\frac{17}{24},\qquad i=t,b
\]
for the case where~only the strong gauge and the top and bottom Yukawa~couplings are active,~while for the case where the other two gauge~and the tau Yukawa couplings are
added as corrections~we obtain
\[
J_t^2=\frac 1{4\pi}\frac{N_t}{D},\quad
J_b^2=\frac 1{4\pi}\frac{N_b}{5D},
\]
where
\[
\begin{split}
D=&
257250 (196000 + 44500 \rho_1 + 2059 \rho_1^2 + 200250 \rho_2 + 22500 \rho_1 \rho_2 +
   50625 \rho_2^2 - \\
&33375 \rho_\tau - 5955 \rho_1 \rho_\tau - 16875 \rho_2 \rho_\tau -
   1350 \rho_\tau^2),
\end{split}
\]

\[
\begin{split}
N_t=&
-(-35714875000 - 10349167500 \rho_1 + 21077903700 \rho_1^2 +9057172327 \rho_1^3 +\\
&   481651575 \rho_1^4 - 55566000000 \rho_2 +2857680000 \rho_1 \rho_2 + 34588894725 \rho_1^2 \rho_2 +\\
&   5202716130 \rho_1^3 \rho_2 +3913875000 \rho_2^2 + 8104595625 \rho_1 \rho_2^2 + 11497621500 \rho_1^2 \rho_2^2 +\\
&   27047671875 \rho_2^3 + 1977918750 \rho_1 \rho_2^3 + 7802578125 \rho_2^4 +3678675000 \rho_\tau +\\
&   1269418500 \rho_1 \rho_\tau - 2827765710 \rho_1^2 \rho_\tau -1420498671 \rho_1^3 \rho_\tau +7557637500 \rho_2 \rho_\tau -\\
&   2378187000 \rho_1 \rho_2 \rho_\tau - 4066909425 \rho_1^2 \rho_2 \rho_\tau -1284018750 \rho_2^2 \rho_\tau - 1035973125 \rho_1 \rho_2^2 \rho_\tau -\\
&   2464171875 \rho_2^3 \rho_\tau + 1230757500 \rho_\tau^2 + 442136100 \rho_1 \rho_\tau^2 -186425070 \rho_1^2 \rho_\tau^2 +\\
&   1727460000 \rho_2 \rho_\tau^2 +794232000 \rho_1 \rho_2 \rho_\tau^2 + 973518750 \rho_2^2 \rho_\tau^2 -\\
&   325804500 \rho_\tau^3 - 126334800 \rho_1 \rho_\tau^3 - 412695000 \rho_2 \rho_\tau^3 -
   32724000 \rho_\tau^4),
\end{split}
\]

\[
\begin{split}
N_b=&
-(-178574375000 - 71734162500 \rho_1 + 36055498500 \rho_1^2 +13029194465 \rho_1^3 +\\
&   977219931 \rho_1^4 - 277830000000 \rho_2 -69523650000 \rho_1 \rho_2 + 72621383625 \rho_1^2 \rho_2 +\\
&   10648126350 \rho_1^3 \rho_2 +19569375000 \rho_2^2 + 13062459375 \rho_1 \rho_2^2 + 25279672500 \rho_1^2 \rho_2^2 +\\
&   135238359375 \rho_2^3 + 16587281250 \rho_1 \rho_2^3 + 39012890625 \rho_2^4 +58460062500 \rho_\tau +\\
&   35924411250 \rho_1 \rho_\tau - 13544261325 \rho_1^2 \rho_\tau -2152509435 \rho_1^3 \rho_\tau - 13050843750 \rho_2 \rho_\tau +\\
&   45805646250 \rho_1 \rho_2 \rho_\tau - 75889125 \rho_1^2 \rho_2 \rho_\tau -24218578125 \rho_2^2 \rho_\tau + 17493046875 \rho_1 \rho_2^2 \rho_\tau -\\
&   1158046875 \rho_2^3 \rho_\tau - 36356775000 \rho_\tau^2 -26724138000 \rho_1 \rho_\tau^2 - 4004587050 \rho_1^2 \rho_\tau^2 -\\
&   97864200000 \rho_2 \rho_\tau^2 - 22359847500 \rho_1 \rho_2 \rho_\tau^2 -39783656250 \rho_2^2 \rho_\tau^2 + 25721797500 \rho_\tau^3 +\\
&   3651097500 \rho_1 \rho_\tau^3 + 11282287500 \rho_2 \rho_\tau^3 + 927855000 \rho_\tau^4).
\end{split}
\]
Let us now move to the~dimensionful couplings of the SSB sector of the~Lagrangian, namely~the trilinear couplings $h_{t,b,\tau}$ of the SSB Lagrangian,~\refeq{SSB_L}. Again,
following the pattern in the Yukawa~reduction,
in the first stage~we reduce $h_{t,b}$, while $h_\tau$ will be
treated as a~correction.
\[
h_i=c_i Y_i M_3 = c_i G_i M_3 g_3,\qquad i=t,b,
\]
where $M_3$ is~the gluino mass.~Using the RGE for the two $h$ we~get
\[
c_t=c_b=-1,
\]
where we have~also used the 1-loop relation between the gaugino mass~and the gauge coupling RGE
\[
2M_i\frac {dg_i}{dt}=g_i\frac {dM_i}{dt},\qquad i=1,2,3.
\]
Adding the other~two gauge couplings as well as~the tau Yukawa $h_\tau$~as correction we get
\[
c_t=-\frac{A_A A_{bb} + A_{tb} B_B}{A_{bt} A_{tb} - A_{bb} A_{tt}},\qquad
c_b=-\frac{A_A A_{bt} + A_{tt} B_B}{A_{bt} A_{tb} - A_{bb} A_{tt}},
\]
where
\beq
\label{rhtau}
\begin{split}
A_{tt}& = G_b^2 - \frac{16}3 - 3\rho_2 - \frac{13}{15}\rho_1,\quad
A_A = \frac{16}3 + 3\rho_2^2 + \frac{13}{15}\rho_1^2\\
A_{bb} &= G_t^2 + \rho_\tau - \frac{16}3 - 3\rho_2 - \frac 7{15}\rho_1,\quad
B_B = \frac{16}3 + 3\rho_2^2 + \frac7{15}\rho_1^2 + \rho_{h_\tau}\rho_\tau^{1/2}\\
A_{tb} &= G_b^2,\quad A_{bt} = G_t^2,\quad \rho_{h_\tau}=\frac{h_\tau}{g_3 M_3}.
\end{split}
\eeq
Finally we consider~the soft squared masses $m^2_\phi$ of the SSB~Lagrangian. Their reduction,~according to the discussion in \refse{roc_dim_1-2}, takes~the form
\beq\label{mM_rel}
m_i^2=c_i M_3^2,\quad i=Q,u,d,H_u,H_d.
\eeq
The 1-loop RGE for~the scalar masses reduce to the following algebraic system~(where we have added~the corrections~from the two gauge couplings, the tau Yukawa and~$h_\tau$)
\[
\begin{split}
-12c_Q&=X_t+X_b-\frac{32}3-6\rho_2^3-\frac 2{15}\rho_1^3+\frac 15\rho_1 S,\\
-12c_u&=2X_t-\frac{32}3-\frac{32}{15}\rho_1^3-\frac 45\rho_1 S,\\
-12c_d&=2X_b-\frac{32}3-\frac 8{15}\rho_1^3+\frac 25\rho_1 S,\\
-12c_{H_u}&=3X_t-6\rho_2^3-\frac 65\rho_1^3+\frac 35\rho_1 S,\\
-12c_{H_d}&=3X_b+X_\tau-6\rho_2^3-\frac 65\rho_1^3-\frac 35\rho_1 S,
\end{split}
\]
where
\[
\begin{split}
X_t&=2G_t^2\left(c_{H_u}+c_Q+c_u\right)+2c_t^2G_t^2,\\
X_b&=2G_b^2\left(c_{H_d}+c_Q+c_d\right)+2c_b^2G_b^2,\\
X_\tau&=2\rho_\tau c_{H_d}+2\rho_{h_\tau}^2,\\
S&=c_{H_u}-c_{H_d}+c_Q-2c_u+c_d.
\end{split}
\]
Solving the above~system for the coefficients~$c_{Q,u,d,H_u,H_d}$ we get
\[
\begin{split}
c_Q=&-\frac{c_{Q{\rm Num}}}{D_m},\quad
c_u=-\frac 13\frac{c_{u{\rm Num}}}{D_m},\quad
c_d=-\frac{c_{d{\rm Num}}}{D_m},\\
c_{H_u}=&-\frac 23\frac{c_{Hu{\rm Num}}}{D_m},\quad
c_{H_d}=-\frac{c_{Hd{\rm Num}}}{D_m},
\end{split}
\]
where
\[
\begin{split}
D_m=&
4 (6480 + 6480 G_b^2 + 6480 G_t^2 + 6300 G_b^2 G_t^2 +
\rho_1(1836  + 1836 G_b^2  + 1836 G_t^2  +1785 G_b^2 G_t^2 )+ \\
&     \rho_\tau \left[1080  + 540 G_b^2  + 1080 G_t^2  + 510 G_b^2 G_t^2  +
 252 \rho_1  + 99 G_b^2 \rho_1  +252 G_t^2 \rho_1  + 92 G_b^2 G_t^2 \rho_1 \right]),
\end{split}
\]

\[
\begin{split}
c_{Q{\rm Num}}=&
2160 F_Q +
G_b^2(- 360 F_d - 360 F_{H_d}  + 1800 F_Q) +
G_t^2(- 360 F_{H_u} + 1800 F_Q  - 360 F_u)+\\
&G_b^2 G_t^2(- 300 F_d - 300 F_{H_d} - 300 F_{H_u} + 1500 F_Q  - 300 F_u ) +\\
&\rho_1(- 36 F_d +   36 F_{H_d}  - 36 F_{H_u}  + 576 F_Q  + 72 F_u  )+\\
&G_b^2 \rho_1(- 138 F_d  -66 F_{H_d} - 36 F_{H_u}  + 474 F_Q  + 72 F_u ) +\\
&G_t^2 \rho_1(- 36 F_d + 36 F_{H_d}  - 138 F_{H_u}  + 474 F_Q  - 30 F_u) +\\
&G_b^2 G_t^2 \rho_1(- 120 F_d - 50 F_{H_d}  -  120 F_{H_u}  + 390 F_Q  - 15 F_u ) +\\
&\rho_\tau\left[
360 F_Q  +G_b^2 (- 60 F_d + 120 F_Q )+
G_t^2 (- 60 F_{H_u} + 300 F_Q  - 60 F_u )+
\right.
\\
&G_b^2 G_t^2 (- 50 F_d - 20 F_{H_u}  + 100 F_Q  - 20 F_u )+
\rho_1 (-6 F_d - 6 F_{H_u}  + 78 F_Q  + 12 F_u )+\\
&G_b^2 \rho_1 ( - 11 F_d  + 22 F_Q ) +
G_t^2 \rho_1 ( - 6 F_d  -20 F_{H_u}  + 64 F_Q  - 2 F_u ) +\\
&\left.
G_b^2 G_t^2 \rho_1 ( -9 F_d  - 4 F_{H_u}  +18 F_Q  - 3 F_u )
\right],
\end{split}
\]

\[
\begin{split}
c_{u{\rm Num}}=&
6480 F_u + 6480 F_u G_b^2 +
G_t^2(- 2160 F_{H_u}  - 2160 F_Q  + 4320 F_u ) +\\
& G_b^2 G_t^2(  360 F_d  + 360 F_{H_d}  - 2160 F_{H_u}  -1800 F_Q  + 4140 F_u )+\\
&    \rho_1( 432 F_d  - 432 F_{H_d} +432 F_{H_u}  + 432 F_Q  + 972 F_u )+\\
&    G_b^2 \rho_1( 432 F_d  -432 F_{H_d}  + 432 F_{H_u}  + 432 F_Q  + 972 F_u )+\\
&   G_t^2 \rho_1(432 F_d  - 432 F_{H_d}  - 180 F_{H_u}  - 180 F_Q  +360 F_u )+\\
&    G_b^2 G_t^2 \rho_1( 522 F_d  - 318 F_{H_d}  - 192 F_{H_u}  - 90 F_Q  + 333 F_u ) +\\
&   \rho_\tau \left[
     1080 F_u  +
     540 G_b^2 F_u + G_t^2(  - 360 F_{H_u}   - 360 F_Q   + 720 F_u )+ \right.\\
&      G_b^2 G_t^2 (60 F_d   - 180 F_{H_u}   - 120 F_Q   + 330 F_u )  +
  \rho_1( 72 F_d   + 72 F_{H_u}   + 72 F_Q   + 108 F_u )  +\\
&   G_b^2 \rho_1( 36 F_{H_u}   + 27 F_u ) +
72 G_t^2 \rho_1 (F_d   - 12 F_{H_u}   - 12 F_Q   + 24 F_u )  +\\
&   \left.
G_b^2 G_t^2 \rho_1 (9 F_d   + 4 F_{H_u}   - 18 F_Q   + 3 F_u )
    \right],
\end{split}
\]

\[
\begin{split}
c_{d{\rm Num}}=&
2160 F_d + G_b^2(1440 F_d  - 720 F_{H_d}  - 720 F_Q ) + 2160 F_d G_t^2 +\\
&   G_b^2 G_t^2 (1380 F_d  - 720 F_{H_d}  + 120 F_{H_u}  -  600 F_Q  + 120 F_u ) +\\
&   \rho_1( 540 F_d  + 72 F_{H_d}  - 72 F_{H_u}  - 72 F_Q  + 144 F_u )+\\
&    G_b^2 \rho_1( 336 F_d  -132 F_{H_d}  - 72 F_{H_u}  - 276 F_Q  + 144 F_u ) +\\
&   G_t^2 \rho_1( 540 F_d  + 72 F_{H_d}  - 72 F_{H_u}  - 72 F_Q  + 144 F_u )+\\
&   G_b^2 G_t^2 \rho_1( 321 F_d  - 134 F_{H_d}  -36 F_{H_u}  - 240 F_Q  + 174 F_u ) +\\
&   \rho_\tau\left[
     360 F_d  + G_b^2( 60 F_d   - 120 F_Q )  + 360 F_d G_t^2  +
   G_b^2 G_t^2(50 F_d   + 20 F_{H_u}   - 100 F_Q   +20 F_u )+ \right.\\
&     \rho_1(72 F_d   - 12 F_{H_u}  -12 F_Q   + 24 F_u)   +  G_b^2 \rho_1 ( 11 F_d -22 F_Q)    +\\
&   \left.  G_t^2 \rho_1( 72 F_d   - 12 F_{H_u}   -12 F_Q   + 24 F_u )+
     G_b^2 G_t^2 \rho_1(9 F_d   + 4 F_{H_u}  - 18 F_Q   + 3 F_u )
    \right],
\end{split}
\]

\[
\begin{split}
c_{Hu{\rm Num}}=&
3240 F_{H_u} + 3240 F_{H_u} G_b^2 + G_t^2( 1620 F_{H_u}  - 1620 F_Q  - 1620 F_u )+\\
&    G_b^2 G_t^2( 270 F_d  + 270 F_{H_d}  + 1530 F_{H_u}  - 1350 F_Q  - 1620 F_u )+ \\
&    \rho_1(- 162 F_d  + 162 F_{H_d}  + 756 F_{H_u}  - 162 F_Q  + 324 F_u )+\\
&    G_b^2 \rho_1(- 162 F_d  + 162 F_{H_d}  + 756 F_{H_u}  - 162 F_Q  + 324 F_u )+\\
&   G_t^2 \rho_1(-162 F_d  + 162 F_{H_d}  + 297 F_{H_u}  - 621 F_Q  - 135 F_u )+\\
&   G_b^2 G_t^2 \rho_1 (- 81 F_d  + 234 F_{H_d}  + 276 F_{H_u}  - 540 F_Q  - 144 F_u ) +\\
&  \rho_\tau \left[
  540 F_{H_u}  + 270 F_{H_u} G_b^2   + G_t^2(270 F_{H_u}   - 270 F_Q    - 270 F_u )+ \right.\\
&       G_b^2 G_t^2( 45 F_d    + 120 F_{H_u}    - 90 F_Q    - 135 F_u )+
   \rho_1(-27 F_d    + 99 F_{H_u}    - 27 F_Q    + 54 F_u )   +\\
&   G_b^2 \rho_1( 36 F_{H_u}    + 27 F_u    - 27 F_d )   + G_t^2 \rho_1( 36 F_{H_u}    - 90 F_Q    - 9 F_u )   +\\
&  \left.  G_b^2 G_t^2 \rho_1( 9 F_d    + 4 F_{H_u}    - 18 F_Q    + 3 F_u )
\right],
\end{split}
\]

\[
\begin{split}
c_{Hd{\rm Num}}=&
2160 F_{H_d}+ G_b^2 (- 1080 F_d + 1080 F_{H_d}  - 1080 F_Q ) + 2160 F_{H_d} G_t^2+\qquad\qquad\qquad\qquad\\
&    G_b^2 G_t^2(- 1080 F_d  + 1020 F_{H_d}  +180 F_{H_u}  - 900 F_Q  + 180 F_u )+\\
&    \rho_1( 108 F_d  +504 F_{H_d} + 108 F_{H_u}  + 108 F_Q  - 216 F_u )+\\
&    G_b^2 \rho_1(- 198 F_d  +198 F_{H_d}  + 108 F_{H_u}  - 198 F_Q  - 216 F_u ) +\\
&   G_t^2 \rho_(108 F_d 1 + 504 F_{H_d}  + 108 F_{H_u}  + 108 F_Q  -216 F_u )+\\
&    G_b^2 G_t^2 \rho_1(- 201 F_d  + 184 F_{H_d}  +156 F_{H_u}  - 150 F_Q  - 159 F_u )
\end{split}
\]
and
\[
\begin{split}
F_Q &= 2 c_t^2 G_t^2 + 2 c_b^2 G_b^2 - \frac{32}{3} - 6 \rho_2^3 - \frac{2}{15} \rho_1^3,\\
F_u &= 4 c_t^2 G_t^2 - \frac{32}{3} - \frac{32}{15} \rho_1^3,\\
F_d &= 4 c_b^2 G_b^2 - \frac{32}{3} - \frac{8}{15} \rho_1^3,\\
F_{H_u}& = 6 c_t^2 G_t^2 - 6 \rho_2^3 - \frac{6}{5} \rho_1^3,\\
F_{H_d}& = 6 c_b^2 G_b^2 + 2 \rho_{h_\tau}^2 - 6 \rho_2^3 - \frac{6}{5} \rho_1^3,
\end{split}
\]
while $G_{t,b}^2$,~$\rho_{1,2,\tau}$ and $\rho_{h_\tau}$ has been defined~in Eqs.(\ref{Gt2_Gb2},\ref{r1_r2_rtau},\ref{rhtau})~respectively.~For our completely reduced system, i.e. $g_3,Y_t,Y_b,h_t,h_b$,~the coefficients of the soft masses
become
\[
c_Q=c_u=c_d=\frac 23,\quad c_{H_u}=c_{H_d}=-1/3,
\]
obeying the~celebrated sum rules
\[
\frac{m_Q^2+m_u^2+m_{H_u}^2}{M_3^2}=c_Q+c_u+c_{H_u}=1,\qquad
\frac{m_Q^2+m_d^2+m_{H_d}^2}{M_3^2}=c_Q+c_d+c_{H_d}=1.
\]

\smallskip

Concerning the gaugino~masses, the Hisano-Shiftman relation (\refeq{M-M0})~is applied to each gaugino mass as a boundary condition at the~GUT scale, where the gauge couplings are considered unified.~Thus, at one-loop level, each gaugino mass is only dependent~on the b-coefficients of the gauge $\beta$-functions~and the arbitrary $M_0$:
\beq
M_i=b_iM_0~.
\eeq
This means that we~can make a choice of $M_0$ such that the gluino mass equals~the unified gaugino mass, and the other two gaugino masses are~equal to the gluino mass times the ratio of the appropriate~b coefficients.

In \refse{se:rmssm} we begin~with the~selection of the free parameters. This discussion~is~intimately connected 
to the fermion masses~predictions.

\chapter{Analysis of Phenomenologically Viable Models}\label{roc_pheno}

The analysis of each model reviewed in the previous chapter is presented here. It includes the predictions for quark masses,~the light Higgs boson mass, the SUSY breaking scale (defined~as the geometric mean of stops), $M_S$, the full SUSY~spectrum and the Cold Dark Matter (CDM) relic density (in~the case the lightest neutralino is considered a CDM candidate), as well as the set of experimental~constraints employed. Note that in the examination of all~four models we use the unified gaugino mass $M$ instead~of $M_S$, as a more indicative parameter of scale. It should be noted that for the first part of the analysis (regarding the light Higgs mass and the first attempt to generate a supersymmetric spectrum) a homebrew code and the new {\tt FeynHiggs} 2.16.0 code~\cite{Degrassi:2002fi,BHHW,FeynHiggs,Bahl:2019hmm} are used, while in the second part of the analysis, that focuses on the discovery potential, the supersymmetric spectrum and its corresponding branching ratios are generated by {\tt SPheno} 4.0.4 \cite{Porod:2003um,Porod:2011nf}. The {\tt MicrOMEGAs 5.0} \cite{Belanger:2001fz,Belanger:2004yn,Barducci:2016pcb} code is used to calculate the Cold Dark Matter relic density, while {\tt MadGraph} \cite{Alwall:2014hca} is used for the calculation of all cross sections.

\section{Phenomenological Constraints}\label{se:constraints}

In the phenomenological~analysis  we apply several experimental constraints, which will~be briefly reviewed in this section. It should be noted that the~constraints presented in this section are the ones that were~valid at the time the original research was carried out.

Starting from the~quark~masses, we calculate the~top quark pole mass, while the bottom~quark mass~is evaluated at $M_Z$, in order not to~encounter~uncertainties inherent to its pole~mass. Their experimental~values~are \cite{Tanabashi:2018oca},
\beq
\mb(M_Z) = 2.83 \pm 0.10 \gev ~.
\label{mbexp}
\eeq
and
\beq
\mt^{\rm exp} = (173.1 \pm 0.9) \gev~.
\label{mtexp}
\eeq
The~discovery~of~a~Higgs-like~particle at~ATLAS~and~CMS in~July 2012~\cite{Aad:2012tfa,Chatrchyan:2012ufa}~can~be interpreted~as~the~discovery~of~the light~$\cal
CP$-even Higgs~boson~of~the~MSSM~Higgs spectrum \cite{Heinemeyer:2011aa,Bechtle:2012jw,Heinemeyer:2016mmm}.~The~experimental~average~for~the (SM) Higgs~boson mass is~\cite{Tanabashi:2018oca}%
\beq
M_H^{\rm exp}=125.10\pm 0.14~{\rm GeV}~.\label{higgsexpval}
\eeq
The~theoretical accuracy \cite{Degrassi:2002fi,BHHW,Bahl:2019hmm},
however,~for~the~prediction~of $\Mh$ in the~MSSM, dominates the uncertainty.~In~our following analysis of each of the models described,~we use~the new~{\tt FeynHiggs} code~\cite{Degrassi:2002fi,BHHW,FeynHiggs,Bahl:2019hmm}
(Version 2.16.0)~to~predict the Higgs mass. {\tt FeynHiggs} evaluates the~Higgs~masses using a combination of 
fixed~order~diagrammatic~calculations and~resummation~of
the~(sub)leading~logarithmic contributions at~all orders, and thus~provides a reliable~evaluation~of~$M_h$ even for~large SUSY~scales.~The refinements~in this~combination (w.r.t.~previous versions~\cite{FeynHiggs})~result~in a
downward~shift~of $M_h$ of order ${\cal O}(2~{\rm GeV})$ for
large SUSY masses. This~version of {\tt FeynHiggs} computes the uncertainty of the~Higgs~boson mass point by point. This theoretical~uncertainty~is added linearly to the experimental error~in~\refeq{higgsexpval}.

Furthermore,~recent results from the ATLAS experiment \cite{Aad:2020zxo}~set limits to the mass of the pseudoscalar Higgs boson, $M_A$,~in comparison with $\tan{\beta}$. For models with~$\tan{\beta}\sim 45-55$, as the ones examined here, the~lowest limit for the physical pseudoscalar Higgs mass~is 
\beq M_A\gtrsim  1900 {\rm ~GeV}.\eeq

We also~consider~four types of flavour
constraints,~in~which~SUSY has non-negligible impact, namely the~flavour~observables $\br(b \to s \ga)$, $\br(B_s \to \mu^+ \mu^-)$, $\br(B_u \to \tau \nu)$ and $\Delta M_{B_s}$.
Although we~do~not use the latest experimental values, no major effect would~be~expected. 
\begin{itemize}
\item
For~the~branching~ratio~$\br(b \to s \gamma)$ we~take a~value
from~the~Heavy~Flavor~Averaging~Group (HFAG) \cite{bsgth,HFAG}:
\beq
\frac{\br(b \to s \gamma )^{\rm exp}}{\br(b \to s \gamma )^{\rm SM}} = 1.089 \pm 0.27~.
\label{bsgaexp}
\eeq

\item
For~the~branching~ratio~$\br(B_s \to \mu^+ \mu^-)$ we~use a~combination~of~CMS~and~LHCb data \cite{Bobeth:2013uxa,RmmMFV,Aaij:2012nna,Chatrchyan:2013bka,BsmmComb}:
\beq
\br(B_s \to \mu^+ \mu^-) = (2.9\pm1.4) \times 10^{-9}~.
\eeq

\item
For~the $B_u$~decay~to~$\tau\nu$ we~use the~limit~\cite{SuFla,HFAG,PDG14}:
\beq
\frac{\br(B_u\to\tau\nu)^{\rm exp}}{\br(B_u\to\tau\nu)^{\rm SM}}=1.39\pm 0.69~.
\eeq

\item
For~$\Delta M_{B_s}$~we~use~\cite{Buras:2000qz,Aaij:2013mpa}:
\beq
\frac{\Delta M_{B_s}^{\rm exp}}{\Delta M_{B_s}^{\rm SM}}=0.97\pm 0.2~.
\eeq
\end{itemize}

We finally~consider~Cold Dark Matter (CDM) constraints.
Since the~lightest~neutralino, being the Lightest
Supersymmetric~Particle (LSP),~is a~very promising candidate for CDM~\cite{EHNOS},~we demand that our LSP is indeed the lightest neutralino~and~we discard parameters~leading to different LSPs.~The~current bound on the CDM relic density at
$2\,\sigma$~level~is~given~by~\cite{Komatsu:2010fb,Komatsu:2014ioa}
\beq
\Omega_{\rm CDM} h^2 = 0.1120 \pm 0.0112~.
\label{cdmexp}
\eeq
For the~calculation~of the relic density of each model we use the \MO~code~\cite{Belanger:2001fz,Belanger:2004yn,Barducci:2016pcb})~(Version 5.0).~The calculation of annihilation and  coannihilation~channels~is also included. It should be noted that other CDM~constraints~do not affect our models significantly,~and thus~were~not included~in the analysis.

\section{Computational setup}\label{se:setup}

\begin{figure}
\centering
\includegraphics[width=0.8\textwidth]{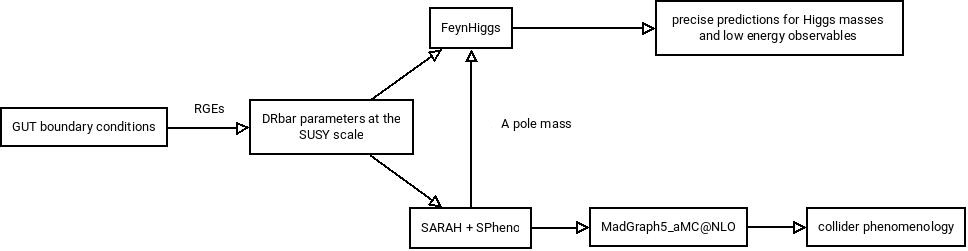}
\caption{\textit{Flow of~information between used computer codes (see text for~details).}\label{fig:flow}}
\end{figure}
The setup for our~phenomenological analysis is as follows.
Starting from an~appropriate set of MSSM boundary conditions at the GUT scale,~parameters are run down to the SUSY scale using a private code.~Two-loop RGEs are used throughout, with the exception of the~soft sector, in which one-loop RGEs are used.~The running parameters are then used as inputs for both {\tt FeynHiggs}~\cite{Degrassi:2002fi,BHHW,FeynHiggs,Bahl:2019hmm} and a
\texttt{SARAH}~\cite{Staub:2013tta} generated, custom MSSM module for~\texttt{SPheno}~\cite{Porod:2003um,Porod:2011nf}. It should be noted~that {\tt FeynHiggs} requires the $m_b(m_b)$ scale,~the \textit{physical} top quark
mass $m_t$ as well~as {\it the physical} pseudoscalar boson mass $\MA$~as input.  The first two  values are calculated by the private code~while $\MA$ is calculated only in \DRbar~scheme.~This single value is obtained from the \texttt{SPheno}~output where it
is calculated at~the two-loop level in the gaugeless limit
\cite{Gabelmann:2018axh,Goodsell:2015ira}.  The flow of information~between codes in our analysis is summarised in \reffi{fig:flow}. 

At this point~both codes contain  a consistent set of all required parameters.~SM-like Higgs boson mass as well as low energy observables~mentioned in Sec.~\ref{se:constraints} are evaluated using {\tt FeynHiggs}.~To obtain collider predictions we use \texttt{SARAH}~to generate \texttt{UFO}~\cite{Degrande:2011ua,Staub:2012pb} model for \texttt{MadGraph} event generator.~Based on SLHA spectrum files generated by \texttt{SPheno},~we use \texttt{MadGraph5\_aMC@NLO}~\cite{Alwall:2014hca} to~calculate cross sections for Higgs boson and SUSY particle~production at the HL-LHC and a 100 TeV FCC-hh.~Processes are generated at the leading order, using~\texttt{NNPDF31\_lo\_as\_0130} \cite{Ball:2017nwa}~structure functions interfaced through \texttt{LHAPDF6} \cite{Buckley:2014ana}.
Cross sections~are computed using dynamic scale choice, where the scale is set~equal to the transverse mass of an event, in 4~or 5-flavour scheme~depending on the presence or not of $b$-quarks in~the final state. The results are 
given in Sec. \ref{sec:minimalsu5}, \ref{sec:finitesu5} and \ref{sec:su33}.

\section{The Minimal \texorpdfstring{$N=1$}~ \texorpdfstring{$SU(5)$}~ }\label{se:minimal}

Let us begin by analysing the particle~spectrum predicted by the Minimal $N=1$ SUSY $SU(5)$ as~discussed~in~\refse{sec:minimalsu5}
for $\mu < 0$ as the only phenomenologically acceptable choice
(in the $\mu> 0$ case the quark masses do not match the experimental measurements).~Below $M_{\rm GUT}$ all couplings and~masses of the theory~run according~to the RGEs of the MSSM.  Thus we examine~the evolution of these~parameters~according to their RGEs up~to two loops for~dimensionless~parameters and at one-loop~for~dimensionful ones imposing the~corresponding boundary~conditions.

\begin{figure}[H]
\centering
\includegraphics[width=0.495\textwidth]{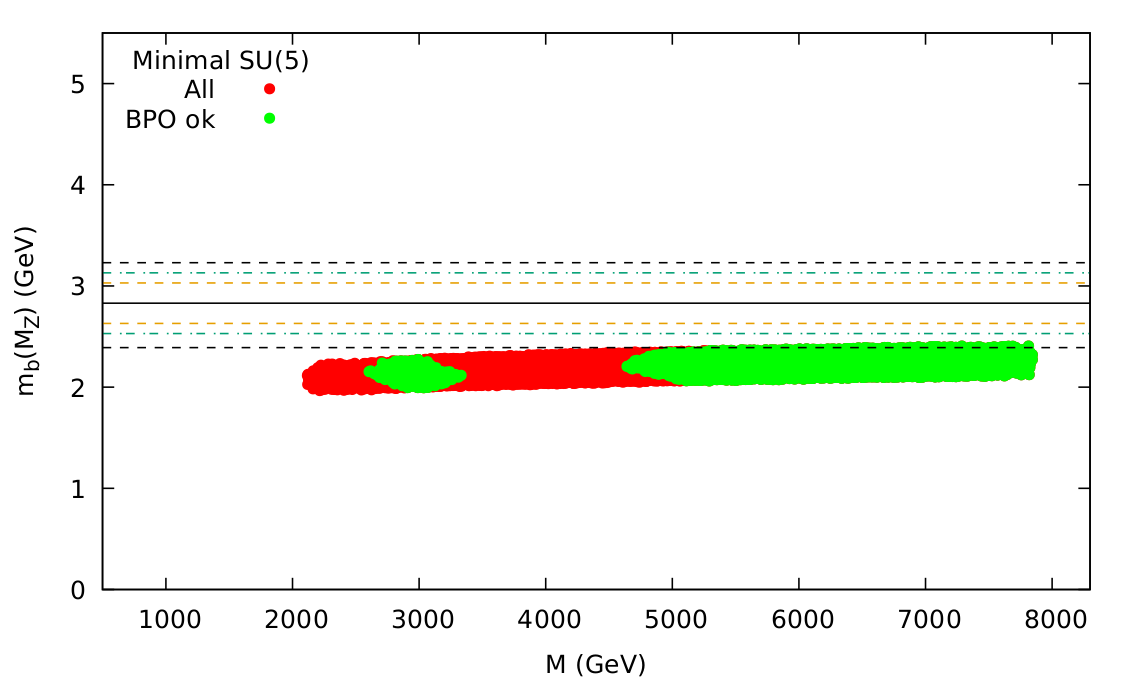}
\includegraphics[width=0.495\textwidth]{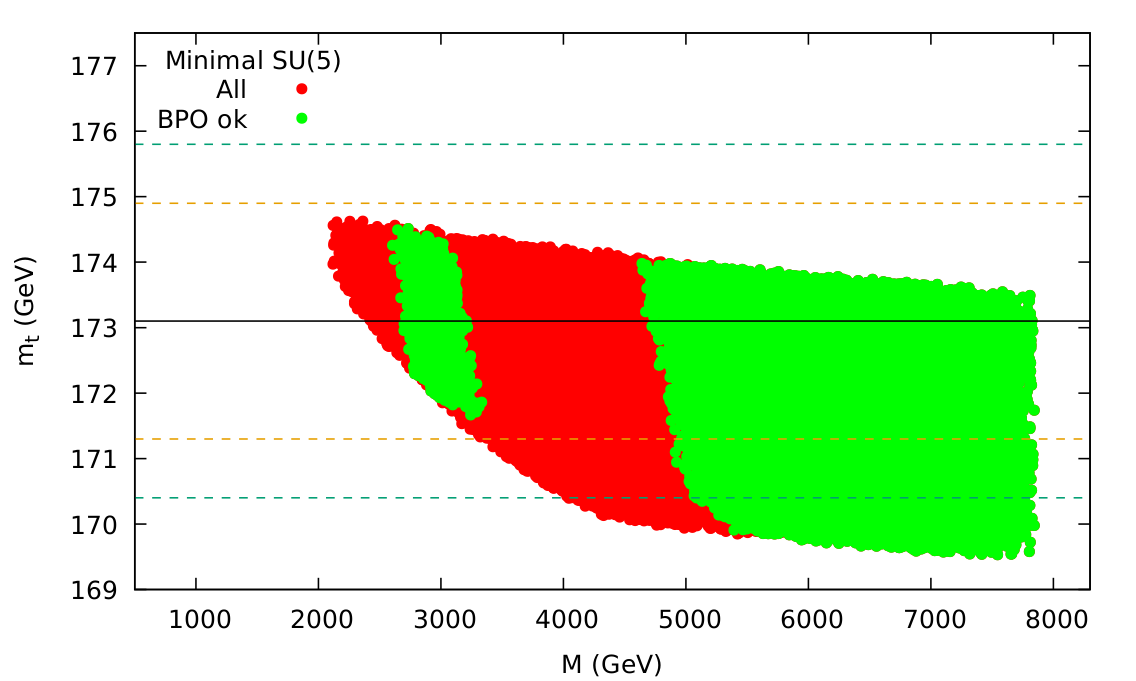}
\caption{\textit{The bottom quark~mass at the $Z$~boson scale (left)
and top quark pole mass~(right) are shown
as a function of $M$ for~the Minimal $N=1$ $SU(5)$. The green points are~the ones that satisfy the B-physics constraints.
The orange~(blue) dashed lines denote the 2$\sigma$ (3$\sigma$) experimental~uncertainties, while the black dashed lines in the left plot~add a $\sim 6 \mev$ theory uncertainty to that.}}
\label{fig:mintopbotvsM}
\end{figure}

In \reffi{fig:mintopbotvsM}, we~show~the predictions for $\mb (M_Z)$ and $\mt$ as a function~of the~unified gaugino mass $M$. The green points include~the $B$-physics~constraints. The $\Delta M_{B_s}$ channel~is responsible for the gap at
the $B$-physics~allowed points.  One can see that, once more, the model~(mostly) prefers the higher energy region of the spectrum (especially~with the admission of $B$-physics constraints).~The orange (blue) lines denote the 2$\sigma$ (3$\sigma$) experimental~uncertainties, while the black dashed lines in the left plot add a~$\sim 6 \mev$ theory~uncertainty to that. The uncertainty~for the boundary conditions of the
Yukawa couplings is~taken to be $7\%$, which is included in the spread~of the points shown. ~In the~evaluation of the bottom mass we~have included the~corrections~coming from bottom squark-gluino loops and~top~squark-chargino loops~\cite{Carena:1999py}. ~One can see in the left plot of \reffi{fig:mintopbotvsM} that only by~taking all uncertainties to their limit, some points at very high~$M$~are within these bounds. I.e.\ confronting the Minimal $N=1$ SUSY~$SU(5)$ with the quark mass~measurements ``nearly'' excludes this model,
and only a very~heavy spectrum might be in agreement with the
experimental~data.

\begin{figure}[H]
\centering
\includegraphics[width=0.495\textwidth]{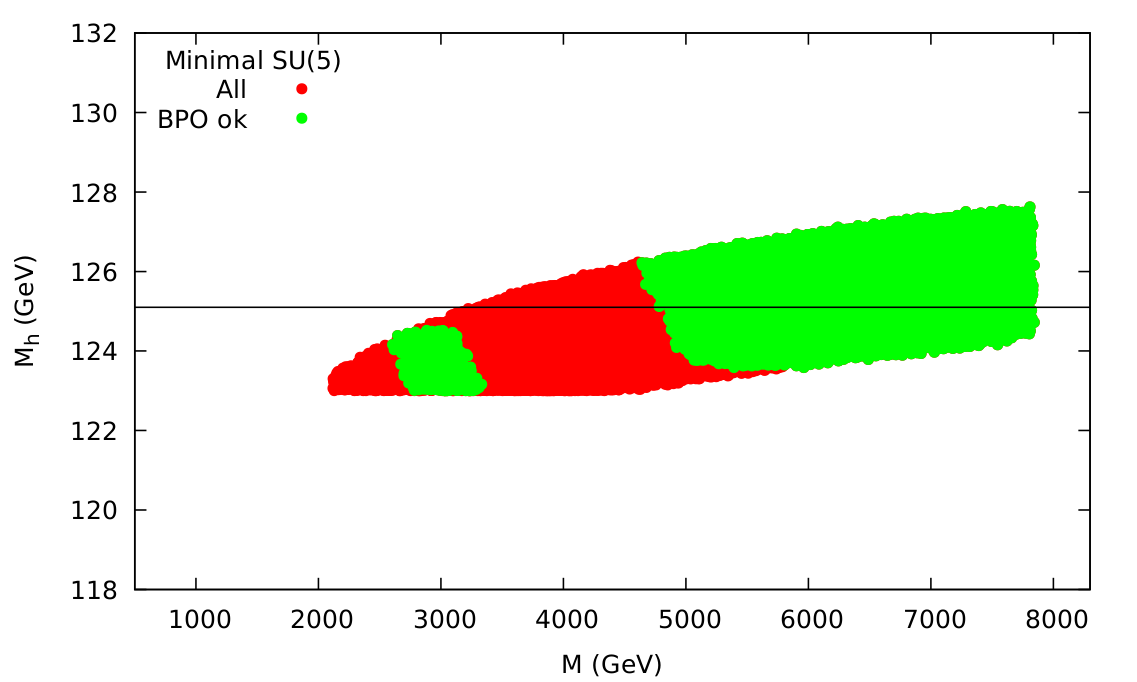}
\includegraphics[width=0.495\textwidth]{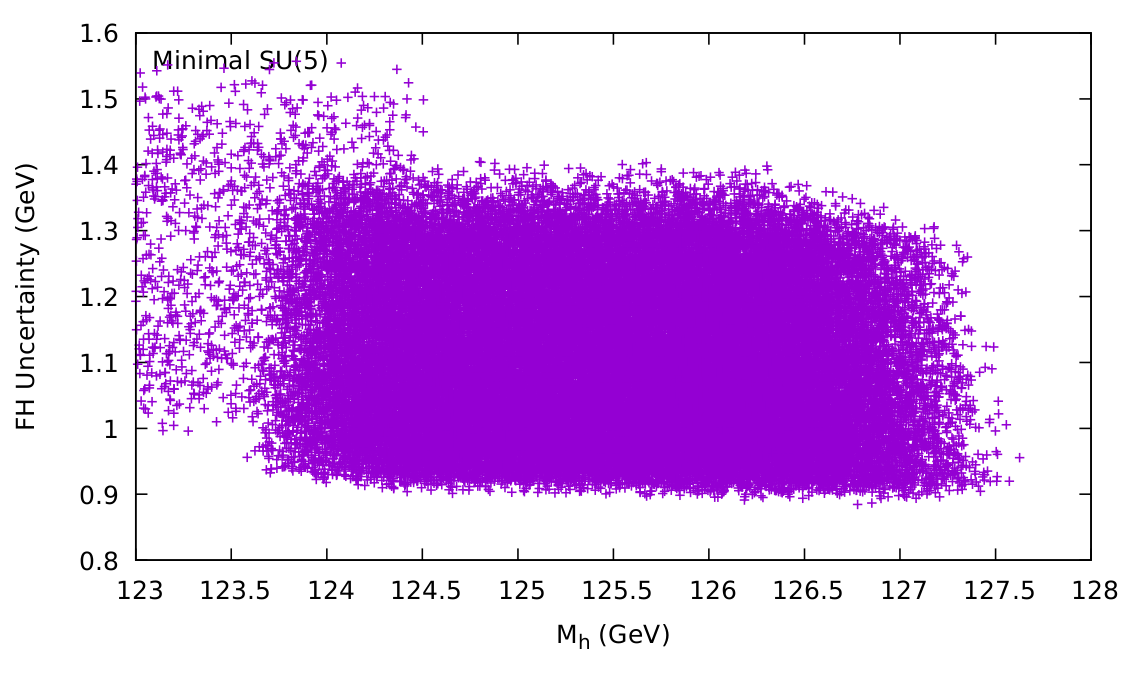}
\caption{\textit{Left: The lightest~Higgs mass, $M_h$, as~a function of $M$ for the~Minimal $N=1$ $SU(5)$ model. The $B$-physics constraints~allow (mostly) higher scale points (with green colour).~Right: The lightest Higgs mass theoretical uncertainty \cite{Bahl:2019hmm}.}}
\label{fig:minhiggsvsM}
\end{figure}

The prediction for $M_h$~with $\mu < 0$ is given in
\reffi{fig:minhiggsvsM}~(left), for a~unified gaugino mass between~$2 \tev$ and $8 \tev$, where~again the green points satisfy $B$-physics~constraints.~\reffi{fig:minhiggsvsM} (right) gives the theoretical~uncertainty of the Higgs mass for each point, calculated with~{\tt FeynHiggs} 2.16.0~\cite{Bahl:2019hmm}.~There is substantial
improvement to the Higgs~mass uncertainty compared to past analyses,~since it has dropped by more than $1 \gev$. 
 
Large parts of the predicted particle spectrum
are in agreement~with~the
$B$-physics~observables~and the lightest Higgs boson~mass
measurement~and its theoretical uncertainty.
In order to test the models discovery potential, three benchmarks are selected, marking the points~with the lightest SUSY particle (LSP)~above $1200$~GeV (MINI-1),~$1500$~GeV
(MINI-2) and $2200$~GeV (MINI-3),~respectively.
The mass of the LSP can go as high~as $\sim 3800 \gev$, but the cross sections calculated below~will then be negligible and we restrict ourselves here to~the low-mass region.
The values presented in \refta{tab:mininput} were used as input to~get the full supersymmetric spectrum from {\tt
  SPheno\,4.0.4}~\cite{Porod:2003um,Porod:2011nf}. $M_i$ are the~gaugino~masses and the rest are~squared soft  sfermion masses which are diagonal~($\mathbf{m^2}=\rm diag(m_1^2,m_2^2,m_3^2)$), and  soft~trilinear
couplings (also diagonal~$\mathbf{A_i}=\mathbb{1}_{3\times3}A_i$). 

\small
\begin{table}[htb!]
\renewcommand{\arraystretch}{1.3}
\centering\small
\begin{tabular}{|c|rrrrrrrrrr|}
\hline
 & $M_1$ & $M_2$ & $M_3$ & $|\mu|$ & $b~~~$ & $A_u$ & $A_d$ & $A_e$ & $\tan{\beta}$ & $m_{Q_{1,2}}^2$  \\
\hline
MINI-1  & 1227 & 2228 & 5310 & 4236 & $401^2$ & 4325 & 4772 & 1732 & 50.3 & $6171^2$  \\
MINI-2  & 1507 & 2721 & 6376 & 5091 & $496^2$  & 5245 & 5586 & 2005 & 52.0 & $7445^2$ \\
MINI-3  & 2249 & 4019 & 9138 & 7367 & $1246^2$  & 7571 & 8317 & 3271 & 50.3 & $10762^2$ \\
\hline
& $m_{Q_{3}}^2$ &  $m_{L_{1,2}}^2$ & $m_{L_{3}}^2$  & $m_{\overline{u}_{1,2}}^2$ & $m_{\overline{u}_{3}}^2$ & $m_{\overline{d}_{1,2}}^2$ & $m_{\overline{d}_{3}}^2$ & $m_{\overline{e}_{1,2}}^2$  & $m_{\overline{e}_{3}}^2$ & \\
\hline
MINI-1 & $4548^2$ & $3714^2$ & $2767^2$ & $5974^2$ & $4181^2$ & $5478^2$ & $4177^2$ & $4160^2$  & $2491^2$   & \\
MINI-2 & $5469^2$ & $4521^2$ & $3358^2$ & $7206^2$ & $5039^2$ & $5478^2$ & $4994^2$ & $5070^2$ & $3019^2$   & \\
MINI-3 & $7890^2$ & $6639^2$ & $4934^2$ & $10412^2$ & $7233^2$ & $9495^2$ & $7211^2$ & $7459^2$ & $4464^2$   & \\
\hline
\end{tabular}
\caption{\textit{
Minimal $N=1$ $SU(5)$ predictions~that are used as input to {\tt SPheno}.
Mass parameters are in~$\gev$ and rounded to $1 \gev$.}}
\label{tab:mininput}
\renewcommand{\arraystretch}{1.0}
\end{table}
\normalsize

The resulting~masses of all~the particles that will be relevant for our~analysis can~be found in \refta{tab:minispheno}.~The three first values
are the heavy Higgs masses.~The~gluino mass is $M_{\tilde{g}}$, the~neutralinos and the~charginos are denoted as $M_{\tilde{\chi}_i^0}$ and~$M_{\tilde{\chi}_i^{\pm}}$, while the slepton and~sneutrino masses
for all three generations~are given as
$M_{\tilde{e}_{1,2,3}},~M_{\tilde{\nu}_{1,2,3}}$. Similarly, the squarks~are denoted as $M_{\tilde{d}_{1,2}}$ and $M_{\tilde{u}_{1,2}}$ for the
first two~generations. The third generation masses are given by~$M_{\tilde{t}_{1,2}}$ for stops and $M_{\tilde{b}_{1,2}}$ for sbottoms.

\begin{center}
\begin{table}[ht]
\begin{center}
\small
\begin{tabular}{|l|r|r|r|r|r|r|r|r|r|r|r|}
\hline
 & $M_{H}$ & $M_A$ & $M_{H^{\pm}}$ & $M_{\tilde{g}}$ & $M_{\tilde{\chi}^0_1}$ & $M_{\tilde{\chi}^0_2}$ & $M_{\tilde{\chi}^0_3}$  & $M_{\tilde{\chi}^0_4}$ &  $M_{\tilde{\chi}_1^\pm}$ & $M_{\tilde{\chi}_2^\pm}$ \\\hline
MINI-1 & 2.660 & 2.660 & 2.637 & 5.596 & 1.221 & 2.316 & 4.224 & 4.225 & 2.316 & 4.225  \\\hline
MINI-2 & 3.329 & 3.329 & 3.300 & 6.717 & 1.500 & 2.827 & 5.076 & 5.077 & 2.827 & 5.078  \\\hline
MINI-3 & 8.656 & 8.656 & 8.631 & 9.618 & 2.239 & 4.176 & 7.357 & 7.358 & 4.176 & 7.359  \\\hline
 & $M_{\tilde{e}_{1,2}}$ & $M_{\tilde{\nu}_{1,2}}$ & $M_{\tilde{\tau}}$ & $M_{\tilde{\nu}_{\tau}}$ & $M_{\tilde{d}_{1,2}}$ & $M_{\tilde{u}_{1,2}}$ & $M_{\tilde{b}_{1}}$ & $M_{\tilde{b}_{2}}$ & $M_{\tilde{t}_{1}}$ & $M_{\tilde{t}_{2}}$ \\\hline
MINI-1 & 3.729 & 3.728 & 2.445 & 2.766 & 5.617 & 6.100 & 4.332 & 4.698 & 4.312 & 4.704 \\\hline
MINI-2 & 4.539 & 4.538 & 2.968 & 3.356 & 6.759 & 7.354 & 5.180 & 5.647 & 5.197 & 5.652 \\\hline
MINI-3 & 6.666 & 6.665 & 4.408 & 4.935 & 9.722 & 10.616 & 7.471 & 8.148 & 7.477 & 8.151 \\\hline
\end{tabular}
\caption{\textit{Masses of~Higgs bosons and some of the SUSY particles for each benchmark~of the Minimal $N=1$ $SU(5)$ (in TeV).}}\label{tab:minispheno}
\end{center}
\end{table}
\end{center}

Table \ref{minSU5xsec} shows~the expected production cross section for~selected channels at the  100 TeV future FCC-hh collider.~We do not show any cross sections for $\sqrt{s} = 14$~TeV, since the~prospects for discovery of MINI scenarios at the HL-LHC are very~dim. SUSY particles are too heavy to be produced with cross sections~greater that 0.01\,fb.
Concerning the heavy Higgs~bosons, the main
search channels~will be $H/A \to \tau^+\tau^-$. Our heavy Higgs-boson~mass scale shows values $\gsim 2500$~GeV with $\tb \sim 50$. The~corresponding reach of the HL-LHC has been estimated~in~\cite{HAtautau-HL-LHC}. In comparison with our benchmark points we~conclude that they will not be accessible at the HL-LHC.\footnote{The analysis presented in~\cite{HAtautau-HL-LHC} only
  reaches $\MA \le 2000$~GeV, where an~exclusion down to
  $\tb \sim 30$ is expected. An~extrapolation to $\tb \sim 50$ reaches~  Higgs-boson mass scales of $\sim 2500$~GeV.}

The situation changes~for the FCC-hh. Theory
analyses~\cite{HAtautau-FCC-hh,Hp-FCC-hh} have shown that for large
$\tb$ heavy Higgs-boson~mass scales up to $\sim 8$~TeV may be
accessible, both for~neutral as well as for charged Higgs bosons. The
relevant decay channels~are $H/A \to \tau^+\tau^-$ and
$H^\pm \to \tau\nu_\tau, tb$.~This places our three benchmark points well~within the covered region (MINI-1 and MINI-2) or at the border~of the
parameter space that~can be probed (MINI-3).

The energy of 100 TeV is~big enough to produce SUSY particles in
pairs. However, the cross~sections remain relatively small. Only for the
MINI-1 scenario the squark pair and~squark-gluino  (summed over all~squarks) production cross sections can reach tens of fb. For~MINI-2 and
MINI-3~scenarios the cross sections are significantly smaller.   In~these scenarios squarks decay preferentially into a quark+LSP (with $\rm
BR\sim 0.95$), gluino~into $\tilde t\bar t$ and $\tilde b\bar b$ +$h.c$~with $\rm BR\sim 0.33$ each.   

The SUSY~discovery reach at the FCC-hh with $3\,\iab$ was evaluated~in~\cite{SUSY-FCC-hh} for a certain set of simplified~models. In the
following we~will compare these simplified model limits with our~benchmark points to get an idea, which part of the spectrum can be~covered at the FCC-hh. A more detailed evaluation with~the future limits
implemented into~proper recasting tools would be necessary to obtain a~firmer statement. However, such a detailed analysis goes beyond the~scope of our paper and we restrict ourselves to the simpler~direct
comparison of~the simplified model limits with our benchmark
predictions.

\begin{center}
\begin{table}[ht]
\begin{center}
\small
\begin{tabular}{|c|c|c|c||c|c|c|c|} \hline
scenarios & MINI-1  & MINI-2 & MINI-3  &  scenarios  &  MINI-1 & MINI-2 & MINI-3\\
$\sqrt{s}$ & 100 TeV & 100 TeV &  100 TeV  &  $\sqrt{s}$  & 100 TeV & 100 TeV &  100 TeV\\ \hline
$\tilde{\chi}^0_1 \tilde{\chi}^0_1 $  &   0.04 & 0.02 & &$\tilde{u}_i \tilde{\chi}^-_1, \tilde{d}_i \tilde{\chi}^+_1 + h.c.$  &   1.00 & 0.35 & 0.03 \\
$\tilde{\chi}^0_1 \tilde{\chi}^0_3 $  &   0.02 & 0.01 & &$\tilde{u}_i \tilde{\chi}^-_2, \tilde{d}_i \tilde{\chi}^+_2 + h.c.$  &   0.07 & 0.02 &     \\
$\tilde{\chi}^0_2 \tilde{\chi}^0_2 $  &   0.06 & 0.02 & &$\tilde{q}_i \tilde{\chi}^0_1, \tilde{q}_i^* \tilde{\chi}^0_1$  &   0.38 & 0.14 & 0.02 \\
$\tilde{\chi}^0_2 \tilde{\chi}^0_3 $  &   0.03 & 0.01 & &$\tilde{q}_i \tilde{\chi}^0_2, \tilde{q}_i^* \tilde{\chi}^0_2$  &   0.51 & 0.17 & 0.02 \\
$\tilde{\chi}^0_2 \tilde{\chi}^0_4 $  &   0.02 & 0.01 & &$\tilde{\nu}_i \tilde{e}_j^*, \tilde{\nu}_i^* \tilde{e}_j$  &   0.06 & 0.02 &     \\
$\tilde{\chi}^0_3 \tilde{\chi}^0_4 $  &   0.05 & 0.02 & & $H  b  \bar{b} $ & 84.04 & 30.10 & 0.17 \\
$\tilde{\chi}^0_2 \tilde{\chi}_1^+ $  &   2.20 & 0.98 & 0.18 & $A b  \bar{b} $ & 84.79 & 29.79 & 0.18\\
$\tilde{\chi}^0_3 \tilde{\chi}_2^+ $  &   0.10 & 0.04 & 0.01 & $H^+ b  \bar{t} + H^- t  \bar{b}$ & 33.24 & 12.76 & 0.1 \\
$\tilde{\chi}^0_4 \tilde{\chi}_2^+ $  &   0.10 & 0.04 & 0.01 & $H^- b  \bar{b} $ & 0.04 & 0.02 & \\
$ \tilde{g}  \tilde{g} $  &   7.76 & 2.02 & 0.11 & $H  t  \bar{t} $ & 0.03 & 0.01 & \\
$ \tilde{g} \tilde{\chi}^0_1 $  &   0.28 & 0.11 & 0.01 & $A t  \bar{t} $ & 0.02 & 0.01 & \\
$ \tilde{g} \tilde{\chi}^0_2 $  &   0.34 & 0.12 & 0.01 & $H  t  b $ & 0.01 &  & \\
$ \tilde{g} \tilde{\chi}_1^+ $  &   0.70 & 0.27 & 0.03 &$ H  A$ & 0.03 & 0.01 &    \\
$\tilde{q}_i \tilde{q}_j, \tilde{q}_i \tilde{q}_j^*$  &   21.15 & 7.44 & 0.74 &$ H  H^+$ & 0.06 & 0.02 &    \\
$\tilde{\chi}_1^+ \tilde{\chi}_1^- $  &   1.19 & 0.54 & 0.09 &$H^+W^-$ & 6.50 & 2.96 & 0.03 \\
$\tilde{\chi}_1^+ \tilde{\chi}_2^- $  &   0.05 & 0.02 & &$ H  W^+$ & 0.02 & 0.01 &    \\
$\tilde{\chi}_2^+ \tilde{\chi}_1^- $  &   0.05 & 0.02 & &$H^+H^-$ & 0.04 & 0.01 &    \\
$\tilde{\chi}_2^+ \tilde{\chi}_2^- $  &   0.06 & 0.02 & &$AH^+$ & 0.06 & 0.02 &    \\
$\tilde{e}_i \tilde{e}_j^*$  &   0.16 & 0.08 & 0.01 &$AW^+$ & 0.02 & 0.01 &    \\
$\tilde{q}_i \tilde{g}, \tilde{q}_i^* \tilde{g}$  &   30.57 & 9.33 & 0.66 &$ H  Z$ & 1.38 & 0.58 & 0.01 \\
$\tilde{\nu}_i \tilde{\nu}_j^*$  &   0.04 & 0.02 & &$AZ$ & 1.20 & 0.52 & 0.01 \\
\hline
\end{tabular}
\caption{\textit{Expected production cross sections (in fb) for SUSY particles in the~MINI scenarios.  There are no channels with cross sections exceeding  0.01 fb at $\sqrt{s}=14$ TeV.}
}
\label{minSU5xsec}
\end{center}
\end{table}
\end{center}

Concerning the scalar tops,~the mass predictions of MINI-1 and MINI-2~are well within the anticipated reach of the FCC-hh, while~MINI-3
predicts~a too heavy stop mass. On the other hand, even for MINI-1 and~MINI-2 no $5\sigma$ discovery can be expected.
The situation~looks more favorable for the first and second
generation~squarks. All the predicted masses can be excluded at the~FCC-hh, whereas a $5\sigma$ discovery will be difficult,~but potentially
possible~(see Fig.~19 in \cite{SUSY-FCC-hh}).
Even more~favorable appear the prospects for gluino searches at the~FCC-hh. All three benchmark points may lead to a $5\sigma$~discovery
(see Fig.~13 in \cite{SUSY-FCC-hh}). On the other hand, chances for~chargino/neutralino searches are slim at the FCC-hh. The~Next-to LSP
(NLSP) can~only be accessed for $\mneu1 \lsim 1$~TeV (see Fig.~21 in~\cite{SUSY-FCC-hh}), where all our benchmark points have~$\mneu1 > 1$~TeV.
Taking~into account that our three benchmark points represent only~the lower part of the possible mass spectrum (with LSP masses of up to~$\sim 1.5 \tev$ higher), we conclude that even at the FCC-hh~large parts
of the possible SUSY spectrum will remain elusive.

\begin{figure}[htb!]
\centering
\includegraphics[width=0.5\textwidth]{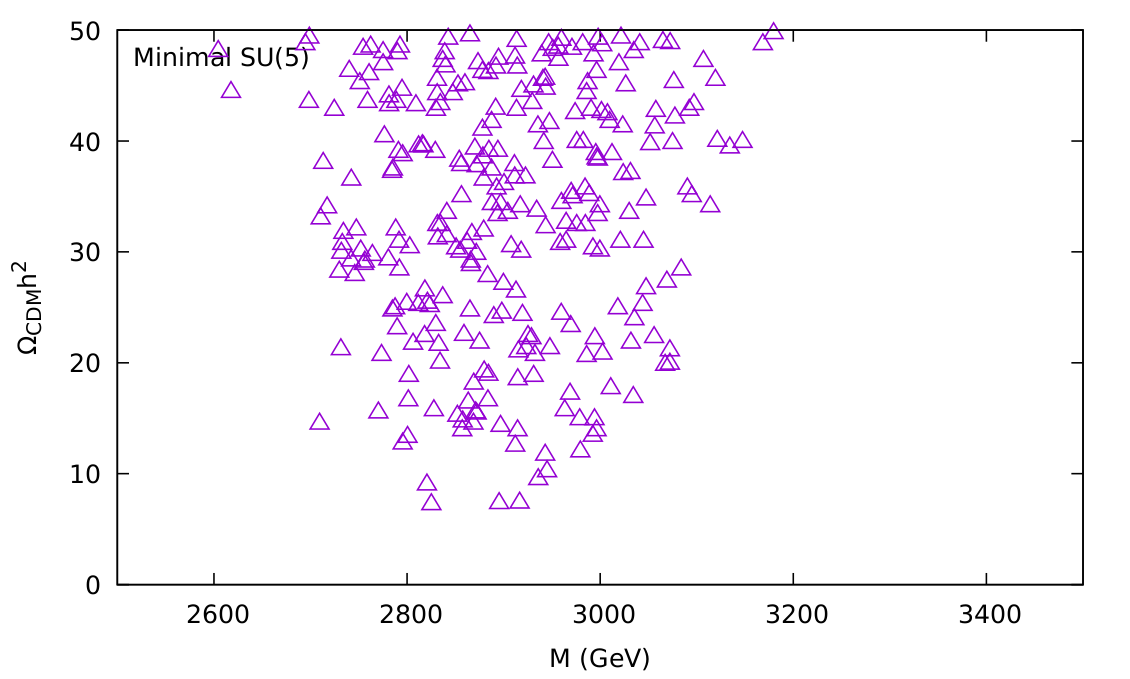}
\caption{\textit{The plot shows~the CDM relic density of
    the Minimal $N=1$ $SU(5)$~model for points with Higgs mass within its calculated~uncertainty. Only the points with the lowest relic density are displayed.}}
\label{fig:mincdm}
\end{figure}

No point~fulfills~the
strict bound of~\refeq{cdmexp}, since the relic~abundance turns out to be too~high, as it can be seen in \reffi{fig:mincdm}. The model features a Bino-like neutralino,  which proves too heavy to give an acceptable relic density. It is $\lesssim 5\%$ Higgsino-like and the Wino contribution is negligible.~Thus, our model needs a mechanism that can
reduce the CDM~abundance~in the early~universe. This~issue could~be related~to~the~problem of neutrino~masses.
These~masses cannot be~generated~naturally in this particular model,~although~a non-zero value for~neutrino masses has been
established~\cite{PDG14}. However,~the model
could be, in~principle, extended by~introducing~bilinear R-parity~violating~terms and introduce~neutrino masses \cite{Valle:1998bs,Valle3}.~R-parity violation \cite{herbi}
would remove~the CDM bound of~\refeq{cdmexp} completely.  Other mechanisms, not~involving R-parity~violation,~that~could be invoked if the~amount of~CDM appears to be~too~large, concern~the cosmology of~the early~universe. For~example, ``thermal inflation'' \cite{thermalinf} or ``late time~entropy
injection''~\cite{latetimeentropy} can~bring the~CDM density~into~agreement~with WMAP~measurements.

\newpage

\section{The Finite \texorpdfstring{$N=1$}~ \texorpdfstring{$SU(5)$}~ }\label{se:futb}

This section contains the full particle spectrum predicted in the~Finite $N=1$  SUSY $SU(5)$ model, as discussed in \refse{sec:finitesu5}. The~gauge symmetry breaks spontaneously below the~GUT scale, so~conditions set by finiteness do~not restrict the ~renormalization~properties at low~energies. We are left with boundary conditions~on the~gauge and Yukawa~couplings
(\ref{zoup-SOL52}),~the~$h=-MC$ 
relation~and the soft~scalar-mass~sum rule at $M_{\rm GUT}$. Again, the uncertainty for the~boundary conditions of the Yukawa couplings is at $7\%$, which again is~included in the spread of the points. 

\begin{figure}[H]
\centering
\includegraphics[width=0.495\textwidth]{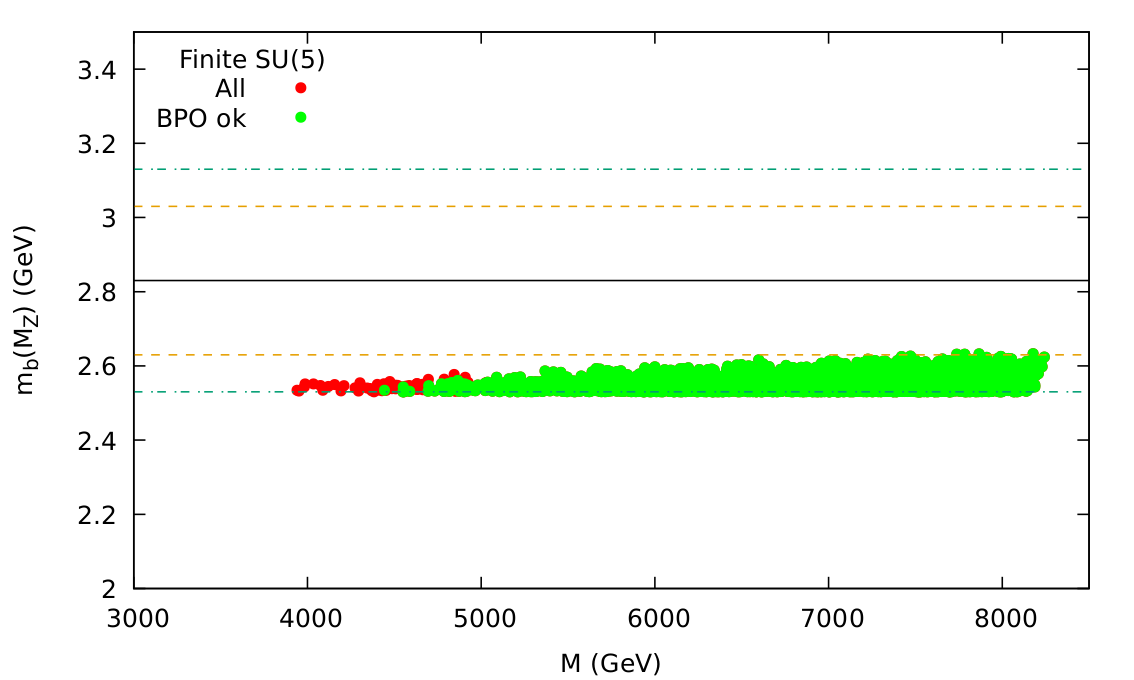}
\includegraphics[width=0.495\textwidth]{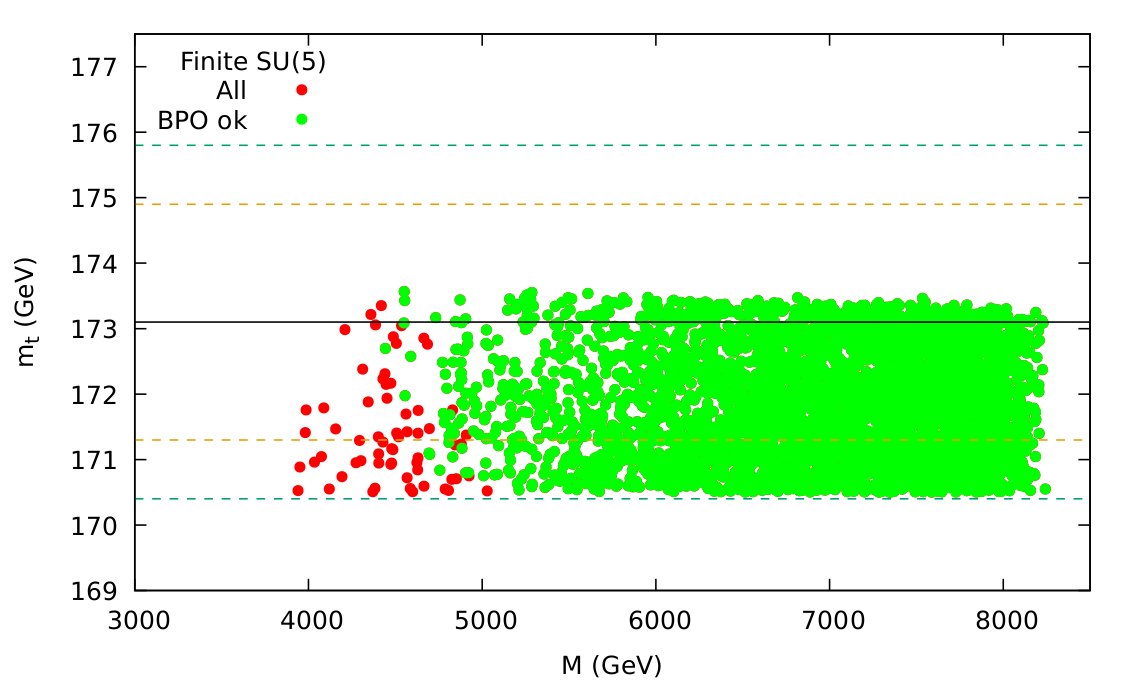}
\caption{\textit{$\mb (M_Z)$ ({left})~and $\mt$ ({right}) as a function of $M$ for the Finite $N=1$ $SU(5)$,~with the colour coding as in \protect\reffi{fig:mintopbotvsM}.}}
\label{fig:futtopbotvsM}
\end{figure}

In \reffi{fig:futtopbotvsM}, $\mb (M_Z)$ and $\mt$ are~shown
as~functions of~the unified~gaugino mass $M$, where the green points~satisfy the B-physics constraints with the same colour coding as in~\reffi{fig:mintopbotvsM}. Here we omitted the additional theoretical~uncertainty of $\sim 6 \mev$. The only phenomenologically viable~option is to consider $\mu < 0$, as is shown in earlier~work \cite{Heinemeyer:2012yj,Heinemeyer:2012ai,Heinemeyer:2012sy}. The~experimental~values are indicated by the~horizontal lines with~the~uncertainties at the $2\,\sigma$ and  $3\,\sigma$ level. The value of~the bottom mass is lower than in past analyses, sending the allowed energy~scale higher. Also the top-quark mass turns out slightly lower than in~previous analyses. 

\begin{figure}[htb!]
\centering
\includegraphics[width=0.495\textwidth]{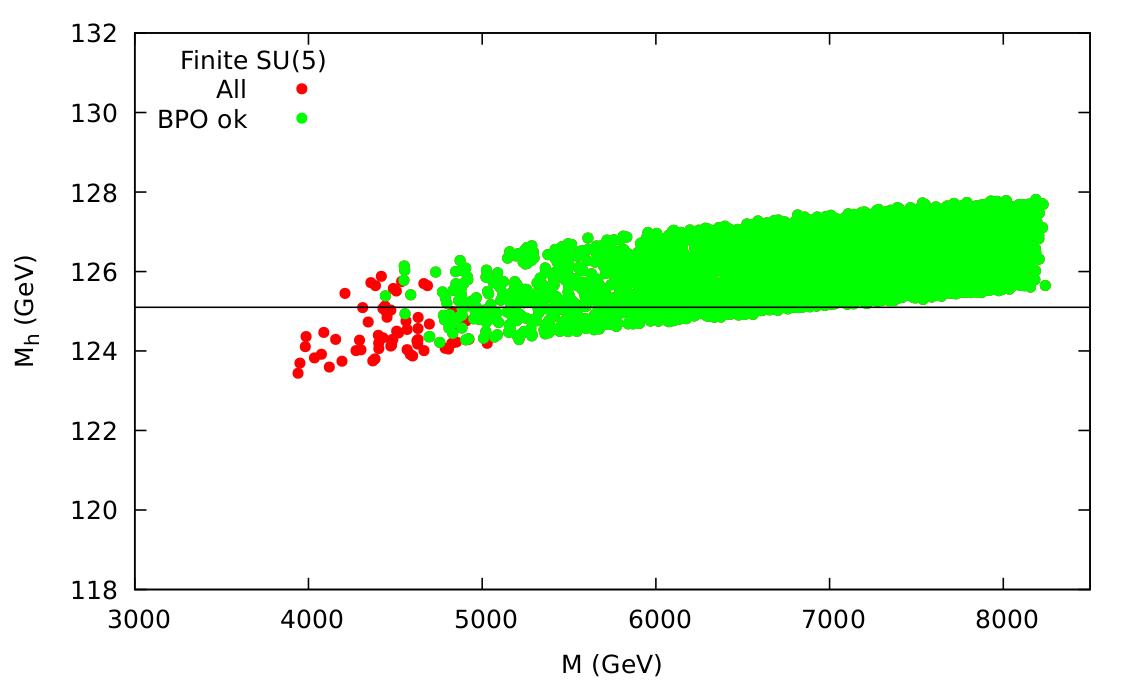}
\includegraphics[width=0.495\textwidth]{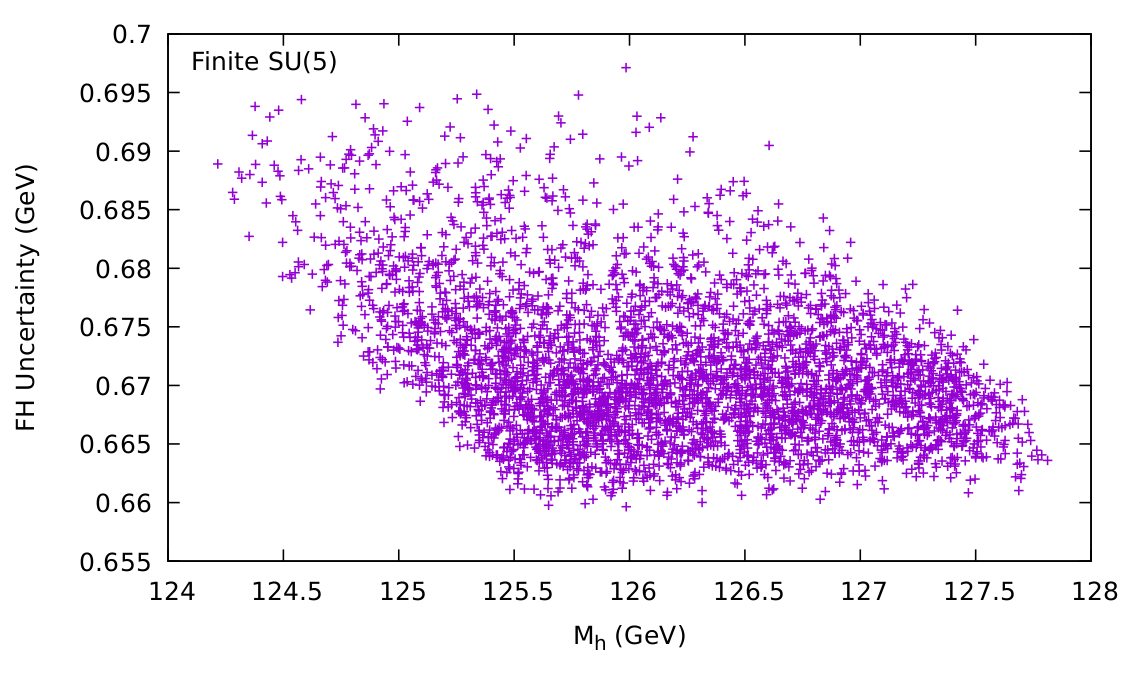}
\caption{\textit{Left: $M_h$ as~a~function of $M$. Green points comply with $B$-physics~constraints.~Right: The lightest Higgs mass theoretical uncertainty~calculated with {\tt FeynHiggs} 2.16.0 \cite{Bahl:2019hmm}.}}
\label{fig:futhiggsvsM}
\end{figure}

The light Higgs~boson mass~is given in \reffi{fig:futhiggsvsM} (left) as~a function~of the unified gaugino mass. Like in the previous section,~these predictions are subject to a theory~uncertainty~\cite{Bahl:2019hmm} that is given in \reffi{fig:futhiggsvsM}~(right). This point-by-point uncertainty (calculated with~{\tt FeynHiggs}) drops significantly from the flat estimate of 2 and 3~GeV~of past analyses to the much improved $0.65-0.70 \gev$.~The $B$-physics
constraints (green points)~and the smaller Higgs uncertainty drive the~energy scale above $\sim4.5 \tev$.~Older analyses,~including~in particular less~refined evaluations of the~light~Higgs mass, are given in
\citeres{Heinemeyer:2012yj,Heinemeyer:2012ai,Heinemeyer:2012sy}.~It should be noted that, w.r.t.\ previous analyses the top-quark mass~turns out to be slightly lower. Consequently, higher scalar top masses~have to be reached in order to yield the Higgs-boson mass around its~central value of \refeq{higgsexpval}, resulting~in a correspondingly
heavier spectrum. Compared to our previous~analyses 
\cite{Heinemeyer:2010xt,Heinemeyer:2012yj,Heinemeyer:2013nza,Heinemeyer:2012ai,Heinemeyer:2012sy,Heinemeyer:2018roq},
the improved~evaluation of $\Mh$ and its uncertainty, together with a~lower prediction of the top-quark mass prefers a heavier Higgs and SUSY spectrum.~In particular, very heavy coloured SUSY particles are favoured (nearly~independent of the $\Mh$ uncertainty), in agreement with~the non-observation of those~particles~at the LHC~\cite{2018:59}. 

For the purpose of testing~the discovery potential we choose three~benchmarks, each~featuring the LSP above
$2100$~GeV, $2400$~GeV~and $2900$~GeV respectively.  Again, they are chosen from the~low-mass region.  Although  
 the LSP can be as heavy~as $\sim 4000 \gev$, but in such cases the production~cross sections even at the FCC-hh would be too small.~The input and output~of {\tt SPheno} 4.0.4
\cite{Porod:2003um,Porod:2011nf} can be found in
\refta{tab:futbinput}~and \refta{tab:futbspheno} (with the notation as~in \refse{sec:minimalsu5}).

\small
\begin{table}[htb!]
\renewcommand{\arraystretch}{1.3}
\centering\small
\begin{tabular}{|c|rrrrrrrrrr|}
\hline
& $M_1$ & $M_2$ & $M_3$ & $|\mu|$ &
  $b~~~$ & $A_u$ & $A_d$ & $A_e$ & $\tan{\beta}$ & $m_{Q_{1,2}}^2$ \\
\hline
FUTSU5-1  & 2124 & 3815 & 8804 & 4825 & $854^2$  & 7282 & 7710 & 2961 & 49.9 & $8112^2$ \\
FUTSU5-2  & 2501 & 4473 & 10198 & 5508 & $1048^2$ & 8493 & 9023 & 3536 & 50.1 & $9387^2$ \\
FUTSU5-3  & 3000 & 5340 & 11996 & 6673 & $2361^2$ & 10086 & 10562 & 4243 & 49.9 & $11030^2$ \\
\hline
 & $m_{Q_{3}}^2$ &
  $m_{L_{1,2}}^2$ & $m_{L_{3}}^2$ & $m_{\overline{u}_{1,2}}^2$ & $m_{\overline{u}_{3}}^2$ & $m_{\overline{d}_{1,2}}^2$ & $m_{\overline{d}_{3}}^2$ & $m_{\overline{e}_{1,2}}^2$  & $m_{\overline{e}_{3}}^2$ & \\
\hline
FUTSU5-1  & $6634^2$ & $3869^2$ & $3120^2$ & $7684^2$ & $5053^2$ & $7635^2$ & $4177^2$ & $3084^2$  & $2241^2$ &  \\
FUTSU5-2  & $7669^2$ & $4521^2$ & $3747^2$ & $8887^2$ & $6865^2$ & $8826^2$ & $6893^2$ & $3602^2$  & $2551^2$ &  \\
FUTSU5-3  & $9116^2$ & $5355^2$ & $3745^2$ & $10419^2$ & $8170^2$ & $10362^2$ & $7708^2$ & $4329^2$  & $3403^2$ &  \\
\hline
\end{tabular}
\caption{\textit{
Finite $N=1$ $SU(5)$  predictions that are used as input to {\tt SPheno}.~Mass parameters are in~$\gev$ and rounded to $1 \gev$.}}
\label{tab:futbinput}
\renewcommand{\arraystretch}{1.0}
\end{table}
\normalsize

The expected production~cross sections for various final states are~listed in Table \ref{futSU5xsec}. 
At 14 TeV~HL-LHC none of the  Finite $N=1$ $SU(5)$ scenarios listed in~Table \ref{tab:futbinput} has a SUSY production cross~section above 0.01
fb,~and thus will (likely) remain unobservable.  All  superpartners are~too heavy to be produced in pairs.  
Also the heavy Higgs~bosons are far outside the reach of the
HL-LHC~\cite{HAtautau-HL-LHC}. 

\begin{center}
\begin{table}[ht]
\begin{center}
\footnotesize
\begin{tabular}{|l|r|r|r|r|r|r|r|r|r|r|r|r|}
\hline
  & $M_{H}$ & $M_A$ & $M_{H^{\pm}}$  & $M_{\tilde{g}}$ & $M_{\tilde{\chi}^0_1}$ & $M_{\tilde{\chi}^0_2}$ & $M_{\tilde{\chi}^0_3}$  & $M_{\tilde{\chi}^0_4}$ &  $M_{\tilde{\chi}_1^\pm}$ & $M_{\tilde{\chi}_2^\pm}$  \\\hline
FUTSU5-1 & 5.688 & 5.688 & 5.688  & 8.966 & 2.103 & 3.917 & 4.829 & 4.832 & 3.917 & 4.833  \\\hline
FUTSU5-2 & 7.039 & 7.039 &  7.086 & 10.380 & 2.476 & 4.592 & 5.515 & 5.518 & 4.592 & 5.519  \\\hline
FUTSU5-3 & 16.382 & 16.382 &  16.401 & 12.210 & 2.972 & 5.484 & 6.688 & 6.691 & 5.484 & 6.691  \\\hline
 & $M_{\tilde{e}_{1,2}}$ & $M_{\tilde{\nu}_{1,2}}$ & $M_{\tilde{\tau}}$ & $M_{\tilde{\nu}_{\tau}}$ & $M_{\tilde{d}_{1,2}}$ & $M_{\tilde{u}_{1,2}}$ & $M_{\tilde{b}_{1}}$ & $M_{\tilde{b}_{2}}$ & $M_{\tilde{t}_{1}}$ & $M_{\tilde{t}_{2}}$ \\\hline
FUTSU5-1 & 3.102 & 3.907 & 2.205 & 3.137 & 7.839 & 7.888 & 6.102 & 6.817 & 6.099 & 6.821 \\\hline
FUTSU5-2 & 3.623 & 4.566 & 2.517 & 3.768 & 9.059 & 9.119 & 7.113 & 7.877 & 7.032 & 7.881 \\\hline
FUTSU5-3 & 4.334 & 5.418 & 3.426 & 3.834 & 10.635 & 10.699 & 8.000 & 9.387 & 8.401 & 9.390 \\\hline
\end{tabular}
\caption{\textit{Masses~for each benchmark of the Finite $N=1$ $SU(5)$ (in TeV).}}\label{tab:futbspheno}
\end{center}
\end{table}
\end{center}

At the FCC-hh the~discovery prospects for the heavy Higgs-boson~spectrum is significantly better. With $\tb \sim 50$ the first two~benchmark points, FUTSU5-1 and FUTSU5-2, are well within the~reach of
the FCC-hh. The~third point, FUTSU5-3, however, with $\MA \sim 16$~TeV~will be far outside the reach of the FCC-hh.
Prospects~for  detecting production
of squark~pairs and squark-gluino pairs are also very dim since their~production cross section is also at the level of a few fb.~This is as a
result~of a heavy spectrum in this class of models
(see \cite{SUSY-FCC-hh} with the same Figures~as discussed
in Sec.~\ref{sec:minimalsu5}).

\begin{center}
\begin{table}[ht]
\begin{center}
\footnotesize
\begin{tabular}{|c|c|c|c||c|c|c|c|}\hline
scenarios &  FUTSU5-1 & FUTSU5-2 & FUTSU5-3  &  scenarios  & FUTSU5-1 & FUTSU5-2 & FUTSU5-3\\
$\sqrt{s}$ &  100 TeV & 100 TeV& 100 TeV  &  $\sqrt{s}$  & 100 TeV & 100 TeV& 100 TeV\\ \hline
$\tilde{\chi}^0_2 \tilde{\chi}^0_3 $ & 0.01 & 0.01 &  & $\tilde{\nu}_i \tilde{\nu}_j^*$ & 0.02 & 0.01 & 0.01 \\
$\tilde{\chi}^0_3 \tilde{\chi}^0_4 $ & 0.03 & 0.01 & & $\tilde{u}_i \tilde{\chi}^-_1, \tilde{d}_i \tilde{\chi}^+_1 + h.c.$ & 0.15 & 0.06 & 0.02 \\
$\tilde{\chi}^0_2 \tilde{\chi}_1^+ $ & 0.17 & 0.08 & 0.03 & $\tilde{q}_i \tilde{\chi}^0_1, \tilde{q}_i^* \tilde{\chi}^0_1$ & 0.08 & 0.03 & 0.01 \\
$\tilde{\chi}^0_3 \tilde{\chi}_2^+ $ & 0.05 & 0.03 & 0.01 & $\tilde{q}_i \tilde{\chi}^0_2, \tilde{q}_i^* \tilde{\chi}^0_2$ & 0.08 & 0.03 & 0.01 \\
$\tilde{\chi}^0_4 \tilde{\chi}_2^+ $ & 0.05 & 0.03 & 0.01 & $\tilde{\nu}_i \tilde{e}_j^*, \tilde{\nu}_i^* \tilde{e}_j$ & 0.09 & 0.04 & 0.01 \\
$ \tilde{g}  \tilde{g} $ & 0.20 & 0.05 & 0.01 & $H  b  \bar{b} $ & 2.76 & 0.85 &  \\
$ \tilde{g} \tilde{\chi}^0_1 $ & 0.03 & 0.01 & & $A b  \bar{b} $ & 2.73 & 0.84 &  \\
$ \tilde{g} \tilde{\chi}^0_2 $ & 0.03 & 0.01 & & $H^+ b  \bar{t} + h.c.$ & 1.32 & 0.42 & \\
$ \tilde{g} \tilde{\chi}_1^+ $ & 0.07 & 0.03 & 0.01 & $H^+W^-$ & 0.38 & 0.12 &    \\
$\tilde{q}_i \tilde{q}_j, \tilde{q}_i \tilde{q}_j^*$ & 3.70 & 1.51 & 0.53 & $H Z$ & 0.09 & 0.03 &    \\
$\tilde{\chi}_1^+ \tilde{\chi}_1^- $ & 0.10 & 0.05 & 0.02 & $AZ$ & 0.09 & 0.03 &    \\
$\tilde{\chi}_2^+ \tilde{\chi}_2^- $ & 0.03 & 0.02 & 0.01 & & & &\\
$\tilde{e}_i \tilde{e}_j^*$ & 0.23 & 0.13 & 0.05 & & & &\\
$\tilde{q}_i \tilde{g}, \tilde{q}_i^* \tilde{g}$ & 2.26 & 0.75 & 0.20 &     &  &  &   \\
 \hline
\end{tabular}
\caption{\textit{Expected production~cross sections (in fb) for SUSY particles
  in the FUTSU5 scenarios.}
}
\label{futSU5xsec}
\end{center}
\end{table}
\end{center}

Concerning the stops,~the lighter one might be accessible in
FUTSU5-1. For the~squarks of the first two generations the prospects~of testing the model are somewhat better. All three benchmark~models
could possibly be excluded at the $2\sigma$ level, but no discovery at~the $5\sigma$ can be expected. The same holds for the~gluino. Charginos and neutralinos will remain unobservable due to the~heavy LSP. As in the previous section, since only the lower part of~the possible mass spectrum has been considered (with~LSP masses higher
by up to $\sim 1 \tev$), we have to conclude that again large parts of~the possible mass spectra will not be observable at the FCC-hh.

\begin{figure}[htb!]
\centering
\includegraphics[width=0.5\textwidth]{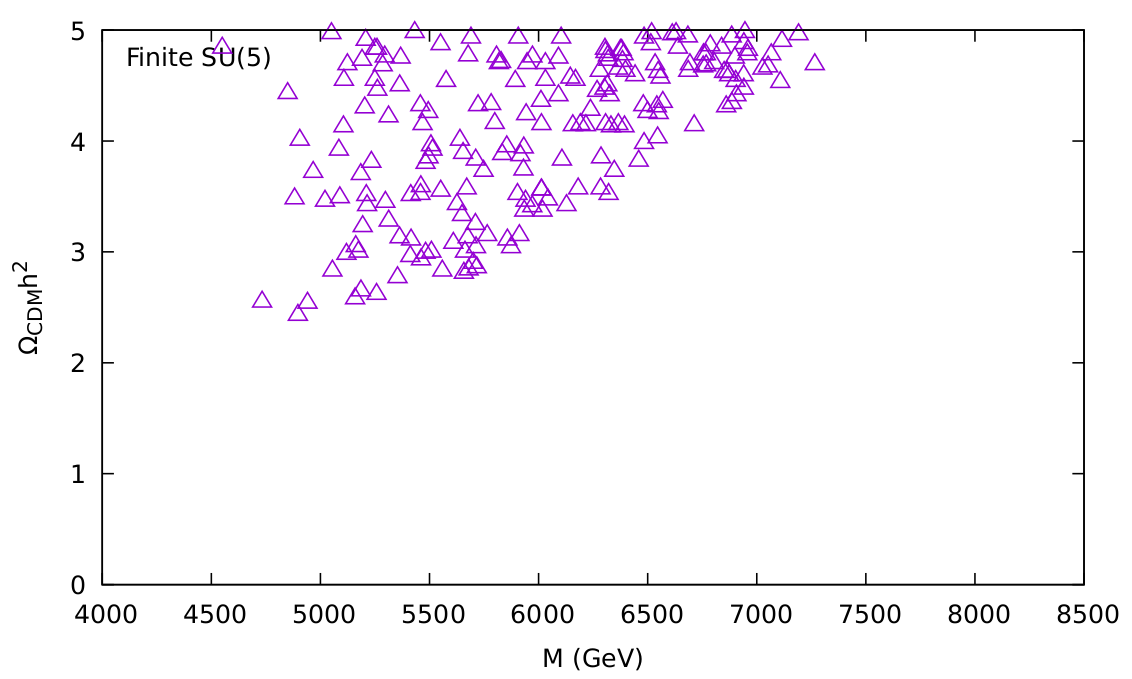}
\caption{\textit{The plot shows~the CDM relic density of
    the Finite $N=1$ $SU(5)$~model for points with Higgs mass within its calculated~uncertainty. Only the points with the lowest relic density are displayed.}}
\label{fig:futbcdm}
\end{figure}

Concerning DM, the model~exhibits a high relic abundance for CDM, as it can be seen in \reffi{fig:futbcdm}. Again, the neutralino is strongly Bino-like, not allowing for a relic density within experimental limits. The~CDM alternatives proposed for the Minimal $SU(5)$ model can also be~applied here. It should be noted that the bilinear R-parity violating~terms proposed in the previous section preserve finiteness,~as well.  

\section{The Two-Loop Finite \texorpdfstring{$N=1$}~ \texorpdfstring{$SU(3)\otimes SU(3)\otimes SU(3)$}~ }\label{se:trinification}

The analysis of the two-loop~finite $N=1$ SUSY $SU(3)\otimes SU(3)\otimes SU(3)$ model,~as described  in the \refse{sec:su33}, is the focus of this section. Again,~below $M_{\rm GUT}$ we get the MSSM. We further~assume a~unique
SUSY breaking scale~$M_{\rm SUSY}$ and~below that scale the
effective theory~is~just the SM. The boundary condition uncertainty is at~$5\%$ for the Yukawa couplings and at $1\%$ for the strong gauge~coupling and the soft parameters.

We take into~account~two new thresholds for the masses of the new exotic particles~at~$\sim10^{13} \gev$ and $\sim 10^{14} \gev$. This results~in a~wider phenomenologically viable parameter space~\cite{Mondragon:2011zzb}.~Specifically, one of the down-like~exotic particles~decouples at
$10^{14} \gev$~ while the~rest  decouple at $10^{13} \gev$.

\begin{figure}[htb!]
\begin{center}
\includegraphics[width=0.495\textwidth]{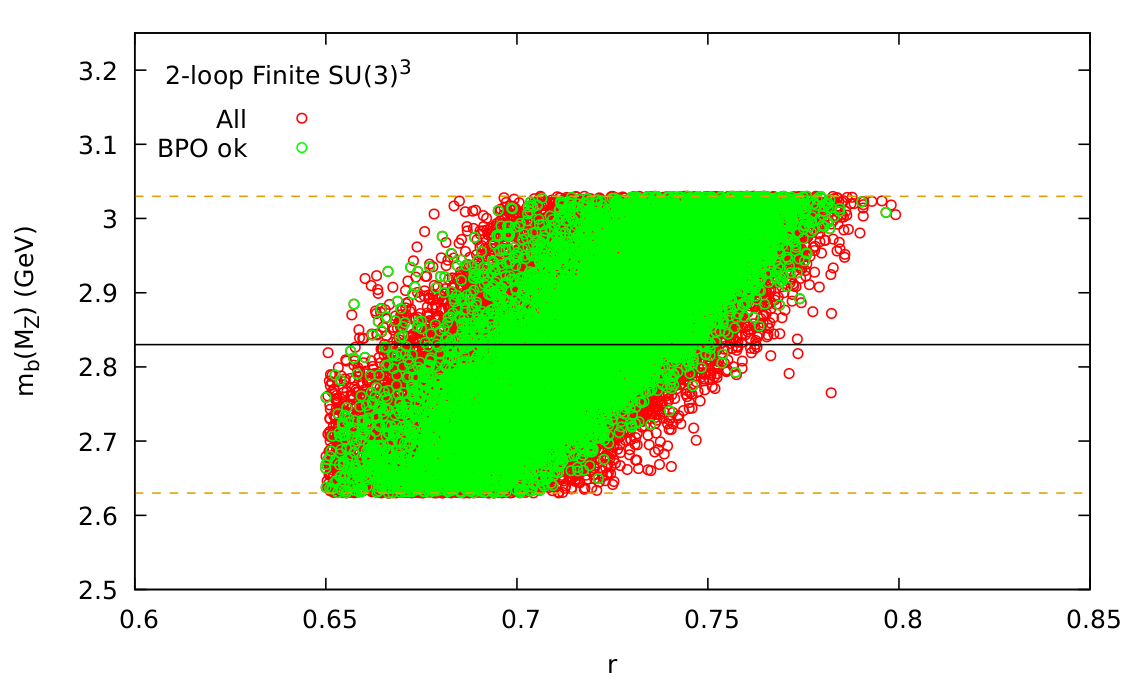}
\includegraphics[width=0.495\textwidth]{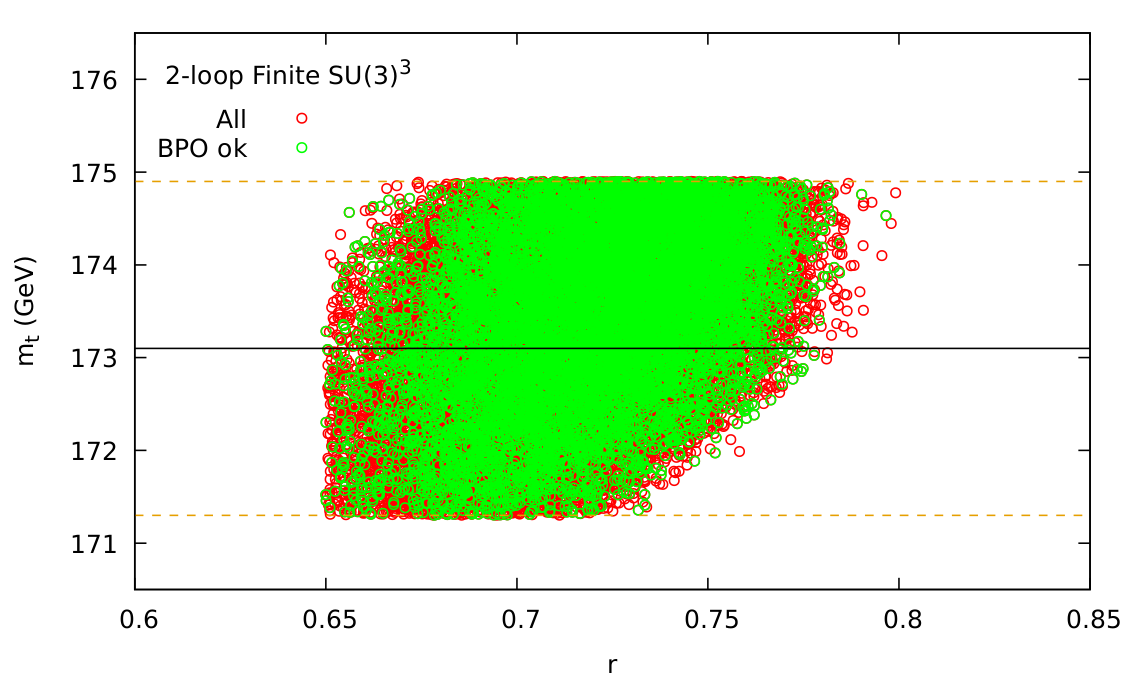}
\caption{\textit{Bottom and~top quark~  masses for the~Finite $N=1$ $SU(3)\otimes SU(3)\otimes SU(3)$~model,~with $\mu <0 $, as functions of~$r$. The~colour coding is as in~\protect\reffi{fig:mintopbotvsM}.}}
\label{fig:su33tbr}
\end{center}
\end{figure}

We compare~our~predictions with the experimental value of $m_t^{\rm exp}$,~while in~the case of the bottom quark we take again the value~evaluated at~$M_Z$, see \refeq{mbexp}.~We single out the $\mu < 0$ case as~the most promising model.~With the inclusion of thresholds~for the decoupling of the~exotic particles, the parameter~space allowed predicts a top~quark mass in agreement with~experimental bounds (see \refeq{mtexp}),~which is an important~improvement from past versions of the~model \cite{Ma:2004mi,Heinemeyer:2009zs,Heinemeyer:2010zzb,Heinemeyer:2010zza}.~Looking for~the values of the~parameter
$r$ (see \refse{sec:su33}) which~comply with the~experimental
limits (see \refse{se:constraints})~for $m_b(M_Z)$  and $m_t$, we~find, as shown in \reffi{fig:su33tbr}, that both masses are in~the experimental~range for the same value of $r$ between $0.65$ and $0.80$.~It is important to note that the two~masses are simultaneously~within two~sigmas of the experimental bounds.

\begin{figure}[htb!]
\centering
\includegraphics[width=0.495\textwidth]{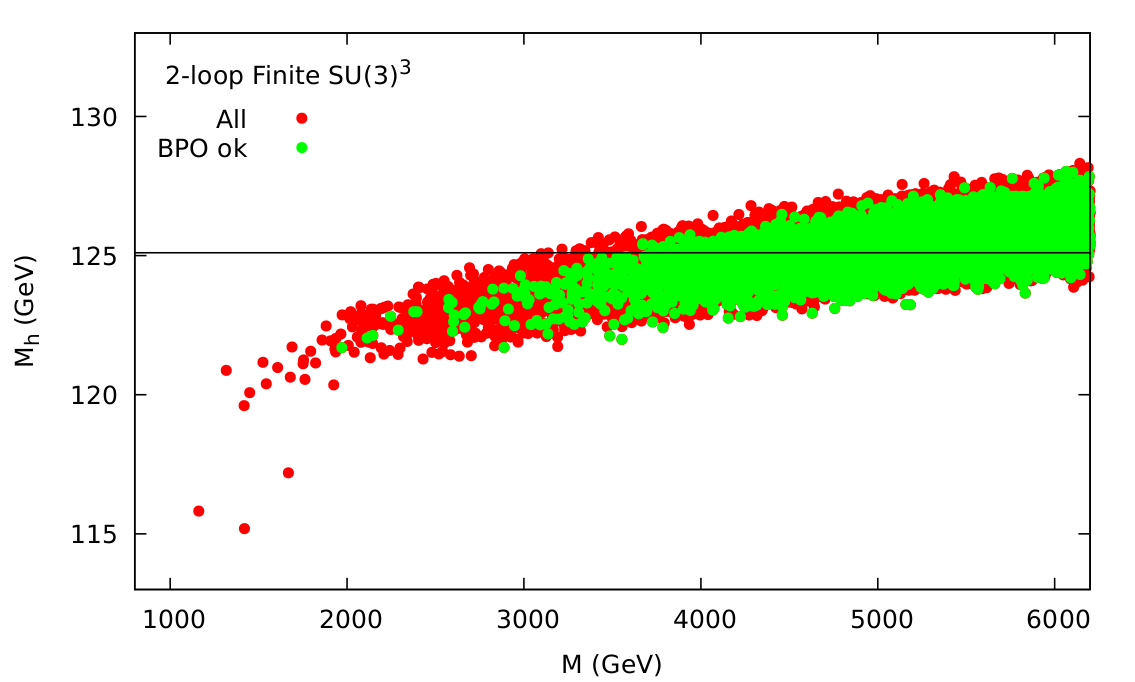}
\includegraphics[width=0.495\textwidth]{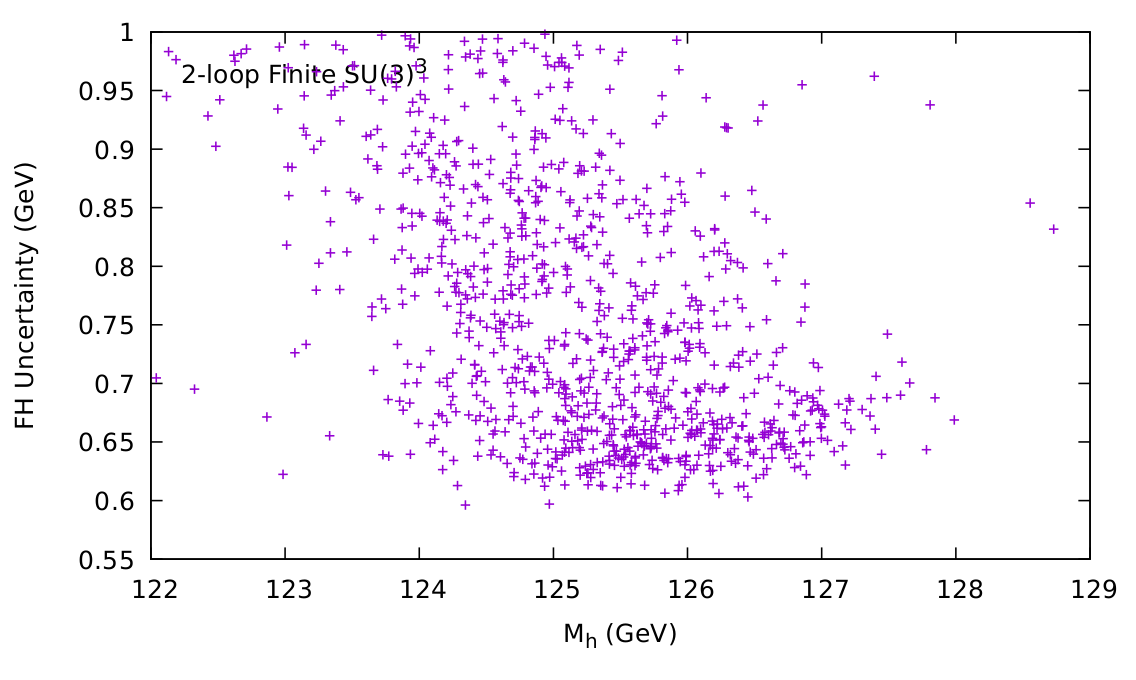}
\caption{\textit{Left: $M_h$ as~a~function of $M$ for the Finite $N=1$ $SU(3)\otimes SU(3)\otimes SU(3)$.~Right: The Higgs mass theoretical~uncertainty \cite{Bahl:2019hmm}.}}
\label{fig:su33higgsvsM}
\end{figure}

In \reffi{fig:su33higgsvsM} (left)~the light Higgs boson mass is shown~as a function of the unified gaugino mass, while with the point-by-point~calculated theoretical uncertainty drops below 1~GeV \cite{Bahl:2019hmm}~(\reffi{fig:su33higgsvsM} (right)).  As in the previous models~examined,
the $B$-physics constraints~(green points in \reffi{fig:su33higgsvsM}~(left) satisfy them) and the new, more restrictive Higgs mass~uncertainty exclude most of the low range of $M$, pushing the particle~spectrum to higher values.

\small
\begin{table}[htb!]
\renewcommand{\arraystretch}{1.3}
\centering\small
\begin{tabular}{|c|rrrrrrrrrr|}
\hline
  & $M_1$ & $M_2$ & $M_3$ & $|\mu|$ &
  $b~~~$ & $A_u$ & $A_d$ & $A_e$ & 
$\tan{\beta}$ & $m_{Q_{1,2}}^2$ \\
\hline
FSU33-1  & 1522 & 2758 & 6369 & 6138 & $1002^2$ & 4520 & 4413 & 1645 & 46.2 & $5574^2$  \\
FSU33-2  & 2070 & 3722 & 8330 & 7129 & $1083^2$ & 5841 & 5734 & 2357 & 45.5 & $7255^2$  \\
FSU33-3  & 2500 & 4484 & 10016 & 6790 & $972^2$ & 7205 & 7110 & 2674 & 49.7 & $8709$  \\
\hline
 & $m_{Q_{3}}^2$ &
  $m_{L_{1,2}}^2$ & $m_{L_{3}}^2$ & $m_{\overline{u}_{1,2}}^2$ & $m_{\overline{u}_{3}}^2$ & $m_{\overline{d}_{1,2}}^2$ & $m_{\overline{d}_{3}}^2$ & $m_{\overline{e}_{1,2}}^2$  & $m_{\overline{e}_{3}}^2$ & \\
\hline
FSU33-1  & $4705^2$ & $2382^2$ & $3754^2$ & $5234^2$ & $5548^2$ & $5197^2$ & $7043^2$ & $1558^2$  & $3095^2$ &  \\
FSU33-2  & $7255^2$ & $3136^2$ & $4131^2$ & $6749^2$ & $7225^2$ & $6745^2$ & $8523^2$ & $2238^2$  & $3342^2$ &  \\
FSU33-3  & $9074^2$ & $3831^2$ & $5483^2$ & $8152^2$ & $7207^2$ & $2558^2$ & $8600^2$ & $2507^2$  & $4000^2$ &  \\
\hline
\end{tabular}
\caption{\textit{
Finite $N=1$ $SU(3)^3$ predictions~that are used as input to {\tt SPheno}.
Mass parameters are in~$\gev$  and rounded to $1 \gev$.}}
\label{tab:su33input}
\renewcommand{\arraystretch}{1.0}
\end{table}
\normalsize

Again, we choose three~benchmarks,~each featuring the  LSP above~$1500$~GeV, $2000$~GeV and $2400$~GeV~respectively (but the LSP~can go as high as $\sim 4100 \gev$, again with too small~cross sections).
The~input~and output of {\tt SPheno} 4.0.4
\cite{Porod:2003um,Porod:2011nf} can be found in
\refta{tab:su33input}~and \refta{tab:su33spheno}  respectively
(with~the notation as in \refse{sec:minimalsu5}). 

It should~be noted that in this model the scale of the heavy Higgs~bosons does not vary monotonously with $\mneu1$, as in the previously~considered models. This can be understood as follows.~ The Higgs bosons masses are determined by a combination of the
sum rule at~the unification scale, and the requirement of successful~electroweak
symmetry~breaking at the low scale.
Like in~the finite scenario of
the previous~section, there are no direct relations between the soft~scalar masses and the unified gaugino mass,  but they are related~through the corresponding sum rule and thus vary correlatedly,~a fact that makes the dependence on
the boundary~values more restrictive. Furthermore (and even more~importantly), the fact that we took into account the two thresholds~at $\sim 10^{13} \gev$ and $\sim 10^{14} \gev$ (as mentioned above), 
allows the new particles, mainly the~Higgsinos of the two other families (that were considered decoupled at
the unification scale in previous analyses) and the down-like exotic~quarks (in a lower degree), to affect the running of the (soft) RGEs
in a~non-negligible way. 
Thus,~since at low energies the heavy
Higgs~masses depend mainly on 
the values of $m^2_{H_u}$, $m^2_{H_d}$, $|\mu|$ and $\tb$,
they~are substantially less connected to $\mneu1$ than
  in~the other models, leading to a different exclusion potential, as~  will be discussed in the following.

\begin{center}
\begin{table}[ht]
\begin{center}
\small
\begin{tabular}{|l|r|r|r|r|r|r|r|r|r|r|}
\hline
   & $M_{H}$ & $M_A$ & $M_{H^{\pm}}$ & $M_{\tilde{g}}$ & $M_{\tilde{\chi}^0_1}$ & $M_{\tilde{\chi}^0_2}$ & $M_{\tilde{\chi}^0_3}$  & $M_{\tilde{\chi}^0_4}$ &  $M_{\tilde{\chi}_1^\pm}$ & $M_{\tilde{\chi}_2^\pm}$ \\\hline
FSU33-1 & 7.029 & 7.029 & 7.028  & 6.526 & 1.506 & 2.840 & 6.108 & 6.109 & 2.839 & 6.109  \\\hline
FSU33-2 & 6.484 & 6.484 & 6.431  & 8.561 & 2.041 & 3.817 & 7.092 & 7.093 & 3.817 & 7.093  \\\hline
FSU33-3 & 6.539 & 6.539 & 6.590  & 10.159 & 2.473 & 4.598 & 6.780 & 6.781 & 4.598 & 6.781  \\\hline
 & $M_{\tilde{e}_{1,2}}$ & $M_{\tilde{\nu}_{1,2}}$ & $M_{\tilde{\tau}}$ & $M_{\tilde{\nu}_{\tau}}$ & $M_{\tilde{d}_{1,2}}$ & $M_{\tilde{u}_{1,2}}$ & $M_{\tilde{b}_{1}}$ & $M_{\tilde{b}_{2}}$ & $M_{\tilde{t}_{1}}$ & $M_{\tilde{t}_{2}}$ \\\hline
FSU33-1 & 2.416 & 2.415 & 1.578 & 2.414 & 5.375 & 5.411 & 4.913 & 5.375 & 4.912 & 5.411 \\\hline
FSU33-2 & 3.188 & 3.187 & 2.269 & 3.186 & 7.026 & 7.029 & 6.006 & 7.026 & 6.005 & 7.029 \\\hline
FSU33-3 & 3.883 & 3.882 & 2.540 & 3.882 & 8.334 & 8.397 & 7.227 & 8.334 & 7.214 & 7.409 \\\hline
\end{tabular}
\caption{\textit{Masses for~each benchmark of the Finite $N=1$ $SU(3)^3$ (in TeV).}}\label{tab:su33spheno}
\end{center}
\end{table}
\end{center}

Scenarios of~Finite $SU(3)^3$  are beyond the reach of the HL-LHC. Not~only superpartners are too heavy, but also  heavy Higgs bosons~with a mass scale of $\sim 7$~TeV cannot be detected at the HL-LHC.
At 100 TeV~collider (see \refta{fSU33xsec}),
on the other~hand, all three benchmark points are well within the~reach of the $H/A \to \tau^+\tau^-$ as well as the
$H^\pm \to \tau\nu_\tau, tb$ searches \cite{HAtautau-FCC-hh,Hp-FCC-hh},
despite~the slightly smaller values of $\tb \sim 45$. This is
particularly~because of the different dependence of the heavy
Higgs-boson~mass scale on $\mneu1$, as discussed above.
However, we~have checked that $\MA$ can go up to to $\sim 11 \tev$,~and thus the heaviest part of the possible spectrum would~escape the
heavy~Higgs-boson searches at the FCC-hh.

Interesting~are also the prospects for production of squark
pairs and~squark-gluino, which 
can reach~$\sim 20$ fb for the FSU33-1 case, going down to a few fb for~FSU33-2 and FSU33-3 scenarios.  The lightest squarks~decay almost
exclusively~ to the third generation quark and chargino/neutralino,~while gluino enjoys many possible decay channels~to quark-squark pairs
each one~with branching fraction of the order of a percent, with the~biggest one $\sim 20\%$  to $t\tilde{t}_1 +h.c.$.

We briefly~discuss the SUSY discovery potential at the FCC-hh,
referring~again to \cite{SUSY-FCC-hh} with the same Figures as discussed~in Sec.~\ref{sec:minimalsu5}.
Stops~in FSU33-1 and FSU33-2 can be tested at the FCC-hh, while~the
masses~turn out to be too heavy in FSU33-3. The situation is better~for scalar quarks, where all three scenarios can be tested,~but will
not allow~for a $5\sigma$ discovery. Even more favorable are the~prospects for gluino. Possibly all three scenarios can be tested at~the $5\sigma$ level. As in the previous scenario, the charginos and~neutralinos will not be accessible, due to the too heavy LSP.~Keeping in mind that only the lower part of possible mass spectrum~  is represented by the three benchmarks (with the LSP up to $\sim 1.5 \tev$
heavier), we~conclude that as before large parts of the parameter space~will not be testable at the FCC-hh. The only partial exception here is the
Higgs-boson~sector, where only the the part with the highest possible~Higgs-boson mass spectra would escape the FCC-hh searches.

\begin{center}
\begin{table}[ht]
\begin{center}
\footnotesize
\begin{tabular}{|c|c|c|c||c|c|c|c|}\hline
scenarios  & FSU33-1 & FSU33-2 & FSU33-3  &  scenarios &  FSU33-1 & FSU33-2 & FSU33-3\\
$\sqrt{s}$  & 100 TeV & 100 TeV& 100 TeV  &  $\sqrt{s}$ &  100 TeV & 100 TeV& 100 TeV\\ \hline
$\tilde{\chi}^0_1 \tilde{\chi}^0_1 $  & 0.04 & 0.01 & 0.01 & $\tilde{q}_i \tilde{g}, \tilde{q}_i^* \tilde{g}$  & 22.12 & 3.71 & 1.05\\
$\tilde{\chi}^0_2 \tilde{\chi}^0_2 $  & 0.04 & 0.01 &  & $\tilde{\nu}_i \tilde{\nu}_j^*$  & 0.10 & 0.03 & 0.01\\
$\tilde{\chi}^0_2 \tilde{\chi}_1^+ $  & 0.58 & 0.16 & 0.07 & $\tilde{u}_i \tilde{\chi}^-_1, \tilde{d}_i \tilde{\chi}^+_1 + h.c.$  & 1.22 & 0.25 & 0.08\\
$\tilde{\chi}^0_3 \tilde{\chi}_2^+ $  & 0.02 & 0.01 & 0.01 & $\tilde{q}_i \tilde{\chi}^0_1, \tilde{q}_i^* \tilde{\chi}^0_1$  & 0.55 & 0.13 & 0.05\\
$\tilde{\chi}^0_4 \tilde{\chi}_2^+ $  & 0.02 & 0.01 & 0.01 & $\tilde{q}_i \tilde{\chi}^0_2, \tilde{q}_i^* \tilde{\chi}^0_2$  & 0.60 & 0.13 & 0.04\\
$ \tilde{g}  \tilde{g} $  & 2.61 & 0.30 & 0.07 & $\tilde{\nu}_i \tilde{e}_j^*, \tilde{\nu}_i^* \tilde{e}_j$  & 0.36 & 0.12 & 0.04\\
$ \tilde{g} \tilde{\chi}^0_1 $  & 0.20 & 0.05 & 0.02 & $H  b  \bar{b} $ & 0.71 & 1.23 & 1.19 \\
$ \tilde{g} \tilde{\chi}^0_2 $  & 0.20 & 0.04 & 0.01 & $A b  \bar{b} $ & 0.72 & 1.23 & 1.18 \\
$ \tilde{g} \tilde{\chi}_1^+ $  & 0.42 & 0.09 & 0.03 & $H^+ b  \bar{t} + h.c.$ & 0.37 & 0.75 & 0.58 \\
$\tilde{q}_i \tilde{q}_j, \tilde{q}_i \tilde{q}_j^*$  & 25.09 & 6.09 & 2.25 & $H^+W^{-}$ & 0.10 & 0.25 & 0.19 \\
$\tilde{\chi}_1^+ \tilde{\chi}_1^- $  & 0.37 & 0.10 & 0.04 & $H Z$ & 0.02 & 0.04 & 0.04 \\
$\tilde{e}_i \tilde{e}_j^*$  & 0.39 & 0.12 & 0.06 & $AZ$ & 0.02 & 0.04 & 0.04 \\
\hline
\end{tabular}
\caption{\textit{Expected production cross sections~(in fb) for SUSY particles in the FSU33 scenarios.}
}
\label{fSU33xsec}
\end{center}
\end{table}
\end{center}

\begin{figure}[htb!]
\centering
\includegraphics[width=0.5\textwidth]{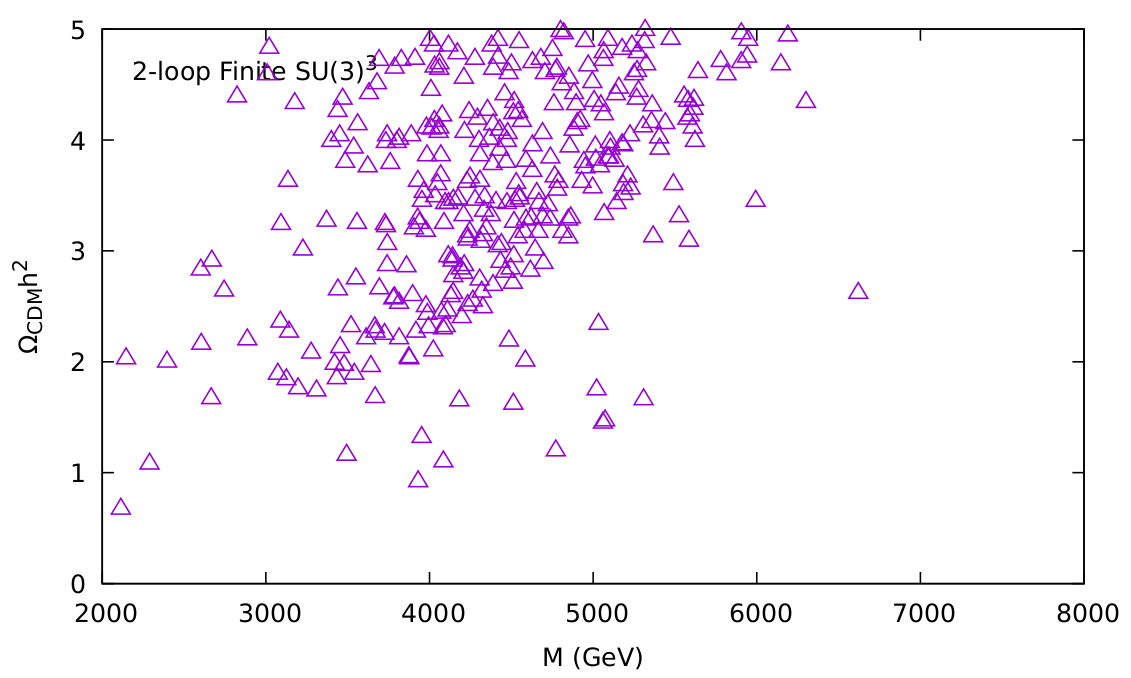}
\caption{\textit{The plot shows~the CDM relic density of
    the two-loop Finite $N=1$ $SU(3)\otimes SU(3)\otimes SU(3)$ model~model for points with Higgs mass within its calculated~uncertainty.}}
\label{fig:su33cdm}
\end{figure}

The only observable~that fails to comply with the
experimental bounds is the CDM~relic density (see \refeq{cdmexp}), as evident in \reffi{fig:su33cdm}. The~lightest neutralino is the LSP and considered as a CDM~candidate, but
its relic density~does not go below $0.15$, since it is strongly
Bino-like and would~require a lower scale of the particle spectrum. It~should be noted that if the B-physics constraints allowed for a unified~gaugino mass $\sim 0.5 \tev$ lower, then agreement with the CDM bounds~as well could be achieved.

\section{The Reduced MSSM}\label{se:rmssm}

The last analysis is the one for the reduced version of the MSSM. The relations~among reduced~parameters in terms
of the~fundamental ones~derived in  \refse{sec:mssm}
have an RGI part and a~part that originates from the corrections,~and thus~scale dependent. In the present
analysis~we choose~the unification~scale to apply the~corrections~to all these RGI~relations.
As noted~earlier, the Hisano-Shiftman relation sets a hierarchy among the gaugino~masses, rendering Wino the lightest of them. As~such, we have~a Wino-like lightest neutralino.

In~the dimensionless~sector of~the
theory,~since~$Y_\tau$ is not reduced in favour of the fundamental~parameter $\al_3$, the tau lepton mass is an input parameter and,~consequently, $\rho_\tau$ is an independent~parameter, too.~At low
energies we fix~$\rho_{\tau}$ and $\tan\beta$ using
the~mass of the~tau~lepton $m_{\tau}(M_Z)=1.7462$ GeV.
Then, we~determine the top and bottom  masses using the value found~for $\tan\beta$ together with $G_{t,b}$, as obtained~from the~REs and their corrections. 

\begin{figure}[htb!]
\begin{center}
\includegraphics[width=0.495\textwidth]{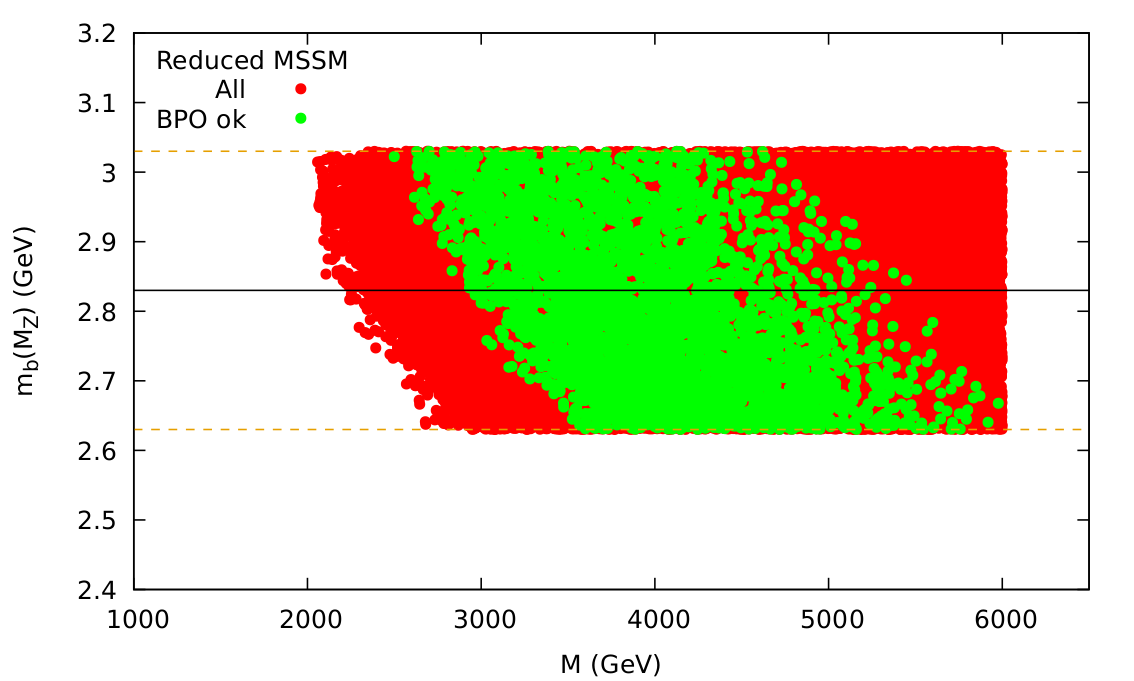}
\includegraphics[width=0.495\textwidth]{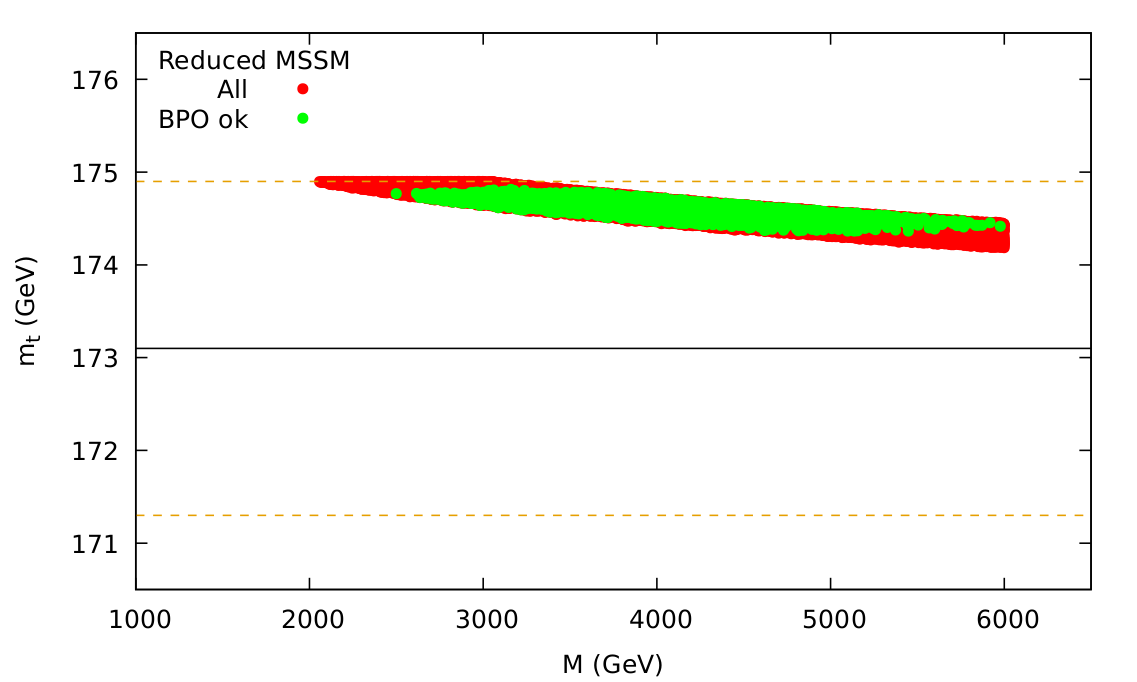}
\caption{\textit{The left (right)~plot shows the bottom (top)~quark mass for the Reduced~MSSM, with the colour coding~as in \protect\reffi{fig:mintopbotvsM}.}}
\label{fig:rmssmtopbotvsM}
\end{center}
\vspace{-1em}
\end{figure}

Correspondingly,~concerning the dimensionful sector,  $h_\tau$ cannot be expressed~in terms of the unified gaugino~mass scale, leaving  $\rho_{h_\tau}$ a~free parameter.
  $\mu$ is a free parameter~as well, as it cannot~be reduced in favour of $M_3$~as discussed~above. On the other hand, $m_3^2$ could be~reduced,~but here
we choose to leave it~free.
However, $\mu$~and $m_3^2$ are~restricted from the requirement of~EWSB, and only $\mu$ is taken~as an independent~parameter.
Finally,~the~other parameter in the Higgs-boson sector,~the $\cp$-odd~Higgs-boson mass $\MA$ is evaluated~from $\mu$, as well as~from $m_{H_u}^2$~and
$m_{H_d}^2$,~which are obtained from the REs.
In total, we~vary the~parameters $\rho_\tau$, $\rho_{h_\tau}$, $M$~and~$\mu$.

As it has been already~mentioned, the variation~of $\rho_\tau$ gives the running~bottom~quark mass at the $Z$~boson mass scale and the top~pole~mass, where points not within $2\sigma$ of
the experimental data~are neglected, as it is~shown in
\reffi{fig:rmssmtopbotvsM}.~The experimental~values (see \refse{se:constraints})~are denoted by the
horizontal lines with~the uncertainties at the $2\,\sigma$ level. 
The green dots~satisfy the flavour constraints. One can see that the~scan yields many parameter points that
are in~very good~agreement with the experimental data and give restrictions~in the allowed range of~$M$ (the common gaugino mass at the~unification scale).

\begin{figure}[htb!]
\begin{center}
\includegraphics[width=0.495\textwidth]{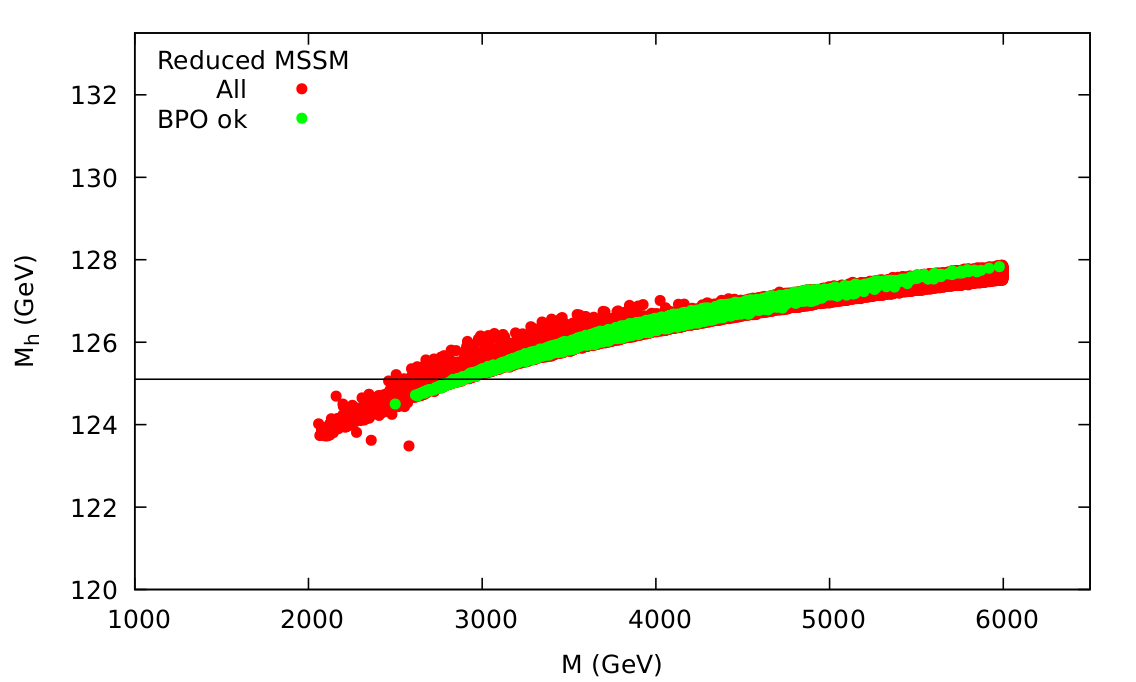}
\includegraphics[width=0.495\textwidth]{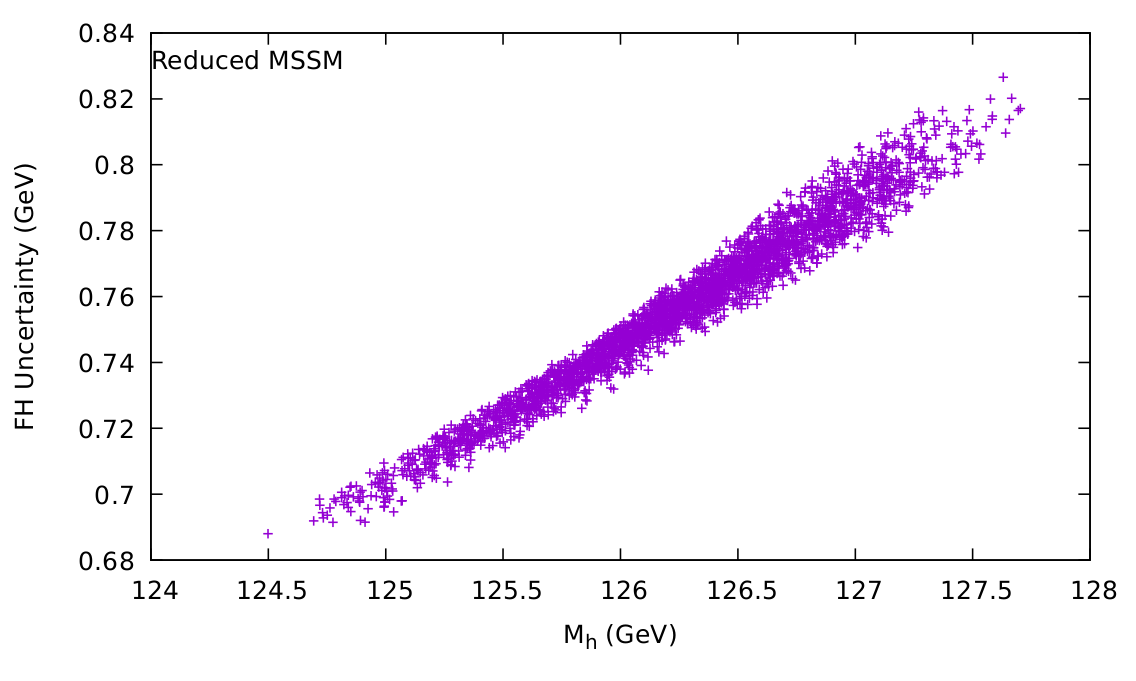}
\caption{\textit{Left: The~lightest Higgs boson mass, $\Mh$ in the Reduced MSSM. The~green points is the full model prediction. Right: the lightest~Higgs mass theoretical uncertainty \cite{Bahl:2019hmm}.}}
\label{fig:rmssmhiggsvsM}
\end{center}
\end{figure}

The prediction~for $\Mh$ is shown in
\reffi{fig:rmssmhiggsvsM}~(left).
Once again, one~should keep in mind that the theory
uncertainty~given in \reffi{fig:rmssmhiggsvsM} (right) has dropped~below 1~GeV \cite{Bahl:2019hmm}.  The Higgs mass predicted~by the model is  
in the~range measured at the LHC, favoring this time relatively small~values of~$M$. This in turn
sets~a limit on the low-energy
SUSY~masses, rendering the Reduced MSSM  highly~predictive and
testable.

The $\Mh$ limits~set a limit on the low-energy
supersymmetric~masses, which we briefly discuss.
The three selected~benchmarks correspond to \htb{\DRbar\ }
pseudoscalar Higgs boson masses above $1900$~GeV, $1950$~GeV and
$2000$~GeV respectively. The input of {\tt SPheno} 4.0.4
\cite{Porod:2003um,Porod:2011nf} can be found in \refta{tab:mssminput}
(notation as in \refse{sec:minimalsu5}). 

\begin{table}[htb!]
\renewcommand{\arraystretch}{1.3}
\centering\small
\begin{tabular}{|c|rrrrrrrrrr|}
\hline
  & $M_1$ & $M_2$ & $M_3$ & $|\mu|$ &
  $b~~~$ & $A_u$ & $A_d$ & $A_e$ & $\tan{\beta}$ & $m_{Q_{1,2}}^2$  \\
\hline
RMSSM-1  & 3711 & 1014 & 7109 & 4897 & $284^2$ & 5274 & 5750 & 20 & 44.9 & $5985^2$  \\
RMSSM-2  & 3792 & 1035 & 7249 & 4983 & $294^2$ & 5381 & 5871 & 557 & 44.6 & $6103^2$  \\
RMSSM-3  & 3829 & 1045 & 7313 & 5012 & $298^2$ & 5427 & 5942 & 420 & 45.3 & $6161^2$ \\
\hline
  & $m_{Q_{3}}^2$ &
  $m_{L_{1,2}}^2$ & $m_{L_{3}}^2$ & $m_{\overline{u}_{1,2}}^2$ & $m_{\overline{u}_{3}}^2$ & $m_{\overline{d}_{1,2}}^2$ & $m_{\overline{d}_{3}}^2$ & $m_{\overline{e}_{1,2}}^2$  & $m_{\overline{e}_{3}}^2$ & \\
\hline
RMSSM-1  & $5545^2$ & $2106^2$ & $2069^2$ & $6277^2$ & $5386^2$ & $5989^2$ & $5114^2$ & $3051^2$  & $4491^2$ &  \\
RMSSM-2  & $5656^2$ & $2122^2$ & $2290^2$ & $6385^2$ & $5476^2$ & $6110^2$ & $5219^2$ & $3153^2$  & $4181^2$ &  \\
RMSSM-3  & $5708^2$ & $2106^2$ & $2279^2$ & $6427^2$ & $5506^2$ & $6172^2$ & $5269^2$ & $3229^2$  & $3504^2$ &  \\
\hline
\end{tabular}

\caption{\textit{
Reduced MSSM~predictions that are used as input to {\tt SPheno}
Mass parameters are in~$\gev$ and rounded to $1 \gev$.}}
\label{tab:mssminput}
\renewcommand{\arraystretch}{1.0}
\end{table}

\begin{figure}[htb!]
\centering
\includegraphics[width=0.5\textwidth]{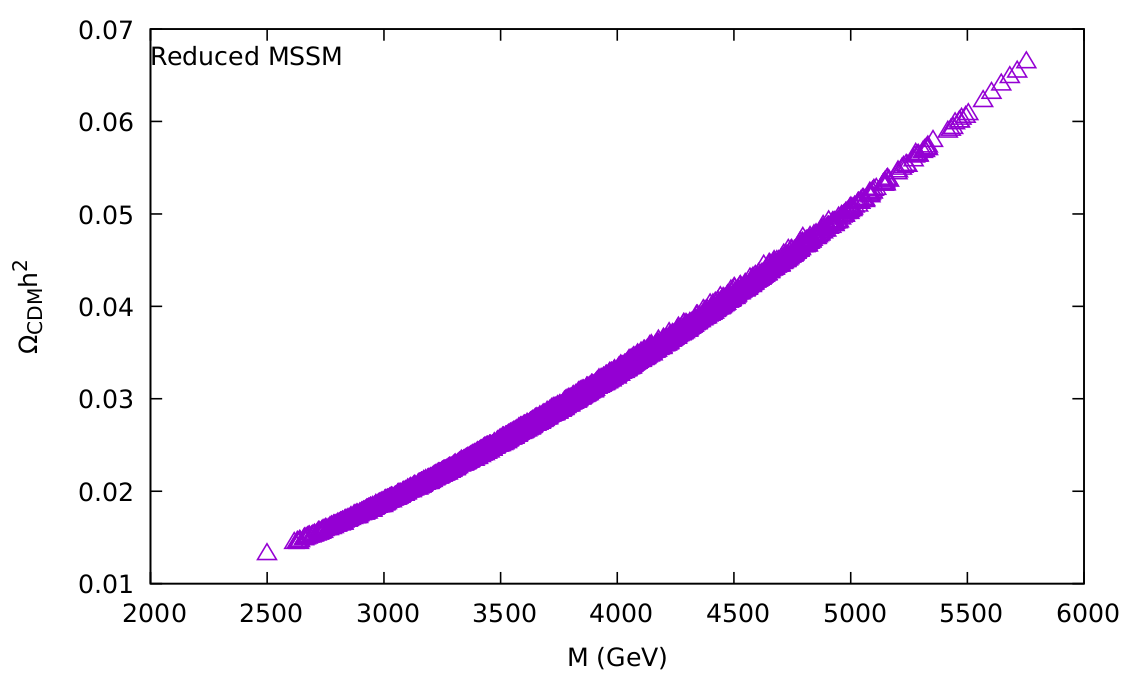}
\caption{\textit{The plot shows~the CDM relic density of
    the Reduced MSSM for points with Higgs mass within its calculated~uncertainty. All points are below the experimental limits.}}
\label{fig:mssmcdm}
\end{figure}

Concerning the DM predictions, it~should be noted that the Hisano-Shiftman~relation imposes a Wino-like~LSP, which unfortunately lowers~the CDM relic density below the
boundaries~of \refeq{cdmexp}. This can be seen in \reffi{fig:mssmcdm}. This could render the model viable if
\refeq{cdmexp} is~applied only as an upper limit and additional sources~of CDM are allowed. This is in contrast to the other three models~discussed previously. 

Table~\ref{tab:mssmspheno} shows the resulting masses of Higgs bosons
and some of the~lightest SUSY particles.
In particular, we find $\MA \lsim 1.5 \tev$ (for large values of~$\tb$ as in the other models), values substantially lower than in the~previously considered models. This can be understood~as follows.

\begin{center}
\begin{table}[ht]
\begin{center}
\small
\begin{tabular}{|l|r|r|r|r|r|r|r|r|r|r|}
\hline
  & $M_{H}$ & $M_A$ & $M_{H^{\pm}}$ & $M_{\tilde{g}}$ & $M_{\tilde{\chi}^0_1}$ & $M_{\tilde{\chi}^0_2}$ & $M_{\tilde{\chi}^0_3}$  & $M_{\tilde{\chi}^0_4}$ &  $M_{\tilde{\chi}_1^\pm}$ & $M_{\tilde{\chi}_2^\pm}$  \\\hline
RMSSM-1 & 1.393 & 1.393 & 1.387 & 7.253 & 1.075 & 3.662 & 4.889 & 4.891 & 1.075 & 4.890  \\\hline
RMSSM-2 & 1.417 & 1.417 & 1.414 & 7.394 & 1.098 & 3.741 & 4.975 & 4.976 & 1.098 & 4.976  \\\hline
RMSSM-3 & 1.491 & 1.491 & 1.492 & 7.459 & 1.109 & 3.776 & 5.003 & 5.004 & 1.108 & 5.004  \\\hline
 & $M_{\tilde{e}_{1,2}}$ & $M_{\tilde{\nu}_{1,2}}$ & $M_{\tilde{\tau}}$ & $M_{\tilde{\nu}_{\tau}}$ & $M_{\tilde{d}_{1,2}}$ & $M_{\tilde{u}_{1,2}}$ & $M_{\tilde{b}_{1}}$ & $M_{\tilde{b}_{2}}$ & $M_{\tilde{t}_{1}}$ & $M_{\tilde{t}_{2}}$ \\\hline
RMSSM-1 & 2.124 & 2.123 & 2.078 & 2.079 & 6.189 & 6.202 & 5.307 & 5.715 & 5.509 & 5.731 \\\hline
RMSSM-2 & 2.297 & 2.139 & 2.140 & 2.139 & 6.314 & 6.324 & 5.414 & 5.828 & 5.602 & 5.842 \\\hline
RMSSM-3 & 2.280 & 2.123 & 2.125 & 2.123 & 6.376 & 6.382 & 5.465 & 5.881 & 5.635 & 5.894 \\\hline
\end{tabular}
\caption{\textit{Masses for each benchmark~of the Reduced MSSM (in TeV).}}\label{tab:mssmspheno}
\end{center}
\end{table}
\end{center}

In this model, we have~direct relations between the soft scalar masses and the~unified gaugino mass, which receive corrections from the~two gauge couplings $g_1$ and $g_2$ and the Yukawa coupling~of the $\tau$ lepton. As mentioned above, in the absence of~these corrections the relations obey the soft scalar mass~sum rule. However, unlike all the previous models, these~corrections make the sum rule only approximate. Thus, these~unique boundary conditions
result in very~low values for the masses of the heavy
Higgs bosons~(even compared to the minimal $SU(5)$ case presented~above, which also exhibits direct relations which however obey the sum rule).~A relatively light spectrum is also favored by the prediction for~the light CP-even Higgs boson mass,~which turns out to be relatively high in this model and does~not allow us to consider heavier spectra.
Thus, in this~model, contrary 
to the models~analyzed before,  because of the
large $\tan\beta\sim 45$ found~here, the \textit{physical} mass of~the pseudoscalar Higgs boson, $M_A$, is excluded by the searches~$H/A
\rightarrow \tau \tau$ at ATLAS with 139/fb \cite{Aad:2020zxo} for all~three benchmarks. One could try considering a heavier spectrum, in~which we would have $M_A\gtrsim 1900\gev$, but in that case the light~Higgs mass would be well above its acceptable region.~Particularly, it would be above 128 GeV, a value that is~clearly excluded, especially given the improved (much smaller)~uncertainty calculated by the new {\tt FeynHiggs} code).~Thus, the
current version of this model has been ruled out~experimentally. Consequently, we do not show any SUSY or Higgs~production cross sections.

\newpage
\mbox{} 
\newpage

\part{Trinification from Coset Space Dimensional Reduction}\label{part2}

\newpage
\mbox{} 
\newpage

\chapter{The Coset Space Dimensional Reduction}\label{csdr}

Consider a unified theory~that is built on a spacetime in higher than four dimensions.~Such a higher-dimensional theory can lead to a~four-dimensional theory through the dimensional reduction~procedure. 
The very first approach~of such reduction of a unified higher-dimensional theory~to a four-dimensional one was made by Kaluza and Klein~\cite{Kaluza:1921tu,Klein:1926tv}. The main idea was to begin by~considering a pure five-dimensional gravitational theory~and perform a dimensional reduction which would lead to a~four-dimensional theory, specifically a theory of electromagnetism~coupled to gravity. Moreover, an appealing aspect of the whole~dimensional reduction scheme is that the extension of the~spacetime~naturally leads to non-Abelian group structures \cite{Kerner:1988fn,Cho:1975sw,Cho:1975sf} (for reviews~see \cite{Mecklenburg:1983uk,Bailin:1987jd,Salam:1981xd}). 

The initial spacetime in high dimensions, $M^D$, can be~considered as $M^D=M^4\times B$, where $B$ is a Riemannian~manifold which is compact and its isometries are described~by a group $S$. If a dimensional reduction is applied on~the higher-dimensional theory, the resulting theory would~be a four-dimensional one describing gravity and a Yang-Mills~theory. The non-Abelian gauge group of the Yang-Mills part~of the four-dimensional theory would include the isometry~group of $B$, that is $S$, accompanied by scalar fields. Besides the main and~most important aspect of the above scheme, namely the unification~of the gravitational interaction with others that are described~by gauge theories in a geometric way, it is also very~welcome that the above scheme sheds light on the fact that~some interactions admit a description of a gauge theory.

On the other hand, there are also some unwelcome features that emerge after performing the above dimensional reduction.~First, expressing the higher-dimensional space~$M^D$ as a product does not allow for a classical ground state.~Although the Ricci tensor of a space like the one considered,~$B$, is non-vanishing, the field equations of the initial,~pure $D$-dimensional gravitational theory yield that the~Ricci tensor is equal to zero. Thus, it is deduced that the~extra dimensions cannot be interpreted in a physical way. Additionally,~another serious drawback in the attempt to result in realistic four-dimensional models is that performing a dimensional~reduction of a higher-dimensional theory in which fermions~are accommodated does not allow to end up with chiral fermions~in the resulting four-dimensional theory \cite{Witten:1983ux}. Nevertheless,~it is possible to overcome the above problems by~choosing an appropriate spectrum of particles in the~higher-dimensional theory, specifically involving Yang-Mills~fields~\cite{Horvath:1977st,Chapline:1982wy}. 

Putting aside gravitational~phenomena in the low-energy regime implies the consideration of~the initial higher-dimensional theory to be purely~Yang-Mills (Y-M). Performing a dimensional reduction could lead~to a (probably unified) four-dimensional Yang-Mills theory~accompanied by a Higgs sector. A naive way way to~perform such a reduction on a Y-M theory defined on $M^D=M^4\times B$ is~to impose the condition that every particle of the~spectrum depends~only on the $M^4$ spacetime coordinates.~Therefore, considering~every particle independent of the extra coordinates implies that the Lagrangian density would also not depend on~them. A less crude approach is to consider that the part of~the space related to the extra dimensions is described by~a coset space $B=S/R$. The elegance of this approach lies in~the fact that one does not have to impose any independence of the fields on the extra dimensions. More specifically, the~fields are allowed to depend on the coordinates of the extra~dimensions in such a way that an application of an S-transformation~would transform the Lagrangian by some quantity that~would be cancelled by a simple gauge transformation.~In other words, the Lagrangian ends up to be independent of the~extra (coset) coordinates because of its gauge invariance, which~is a built-in property, and not because of an arbitrary demand. This~is the most fundamental argument that renders the Coset Space~Dimensional Reduction (CSDR) a very appealing scheme for~the treatment of higher-dimensional theories and the construction~of promising four-dimensional theories \cite{Forgacs:1979zs,Kapetanakis:1992hf,Kubyshin:1989vd}. 


\section{Coset Space Geometry}\label{geometry}

Before we proceed to the description of the CSDR scheme, it is  useful to recall some basic information~about the geometry of coset~spaces. This section includes the prerequisite geometrical aspects of coset~spaces needed in order to understand the CSDR scheme. For an extended (and more complete) discussion of the geometry of coset spaces see \cite{Kapetanakis:1992hf,Castellani:1999fz}.

In the CSDR~scheme one starts by considering a Yang-Mills-Dirac (YMD) action with a gauge~group $G$. The action is defined on a spacetime of $D$ dimensions, $M^{D}$ and its metric is $g^{MN}$. The spacetime $M^D$ gets compactified to the spacetime $M^{4}\times S/R$, with $S/R$ a coset space. The metric of the compactified space has the~following block diagonal form:
\begin{equation}\label{pan:1}
g^{MN}=
\left[\begin{array}{cc}\eta^{\mu\nu}&0\\0&-g^{ab}\end{array}
\right],
\end{equation}
where $\eta^{\mu\nu}= \text{diag}(1,-1,-1,-1)$ and $g^{ab}$ is the metric of the coset
space. The above compactification of the initial spacetime points to further exploration of the geometry~of coset spaces $S/R$, where $S$ is a compact Lie~group and $R$ is compact Lie subgroup~of $S$. 

It is useful to divide the set of the generators of the group $S$, $Q_{A}$, into two subsets, one including the generators of $R$, $Q_{i}$ $(i=1, \ldots,\text{dim}R)$  and another including the generators of $S/R$, $Q_{a}$, where $a=\text{dim}R+1 \ldots,\text{dim}S$, with $\text{dim}(S/R)=\text{dim}S-\text{dim}R = d$. Therefore, their commutation relations can be written in~the following form:
\begin{eqnarray}\label{pan:2}
\left[ Q_{i},Q_{j} \right] &=& f_{ij}^{\ \ k} Q_{k} \ ,\nonumber \\
\left[ Q_{i},Q_{a} \right]&=& f_{ia}^{\ \ b}Q_{b} \ ,\nonumber\\
\left[ Q_{a},Q_{b} \right]&=& f_{ab}^{\ \ i}Q_{i}+f_{ab}^{\ \
c}Q_{c}\ .
\end{eqnarray}
The second commutation relation demonstrates the fact that the coset space $S/R$ is reductive (since $R$ is compact) and, in general, non-symmetric\footnote{In case $f_{ab}^{\ \ c}=0$ in \refeq{pan:2}, the coset space is called symmetric.}. 

Now, let us denote the coordinates~of the compactified space $M^{4} \times S/R$ as $z^{M}= (x^{m},y^{\alpha})$, where $\alpha$ is a curved index of the~coset space and $y$ is a parametrization of an element of~$S$, which is actually a coset representative, $L(y)$. Latin indices $a,b,\ldots$ will denote the coordinates of the tangent space. The vielbein and the $R$-connection are defined by the~Maurer-Cartan form which takes its values~in the Lie algebra of $S$ :
\begin{equation}\label{pan:3}
L^{-1}(y)dL(y) = e^{A}_{\alpha}Q_{A}dy^{\alpha}~.
\end{equation}
At the origin it holds that $y = 0$, $e^{~a}_{\alpha} = \delta^{~a}_{\alpha}$ and $e^{~i}_{\alpha} = 0$. In general, a connection form $\theta^{a}_{\ b}$ on $S/R$ consists of both curvature and torsion. In the case that the torsion vanishes, the calculation of the torsionless part $\omega^{a}_{\ b}$ takes place by setting the torsion form, $T^{a}$, to zero:
\begin{equation}
T^{a} = de^{a} + \omega^{a}_{\ b} \wedge e^{b} = 0\ ,
\end{equation}
while using the Maurer-Cartan equation
\begin{equation}
de^{a} = \frac{1}{2}f^{a}_{\ bc}e^{b}\wedge e^{c} +f^{a}_{\
bi}e^{b}\wedge e^{i}\ ,
\end{equation}
one obtains the following expression for the connection
\begin{equation}
\omega^{a}_{\ b}= -f^{a}_{\ ib}e^{i}-\frac{1}{2}f^{a}_{\ bc}e^{c}-
\frac{1}{2}K^{a}_{\ bc}e^{c}\ ,\label{connectiontorsionless}
\end{equation}
where $K^{a}_{\ bc}$ is symmetric under the exchange of the lower indices $b,c$, thus
$K^{a}_{\ bc}e^{c} \wedge e^{b}=0$. The expression of the $K^{a}_{\ bc}$ can be
obtained by taking into consideration that $\omega^{a}_{\ b}$ is antisymmetric, that is $\omega^{a}_{\ b}g^{cb}=-\omega^{b}_{\ c}g^{ca}$:
\begin{equation}
K_{\ \ bc}^{a}=g^{ad}(g_{be}f_{dc}^{\ \ e}+g_{ce}f_{db}^{\ \ e})\ .
\end{equation}
In turn, replacing the above expression for $K^{a}_{\ bc}$ in the expression for $\omega^{a}_{\ b}$, \eqref{connectiontorsionless}, one obtains the following form of the connection:
\begin{equation}
\omega^{a}_{\ b}= -f^{a}_{\ ib}e^{i}-D_{\ bc}^{a}e^{c},
\end{equation}
where
\begin{equation} 
D_{\ bc}^{a}=\frac{1}{2}g^{ad}[f_{db}^{\ \
e}g_{ec}+f_{cb}^{\ \ e} g_{de}- f_{cd}^{\ \ e}g_{be}]\ .\label{TheDs}
\end{equation}
According to \cite{Kapetanakis:1992hf}, the $D$'s can be related to the structure constants by a rescaling: $$ D^{a}_{\
bc}=(\lambda^{a}\lambda^{b}/\lambda^{c})f^{a}_{\ bc}\ ,$$ where the $\lambda$'s depend only on the~radii of the coset. It should be noted that, in general, the~rescaling factors affect the antisymmetry properties of the~structure constants. Also, in the case that the radii are equal, it holds:
\begin{equation}
D^{a}_{\ bc}=\frac{1}{2}f^{a}_{\ bc}\ .\label{Dequalradii}
\end{equation}
Moreover, the connection form $\omega^{a}_{\ b}$ is $S$-invariant, meaning that a parallel transport commutes with the $S$ action \cite{Castellani:1999fz}. Consequently, the most general expression of an $S$-invariant~connection on $S/R$ is: 
\begin{equation}
\omega^{a}_{\ b} = f_{\ \ ib}^{a}e^{i}+J_{\ \ cb}^{a}e^{c}\ ,
\end{equation}
with $J$ being a tensor that~is $R$-invariant, namely: $$ \delta J_{cb}^{\ \ a}
=-f_{ic}^{\ \ d}J_{db}^{\ \ a}+ f_{id}^{\ \ a}J_{cb}^{\ \
d}-f_{ib}^{\ \ d}J_{cd}^{\ \ a}=0. $$ It can be easily shown -using the Jacobi identity- that~the $D$ tensor, given in \refeq{TheDs}, satisfies the above~condition.

\noindent Then, for the case of non-vanishing torsion the expression is: 
\begin{equation}
T^{a} = de^{a} + \Omega^{a}_{\ b} \wedge e^{b}\ ,
\end{equation}
where the~connection $\Omega^{a}_{\ b}$ is expressed as a decomposition~of two parts: $$\Omega^{a}_{\ b}=\omega^{a}_{\ b}+\tau^{a}_{\ b}\ .$$ The first is the connection of the torsionless~case  mentioned above and the second is given by:  
\begin{equation}
\tau^{a}_{\ b} = - \frac{1}{2} \Sigma^{a}_{\ bc}e^{c},
\end{equation}
where $ \Sigma^{a}_{\ \ bc} $ is the~contorsion, expressed as
\begin{equation}
\Sigma^{a}_{\ bc} = T^{a}_{\ \ bc}+T_{bc}^{\ \ a}-T_{cb}^{\ \ a}
\end{equation}
in terms of the torsion~components $T^{a}_{\ bc}$. Therefore, putting all the expressions of the connections together, the general connection form $\theta^{a}_{\ b}$ is~written as:
\begin{equation}
\Omega^{a}_{\ b} = -f^{a}_{\ ib}e^{i} -\left(D^{a}_{\
bc}+\frac{1}{2}\Sigma^{a}_{\ bc}\right)e^{c}= -f^{a}_{\
ib}e^{i}-G^{a}_{\ bc}e^{c}\ .\label{definitionofG}
\end{equation}
In the case of equal radii, it is possible to~express $T^{a}_{\ bc}$ as $T^{a}_{\ bc}=\eta f^{a}_{\ bc}$, where $\eta$ is a parameter \cite{Lust:1986ix,Castellani:1986rg,Gavrilik:1999xr,MuellerHoissen:1987cq,Batakis:1989gb}. A natural~generalization of the expression of  $T^{a}_{\ bc}$ in the case of unequal radii would be $T^{a}_{\ bc}=2\tau D^{a}_{\ bc}$, where $\tau$ is another parameter.~The latter was obtained after taking into consideration and generalizing the expression~of $D^a_{\ bc}$ for equal radii, \refeq{Dequalradii}. However,~generalizing the case of equal radii to the relation $T^{a}_{\ bc}=2\tau D^{a}_{\ bc}$ for different radii turns out to be insufficient, as the $D^{a}_{\ bc}$ fail to have the appropriate~symmetry properties. In order to ameliorate this~drawback, the $\Sigma$ is defined as~a combination of $D^{a}_{\ bc}$ in such a way that it is rendered~completely antisymmetric and, at the same time,~$S$-invariant:
\begin{equation}
\Sigma_{abc} \equiv 2\tau(D_{abc}+D_{bca}-D_{cba})\ .
\end{equation}
Furthermore, the curvature two-form is given by the~following expression \cite{Kapetanakis:1992hf}, \cite{MuellerHoissen:1987cq,Batakis:1989gb}:
\begin{equation}
R^{a}_{\ b}=\left[-\frac{1}{2}f_{ib}^{\ a}f_{de}^{~~i}-
\frac{1}{2}G_{cb}^{~~a}f_{de}^{~~c}+ \frac{1}{2}(G_{dc}^{~~
a}G_{eb}^{~~c}-G_{ec}^{~~a}G_{db}^{~~c})\right]e^{d} \wedge e^{e},
\end{equation}
while the Ricci~tensor, $R_{ab}=R^{d}_{\ adb}$, is obtained as:
\begin{equation}
R_{ab}=G_{ba}^{~~c}G_{dc}^{~~d}-G_{bc }^{~~d}G_{da }^{~~c}-G_{ca
}^{~~d}f_{db }^{~~c}-f_{ia }^{~~d}f_{db }^{~~i}\ .
\end{equation}
If we choose $\tau$ to vanish, then the corresponding connection is called \textit{Riemannian connection}, $\Omega_{R~b}^{~a}$. Another special case worth mentioning is the \textit{canonical connection}, which is obtained by adjusting both the radii and $\tau$ in such a way that the connection form is $\Omega_{C~~b}^{~a}=-f^{a}_{~bi}e^{i}$, that is an $R$-gauge field \cite{Lust:1986ix,Castellani:1986rg}. Also, the adjustments of the parameters are such that satisfy $G_{abc}=0$. In the case of
$SU(3)/(U(1) \times
U(1))$ with metric $g_{ab}=\text{diag}(a,a,b,b,c,c)$, one needs to
set $a=b=c$ and therefore consider that $\tau=-\frac{1}{3}$. Similar adjustments can cause the Ricci tensor \cite{Lust:1986ix,Castellani:1986rg} to vanish, with the corresponding connection called \textit{Ricci flattening connection}.

\section{Main Features and Constraints of CSDR}\label{featurescsdr}

We continue with the description of the main aspects of the CSDR that will be necessary in the next sections. 
Consider a Y-M theory in $D$ dimensions with gauge group by $G$. The action will be:
\begin{align}
S&=\int d^4xd^dy\sqrt{-g}\left[-\frac{1}{4}\text{Tr}(F_{MN}F_{K\Lambda})g^{MK}g^{N\Lambda}+\frac{i}{2}\bar{\psi} \Gamma^MD_{M}\psi\right]\,,\label{actioncsdr}
\end{align}
in which it is~evident from the last term that~fermionic fields have been introduced in the theory. the metric tensor of the $D$-dimensional spacetime is denoted $g_{MN}$\,\footnote{Upper case latin middle-alphabet~letters refer to the $D$-dimensional spacetime.} and $D_M$ is the covariant derivative expressed as
\begin{equation}
    D_M=\partial_M-\theta_M-igA_M\,, \,\text{where}\,\,\, \theta_M=\frac{1}{2}\theta_{MN\Lambda}\Sigma^{N\Lambda}\,
\end{equation}
is the spin connection~of the spacetime. Furthermore, $\psi$ is a spinor representing~the fermionic fields of the theory, which can be assigned~into an arbitrary representation $F$ of $G$. Nevertheless,~in a supersymmetric version of the theory, the fermions should~be accommodated in the adjoint representation along with $A_M$, which~is the gauge field related to $G$, forming a vector~supermultiplet. Subsequently, $F_{MN}$ is the field strength~tensor of $A_M$:
\begin{equation}
    F_{MN}=\partial_M A_N-\partial_NA_M-ig[A_M,A_N]\,.
\end{equation}
Now, let the Killing~vectors of $S/R$ be denoted as $\xi^{\alpha}_{A}$,~where $A=1,\ldots,\text{dim}S$ and $\alpha=\text{dim}R+1,\ldots,\text{dim}S$. Let us also consider a gauge~transformation, $W_A$, which is related to the Killing vector~$\xi_A$. The condition that all kinds of fields (scalars $\phi$,~spinors  $\psi$ and vectors $A_\alpha$) that live on the coset~space are symmetric is translated to the following constraints:
\begin{align}
\delta_A\phi&=\xi^{~\alpha}_{A}\partial_\alpha\phi=D(W_A)\phi,\label{ena} \\
\delta_AA_{\alpha}&=\xi^{~\beta}_{A}\partial_{\beta}A_{\alpha}+\partial_{\alpha}\xi^{~\beta}_{A}A_{\beta}=\partial_{\alpha}W_A-[W_A,A_{\alpha}], \label{duo}\\
\delta_A\psi&=\xi^{~\alpha}_{A}\partial_\alpha\psi-\frac{1}{2}G_{Abc}\Sigma^{bc}\psi=D(W_A)\psi\,.
\label{tria}
\end{align}
In the above, $D(W_A)$ is the gauge~transformation $W_A$ in the appropriate representation of the~group according to the representation in~which the various fields are assigned. The transformation~$W_A$, no matter the representation, depends only on~the coordinates of the coset space. The above equations of the~constraints \eqref{ena}-\eqref{tria} are particularly important, since their solutions determine both the gauge group and the~spectrum of the reduced, four-dimensional theory\footnote{After the reduction the fields are no longer considered constrained.} \cite{Forgacs:1979zs,Kapetanakis:1992hf}. Let us~now work through the above constraints in some detail.~First, the gauge field $A_M$ defined~on the $D$-dimensional spacetime can be written explicitly~in the split form $A_M=(A_\mu,A_\alpha)$, since $M^D=M^4\times S/R$.~Under an $S$-transformation the four-dimensional part $A_\mu$ follows the transformation~rule of a scalar field, which is~obviously accommodated in the adjoint representation~of the initial gauge group, $G$. The first constraint,~\eqref{ena}, leads to two enlightening results: 
\begin{itemize}
\item The gauge group of the four-dimensional theory is $H=C_G(R_G)$, that is the centralizer~of $R$ in $G$.
\item The first part of~the $(A_\mu, A_\alpha)$ is identified as the four-dimensional~gauge field $A_\mu$ and does not depend on the coordinates of~the coset space. Regarding the second component of the initial~gauge field, $A_\alpha$, it is identified as a set of~scalar fields in the four-dimensional theory, assuming~the role of intertwining operators among the representations~of $R$ in $G$ and in $S$. The latter is obtained by~solving the second constraint, \eqref{duo}. The following~decomposition of the initial gauge group $G$, along with the~decomposition of $S$ under its subgroup $R$, give insight~on the representation of the produced scalars in the four-dimensional~theory:
\begin{equation}\label{tria ena}
G\supset R_G\times H\,,\qquad
\mathrm{adj}G=(\mathrm{adj}R,1)+(1,\mathrm{adj}H)+\sum(r_i,h_i)\,,
\end{equation}
\begin{equation}\label{tria dyo}
S\supset R\,,\qquad
\mathrm{adj}S=\mathrm{adj}R+\sum s_i\,.
\end{equation}
Given that the~irreducible representations $r_i$ and $s_i$ of $R$ are identical,~the representation $h_i$ of $H$ accounts for a scalar multiplet.~Any scalars that~are not included in $h_i$~vanish.
\end{itemize}
In turn, the third~constraint in \eqref{tria} is related to the spinorial part~and its form and properties in the four-dimensional~theory. The line~of reasoning is similar to the one described above about~the scalars \cite{Kapetanakis:1992hf,Manton:1981es,Chapline:1982wy,
Wetterich:1982ed,Palla:1983re,Pilch:1984xx,Forgacs:1985vp}. By solving the third constraint it becomes apparent that the four-dimensional spinors, $\psi$, depend~exclusively on the coordinates of the four-dimensional~part. In addition, similarly to the~scalar fields, $\psi$ relates the induced representations~of $R$ between $SO(d)$ and $G$ as an intertwining~operator. The representation of the four-dimensional~gauge group$H$ in which fermions are assigned, denoted by $f_i$,~is determined by the decomposition rule of the representation~$F$ of $G$, that is the fermionic representation in~the higher-dimensional theory, with respect to $R_G\times H$~and the spinorial representation of $SO(d)$ under $R$:
\begin{equation}
 G\supset R_G\times H\,,\qquad F=\sum (r_i,f_i),
\end{equation}
\begin{equation}
SO(d)\supset R\,,\qquad \sigma_d=\sum \sigma _j\,.
\end{equation}
Similarly to the~scalar case examined above, the $r_i$ and $\sigma_i$~representations are effectively the same and for each pair~of those an~$f_i$ multiplet of four-dimensional spinors exists. 

At this point a few~comments on the nature of the fermions before and after~the dimensional reduction are in order. If the fermions~introduced in the theory in higher dimensions are considered to be Dirac-type, then the surviving~ones in the four-dimensional~theory will not be chiral, which is surely a most unwelcome feature. However,~considering the higher-dimensional theory~defined on an even-dimensional spacetime and imposing the Weyl condition,~the fermions of the resulting four-dimensional~theory will have the property of chirality.~In particular, fermions that are assigned to the adjoint~representation of a gauge group of a~higher-dimensional theory defined on a~spacetime of $D=2n+2$ dimensions and also obey~the Weyl constraint, eventually result to two~sets of chiral fermions in the four-dimensional theory with~the quantum numbers being identical for the components~of the two sets. If the condition imposed on the fermions of the higher-dimensional theory is extended to include the Majorana condition, then the doubling of the fermionic spectrum does not~appear in the four-dimensional theory. Imposing both the Weyl and the Majorana conditions is possible when the dimensions of~the initial theory are $D=4n+2$.

\section{The action of the reduced four-dimensional theory}\label{4daction}

Let us now focus~on the action of the four-dimensional theory, obtained after~the dimensional reduction. First, consider that the higher-dimensional~spacetime, $M^D$, is compactified as $M^4\times S/R$, where $S/R$ is a compact coset space. This kind of~compactification induces the following modification on the form~of the metric:
\begin{equation}
g^{MN}=\left(
         \begin{array}{cc}
           \eta^{\mu \nu} & 0 \\
           0 & -g^{ab} \\
         \end{array}
       \right)\,,\label{metric_comp}
\end{equation}
where the upper~left block is the (mainly negative) four-dimensional~Minkowski metric, while the lower right is the coset space~metric tensor. Introducing the above metric in the Y-M action~of the higher-dimensional theory given in \refeq{actioncsdr} and~taking into account the constraints \eqref{ena}-\eqref{tria},~the reduced action of the four-dimensional~theory is:
\begin{align}
    S=C\int d^4x\,\mathrm{tr}\left[-\frac{1}{8}F_{\mu\nu}F^{\mu\nu}-\frac{1}{4}(D_\mu\phi_a)(D^\mu\phi^a)\right]+V(\phi)+\frac{i}{2}\bar{\psi}\Gamma^\mu D_\mu\psi-\frac{i}{2}\bar{\psi}\Gamma^aD_a\psi \,,\label{tessera}
\end{align}
in~which:~
\begin{itemize}
    \item $C$ denotes the~coset space volume
    \item $D_\mu = \partial_{\mu}-igA_\mu$ is the covariant~derivative in four dimensions
    \item $D_a = \partial_{a}-\theta_a-ig\phi_a$ is the covariant~derivative of the coset space, $\theta_a = \tfrac{1}{2}\theta_{abc}\Sigma^{bc}$ its connection~and $\phi_a\equiv A_a$
    \item $V(\phi)$ is the four-dimensional~scalar potential which can be explicitly written~as:  
    \begin{align}
    V(\phi)=-\frac{1}{8}g^{ac}g^{bd}\mathrm{Tr}(f^{~~C}_{ab} \phi_C -ig[\phi_a,
    \phi_b])(f^{~~D}_{cd} \phi_D - ig[\phi_c, \phi_d])\,,\label{scalarpotentialafterreductiongeneralcase}
\end{align}
in which~$A=1,...,\mathrm{dim}S$ and $f^{~~C}_{ab}, f^{~~D}_{cd}$ are the~structure constants of the~algebra~of $S$, which is the group of isometries~of the coset.
\end{itemize}
Furthermore, the constraints \eqref{ena} - \eqref{tria} imply that for~the scalars $\phi_a$ we have:
\begin{align}
f^{~~D}_{ai}\phi_D-ig[\phi_a, \phi_i]=0\,, \label{tessera_ena}
\end{align}
in which~$\phi_i$ denote the generators of~the subgroup $R_G$. The scalar fields that survive the dimensional~reduction are identified as the Higgs sector of the four-dimensional theory. The vacuum~of the theory is obtained by the minimization of the scalar~potential $V(\phi)$ written down in \eqref{scalarpotentialafterreductiongeneralcase}~and eventually the~corresponding gauge symmetry is obtained \cite{Chapline:1980mr,Bais:1985yd,Kubyshin:1987cv,Harnad:1978vj,Harnad:1979in,Harnad:1980fz,Farakos:1986sm,Farakos:1986cj}. In general, the procedure~above is not straightforward and is indeed difficult to~carry out. Nevertheless, in the very special case in~which the isometry group of the coset space, $S$, has an~isomorphic image (denoted as $S_G$) into $G$, then the four-dimensional~gauge group $H$ breaks~spontaneously in a subgroup of $H$,~denoted $K$. The subgroup $K$ is obtained~by the centralizer of the~isomorphic image of $S$ in $G$ in the gauge group~$G$ of the higher-dimensional theory \cite{Kapetanakis:1992hf,Chapline:1980mr,Bais:1985yd,Kubyshin:1987cv,Harnad:1978vj,Harnad:1979in,Harnad:1980fz,Farakos:1986sm,Farakos:1986cj}.~The latter gets well-understood by the following~scheme:
\begin{align}
  G\supset &\,S_G\times K\nonumber\\
  &\,\cup\quad\,\,\,\cap    \nonumber\\
  G\supset&\,R_G\times H~.
\end{align}
In case of a~symmetric coset space (this means $f_{ab}^{~~c}=0$) the four-dimensional scalar~potential always takes a form which allows the~spontaneous symmetry breaking \cite{Kapetanakis:1992hf}. Although an unwelcome~feature of the application of the above~theorem is that the fermions of the four-dimensional~theory appear to become supermassive \cite{Barnes:1986ea}~(similarly to the Kaluza-Klein case), there is an~exceptional case in which fermions are prevented form acquiring superheavy masses. 

Examining~the fermionic part of the effective four-dimensional action, one~has to focus on the last two terms of \eqref{tessera}.~The first term is identified as the fermionic kinetic term,~while the second is related to Yukawa~interactions. Since the fermions of the~higher-dimensional theory are considered to be satisfying the~Weyl and Majorana conditions, then they it is possible to assign~them in a real representation of the initial gauge group.~In particular, the last term (Yukawa term) of the Lagrangian~of \refeq{tessera} can be expressed as:
\begin{align}
L_Y = -\frac{i}{2}\bar \psi \Gamma^{a}(\partial_a
-\frac{1}{2}f_{ibc}e_\Gamma^i e_a^\Gamma \Sigma^{bc} -
\frac{1}{2}G_{abc}\Sigma^{bc}-\phi_a)\psi=\frac{i}{2} \bar \psi
\Gamma^a \nabla_a \psi +\bar \psi V\psi\,,\label{pente}
\end{align}
in which it is~implied that
\begin{align}
\nabla_a &=-\partial_a +\frac{1}{2}f_{ibc}e_\Gamma^i e_a^\Gamma \Sigma^{bc}+\phi_a, \label{exi} \\
V &=\frac{i}{4}\Gamma^aG_{abc}\Sigma^{bc},\label{forgaugini}
\end{align}
making use of the~complete connection, that is the one including torsion~\cite{MuellerHoissen:1987cq,Batakis:1989gb}. As mentioned in~\refse{featurescsdr}, the third constraint \refeq{tria} implies~that they do not depend on the coordinates of the~coset space. Moreover, due to the invariance of the~Lagrangian under $S$-transformations, it is equivalent to work~with the Lagrangian in the special case in which $y=0$ and~$e_\Gamma^{~i}$. Under these conditions, \eqref{exi} is~reduced to the following form:
\begin{equation}
    \nabla_a=\phi_a
\end{equation}
The fermionic~fields are independent of the coset space coordinates~(i.e.~$\partial_a\psi=0$) since they are symmetric fields, satisfying the~constraint equation \refeq{tria}. Furthermore,~we can consider the~Lagrangian at the point $y=0$ due to its invariance under~$S$-tranformations and $e^i_{\Gamma}=0$ at that~point.~Therefore,
\refeq{exi} becomes just~$\nabla_a=\phi_a$ and the term
$\frac{i}{2} \bar \psi \Gamma^a \nabla_a \psi$ of~\refeq{pente} is identified exactly as a~Yukawa term. 

Further~examining the last term of \refeq{pente}, one can easily show~that $V$ and the six-dimensional operator of helicity are~anticommutative \cite{Kapetanakis:1992hf}, while, on the other~hand, $V$~and $T_i=-\frac{1}{2}f_{ibc}\Sigma^{bc}$\footnote{The $T_i$ generators~close
the $R$-subalgebra of $SO(6)$.} are commutative~quantities. Taking into consideration~Schur's lemma, it is deduced that the surviving terms of $V$~are those that are present in the decomposition of $4$ and~$\bar{4}$ irreducible representations of $SO(6)$, namely the~singlets. Due to their geometric origin the singlets~will obtain masses in a very large scale, a fact that most likely leads to problematic phenomenology. In particular, the gauginos that appear in the four-dimensional theory after the application of the CSDR theme on a supersymmetric higher-dimensional theory will obtain masses at the order of the compactification scale.
However, such an outcome is prevented by the inclusion of the torsion. This will be the case considered in the following construction.

\section{CSDR and the Einstein-Yang-Mills Theory}\label{einsteinYM}
Finally, consider the case of~the Einstein-Yang-Mills (EYM) theory in $4+d$ dimensions,~in which a cosmological constant is present. Such a system~is described by the following Lagrangian~density:
\begin{equation}
{\cal L}  = -\frac{1}{16 \pi G} \sqrt{-g}
R^{D}-\frac{1}{4e^{2}}\sqrt{-g}F^{a}_{\ MN}F^{aMN}- \sqrt{-g}
\Lambda\,.
\end{equation}
The corresponding~equations of motion are obtained:
\begin{eqnarray}\label{eq:Tmn}
R_{MN}-\frac{1}{2}Rg_{MN}&=&-8 \pi G T_{MN}\,,\nonumber\\
D_{M}F^{MN}&=&0\,.
\end{eqnarray}
in which the~energy-momentum tensor, $T_{MN}$, of \refeq{eq:Tmn}, has been~identified as:
\begin{equation}
T_{MN}=\frac{1}{e^{2}}F^{a}_{\ MP}F^{a\ P}_{\
N}-g_{MN}\left(\frac{1}{e^{2}}F^{a}_{~PS}F^{aPS}+\Lambda\right)\,.
\end{equation}
At first, Cremmer~and Scherk suggested that the above theoretical system~admits a spontaneous compactification, in other words, they~observed that there exist solutions if the EYM equations~that correspond to manifolds of the form $M^4\times B$.~Regarding the cosmological constant included in the $4+d$-dimensional~theory, it ceases to appear in the four-dimensional~theory since its role is restricted on counterbalancing~the vacuum energy of the gauge fields. In the specific case in~which the higher-dimensional theory is defined on a~spacetime of the $M^4\times S/R$ form, meaning that the extra~dimensions are described by a coset space, the corresponding~metric is of the following form:
\begin{eqnarray}
g_{MN}(x,y)=
\left(\begin{array}{cc}g_{\mu\nu}(x^{\mu})&0\\0&g_{mn}(y^{\alpha})\end{array}
\right),
\end{eqnarray}
where the coordinate~$x^M$ of the higher-dimensional spacetime has been considered~as a decomposition on the coordinates $x^\mu$ of the~four-dimensional Minkowski spacetime and the extra-dimensional~coset space ones, $y^\alpha$, namely $x^{M}=(x^{\mu},y^{\alpha})$.~In general, the gauge symmetry of the emerging four-dimensional~theory is at least $S \times H$, where $H = C_{G} (R)$. 

Nevertheless, the gauge~group of the four-dimensional theory cannot include the whole~isometry of the coset space \cite{Chatzistavrakidis:2007by}, \cite{Chatzistavrakidis:2007pp}.~Also, particularly for non-symmetric coset spaces,~the modification -due to the extra scalar fields- on the potential of the~four-dimensional theory is already known~\cite{Chatzistavrakidis:2007pp}.

Let us consider a theory~that is defined on a higher-dimensional~spacetime and compactification leads to a spacetime in which~the extra dimensions form a coset manifold, $S/R$. In such a~case, by making use of $S$-symmetric gauge fields it is~possible to result with exact solutions of the field equations.~Due to the product structure of the spacetime $M^D=M^4\times S/R$ the~expression of the curvature tensor, $R^D_{MN}$, comprises~of $R_{\mu \nu}$ and $R_{\alpha \beta}$ (the four-dimensional~and the coset space curvature tensors respectively) and is~thus expressed as $R^{D}=R^{4}+R^{d}$ (up to some index regulation~through vielbeins). Making the assumptions that the~four-dimensional part of the gauge fields is independent of the~coordinates of the coset space, i.e. $A_{M}=(A_{\mu}(x),A_{\alpha}(x,y))$~and that the coset space component of the gauge~field, $A_{\alpha}(x,y)$, is symmetric, namely:  
\begin{equation} \label{eq:58}
A_{\alpha}(x,y)=e^{A}_{\alpha}(y)\phi_{A}(x)\,,
\end{equation}
in which $A$ is an~$S$-algebra index and the generalized vielbein is written~as $e^{A}_{\alpha}\rightarrow(e^{a}_{\alpha}, e^{i}_{\alpha})$, the expression of~the mixed component of the field strength tensor is found to~be:
\begin{equation}\label{eq:kinsc}
F_{\mu \alpha}=e_{\alpha}^{~A}D_{\mu}\phi_{A}\,,
\end{equation}
from which, evidently,~the kinetic term of $\phi_{A}$ will emerge. Moreover, the~corresponding component related to the coset space will~be:
\begin{equation}\label{eq:pot}
\tilde{F}_{\alpha \beta}=e^{A}_{\alpha}e^{B}_{\beta}(f_{A B}^{\ \
C}\phi_{C}-[\phi_{A},\phi_{B}])=e^{A}_{\alpha}e^{B}_{\beta}F_{A
B}\,,
\end{equation}
an expression that~will be associated to the potential. As mentioned above,~in the case of interest the field equations can be obtained. Their~explicit form is:
\begin{equation}
e(y)D_{\mu}(\sqrt{-g}F^{\mu
\nu})+\sqrt{-g(x)}D_{\alpha}(e(y)F^{\alpha \nu})=0
\end{equation}
\begin{equation}\label{eq:gfe2}
e(y)D_{\mu}(\sqrt{-g}F^{\mu
\beta})+\sqrt{-g(x)}D_{\alpha}(e(y)C^{\alpha \beta})=0\,,
\end{equation}
in which $e(y)$ is~the determinant of the veilbeins of the coset space. Now,~replacing \refeqs{eq:kinsc} and \eqref{eq:pot} into~\refeq{eq:gfe2}, it is found that: 
\begin{equation}
e(y)D_{\mu}(\sqrt{-g}F^{\mu
\nu})+\sqrt{-g(x)}\{\partial_{\alpha}(e(y)e^{\alpha}_{A})D^{\nu}\phi^{A}
+e(y)e_{\alpha}^{A}e^{B}_{\beta}g^{\alpha \beta}[\phi_{A},
D^{\mu}\phi_{B}]\}=0\,.
\end{equation}
Demanding that~the Y-M equations in the vacuum: 
\begin{equation}\label{eq:Maxeq}
D_{\mu}(\sqrt{-g}F^{\mu \nu})=0
\end{equation}
are in effect and~also that $\phi_{A}$ have constant classical values, it follows~that:
\begin{equation}\label{eq:centr}
[A_{\mu},\phi_{A}]=0\,.
\end{equation}
Moreover, \refeq{eq:pot} leads~to the following requirements:
\begin{eqnarray}\label{eq:cons}
F_{ib}=f_{ib}^{\ \ c}\phi_{c}-[\phi_{i},\phi_{b}]=0,\nonumber\\
F_{ij}=f_{ij}^{\ \ k}\phi_{k}-[\phi_{i},\phi_{j}]=0,\nonumber\\
\frac{1}{2}f_{ab}^{\ \ c}F^{ab}-[\phi_{a},F^{ac}]=0\,.
\end{eqnarray}
The obtained solutions~above, \refeq{eq:cons}, lead to the conclusion that~consistency of the solutions of \refeq{eq:gfe2}~and the CSDR constraints is achieved. 
Next, let us turn~to the Einstein equations of the $D$-dimensional~spacetime after the compactification takes place. Making use of~\refeqs{eq:Maxeq}, \eqref{eq:centr} and \eqref{eq:cons} the~equations become:
\begin{equation}\label{eq:ext}
R_{\mu \nu}-\frac{1}{2}g_{\mu \nu}R^{(4)}=g_{\mu
\nu}\left[\frac{1}{2}R^{(d)}+6 \pi G \left(F_{\alpha
\beta}F^{\alpha \beta}\right)+ \Lambda \right]\,,
\end{equation}
\begin{equation}\label{eq:int}
R_{\alpha \beta}-\frac{1}{2}R^{(d)}=g_{\alpha \beta}\left[
\frac{1}{9}R^{(4)}+8 \pi G \left(\frac{1}{4}\mathrm{Tr}(F_{\alpha
\beta}F^{\alpha \beta})+\Lambda\right)\right]-8 \pi G
\mathrm{Tr}(F_{\alpha \gamma}F_{\beta}^{\ \gamma}).
\end{equation}
Considering the~Minkowski spacetime, $M^4$, as the four-dimensional~solution, the \refeq{eq:ext} yields: 
\begin{equation}\label{eq:Rd}
R^{(d)}=-16 \pi G\left[\frac{9}{4}\mathrm{Tr}(F_{\alpha \beta}F^{\alpha
\beta})+\Lambda \right]
\end{equation}
and the \refeq{eq:int} becomes:
\begin{equation}\label{eq:int1}
R_{\alpha \beta}=-8 \pi G \mathrm{Tr}(F_{\alpha \gamma}F_{\beta}^{\
\gamma})\,.
\end{equation}
Last, calculating~and then replacing the trace of \refeq{eq:int1}~into $R^{(d)}$ of \eqref{eq:Rd}, it is obtained~that:
\begin{equation}
\Lambda=\frac{1}{4}\mathrm{Tr}(F_{\alpha \beta}F^{\alpha \beta})\,.
\end{equation}
Concluding,~consideration of a suitable cosmological constant in the initial theory~leads to a compactified classical solution for the EYM~system of a $M^{4}\times S/R$ spacetime, where the four-dimensional~fields obey the CSDR constraints.
The cosmological constant~has to be equal to the minimum of the potential~of the theory, which in general is not zero, except in the case~where $S$ is in $G$ which is the case that will be examined in~the following. As a final remark, it should be noted~that the gauge couplings of a four-dimensional theory produced~by the reduction of a higher-dimensional one should depend,~in general, on spacetime coordinates. For instance, the~gravitational and gauge couplings are of the following~form:
\begin{equation}
    G^{(4)}=\frac{G}{V_B}\,,\quad g^{(4)}=\frac{g}{V_B}\,,
\end{equation}
where $V_B$ denotes~the volume of the space $B$.

\chapter{Dimensional Reduction of \texorpdfstring{$E_8$}{Lg} over \texorpdfstring{$SU(3)/U(1)\times U(1)$}{Lg}}\label{e8_coset}

An illustrative example of the CSDR scheme is the case of~an $\mathcal{N}=1$ supersymmetric $E_8$ YM theory that~is dimensionally reduced over the non-symmetric~coset space  $SU(3)/U(1)\times U(1)$ \cite{Lust:1985be,
Kapetanakis:1990xx,Kapetanakis:1990rz}.~The four-dimensional gauge group is determined  -in~accordance with \refeq{tria ena},~\refeq{tria dyo}- by the way~the $R=U(1)\times U(1)$ is embedded in $E_8$ according~to the following~decomposition:
\begin{align}
E_8\supset E_6\times SU(3)\supset E_6\times U(1)_A \times U(1)_B\,.
\label{efta}
\end{align}
As mentioned earlier~in \refse{featurescsdr} (first CSDR constraint), the~gauge group of the four-dimensional theory after the coset~space dimensional reduction of
$E_8$ under~$SU(3)/U(1)\times U(1)$ is obtained as the centralizer~of $R=U(1)\times U(1)$ in $G=E_8$, that is:
\begin{align}
H=C_{E_8}(U(1)_A\times U(1)_B)=E_6\times U(1)_A\times U(1)_B\,.
\end{align}
Moreover, the~second and third constraints of \refse{featurescsdr}~imply that the scalar and fermion fields that remain in the~four-dimensional theory are obtained by the decomposition of~the representation 248 -adjoint representation- of~$E_8$ under $U(1)_A\times U(1)_B$:
\begin{gather}
248=1_{(0,0)}+1_{(0,0)}+1_{(3,\tfrac{1}{2})}+1_{(-3,\tfrac{1}{2})}+1_{(0,-1)}+1_{(0,1)}+1_{(-3,-\tfrac{1}{2})}+1_{(3,-\tfrac{1}{2})}+\nonumber\\
78_{(0,0)}+27_{(3,\tfrac{1}{2})}+27_{(-3,\tfrac{1}{2})}+27_{(0,-1)}+
\overline {27}_{(-3,-\tfrac{1}{2})}+
\overline{27}_{(3,-\tfrac{1}{2})}+\overline{27}_{(0,1)}\,.
\label{okto}
\end{gather}
In order to result~with the representations of the surviving fields of the~four-dimensional theory, it is necessary to examine the~decompositions of the vector and spinor
representations~of $SO(6)$ under $R=U(1)_A\times U(1)_B$: 
\begin{align*}
(3,\tfrac{1}{2})+(-3,\tfrac{1}{2})+(0,-1)+(-3,-\tfrac{1}{2})+(3,-\tfrac{1}{2})+(0,1)
\end{align*}
and
\begin{align*}
(0,0)+(3,\tfrac{1}{2})+(-3,\tfrac{1}{2})+(0,-1)\,,
\end{align*}
respectively.~Therefore, the CSDR rules imply that the surviving~gauge fields of the initial gauge theory ($E_6\times
U(1)_A\times U(1)_B$) are~accommodated in three $\mathcal{N}=1$ vector~supermultiplets in the four-dimensional theory. Also, the~matter fields of the four-dimensional theory end up in~six chiral multiplets. Three of them are
$E_6$ singlets~carrying $U(1)_A\times U(1)_B$ charges, while the~rest are chiral multiplets, let us denote them as $A^i, B^i$ and~$C^i$, with $i=1,\ldots,27$~an $E_6$ index.

In turn, the~decomposition $S\supset R$, where $S=SU(3)$ and $R=U(1)\times U(1)$,~namely $SU(3)\supset U(1)\times U(1)$:
\begin{equation}
8=(0,0)+(0,0)+(3,\tfrac{1}{2})+(-3,\tfrac{1}{2})+(0,-1)+(-3,-\tfrac{1}{2})+(3,-\tfrac{1}{2})+(0,1)\,.
\end{equation}
implies the introduction~of the following set of generators for the $SU(3)$:
\begin{align}
Q_{SU(3)}=\big\{Q_0,Q_0^\prime,Q_1,Q_2,Q_3,Q^1,Q^2,Q^3 \big\}.
\label{okto_duo}
\end{align}
The above generators~obey non-trivial commutation relations which can be found~in \cite{Manousselis:2001re}. The above notation of the~generators suggests to denote the scalar fields as:
\begin{align}
(\phi_I, I=1,...,8)\longrightarrow
(\phi_0,\phi_0^\prime,\phi_1,\phi^1,\phi_2,\phi^2,\phi_3,\phi^3).
\label{ennia}
\end{align}
Adopting the above~notation of the scalar fields, the scalar potential of any~theory that results from the reduction under the $SU(3)/U(1)\times U(1)$~is: 
\begin{align}
\frac{2}{g^2}V(\phi)=&(3\Lambda^2+\Lambda^{'2})\left(\frac{1}{R_1^4}+\frac{1}{R_2^4}\right)+\frac{4\Lambda^{'2}}{R_3^4}\nonumber\\
&+\frac{2}{R_2^2R_3^2}\mathrm{Tr}(\phi_1\phi^1)+\frac{2}{R_1^2R_3^2}\mathrm{Tr}(\phi_2\phi^2)+\frac{2}{R_1^2R_2^2}\mathrm{Tr}(\phi_3\phi^3)\nonumber\\
&+\frac{\sqrt{3}\Lambda}{R_1^4}Tr(Q_0[\phi_1,\phi^1])-\frac{\sqrt{3}\Lambda}{R_2^4}\mathrm{Tr}(Q_0[\phi_2,\phi^2])-\frac{\sqrt{3}\Lambda}{R_3^4}\mathrm{Tr}(Q_0[\phi_3,\phi^3]) \nonumber\\
&+\frac{\Lambda^{'}}{R_1^4}\mathrm{Tr}(Q_0^{'}[\phi_1,\phi^1])+\frac{\Lambda^{'}}{R_2^4}\mathrm{Tr}(Q_0^{'}[\phi_2,\phi^2])-\frac{2\Lambda^{'}}{R_3^4}\mathrm{Tr}(Q_0^{'}[\phi_3,\phi^3]) \nonumber\\
&+\Big[\frac{2\sqrt{2}}{R_1^2R_2^2}\mathrm{Tr}(\phi_3[\phi_1,\phi_2])+\frac{2\sqrt{2}}{R_1^2R_3^3}\mathrm{Tr}(\phi_2[\phi_3,\phi_1])+\frac{2\sqrt{2}}{R_2^2R_3^2}\mathrm{Tr}(\phi_1[\phi_2,\phi_3])+h.c\Big] \nonumber\\
&+\frac{1}{2}\mathrm{Tr}\Big(\frac{1}{R_1^2}([\phi_1,\phi^1])+\frac{1}{R_2^2}([\phi_2,\phi^2])+\frac{1}{R_3^2}([\phi_3,\phi^3])\Big)^2\nonumber\\
&-\frac{1}{R_1^2R_2^2}\mathrm{Tr}([\phi_1,\phi_2][\phi^1,\phi^2])-\frac{1}{R_1^2R_3^2}\mathrm{Tr}([\phi_1,\phi_3][\phi^1,\phi^3])\nonumber\\
&-\frac{1}{R_2^2R_3^2}\mathrm{Tr}([\phi_2,\phi_3][\phi^2,\phi^3]),
\label{ennia_duo}
\end{align}
where $R_1$, $R_2$, $R_3$ are the~radii of the coset space\footnote{The potential~is expressed in this form making use of (A.22)~of \cite{Kapetanakis:1992hf} but also the equations (7),(8)~of \cite{Witten:1985xb}.}. The real metric tensor\footnote{The complex~metric tensor
that was used~is $g^{1\bar{1}}=\frac{1}{R_1^2},g^{2\bar{2}}=\frac{1}{R_2^2},g^{2\bar{3}}=\frac{1}{R_3^2}$.}
of the coset~space with respect to the radii is:
\begin{align}
g_{ab}=diag(R_1^2,R_1^2,R_2^2,R_2^2,R_3^2,R_3^2)\,.
\end{align}
It is now meaningful to write down the generators of~the initial group, $E_8$, in accordance to the~decomposition of \refeq{okto}, as~follows:
\begin{align}
Q_{E_8}=\big\{Q_0,Q_0^{'}
,Q_1,Q_2,Q_3,Q^1,Q^2,Q^3,Q^{\alpha},Q_{1i},Q_{2i},Q_{3i},
Q^{1i},Q^{2i},Q^{3i} \big\}\,,\label{E8generators}
\end{align}
where the indices~take the values $\alpha =1,...,78$ and $i=1,...,27$. According~to the~\refeq{tessera_ena} the redefined fields obey~the following constraints:
\begin{align}
[\phi_1,\phi_0] &= \sqrt{3}\phi_1, & [\phi_3,\phi_0]&=0, & [\phi_1,\phi_0^{'}]&=\phi_1, \nonumber\\
[\phi_2,\phi_0] &= -\sqrt{3}\phi_2, & [\phi_2,\phi_0^{'}]&=\phi_2, &
[\phi_3,\phi_0^{'}]&=-2\phi_3\,.
\end{align}
Solving the above~constraints in terms of the genuine Higgs scalar fields and~$E_8$ generators, \eqref{E8generators} corresponding to~the embedding of $R=U(1)\times U(1)$ in the $E_8$, \eqref{okto} are~$\phi_0=\Lambda Q_0$ and $\phi_0^{'}=\Lambda^{'}  Q_0^{'}$, with~$ \Lambda =\Lambda^{'}=\frac{1}{\sqrt{10}}$~and
\begin{align*}
\phi_1 &= R_1 \alpha^i Q_{1i}+R_1 \alpha Q_1, & \phi_2 &= R_2
\beta^i Q_{2i}+R_2\beta Q_2, & \phi_3 &= R_3\gamma^iQ_{3i}+R_3\gamma
Q_3,
\end{align*}
with the unconstrained~fields transforming under $E_6\times
U(1)_A\times U(1)_B$ as:
\begin{align*}
\alpha_i \sim 27_{(3,\frac{1}{2})}, \quad  \beta_i \sim
27_{(-3,\frac{1}{2})}, \quad  \gamma_i \sim 27_{(0,-1)}, \quad
\alpha \sim 1_{(3,\frac{1}{2})}, \quad  \beta \sim
1_{(-3,\frac{1}{2})}, \quad  \gamma \sim 1_{(0,-1)}\,,
\end{align*}
the potential~of \refeq{ennia_duo} is expressed as:
\begin{align}
\frac{2}{g^2}V&(\alpha^i,\alpha,\beta^i,\beta,\gamma^i,\gamma)=\nonumber\\ &~~\frac{2}{5}\left(\frac{1}{R_1^4}+\frac{1}{R_2^4}+\frac{1}{R_3^4}\right)\nonumber\\
&+\bigg(\frac{4R_1^2}{R_2^2 R_3^2}-\frac{8}{R_1^2}\bigg)\alpha^i \alpha_i + \bigg(\frac{4R_1^2}{R_2^2 R_3^2}- \frac{8}{R_1^2}\bigg)\bar{\alpha} \alpha \nonumber\\
&+\bigg(\frac{4R_2^2}{R_1^2 R_3^2}-\frac{8}{R_2^2}\bigg)\beta^i \beta_i +\bigg(\frac{4R_2^2}{R_1^2 R_3^2}-\frac{8}{R_2^2}\bigg)\bar{\beta} \beta \nonumber\\
&+\bigg(\frac{4R_3^2}{R_1^2 R_2^2}-\frac{8}{R_3^2}\bigg)\gamma^i \gamma_i +\bigg(\frac{4R_3^2}{R_1^2 R_2^2}-\frac{8}{R_3^2}\bigg)\bar{\gamma} \gamma \nonumber\\
&+\bigg[\sqrt{2}80 \bigg(\frac{R_1}{R_2 R_3}+\frac{R_2}{R_1 R_3}+\frac{R_3}{R_2 R_1}\bigg)d_{ijk}\alpha^i \beta^j \gamma^k
+\sqrt{2}80\bigg(\frac{R_1}{R_2 R_3}+\frac{R_2}{R_1 R_3}+\frac{R_3}{R_2 R_1}\bigg)\alpha \beta \gamma +h.c \bigg] \nonumber \\
&+ \frac{1}{6}\bigg( \alpha^i(G^\alpha)_i^j\alpha_j+\beta^i(G^\alpha)_i^j\beta_j+\gamma^i(G^\alpha)_i^j\gamma_j\bigg)^2 \nonumber\\
&+\frac{10}{6}\bigg( \alpha^i(3\delta_i^j)\alpha_j+\bar{\alpha}(3)\alpha+\beta^i(-3\delta_i^j)\beta_j+\bar{\beta}(-3)\beta \bigg)^2 \nonumber\\
&+\frac{40}{6}\bigg( \alpha^i(\tfrac{1}{2}\delta_i^j)\alpha_j+\bar{\alpha}(\tfrac{1}{2})\alpha+\beta^i(\tfrac{1}{2}\delta_i^j)\beta_j+\bar{\beta}(\tfrac{1}{2})\beta+\gamma^i(-1\delta_i^j)\gamma^j+\bar{\gamma}(-1)\gamma \bigg)^2 \nonumber\\
&+40\alpha^i \beta^j d_{ijk}d^{klm} \alpha_l \beta_m+40\beta^i
\gamma^j d_{ijk}d^{klm} \beta_l
\gamma_m+40 \alpha^i \gamma^jd_{ijk} d^{klm} \alpha_l \gamma_m \nonumber\\
&+40(\bar{\alpha}\bar{\beta})(\alpha\beta)+40(\bar{\beta}\bar{\gamma})(\beta\gamma)+40(\bar{\gamma}\bar{\alpha})(\gamma
\alpha)\,,\label{deka}
\end{align}
which is positive definite as well. In the above expression~of the scalar potential~the $F-, D-$ and soft~supersymmetry breaking (SSB) terms are~identified. The $F$-terms emerge~from the superpotential:
\begin{align}
\mathcal{W}(A^i,B^j,C^k,A,B,C)=\sqrt{40}d_{ijk}A^iB^jC^k+\sqrt{40}ABC\,,
\end{align}
while the $D$-terms are~structured as:
\begin{align}
\frac{1}{2}D^{\alpha}D^{\alpha}+\frac{1}{2}D_1D_1+\frac{1}{2}D_2D_2\,,
\end{align}
where the $D$ quantities~are calculated as:
\begin{align}
D^{\alpha}&=\frac{1}{\sqrt{3}}\Big(\alpha^i(G^{\alpha})_i^j\alpha_j+\beta^i(G^{\alpha})_i^j\beta_j+\gamma^i(G^{\alpha})_i^j\gamma_j\Big),\label{DAinitial}\\
D_1&=\sqrt{\frac{10}{3}}\Big(\alpha^i(3\delta_i^j)\alpha_j+\bar{\alpha}(3)\alpha+\beta^i(-3\delta_i^j)\beta_j+\bar{\beta}(-3)\beta\Big)\label{D1initial}\\
D_2&=\sqrt{\tfrac{40}{3}}\Big(\alpha^i(\tfrac{1}{2}\delta_i^j)\alpha_j+\bar{\alpha}(\tfrac{1}{2})\alpha+\beta^i(\tfrac{1}{2}\delta_i^j)\beta_j+\bar{\beta}(\tfrac{1}{2})\beta+\gamma^i(-1\delta_i^j)\gamma_j+\bar{\gamma}(-1)\gamma\Big)\label{D2initial}\,.
\end{align}
Apart from the terms of~the potential of \refeq{deka} identified as $F$- and $D$- terms, the~remaining ones admit the interpretation of soft scalar masses~and trilinear soft terms. 
Last but not least, in order to complete the description of the (emerging) SSB sector, one has to determine the gaugino mass. This is done by calculating the $V$ operator of \refeq{forgaugini}, using \refapp{appendixB}. Moreover,  using the $\Gamma$ matrices given in \refapp{appendixA} we calculate $\Sigma^{ab}=\frac{1}{4}[\Gamma^a\Gamma^b]$ and then $G_{abc}\Gamma^a\Gamma^{bc}$. The combination~ of all leads~to the gaugino mass:
\begin{equation}
    M=(1+3\tau)\frac{R_1^2+R_2^2+R_3^2}{8\sqrt{R_1^2R_2^2R_3^2}}\,.
\end{equation}\label{gauginomass}
Thus, since there is a contribution from the torsion of the coset space\footnote{It should be noted that the adjustment required to obtain the \textit{canonical connection} leads to vanishing gaugino masses.}, the gaugino mass can in principle be at a different energy scale than the rest of the soft terms, since they do not receive any contribution from torsion. This is due to the fact that~gauge fields, contrary to fermions, do not couple~to torsion.

The next step is the minimization of the potential. This happens if two of the three singlets ($\alpha, \beta, \gamma$) acquire vevs at the compactification scale. For the purposes of the current work we choose the singlets $\alpha$ and $\beta$ to acquire vevs, while $\gamma$ remains massless. The reason is that $\gamma$ is the only singlet that will survive the Wilson breaking described in the next chapter and we will needed in the construction of higher-dimensional operators as we will see in \refcha{selection-breaking}.

This results in the breaking of the two $U(1)$, reducing our gauge group from $E_6\times U(1)_A\times U(1)_B$ to $E_6$. The two abelian groups remain, however, as global symmetries, which, as it will also be discussed in \refcha{selection-breaking}, will be very useful in conserving Baryon number. This does not come as a surprise, as in our case $S \supset G$, so the gauge group $H$ further breaks to $K=SU(3)^3$. 

It is useful at this point to make a \textit{brief recap}:
starting with an $\mathcal{N} = 1$ supersymmetric, $10D$ $E_8$ theory defined on a $M^4\times S/R$ manifold, where $S/R$ is the non-symmetric coset space$SU(3)/U(1)\times U(1)$, the CSDR leads to the content of an $N=1$ supersymmetric $E_6\times U(1)_A\times U(1)_B$ GUT. Subsequently, the two $U(1)$ groups are broken and end up parametrizing~global symmetries, leaving an $N=1$ $E_6$ theory. In order to reduce our gauge symmetry to a simpler one (which will turn out to be $SU(3)\times SU(3)\times SU(3)$), we have to break our theory using the Wilson flux mechanism.

\newpage
\mbox{} 
\newpage

\chapter{Wilson Flux Breaking}\label{wilson}

In the above~section the case in which the CSDR scheme is applied~on a higher-dimensional $E_8$ gauge theory which is reduced~over an $SU(3)/U(1)\times U(1)$ coset space and leads to~a four-dimensional $E_6$ gauge theory was examined in detail. However, the $E_6$ group cannot be broken exclusively by the presence~of the $27$ Higgs multiplet. For this reason, that is to~further reduce the resulting gauge symmetry, the Wilson flux~breaking mechanism is employed \cite{Hosotani:1983xw,Zoupanos:1987wj,Kozimirov:1989kn}. First, some~information about the mechanism is given and then its~application on the above model follows.

\section{Wilson Flux basics}\label{wilsonbasics}

In the above sections the dimensional reduction occurs over a simply connected manifold, i.e. a~coset space, $B_0=S/R$. Instead, there is an alternative~choice in which the manifold can be chosen to be multiply~connected, namely the $B=B_0/F^{S/R}$, where~$F^{S/R}$ is a freely-acting discrete symmetry of the manifold $B_0$. For~an element $g\in F^{S/R}$, an element~$U_g$ in $H$ is corresponded, which may be considered~as the Wilson Loop:
\begin{equation}
U_g={\mathcal{P}}exp\left(-i\oint_{\gamma_g} T^a A^{~a}_M dx^M \right),
\end{equation}
where $A^{~a}_M$ are the~gauge fields, $T^a$ the generators of the group,~$\gamma_g$ is a contour representing the element $g$ of~$F^{S/R}$ and $\mathcal{P}$ standing before the integration denotes~the path ordering. In case that the~considered manifold is simply connected,~considering the field strength tensor to be vanishing implies~that the gauge field could be set to zero by a gauge transformation.~Choosing $\gamma_g$ to be non-contractible to a~point leads to $U[\gamma]\neq 1$ and is also gauge covariant.~In this case, it is understood that the vanishing of the vacuum~field strength does not lead to $U_g=1$ and also the gauge field~cannot be gauge fixed to zero. Therefore, a homomorphism~of the discrete symmetry $F^{S/R}$ into~$H$ is induced, with image denoted as $T^H$, which is actually the subgroup of $H$~generated by {$U_g$}. Moreover, consider~a field $f(x)$ on $B_0$. It is obvious that~$f(x)$ is equivalent to another field~on $B_0$ for which it holds that $f(g(x))=f(x)$ for~every $g\in F^{S/R}$. Nevertheless, the presence of the gauge~group $H$ generalizes this statement to the following:
\begin{align}
f(g(x))=U_gf(x)\,.\label{dwdeka}
\end{align}
Now, about the gauge symmetry that~remains by the vacuum: in the vacuum state it holds~that $A_\mu^a=0$, and also consider a gauge transformation by the~coordinate-dependent matrix $V(x)$ of $H$. In order to~maintain $A_\mu^{~a}=0$ and preserve the invariance of~the vacuum, the $V(x)$~must be chosen to be constant.~Furthermore, $f\to Vf$ is consistent with~\refeq{dwdeka} only in case that:
\begin{equation}
 [V,U_g]=0\, \label{consistenconditionforwilsonflux}    
\end{equation}
for~every $g\in F^{S/R}$. Therefore, the subgroup~of $H$ that remains unbroken is the~centralizer of $T^H$ in $H$. As for the surviving matter fields~of the theory, i.e. the matter fields satisfying~\refeq{dwdeka}, they must be invariant~under the combination:
\begin{align*}
F^{S/R}\oplus T^H.
\end{align*}
Last, the freely-acting discrete symmetries, $F^{S/R}$, of the coset spaces $B_0=S/R$ are~the center of $S$, $Z(S)$ and $W=W_S/W_R$, where $W_S$~and $W_R$ are the Weyl groups of $S$ and $R$, respectively.~In the case examined in the present thesis the coset space is the $B_0=SU(3)/U(1)\times U(1)$, therefore it holds:
\begin{equation}
F^{S/R}=\mathbb{Z}_3  \subseteq W = S_{3}.
\end{equation}

\section{\texorpdfstring{$SU(3)^3$}{Lg} produced by Wilson Flux}\label{fluxsu33}

The Wilson~flux breaking mechanism~projects the theory in such a way that~the surviving fields~are those which remain invariant under~the action of the~freely acting discrete symmetry, the $\mathbf{Z}_3$, on~their gauge~and geometric indices. The non-trivial~action~of the $\mathbf{Z}_3$ group on the gauge indices of the~various~fields is parametrized by the matrix \cite{Chatzistavrakidis:2010xi}:
\begin{equation}
    \gamma_3=\text{diag}\{\mathbf{1}_3,\omega \mathbf{1}_3, \omega^2
\mathbf{1}_3\}\,,
\end{equation}
where 
\begin{equation}
\omega=e^{i\frac{2\pi}{3}}~.
\end{equation}
The latter~acts on the gauge fields of the $E_6$ gauge~theory~and a non-trivial phase acts on the matter fields. First,~the~gauge fields that pass through the filtering of the~projection~are those which satisfy the condition:
\begin{equation}
[A_M,\gamma_3]=0\;\;\Rightarrow A_M=\gamma_3 A_M \gamma_3^{-1}\label{filteringgaugefields}
\end{equation}
and the remaining~gauge symmetry~is 
\begin{equation}
SU(3)_c\times SU(3)_L \times SU(3)_R~.
\end{equation}                     
The matter~counterpart of \refeq{filteringgaugefields} for is:
\begin{equation}
\vec{\alpha}=\omega\gamma_3\vec{\alpha},\;\;\vec{\beta}=\omega^2
\gamma_3\vec{\beta},\;\;\vec{\gamma}=\omega^3 \gamma_3\vec{\gamma}\,,\qquad \alpha=\omega \alpha,\;\;\beta=\omega^2 {\beta},\;\;{\gamma}=\omega^3 {\gamma}\,.
\end{equation}\label{wilsonprojectionmatter}
where~$\vec \alpha,\vec \beta, \vec \gamma$ are the~matter superfields which belong~to the 27
representation and $\alpha,\beta,\gamma$ the singlets~that only carry $U(1)_{A,B}$~charges. 
The representations of~the remnant~group, $SU(3)_c\times SU(3)_L\times SU(3)_R$,~in which~the above fields are accommodated, are obtained~after~considering the decomposition rule of the 27~representation of $E_6$ under the new group and are given as
\begin{equation}
(1,3,\bar 3)\oplus(\bar3,1,3)\oplus(3,\bar 3,1)~.
\end{equation}
Therefore, in~the projected theory we are left with the following matter~content:
\begin{align*}
\alpha_3\equiv \Psi_1\sim (\bar{3},1,3)_{(3,\frac{1}{2})}, \;\;\; \beta_2\equiv
\Psi_2\sim (3,\bar{3},1)_{(-3,\frac{1}{2})}, \;\;\; \gamma_1\equiv \Psi_3\sim
(1,\bar{3},3)_{(0,-1)}, \;\;\; \gamma\equiv \theta_{(0,-1)},
\end{align*}
where the three former~are the leftovers of $\vec \alpha,\vec \beta, \vec \gamma$ and~together they form a 27 representation of $E_6$, that means that~the leftover content can be identified as one generation.~In order to obtain 
 a spectrum consisting of~three generations, one may introduce non-trivial~monopole~charges in the $U(1)$s in $R$, resulting in a total of three replicas~of the above~fields
(where an index $l = 1, 2, 3$ can~be used to specify each of the three~families).

The scalar potential~of~the $E_6$ (plus the global abelian symmetries) that was obtained~after the dimensional~reduction of $E_8$, \refeq{deka}, can now~(that is after the adoption of the~Wilson flux breaking~mechanism and the~projection of~the theory) be rewritten~in~the $SU(3)_c\times SU(3)_L\times SU(3)_R$ language as~\cite{Irges:2011de}:
\begin{align}
V_{sc}=3\cdot \frac{2}{5}\Big(\frac{1}{R_1^4}+\frac{1}{R_2^4}+\frac{1}{R_3^4}\Big)+\underset{l=1,2,3}{\sum}V^{(l)}\,,
\end{align}
in~which:
\begin{align}
V^{(l)}=V_{susy}+V_{soft}=V_D+V_F+V_{soft}\,.
\end{align}
From now on, we~give up on~the generation superscript $(l)$, since our analysis will be~focused on the third generation, and it will only be written~explicitly when required. Regarding the $D$- and $F$-terms,~they are identified as:
\begin{align}
V_D =\frac{1}{2}\underset{A}{\sum}D^AD^A+\frac{1}{2}D_1D_1+\frac{1}{2}D_2D_2, \label{dterm}\\
V_F =\underset{i=1,2,3}{\sum}|F_{\Psi_i}|^2+|F_{\theta}|^2, \;\;
F_{\Psi_i}=\frac{\partial\mathcal{W}}{\partial \Psi_i} ,\;\;
F_{\theta}=\frac{\partial\mathcal{W}}{\partial \theta}\,,\label{fterm}
\end{align}
where the $F$-terms~derive~from the expression:
\begin{align}
\mathcal{W}=\sqrt{40}d_{abc}\Psi_1^a\Psi_2^b\Psi_3^c\,,
\end{align}
while the $D$-terms~are written~explicitly as:
\begin{align}
D^A &=\frac{1}{\sqrt{3}}\big<\Psi_i|G^A|\Psi_i\big>, \label{DA} \\
D_1 &=3\sqrt{\frac{10}{3}}(\big< \Psi_1|\Psi_1\big>-\big<\Psi_2|\Psi_2\big>),\label{D1} \\
D_2 &=\sqrt{\frac{10}{3}}(\big<
\Psi_1|\Psi_1\big>+\big<\Psi_2|\Psi_2\big>-2\big<\Psi_3|\Psi_3\big>-2|\theta|^2)\,. \label{D2}
\end{align}
where
\begin{align*}
\big<\Psi_i|G^A|\Psi_i\big> &= \underset{i=1,2,3}{\sum}\Psi_i^a(G_A)_a^b\Psi_{ib}, \\
\big<\Psi_i|\Psi_i\big> &=  \underset{i=1,2,3}{\sum}\Psi_i^a\delta_a^b\Psi_{ib}.
\end{align*}
The $(G^A)_a^{~b}$ are~the structure~constants of the $SU(3)_c\times SU(3)_L\times SU(3)_R$~and~therefore antisymmetric in $a$ and $b$. Their explicit calculation can be found in \refapp{appendixC}.~The soft~supersymmetry~breaking terms are written down~as:
\begin{align}
V_{soft}=&\left(\frac{4R_1^2}{R_2^2R_3^2}-\frac{8}{R_1^2}\right)\big<\Psi_1|\Psi_1\big>+\left(\frac{4R_2^2}{R_1^2R_3^2}-\frac{8}{R_2^2}\right)\big<\Psi_2|\Psi_2\big>\nonumber \\
&+\left(\frac{4R_3^2}{R_1^2R_2^2}-\frac{8}{R_3^2}\right)(\big<\Psi_3|\Psi_3\big>+|\theta|^2)\nonumber \\
&+80\sqrt{2}\left(\frac{R_1}{R_2R_3}+\frac{R_2}{R_1R_3}+\frac{R_3}{R_1R_2}\right)(d_{abc}\Psi_1^a\Psi_2^b\Psi_3^c+h.c)\\
=& m_1^2\big<\Psi_1|\Psi_1\big>+m_2^2\big<\Psi_2|\Psi_2\big>+m_3^2\Big(\big<\Psi_3|\Psi_3\big>+|\theta|^2\Big)+(\alpha_{abc}\Psi_1^a\Psi_2^b\Psi_3^c+h.c)\,.\label{softterms}
\end{align}
According~to \cite{Kephart:1981gf}, the~vectors of~the $27$ of $E_6$ can~be written in a more~convenient~form in the $SU(3)_c\times SU(3)_L\times SU(3)_R$~language,~that is in complex $3\times 3$ matrices.~Identification of:
\begin{align}
\Psi_1\sim (\bar{3},1,3)\rightarrow (q^c)_p^{~\alpha} , \;\;\Psi_2\sim
(3,\bar{3},1)\rightarrow (Q^{~a}_{\alpha}), \;\; \Psi_3\sim
(1,3,\bar{3})\rightarrow L_a^{~p}\,,\label{multipletcharges}
\end{align}
leads to the~following~relabeling and assignment of the particle content~of the~MSSM (and more) in the~above representation of~the model:
\begin{eqnarray*}
  q^c=\left(\begin{array}{ccc}
 d^{c1}_R & u^{c1}_R & D^{c1}_R \\
 d^{c2}_R & u^{c2}_R & D^{c2}_R \\
 d^{c3}_R & u^{c3}_R & D^{c3}_R
 \end{array}
 \right)\,,\,\, Q=\left(\begin{array}{ccc}
 -d^1_L & -d^2_L & -d^3_L \\
 u^1_L & u^2_L & u^3_L \\
 D^1_L & D^2_L & D^3_L
 \end{array}\right)\,,\,\, L=\left(\begin{array}{ccc}
 H_d^0 & H_u^+ & \nu_L \\
 H_d^- & H_u^0 & e_L \\
 \nu^c_R & e^c_R & S
 \end{array}\right)\,.\label{multiplets}
\end{eqnarray*}
It is evident~from the~above that $d_{L,R},u_{L,R},D_{L,R}$ transform as $3, \bar{3}$~under the
colour group.

\noindent In turn, as~suggested in \cite{Kephart:1981gf}, the following expressions are taken into account:
\begin{align}
(\hat{q}^c)_{\alpha}^{~p}=\frac{1}{3}\frac{\partial I_3}{\partial
(q^{c})_p^{~\alpha}}, \;\;  
\hat{Q}^{~\alpha}_a=\frac{1}{3}\frac{\partial I_3}{\partial
Q^{~a}_{\alpha}},\;\;
\hat{L}_{p}^{~a}=\frac{1}{3}\frac{\partial
I_3}{\partial L^{~p}_a}\,,\label{orf420}
\end{align}
where $I_3$ is~an $E_6 $ trilinear invariant:
\begin{equation}
   I_3=d_{abc}A^{a}A^bA^c\,,\quad \bar{I}_3=d^{abc}A_aA_bA_c\,,
\end{equation}
where $A$ is a vector of~the $\mathbf{27}$ representation of $E_6$. The $I_3$ invariant~quantity can be written as a decomposition under~$SU(3)\times SU(3)\times SU(3)$ as:
\begin{equation}
I_3=\mathrm{det}[Q]+\mathrm{det}[q^c]+\mathrm{det}[L]-\mathrm{tr}(q^c\cdot L\cdot Q)\,.
\end{equation}
The $\langle \Psi_i|\Psi_i\rangle$ that~appear in the calculation of the $D$-terms,~may now be written in terms of the above matrices~as: 
\begin{equation}
    \big< \Psi_1|\Psi_1\big>=\mathrm{tr}(q^{c\dagger}q^c),\quad \big< \Psi_2|\Psi_2\big> =\mathrm{tr}(Q^{\dagger}Q),\quad \big<\Psi_3|\Psi_3\big> =\mathrm{tr}(L^{\dagger}L)
\end{equation} 
and~also:
\begin{align*}
I_3&=d_{abc}\Psi_1^a\Psi_2^b\Psi_3^c=\mathrm{det}q^c+\mathrm{det}Q+\mathrm{det}L-\mathrm{tr}(q^cQL)\\
\bar{I}_3&=d^{abc}\Psi_{1a}\Psi_{2b}\Psi_{3c}=\mathrm{det}q^{c\dagger}+\mathrm{det}Q^{\dagger}+\mathrm{det}L^{\dagger}-\mathrm{tr}((q^c)^{\dagger}Q^{\dagger}L^{\dagger})\,.
\end{align*}
The expression~of the potential due to the $F$-terms in terms of the $3\times 3$~complex matrices is:
\begin{align}
V_F=40d_{abc}d^{cde}(\Psi_1^a\Psi_2^b\Psi_{1d}\Psi_{2e}+\Psi_2^a\Psi_3^b\Psi_{2d}\Psi_{3e}+\Psi_1^a\Psi_3^b\Psi_{1d}\Psi_{3e})\label{orf423}
\end{align}
and may now be written~in terms of the above hatted quantities as: 
\begin{align}
V_F=360
\mathrm{tr}(\hat{q}^{c^{\dagger}}\hat{q}^c+\hat{Q}^{\dagger}\hat{Q}+\hat{L}^{\dagger}\hat{L})\,.\label{ftermspotentialhattedquantities}
\end{align}
The explicit calculation of the above, including $D$-, $F$- and soft terms, can be found in \refapp{appendixD}.

\newpage
\mbox{} 
\newpage

\chapter{Selection of Parameters and GUT breaking}\label{selection-breaking}

With the above-mentioned~theoretical framework fully in place, it is time~to specify~the compactification scale of the theory, as well as other~(resulting) quantities, in order to proceed~to phenomenology.

\section{Choice of radii}\label{radii}
We will examine the case~where the compactification scale is high\footnote{In this case~Kaluza-Klein excitations are irrelevant. Otherwise one~would need the eigenvalues of the Dirac and Laplace operators~in the $6D$ compactification space.}, and more specifically~$M_C=M_{GUT}$. Thus for the radii we have~$R_l\sim \frac{1}{M_{GUT}}~, ~ l=1,2,3$.

Without any special~treatment, this results in soft trilinear couplings and soft~scalar masses around $M_{GUT}$. However, we can select our third~radius slightly different than the other two in a way that~yields:
\begin{equation}
m_3^2\sim-\mathcal{O}(TeV^2),~~~~m_{1,2}^2\sim-\mathcal{O}(M_{GUT}^2),~~~~a_{abc}\gtrsim M_{GUT}~.
\end{equation}
In other words, we have~supermassive squarks and $TeV$-scaled sleptons. Thus, supersymmetry~is softly broken already at the unification scale, in addition~to its breaking by both $D$-terms and $F$-terms.

\section{Further gauge symmetry breaking of $SU(3)^3$}\label{breakings}
The spontaneous breaking~of the $SU(3)_L$ and $SU(3)_R$ can be
triggered by the following~vevs of the two families of $L$'s.
\[
\langle L_s^{(3)}\rangle=\left(\begin{array}{ccc}
0 & 0&0\\
0&0&0\\
0&0&V
\end{array}\right),\;\;
\langle L_s^{(2)}\rangle=\left(\begin{array}{ccc}
0 & 0&0\\
0&0&0\\
V&0&0
\end{array}\right)~,
\]
where the $s$ index denotes~the scalar component of the multiplet. These vevs are~singlets under $SU(3)_c$, so they leave the colour group~unbroken.~If we use only $\langle L_s^{(3)}\rangle$ we~get the breaking
\begin{equation}
SU(3)_c\times SU(3)_L\times SU(3)_R \rightarrow SU(3)_c\times SU(2)_L\times SU(2)_R\times U(1)~,
\end{equation}
while if we use~only $\langle L_s^{(2)}\rangle$ we get the breaking 
\begin{equation}
SU(3)_c\times SU(3)_L\times SU(3)_R \rightarrow SU(3)_c\times SU(2)_L\times SU(2)'_R\times U(1)'~.
\end{equation}
Their combination~gives the desired breaking \cite{Babu:1985gi}:
\begin{equation}\label{desiredbreaking}
SU(3)_c\times SU(3)_L\times SU(3)_R \rightarrow SU(3)_c\times SU(2)_L\times U(1)_Y~.
\end{equation}

\vspace{0.5cm}

\noindent Let us examine this unusual breaking in more detail. The breaking of the gauge~sector of the theory comes from the term $\mathrm{Tr}|D_{\mu}L_0^{(i)}|^2$ , where
\begin{equation}
D_{\mu}L_s=\partial_{\mu}L_0^{(i)}\underbrace{-ig_L(A_{\mu}^{\al}T^{\al})L_0^{(i)}}_{SU(3)_L}-\underbrace{iL_0^{(i)}g_RB_{\mu}^{\al}\overline{T}^{\al}}_{SU(3)_R}~,
\end{equation}
with $T^{\al}$ the Gell-Mann~matrices with $\al=1,...,8$ ($\overline{T}^{\al}=-{T^{\al}}^*$) and $A_{\mu}^{\al}$ and $B_{\mu}^{\al}$ the gauge bosons of each group.\footnote{At the last ($SU(3)_R$) term, $L_0$ is placed before $T^{\alpha}$, as $T^{\alpha}$ acts on the rows and not the columns of $L_0$.} 

\noindent For simplicity we abandon (temporarily) the $\mu$ indices:
\footnotesize
\begin{align}
&A^1T^1=\frac{1}{2}
\begin{pmatrix}
  0  & A^1 &0 \\
 A^1 &   0  &0 \\
   0  &   0  &0 \\
\end{pmatrix}, \pan 
A^2T^2=\frac{1}{2}
\begin{pmatrix}
    0 &-iA^2&0 \\
iA^2 &     0&0 \\
     0&     0&0 \\
\end{pmatrix}, \pan 
A^3T^3=\frac{1}{2}
\begin{pmatrix}
 A^3 &0     &0 \\
     0&-A^3 & 0\\
    0 &    0 & 0\\
\end{pmatrix}, \nonumber\\
&A^4T^4=\frac{1}{2}
\begin{pmatrix}
     0&0     & A^4 \\
     0& 0    &0 \\
 A^4 &     0& 0\\
\end{pmatrix}, \pan 
A^5T^5=\frac{1}{2}
\begin{pmatrix}
     0&     0& -iA^5\\
     0&     0&0 \\
iA^5 &     0&0 \\
\end{pmatrix}, \pan 
A^6T^6=\frac{1}{2}
\begin{pmatrix}
     0&0     &0 \\
     0&0     & A^6\\
     0& A^6 &0 \\
\end{pmatrix}, \nonumber\\
&A^7T^7=\frac{1}{2}
\begin{pmatrix}
     0&     0&0 \\
     0&     0&-iA^7\\
     0&iA^7 &0 \\
\end{pmatrix}, \pan
A^8T^8=\frac{1}{2\sqrt{3}}
\begin{pmatrix}
 A^8 &0     &0 \\
    0 & A^8 &0 \\
    0 & 0    &-2A^8\\
\end{pmatrix}
\end{align}\normalsize
and similarly for $B^{\al}T^{\al}$. Thus:
\begin{align}
A_{\mu}\equiv A_{\mu}^{\al}T^{\al}&=\frac{1}{2}
\begin{pmatrix}
A^3+\frac{1}{\sqrt{3}}A^8 & A^1-iA^2 & A^4-iA^5 \\
A^1+iA^2 & \frac{1}{\sqrt{3}}A^8-A^3 & A^6-iA^7 \\
A^4+iA^5 & A^6+iA^7 & -\frac{2}{\sqrt{3}}A^8 \\
\end{pmatrix}\\
B_{\mu}\equiv B_{\mu}^{\al}{T^{\al}}^*&=\frac{1}{2}
\begin{pmatrix}
B^3+\frac{1}{\sqrt{3}}B^8 & B^1+iB^2 & B^4+iB^5 \\
B^1-iB^2 & \frac{1}{\sqrt{3}}B^8-B^3 & B^6+iB^7 \\
B^4-iB^5 & B^6-iB^7 & -\frac{2}{\sqrt{3}}B^8 \\
\end{pmatrix}
\end{align}
and we (temporarily) redefine
\begin{equation}\label{uno}
-ig_LA_{\mu}L_0^{(i)}+iL_0^{(i)}g_RB_{\mu}\equiv -i\tilde{A}_{\mu}L_0^{(i)}+iL_0^{(i)}\tilde{B}_{\mu}~.
\end{equation}

\subsection*{First Breaking}

First, we attempt to use only one of the two multiplets that get a vev.
For simplicity we drop the Lorentz indices, as they 
We use $L_0^{(3)}$ first:
\begin{align*}
-i\tilde{A}_{\mu}\langle L_0^{(3)}\rangle &+i\langle L_0^{(3)}\rangle\tilde{B}_{\mu}=\\
&=-i\frac{V}{2}
\begin{pmatrix}
\phantom{\tal^4}0\phantom{i\tal^5}&\phantom{\tal^4}0\phantom{i\tal^5}&\tal^4-i\tal^5 \\
\phantom{\tal^4}0\phantom{i\tal^5}&\phantom{\tal^4}0\phantom{i\tal^5}&\tal^6-i\tal^7\\
\phantom{\tal^4}0\phantom{i\tal^5}&\phantom{\tal^4}0\phantom{i\tal^5}&-\frac{2}{\sqrt{3}}\tal^8\\
\end{pmatrix}\\
&\pan +i\frac{V}{2}
\begin{pmatrix}
0&0&0\\
0&0&0\\
\tbl^4-i\tbl^5 & \tbl^6-i\tbl^7 & -\frac{2}{\sqrt{3}}\tbl^8\\
\end{pmatrix}\\
&=-i\frac{V}{2}
\begin{pmatrix}
0&0&\tal^4-i\tal^5 \\
0&0&\tal^6-i\tal^7\\
-(\tbl^4-i\tbl^5) & -(\tbl^6-i\tbl^7) &\frac{2}{\sqrt{3}}(\tbl^8-\tal^8)\\
\end{pmatrix}
\end{align*}
Then the relevant part of its square is:
\begin{align*}
&|D_{\mu}L_0^{(3)}|^2=\frac{V^2}{4}
\begin{pmatrix}
0&0& -(\tbl^4-i\tbl^5) \\
0&0& -(\tbl^6-i\tbl^7) \\
 \tal^4-i\tal^5 & \tal^6-i\tal^7 &\frac{2}{\sqrt{3}}(\tbl^8-\tal^8)\\
\end{pmatrix}\\
&\phantom{aaaaaaaaa}\times\begin{pmatrix}
0&0&\tal^4-i\tal^5 \\
0&0&\tal^6-i\tal^7\\
-(\tbl^4-i\tbl^5) & -(\tbl^6-i\tbl^7) &\frac{2}{\sqrt{3}}(\tbl^8-\tal^8)\\
\end{pmatrix}\\
\phantom{aaaa}=&\frac{V^2}{4}
\begin{pmatrix}
|\tbl^4-i\tbl^5|^2 & 
\begin{matrix}
(\tbl^6-i\tbl^7)\times \\ \times(\tbl^4-i\tbl^5) \\ \\
\end{matrix} & 
\begin{matrix}
(\tbl^6-i\tbl^7)\times \\ \times(\tbl^4-i\tbl^5) \\ \\
\end{matrix} \\
\begin{matrix}
(\tbl^6+i\tbl^7)\times \\ \times(\tbl^4+i\tbl^5) \\ \\
\end{matrix} & 
|\tbl^6-i\tbl^7|^2 & 
\begin{matrix}
(\tbl^6+i\tbl^7)\times \\ \times(\tbl^4+i\tbl^5) \\ \\
\end{matrix} \\
\begin{matrix}
\frac{2}{\sqrt{3}}(\tal^8-\tbl^8)\times \\ \times(\tbl^4-i\tbl^5) \\ 
\end{matrix} & 
\begin{matrix}
\frac{2}{\sqrt{3}}(\tal^8-\tbl^8)\times \\ \times(\tbl^6-i\tbl^7) \\ 
\end{matrix} & 
\begin{matrix}
|\tal^4+i\tal^5|^2+ \\ +|\tal^6-i\tal^7|^2+ \\ +\frac{4}{3}(\tbl^8-\tal^8)^2 \\ 
\end{matrix}\\
\end{pmatrix}
\end{align*}
and (the relevant part of) its trace will be 
\begin{align}\label{trace1}
Tr|D_{\mu}L_0^{(3)}|^2=&\frac{V^2}{4}\Big((\tal^4)^2+(\tal^5)^2+(\tal^6)^2+(\tal^7)^2\nonumber\\
&+(\tbl^4)^2+(\tbl^5)^2+(\tbl^6)^2+(\tbl^7)^2\\
&+\frac{4}{3}(\tbl^8)^2+\frac{4}{3}(\tal^8)^2-\frac{4}{3}\tal^8\tbl^8-\frac{4}{3}\tbl^8\tal^8\Big)\nonumber
\end{align}
\begin{align}
=&\frac{V^2}{4}g_L^2\Big((A^4)^2+(A^5)^2+(A^6)^2+(A^7)^2\Big)\nonumber\\
&+\frac{V^2}{4}g_R^2\Big((B^4)^2+(B^5)^2+(B^6)^2+(B^7)^2\Big)\\
&+\frac{V^2}{4}\Big(\frac{4}{3}g_R^2(B^8)^2+\frac{4}{3}g_L^2(A^8)^2-\frac{4}{3}g_Lg_R(A^8B^8-B^8A^8)\Big)~.\nonumber
\end{align}
Concerning $A^8$ and $B^8$, we have:
\begin{align}V^2
\begin{pmatrix}
A^8 & B^8
\end{pmatrix}&
\begin{pmatrix}
\frac{1}{3}g_L^2 & -\frac{1}{3}g_Lg_R \\
-\frac{1}{3}g_Lg_R & \frac{1}{3}g_R^2
\end{pmatrix}
\begin{pmatrix}
A^8 \\ B^8
\end{pmatrix}\\=V^2
\begin{pmatrix}
A^8 & B^8
\end{pmatrix}P_1^{-1}P_1&
\begin{pmatrix}
\frac{1}{3}g_L^2 & -\frac{1}{3}g_Lg_R \\
-\frac{1}{3}g_Lg_R & \frac{1}{3}g_R^2
\end{pmatrix}P_1^{-1}P_1
\begin{pmatrix}
A^8 \\ B^8
\end{pmatrix}\\=V^2
\begin{pmatrix}
K^1 & K^2
\end{pmatrix}&
\begin{pmatrix}
0 & 0 \\
0 & \frac{1}{3}(g_L^2+g_R^2)
\end{pmatrix}
\begin{pmatrix}
K^1 \\ K^2
\end{pmatrix}~,\\
\end{align}
where
\begin{equation}
\begin{pmatrix}
K^1 \\ K^2
\end{pmatrix}=\frac{1}{\sqrt{g_L^2+g_R^2}}
\begin{pmatrix}
g_LB^8+g_RA^8 \\ g_RB^8-g_LA^8
\end{pmatrix}~.
\end{equation}
We see that $A^4,~A^5,~A^6,~A^7,~B^4,~B^5,~B^6,~B^7$ and $K^2$ acquired mass. So we are left with:
\begin{align}
A^1,~A^2,~A^3 ~ & ~ \rightarrow ~ SU(2)_L\\
B^1,~B^2,~B^3 ~ & ~ \rightarrow ~ SU(2)_R\\
K^1 ~ & ~ \rightarrow ~ U(1)
\end{align}
Counting in the 8 gauge bosons of $SU(3)_C$ we have 15 massless gauge bosons.

\subsection*{Second Breaking}

We will now use $L_0^{(2)}$ instead:
\begin{align*}
-i&\tilde{A}_{\mu}\langle L_0^{(2)}\rangle +i\langle L_0^{(2)}\rangle\tilde{B}_{\mu}=\\
&=-i\frac{V}{2}
\begin{pmatrix}
\tal^4-i\tal^5 &0& 0 \\
\tal^6-i\tal^7 &0&0\\
-(\tbl^3+\frac{1}{\sqrt{3}}\tbl^8)-\frac{2}{\sqrt{3}}\tal^8 & -(\tbl^1+i\tbl^2) & -(\tbl^4+i\tbl^5)\\
\end{pmatrix}
\end{align*}
and in a similar way:\small
\begin{align*}
&|D_{\mu}L_0^{(2)}|^2=\frac{V^2}{4}\times\\
&\begin{pmatrix}
\begin{matrix}
|\tal^4-i\tal^5|^2+|\tal^6-i\tal^7|^2+\\ +(\tbl^3+\frac{1}{\sqrt{3}}\tbl^8+\frac{2}{\sqrt{3}}\tal^8)^2\\ \\
\end{matrix} &
\begin{matrix}
\tbl^1+i\tbl^2)\times \\ \times(\tbl^3++\frac{1}{\sqrt{3}}\tbl^8+\frac{2}{\sqrt{3}}\tal^8) \\ \\
\end{matrix} &
\begin{matrix}
-(\tbl^4+i\tbl^5)\times \\ \times(\tbl^3++\frac{1}{\sqrt{3}}\tbl^8+\frac{2}{\sqrt{3}}\tal^8)\\ \\
\end{matrix} \\ 
\begin{matrix}
-(\tbl^1i\tbl^2)\times \\ \times (\tbl^3++\frac{1}{\sqrt{3}}\tbl^8+\frac{2}{\sqrt{3}}\tal^8)  \\ \\
\end{matrix}& 
\begin{matrix}
 |\tbl^1+i\tbl^2|^2
\end{matrix} &
\begin{matrix}
(\tbl^4+i\tbl^5)\times \\ \times(\tbl^1-i\tbl^2)  \\ \\
\end{matrix} \\
\begin{matrix}
\tbl^4-i\tbl^5)\times \\ \times(\tbl^3++\frac{1}{\sqrt{3}}\tbl^8+\frac{2}{\sqrt{3}}\tal^8) \\ 
\end{matrix} & 
\begin{matrix}
\tbl^1+i\tbl^2)\times \\ \times(\tbl^4-i\tbl^5)  \\ 
\end{matrix} & |\tbl^4+i\tbl^5|^2 \\
\end{pmatrix}~.
\end{align*}\normalsize
Its trace will be:
\begin{align}\label{trace2}
Tr|D_{\mu}L_0^{(2)}|^2=&\frac{V^2}{4}\Big((\tal^4)^2+(\tal^5)^2+(\tal^6)^2+(\tal^7)^2\nonumber\\
&+(\tbl^4)^2+(\tbl^5)^2+(\tbl^1)^2+(\tbl^2)^2\nonumber\\
&+(\tbl^3)^2+\frac{1}{3}(\tbl^8)^2+\frac{4}{3}(\tal^8)^2\nonumber\\
&+\frac{1}{\sqrt{3}}(\tbl^3\tbl^8+\tbl^8\tbl^3)+\frac{2}{\sqrt{3}}(\tbl^3\tal^8+\tal^8\tbl^3)\nonumber\\
&+\frac{2}{3}(\tbl^8\tal^8+\tal^8\tbl^8)\Big)\\
=&\frac{V^2}{4}g_L^2\Big((A^4)^2+(A^5)^2+(A^6)^2+(A^7)^2\Big)\nonumber\\
&+\frac{V^2}{4}g_R^2\Big((\tbl^1)^2+(\tbl^2)^2+(\tbl^4)^2+(\tbl^5)^2\Big)\nonumber\\
&+\frac{V^2}{4}\Big(g_R^2(B^3)^2+\frac{1}{3}g_R^2(B^8)^2+\frac{4}{3}g_L^2(A^8)^2\nonumber\\
&+\frac{1}{\sqrt{3}}g_R^2(B^3B^8+B^8B^3)+\frac{2}{\sqrt{3}}g_Lg_R(B^3A^8+A^8B^3)\nonumber\\
&+\frac{2}{3}g_Lg_R(B^8A^8+A^8B^8)\Big)
\end{align}
Concerning $A^8,~B^8$ and $B^3$, we have:
\begin{align}V^2
\begin{pmatrix}
A^8 & B^8 & B^3
\end{pmatrix}&
\begin{pmatrix}
\frac{1}{3}g_L^2 & \frac{1}{6}g_Lg_R & \frac{1}{2\sqrt{3}}g_Lg_R\\
 \frac{1}{6}g_Lg_R & \frac{1}{12}g_R^2 & \frac{1}{4\sqrt{3}}g_R^2 \\
\frac{1}{2\sqrt{3}}g_Lg_R & \frac{1}{4\sqrt{3}}g_R^2 & \frac{g_R^2}{4}
\end{pmatrix}
\begin{pmatrix}
A^8 \\ B^8 \\ B^3
\end{pmatrix}
\end{align}
\begin{align}
=V^2
\begin{pmatrix}
A^8 & B^8 & B^3
\end{pmatrix}P_2^{-1}P_2&
\begin{pmatrix}
\frac{1}{3}g_L^2 & \frac{1}{6}g_Lg_R & \frac{1}{2\sqrt{3}}g_Lg_R \\
 \frac{1}{6}g_Lg_R & \frac{1}{12}g_R^2 & \frac{1}{4\sqrt{3}}g_R^2 \\
\frac{1}{2\sqrt{3}}g_Lg_R & \frac{1}{4\sqrt{3}}g_R^2 & \frac{g_R^2}{4}
\end{pmatrix}P_2^{-1}P_2
\begin{pmatrix}
A^8 \\ B^8 \\ B^3
\end{pmatrix}\\=V^2
\begin{pmatrix}
K^3 & K^4 & K^5 
\end{pmatrix}&
\begin{pmatrix}
0 & 0 & 0 \\
0 & 0 & 0 \\
0 & 0 & \frac{1}{3}(g_L^2+g_R^2)
\end{pmatrix}
\begin{pmatrix}
K^3 \\ K^4 \\ K^5 
\end{pmatrix}~,
\end{align}
where
\begin{equation}
\begin{pmatrix}
K^3 \\ K^4 \\ K^5
\end{pmatrix}=
\begin{pmatrix}
\frac{g_LB^8}{\sqrt{g_L^2+\frac{3}{4}g_R^2}}-\frac{\sqrt{3}g_RA^8}{2\sqrt{g_L^2+\frac{3}{4}g_R^2}} \\ 
\frac{g_LB^8}{\sqrt{g_L^2+\frac{1}{4}g_R^2}}-\frac{g_RA^8}{2\sqrt{g_L^2+\frac{1}{4}g_R^2}} \\ 
\frac{\sqrt{3}g_RB^3}{2\sqrt{g_L^2+g_R^2}} + \frac{g_RB^8}{2\sqrt{g_L^2+g_R^2}} + \frac{g_LA^3}{\sqrt{g_L^2+g_R^2}}
\end{pmatrix}~.
\end{equation}
We see that $A^4,~A^5,~A^6,~A^7,~B^1,~B^2,~B^4,~B^5$ and $K^5$ acquired mass. So we are left with:
\begin{align}
A^1,~A^2,~A^3 ~ & ~ \rightarrow ~ SU(2)_L\\
B^6,~B^7,~K^3 ~ & ~ \rightarrow ~ SU(2)'_R\\
K^4 ~ & ~ \rightarrow ~ U(1)'
\end{align}
Again, counting the 8 gauge bosons of $SU(3)_C$ as well we have 15 massless gauge bosons.

\subsection*{Combined Breaking}

However, in order for our gauge~group to become $SU(3)_c\times SU(2)_L\times U(1)_Y$ we want~both breakings to happen simultaneously. This way,~$B^1,~B^2,~B^6,~B^7$ all acquire masses, so we do not have~any of the two $SU(2)_R$ groups of the above isolated scenaria.~Furthermore, only one linear combination of $A^8,~B^8,~B^3$~remains massless (and will be identified as $B_{\mu}$ of $U(1)_Y$),~so we can only have one $U(1)$ gauge group.

\noindent Thus, we have the masses
\begin{align}
\frac{1}{2}m_{A^4}^2=\frac{1}{2}m_{A^5}^2=\frac{1}{2}m_{A^6}^2=\frac{1}{2}m_{A^7}^2=&\frac{g_L^2V^2}{2}\\
\frac{1}{2}m_{B^4}^2=\frac{1}{2}m_{B^5}^2=&\frac{g_R^2V^2}{2}\\
\frac{1}{2}m_{B^1}^2=\frac{1}{2}m_{B^2}^2=\frac{1}{2}m_{B^6}^2=\frac{1}{2}m_{B^7}^2=&\frac{g_R^2V^2}{4}~,
\end{align}
while for $A^8,~B^8,~B^3$ we have
\begin{align}V^2
\begin{pmatrix}
A^8 & B^8 & B^3
\end{pmatrix}&
\begin{pmatrix}
\frac{2}{3}g_L^2 & -\frac{1}{6}g_Lg_R & \frac{1}{2\sqrt{3}}g_Lg_R \\
 -\frac{1}{6}g_Lg_R & \frac{5}{12}g_R^2 & \frac{1}{4\sqrt{3}}g_R^2 \\
\frac{1}{2\sqrt{3}}g_Lg_R & \frac{1}{4\sqrt{3}}g_R^2 & \frac{g_R^2}{4}
\end{pmatrix}
\begin{pmatrix}
A^8 \\ B^8 \\ B^3
\end{pmatrix}
\end{align}
\begin{align}
=V^2
\begin{pmatrix}
A^8 & B^8 & B^3
\end{pmatrix}P^{-1}P&
\begin{pmatrix}
\frac{2}{3}g_L^2 & -\frac{1}{6}g_Lg_R & \frac{1}{2\sqrt{3}}g_Lg_R \\
 -\frac{1}{6}g_Lg_R & \frac{5}{12}g_R^2 & \frac{1}{4\sqrt{3}}g_R^2 \\
\frac{1}{2\sqrt{3}}g_Lg_R & \frac{1}{4\sqrt{3}}g_R^2 & \frac{g_R^2}{4}
\end{pmatrix}P^{-1}P
\begin{pmatrix}
A^8 \\ B^8 \\ B^3
\end{pmatrix}\\=V^2
\begin{pmatrix}
B_{\mu} & C^1 & C^2 
\end{pmatrix}&
\begin{pmatrix}
0 & 0 & 0 \\
0 & \frac{g_R^2}{2} & 0 \\
0 & 0 & \frac{1}{6}(4g_L^2+g_R^2)
\end{pmatrix}
\begin{pmatrix}
B_{\mu} \\ C^1 \\ C^2 
\end{pmatrix}~,
\end{align}
where the rotation matrix is
\begin{equation}
P=\frac{1}{\sqrt{4g_L^2+g_R^2}}\begin{pmatrix}
-g_R & -g_L & \sqrt{3}g_L \\
0 & \frac{\sqrt{3}}{2}\sqrt{4g_L^2+g_R^2} & \frac{1}{2}\sqrt{4g_L^2+g_R^2} \\
2g_L & -\frac{1}{2}g_R & \frac{\sqrt{3}}{2}g_R
\end{pmatrix}
\end{equation}
and
\begin{equation}
\begin{pmatrix}
B_{\mu} \\ C^1 \\ C^2 
\end{pmatrix}= P\begin{pmatrix}
A^8 \\ B^8 \\ B^3
\end{pmatrix}
=\frac{1}{\sqrt{4g_L^2+g_R^2}}
\begin{pmatrix}
\sqrt{3}g_LB^3-g_LB^8-g_RA^8\\
B^3\frac{\sqrt{4g_L^2+g_R^2}}{2}+B^8\frac{\sqrt{3}\sqrt{4g_L^2+g_R^2}}{2}\\
\frac{\sqrt{3}}{2}g_RB^3-\frac{1}{2}g_RB^8+2g_LA^8
\end{pmatrix}~.
\end{equation}
It is understood that $C^i$ are two physical massive gauge bosons that do not correspond to any of the unbroken symmetries. It is useful to also write down
\begin{equation}
\begin{pmatrix}
A^8 \\ B^8 \\ B^3
\end{pmatrix}= P^{-1}\begin{pmatrix}
B_{\mu} \\ C^1 \\ C^2 
\end{pmatrix}
=\frac{1}{\sqrt{4g_L^2+g_R^2}}
\begin{pmatrix}
-g_RB_{\mu} + 2g_LC^2 \\
-g_LB_{\mu} - \frac{g_R}{2}C^2 + \frac{\sqrt{3}}{2}\sqrt{4g_L^2+g_R^2}C^1 \\
\sqrt{3}g_LB_{\mu} + \frac{\sqrt{3}}{2}C^2 + \frac{1}{2}\sqrt{4g_L^2+g_R^2}C^1
\end{pmatrix}\label{A8B8B3}
\end{equation}
Thus, the masses of $C^i$ will be
\begin{equation}
\frac{1}{2}m_{C^1}^2=\frac{g_R^2V^2}{2}~~~~~~~~~~~~
\frac{1}{2}m_{C^2}^2=\frac{1}{6}(4g_L^2+g_R^2)V^2~.
\end{equation}
Restoring the Lorentz indices, we see that now we have 12 massless gauge fields:
\begin{itemize}
 \item the 8 $SU(3)_C$ bosons
 \item $A_{\mu}^1,~A_{\mu}^2,~A_{\mu}^3$ that form $SU(2)_L$ and 
 \item $B_{\mu}$ that is identified as the $U(1)_Y$ gauge boson:
\begin{equation}
B_{\mu}=\sqrt{3}\frac{g_LB_{\mu}^3-\frac{g_L}{\sqrt{3}}B_{\mu}^8-\frac{g_R}{\sqrt{3}}A_{\mu}^8}{\sqrt{4g_L^2+g_R^2}}~.\label{bmu-rotation}
\end{equation}
\end{itemize}
\noindent Electroweak (EW) breaking then proceeds by the vevs
\cite{Ma:2004mi}:
\[
\langle L_s^{(3)}\rangle=\left(\begin{array}{ccc}
\upsilon_d& 0&0\\
0&\upsilon_u&0\\
0&0&0
\end{array}\right)\;.
\]

\section{Electroweak Currents and \texorpdfstring{$\sin{\theta_W}$}{Lg}}\label{chargequant}

Now one can determine the relation between the gauge couplings of $SU(3)_L\times SU(3)_R$, $g_L$ and $g_R$, and the gauge coupling of $SU(2)_L\times U(1)_Y$, $g_2$ and $g_Y$. The way to do that is by comparing the currents of the theory with the electroweak currents.
For the sake of simplicity only one of the generations will be considered. The chiral multiplets can be written once more (in a colour-implicit way):
\begin{align}
\Psi_1\equiv q^c =& \begin{pmatrix} d_R^{*1} & u_R^{*1} & D_R^{*1} \\ d_R^{*2} & u_R^{*2} & D_R^{*2} \\ d_R^{*3} & u_R^{*3} & D_R^{*3} \end{pmatrix} \equiv \begin{pmatrix} d_R^* & u_R^* & D_R^* \end{pmatrix}  ~~~ \sim (\overline{3},1,3)_{(3,1/2)}\label{Q}\\
\Psi_2\equiv Q =& \begin{pmatrix} -d_L^1 & -d_L^2 & -d_L^3 \\ u_L^1 & u_L^2 & u_L^3 \\ D_L^1 & D_L^2 & D_L^3
\end{pmatrix} \equiv \begin{pmatrix} -d_L \\ u_L \\ D_L \end{pmatrix}  ~~~ \sim (3,\overline{3},1)_{(-3,1/2)}\label{qc}\\
\Psi_3\equiv L =& \begin{pmatrix} H_d^0 & H_u^+ & \nu_L \\ H_d^- & H_u^0 & e_L \\ \nu_R^* & e_R^* & S
\end{pmatrix} ~~~ \sim (1,3,\overline{3})_{(0,-1)}~,\label{L}
\end{align}
where the first parenthesis has the representation each one belongs under the three $SU(3)$'s and  the index parenthesis the charges under the two $U(1)$'s.

\noindent We consider the generic Dirac kinetic term
\begin{equation}
i\overline{\psi}\gamma^{\mu}(\partial_{\mu}-igG_{\alpha}W_{\mu}^{\alpha})\psi
\end{equation}
where $g$ is the coupling constant, $G_{\alpha}$ are the normalized generators of the gauge group and $W_{\mu}^{\alpha}$ are the corresponding gauge bosons.
Ignoring $SU(3)_C$, we have the following expressions\footnote{We use $\overline{T}_i=-T_i^*$ instead of $T_i$ when we are in the $\overline{\mathrm{3}}$ rep instead of $\mathrm{3}$.} (we discard the $\partial_{\mu}$ term as it is not of interest at the moment):
\begin{align}
\bullet ~~&(-i^2)\overline{Q}\gamma^{\mu}g_L\overline{T}_{i}A_{\mu}^iQ~,~~~~i=1,2,3\label{a}\\
\bullet ~~&(-i^2)Tr[\overline{L}\gamma^{\mu}g_LT_iA_{\mu}^iL]~,~~~~i=1,2,3\label{b}\\
\bullet ~~&(-i^2)\overline{Q}\gamma^{\mu}g_L\overline{T}_8A_{\mu}^8Q\label{c}\\
\bullet ~~&(-i^2)Tr[\overline{q}^c\gamma^{\mu}g_Rq^cT_iB_{\mu}^i]~,~~~~i=3,8\label{d}\\
\bullet ~~&(-i^2)Tr[\overline{L}\gamma^{\mu}g_RL\overline{T}_iB_{\mu}^i]~,~~~~i=3,8\label{e}\\
\bullet ~~&(-i^2)Tr[\overline{L}\gamma^{\mu}g_LT_8A_{\mu}^8L]~.\label{st}
\end{align}

\subsection*{Weak Currents}

Expanding  \refeq{a} and \refeq{b} we get:
\begin{align}
\bullet ~~& \frac{g_L}{2}\Big[(\overline{d}_L\gamma^{\mu}u_L+\overline{u}_L\gamma^{\mu}d_L)A_{\mu}^1 + i(\overline{d}_L\gamma^{\mu}u_L-\overline{u}_L\gamma^{\mu}d_L)A_{\mu}^2 + (\overline{u}_L\gamma^{\mu}u_L-\overline{d}_L\gamma^{\mu}d_L)A_{\mu}^3 \Big] \\
\bullet ~~& \frac{g_L}{2}\Big[(\overline{e}_L\gamma^{\mu}\nu_L+\overline{\nu}_L\gamma^{\mu}e_L)A_{\mu}^1 + i(\overline{e}_L\gamma^{\mu}\nu_L-\overline{\nu}_L\gamma^{\mu}e_L)A_{\mu}^2 + (\overline{\nu}_L\gamma^{\mu}\nu_L-\overline{e}_L\gamma^{\mu}e_L)A_{\mu}^3 \Big]~,
\end{align}
where we omit the Higgsino currents as they are irrelevant to the present discussion. Re-expressing the above, we have:
 \begin{equation}
 \mathcal{L}=g_Lj^{\mu 1}A_{\mu}^1+g_Lj^{\mu 2}A_{\mu}^2+g_Lj^{\mu 3}A_{\mu}^3~,
 \end{equation}
where
\begin{align}
j^{\mu 1}=& \frac{1}{2}\Big(\overline{e}_L\gamma^{\mu}\nu_L+\overline{\nu}_L\gamma^{\mu}e_L+\overline{d}_L\gamma^{\mu}u_L+\overline{u}_L\gamma^{\mu}d_L\Big)\\
j^{\mu 2}=& \frac{i}{2}\Big(\overline{e}_L\gamma^{\mu}\nu_L-\overline{\nu}_L\gamma^{\mu}e_L+\overline{d}_L\gamma^{\mu}u_L-\overline{u}_L\gamma^{\mu}d_L\Big)\\
j^{\mu 3}=& \frac{1}{2}\Big(\overline{\nu}_L\gamma^{\mu}\nu_L-\overline{e}_L\gamma^{\mu}e_L+\overline{u}_L\gamma^{\mu}u_L-\overline{d}_L\gamma^{\mu}d_L\Big)~
\end{align}
exactly match the (supersymmetric) SM weak currents.
We can thus really identify $A_{\mu}^{1,2,3}$ as the $W_{\mu}^{1,2,3}$ of the weak interaction and $g_L$ as $g_2$.

\subsection*{$U(1)_Y$ currents}

Having identified the weak currents, \refeqs{c},(\ref{d}),(\ref{e}) and (\ref{st}) must be matched to the $U(1)_Y$ currents:
\begin{align}
\bullet ~~\phantom{-}&\frac{g_Y}{6}\Big(\overline{u}_L\gamma^{\mu}u_L+\overline{d}_L\gamma^{\mu}d_L\Big)B_{\mu}\label{i}\\
\bullet ~~-&\frac{g_Y}{2}\Big(\overline{\nu}_L\gamma^{\mu}\nu_L+\overline{e}_L\gamma^{\mu}e_L\Big)B_{\mu}\label{ii}\\
\bullet ~~-&g_Y\overline{e}_R\gamma^{\mu}e_RB_{\mu}\label{iii}\\
\bullet ~~\phantom{-}&\frac{2}{3}g_Y\overline{u}_R\gamma^{\mu}u_RB_{\mu}\label{iv}\\
\bullet ~~-&\frac{g_Y}{3}\overline{d}_R\gamma^{\mu}d_RB_{\mu}\label{v}
\end{align}
We will substitute $A_{\mu}^8,~B_{\mu}^8,~B_{\mu}^3$ from \refeq{A8B8B3}, keeping only the (relevant) $\sim B_{\mu}$ part. Again, we will ignore Higgsino currents and exotic particle currents as they are not relevant.
Matching \refeq{c} with \refeq{i} we get:
\begin{align}
-\frac{g_L}{2\sqrt{3}}\Big(\overline{u}_L\gamma^{\mu}u_L+\overline{d}_L\gamma^{\mu}d_L\Big)A_{\mu}^8+...=&
-\frac{g_L}{2\sqrt{3}}\Big(\overline{u}_L\gamma^{\mu}u_L+\overline{d}_L\gamma^{\mu}d_L\Big)\Big(\frac{-g_R}{\sqrt{4g_L^2+g_R^2}}\Big)B_{\mu}\nonumber\\
=&\frac{1}{6}\frac{\sqrt{3}g_Lg_R}{\sqrt{4g_L^2+g_R^2}}\Big(\overline{u}_L\gamma^{\mu}u_L+\overline{d}_L\gamma^{\mu}d_L\Big)B_{\mu}\nonumber\\
=&\frac{g_Y}{6}\Big(\overline{u}_L\gamma^{\mu}u_L+\overline{d}_L\gamma^{\mu}d_L\Big)B_{\mu}~.\label{ci}
\end{align}
From the above we identify\large
\begin{equation}\label{gY}
g_Y=\frac{\sqrt{3}g_Lg_R}{\sqrt{4g_L^2+g_R^2}}~.
\end{equation}\normalsize
Matching \refeq{d} with \refeqs{iv},(\ref{v}) and \refeqs{e},(\ref{st}) with \refeqs{ii},(\ref{iii}) we confirm \refeq{gY}:
\begin{align}
\frac{g_R}{2}\Big(\overline{u}_R\gamma^{\mu}u_R-\overline{d}_R\gamma^{\mu}d_R\Big)B_{\mu}^3-&
\frac{g_R}{2\sqrt{3}}\Big(\overline{u}_R\gamma^{\mu}u_R+\overline{d}_R\gamma^{\mu}d_R\Big)B_{\mu}^8+...=\nonumber\\
&=...=\frac{2}{3}\frac{\sqrt{3}g_Lg_R}{\sqrt{4g_L^2+g_R^2}}\Big(\overline{u}_R\gamma^{\mu}u_R\Big)B_{\mu}-\frac{1}{3}\frac{\sqrt{3}g_Lg_R}{\sqrt{4g_L^2+g_R^2}}\Big(\overline{d}_R\gamma^{\mu}d_R\Big)B_{\mu}\label{divv}
\end{align}
\begin{align}
\Big(\overline{\nu}_L\gamma^{\mu}\nu_L+\overline{e}_L\gamma^{\mu}e_L\Big)&\Big[\frac{g_L}{2\sqrt{3}}A_{\mu}^8+\frac{g_R}{\sqrt{3}}B_{\mu}^8\Big]+\Big(\overline{e}_R\gamma^{\mu}e_R\Big)\Big[\frac{g_L}{\sqrt{3}}A_{\mu}^8-\frac{g_R}{2}B_{\mu}^3+\frac{g_R}{2\sqrt{3}}B_{\mu}^8\Big]+...=\nonumber\\
&=...=-\frac{1}{2}\frac{\sqrt{3}g_Lg_R}{\sqrt{4g_L^2+g_R^2}}\Big(\overline{\nu}_L\gamma^{\mu}\nu_L+\overline{e}_L\gamma^{\mu}e_L\Big)B_{\mu}-\frac{\sqrt{3}g_Lg_R}{\sqrt{4g_L^2+g_R^2}}
\Big(\overline{e}_R\gamma^{\mu}e_R\Big)B_{\mu}~.
\end{align}
Now we can use \refeq{gY} to see that at $M_{GUT}$ where $g_C=g_L=g_R$ (remember that $g_L=g_2$):\large
\begin{equation}
sin^2(\theta_W)=\frac{g_Y^2}{g_Y^2+g_L^2}=\frac{\frac{3g_L^2g_R^2}{4g_L^2+g_R^2}}{\frac{3g_L^2g_R^2}{4g_L^2+g_R^2}+g_L^2}\xrightarrow{\text{$M=M_{GUT}$}}\frac{\frac{3}{4+1}}{\frac{3}{4+1}+1}=\frac{3}{8}~.
\end{equation}\normalsize

\section{Lepton Yukawa couplings and $\mu$ terms}\label{higherdimop}

The configuration of the scalar~potential just after the breaking (for details see \refapp{appendixE}) gives vevs to the singlet~of each family (not necessarily to all three). It should be reminded that, from the three singlets (for each family) only one, namely $\gamma$, survived the Wilson breaking. In our case we have~$\langle\theta^{(3)}\rangle\sim\mathcal{O}(TeV)~,~\langle\theta^{(1,2)}\rangle\sim\mathcal{O}(M_{GUT})$. The fact that one of them is at the $TeV$ scale and the other two are superheavy is crucial for the phenomenological viability of the model, as it will become apparent in the following.

Although~the two $U(1)$s were already broken before the Wilson flux breaking,~they still impose global symmetries. As a result, in the~lepton sector we cannot have invariant Yukawa terms. However~ below the unification scale, an effective term can occur from~higher-dimensional operators \cite{Irges:2011de}:
\begin{equation}
L\overline{e}H_d\Big(\frac{\overline{K}}{M}\Big)^3~,\label{leptonyukawa}
\end{equation}
where $\overline{K}$ is the~vacuum expectation value of the  conjugate scalar component~of either $S^{(i)},~\nu_R^{(i)}$ or $\theta^{(i)}$, or any~combination of them, with or without mixing of flavours.~Using similar arguments, one can also have mass terms for $S^{(i)}$~and $\nu_R^{(i)}$, which will then be rendered~supermassive.

Another much needed~quantity that is missing from our model is the $\mu$ term, one~for each family of Higgs doublets. 
In the same way,~we can have:
\begin{equation}
H_u^{(i)}H_d^{(i)}\overline{\theta}^{(i)}\frac{\overline{K}}{M}~.\label{muterm}
\end{equation}
The first two generations~of Higgs doublets will then have supermassive $\mu$ terms,~while the $\mu$ term of the third generation will be at the~$TeV$ scale.\\

\vspace{1cm}

\noindent Before proceeding to the phenomenological analysis of the next chapter, it is useful to sum up the scale~of some important~parameters in \refta{parameterscale} in order to avoid confusion.

\begin{center}
\begin{table}[ht]
\begin{center}
\small
\begin{tabular}{|l|r|}
\hline
 Parameter & Scale \\\hline
soft trilinear couplings & $\mathcal{O}(GUT)$  \\\hline
squark masses & $\mathcal{O}(GUT)$  \\\hline
slepton masses & $\mathcal{O}(TeV)$  \\\hline
$\mu^{(3)}$ & $\mathcal{O}(TeV) $  \\\hline
$\mu^{(1,2)}$ & $\mathcal{O}(GUT) $  \\\hline
unified gaugino mass $M_U$ & $\mathcal{O}(TeV) $  \\\hline
\end{tabular}
\caption{\textit{Approximate scale of parameters.}}
\label{parameterscale}
\end{center}
\end{table}
\end{center}

\newpage
\mbox{} 
\newpage

\chapter{Phenomenological Analysis of \texorpdfstring{$SU(3)^3$}{Lg} with Small Coset Space Radii}\label{pheno-su33}

Like every GUT,~this model considers all gauge couplings to start as one coupling $g$ at $M_{GUT}$. However, since at the $E_8$ level there is~only one coupling, it is clear that the (quark) Yukawa couplings~are equal to $g$ at $M_{GUT}$ as well (for a discussion on Yukawa couplings from CSDR see the last part of \refse{4daction}). This makes the selection of~a large $tan\beta$ necessary. The unified coupling $g$ is used as a~boundary condition for all the above-mentioned couplings at $M_{GUT}$. 

In the following analysis 1-loop $\beta$ functions are used for all parameters included. Below the~unification scale they run according the RGEs of the MSSM~(squarks included) plus the 4 additional Higgs doublets (and~their supersymmetric counterparts) that come from~the two extra $L$ multiplets of the first and second generations,~down to an intermediate scale $M_{int}$. Below this scale,~all supermassive particles and parameters are considered~decoupled, and the RGEs used include only the 2 Higgs doublets~that originate from the third generation (and their~respective Higgsinos), the sleptons and the gauginos. Finally,~below a second intermediate scale that we call $M_{TeV}$, we run~the RGEs of a non-supersymmetric 2HDM.

\section{Constraints}\label{constraints}
In our analysis we apply several experimental constraints, which we briefly review in this subsection\footnote{Some of the experimental values shown here are slightly different from the constraints considered in \refpa{part1}. This happened because, at the time of the original analysis of \refpa{part1}, most of the values of this section were not yet published.}.\\ 
Starting from the strong gauge coupling, we use the experimental value \cite{Zyla:2020zbs}:
\begin{equation}\label{astrong}
a_s(M_Z)=0.1187\pm0.0016\,.
\end{equation}
We calculate the top quark pole mass,~while the bottom quark~mass is evaluated~at $M_Z$, in order not to induce uncertainties that are inherent to~its pole~mass.  Their experimental values are \cite{Zyla:2020zbs}:
\beq
m_t^{\rm exp} = (172.4 \pm 0.7) \gev\,,~~~~~~ m_b(M_Z) = 2.83 \pm 0.10 \gev~.
\label{mtmbexp}
\eeq
We interpret the Higgs-like particle discovered in July 2012 by ATLAS and CMS
\cite{Aad:2012tfa}
as the light $\cal CP$-even Higgs~boson of~the supersymmetric SM.
The (SM) Higgs boson experimental average mass is \cite{Zyla:2020zbs}:
\beq
M_H^{\rm exp}=125.10\pm 0.14~{\rm GeV}~.\label{higgsexp}
\eeq

\section{Gauge unification}\label{unification}
A first challenge for~each unification model is to predict a unification scale,~while maintaining agreement with experimental constraints~on gauge couplings. The 1-loop gauge $\beta$ functions are~given by:
\begin{equation}
2\pi\beta_i=b_i\alpha_i^2~,
\end{equation}
where for our field content in each of  the three~energy regions the $b$ coefficients are given in \refta{bcoeff}~(see \refapp{appendixF} for the calculation of each $\beta$~function).
 
\begin{center}
\begin{table}[ht]
\begin{center}
\small
\begin{tabular}{|l|r|r|r|}
\hline
 Scale & $b_1$ & $b_2$ & $b_3$ \\\hline
 $M_{EW}$-$M_{TeV}$ & $\frac{21}{5}$ & $-3$ & $-7$ \\\hline
 $M_{TeV}$-$M_{int}$ & $\frac{11}{2}$ & $-\frac{1}{2}$ & $-5$ \\\hline
 $M_{int}$-$M_{GUT}$ & $\frac{39}{5}$ & $3$ & $-3$ \\\hline
\end{tabular}
\caption{\textit{$b$ coefficients for gauge RGEs.}}
\label{bcoeff}
\end{center}
\end{table}
\end{center}

The $a_{1,2}$ determine~the unification scale and the $a_3$ is used to confirm that~unification is indeed possible. Using a $0.3\%$ uncertainty~at the unification scale boundary, we predict the different~scales of our model (shown on \refta{scalesuni}), while the strong~coupling is predicted within $2\sigma$ of the experimental value~(\refeq{astrong}):
\begin{equation}
a_s(M_Z)=0.1218~.
\end{equation}

\begin{center}
\begin{table}[ht]
\begin{center}
\small
\begin{tabular}{|l|r|r|r|}
\hline
 Scale & $GeV$~~~~~  \\\hline
 $M_{GUT}$ & $\sim 1.7\times 10^{15}$  \\\hline
 $M_{int}$ & $\sim 9\times 10^{13}$~~ \\\hline
 $M_{TeV}$ & $\sim 1500$~~~~~~  \\\hline
\end{tabular}
\caption{\textit{Scale predicted by gauge unification.}}
\label{scalesuni}
\end{center}
\end{table}
\end{center}

It should be noted that~although the unification scale is somewhat lower than~expected in a supersymmetric theory, there is no fear of fast~proton decay, as the $U(1)_A$ remaining global symmetry can~be immediately recognised\footnote{One can simply compare the charges of each field under $U(1)_A$ with its respective Baryon number, see \refeq{multipletcharges}.} as:
\begin{equation}
U(1)_A=-\frac{1}{9}B~,
\end{equation} 
where $B$ is the baryon number. Therefore, the unification scale could, in principle, lie even lower without such problems.

\section{Higgs potential}\label{2hdm}
We once again turn our focus on the third family. After GUT breaking, the Higgs scalar potential is calculated from the $D$-, $F$- and soft terms of \refse{fluxsu33}. While the soft terms are extracted in a very straightforward way at this point, the $D$- and $F$- terms are somewhat more involved.

First, the $D^A$-terms $\langle\phi|G^A|\phi\rangle$ will eventually vanish identically at the vacuum for our choice of vevs, since the coefficients $(G^A)_a^b$ are antisymmetric in $a$ and $b$, so we can ignore them. The $D_1$-terms are also zero, since they do not contain any of the Higgs fields in their expressions. With help from \refapp{appendixD} the $D_2$- and $F$-terms are calculated:
\begin{align}
V_{F}^{Higgs}=&-20g^2\Big[\overline{H_d^0}H_d^-\overline{H_u^0}H_u^++c.c.\Big]\;\label{ftermhiggs}\\
V_{D_2}^{Higgs}=&\frac{10}{3}g^2\Big[|H_d^0|^4+|H_d^-|^4+|H_u^0|^4+|H_u^+|^4+ \nonumber \\
&~~~~~~~~~~~~2|H_d^0|^2|H_d^-|^2+2|H_d^-|^2|H_u^0|^2+2|H_d^0|^2|H_u^+|^2+2|H_u^0|^2|H_u^+|^2\Big]\nonumber \\
&+\frac{20}{3}g^2\Big[|H_d^0|^2|H_u^0|^2+|H_d^-|^2|H_u^+|^2\Big]\label{d2termhiggs}
\end{align}
Together with the $\mu$ term of \refeq{muterm} (see \refse{higherdimop}) and the soft terms (see \refapp{appendixD}) we have the Higgs potential:
\begin{align}\label{higgspot}
V_{Higgs}=&\Big(3|\mu^{(3)}|^2+m_3^2\Big)\Big(|H_d^0|^2+|H_d^-|^2\Big)+\Big(3|\mu^{(3)}|^2+m_3^2\Big)\Big(|H_u^0|^2+|H_u^+|^2\Big)\nonumber\\
&+b^{(3)}\Big[(H_u^+H_D^--H_u^0H_D^0)+c.c.\Big]\nonumber\\
&+\frac{10}{3}g^2\Big[|H_d^0|^4+|H_d^-|^4+|H_u^0|^4+|H_u^+|^4+ \nonumber \\
&~~~~~~~~~~~~2|H_d^0|^2|H_d^-|^2+2|H_d^-|^2|H_u^0|^2+2|H_d^0|^2|H_u^+|^2+2|H_u^0|^2|H_u^+|^2\Big]\nonumber \\
&+\frac{20}{3}g^2\Big[|H_d^0|^2|H_u^0|^2+|H_d^-|^2|H_u^+|^2\Big]-20g^2\Big[\overline{H_d^0}H_d^-\overline{H_u^0}H_u^++c.c.\Big]~,
\end{align}
where $g$ is the gauge~coupling at the unification scale and it is understood that~the RG running has not yet taken place. One can easily compare~the above potential with the standard 2 Higgs doublet scalar~potential \cite{Gunion:1984yn,Quiros:1997vk,Branco:2011iw} and identify:
\begin{equation}
\lambda_1=\lambda_2=\lambda_3=\frac{20}{3}g^2~,~~~~~~\lambda_4=20g^2~,~~~~~~\lambda_5=\lambda_6=\lambda_7=0~.
\end{equation}\label{lambdas}
The absence of $\lambda_{5,6,7}$ is expected in a supersymmetric theory (even in the case it is a broken one).
The above relations~are used as boundary conditions at GUT scale, then all the~Higgs couplings run using their RGEs (see \refapp{appendixF}~for the full expressions for each energy regime), which in turn change appropriately~for each energy~interval explained~above.

\section{1-loop results}\label{1loopresults}
The Higgs couplings $\lambda_i$ are~evolved from the GUT scale down to the EW scale together with~the gauge couplings, the top, bottom and tau Yukawas, all~at one loop. It is useful to recall that all gauge~and~quark Yukawa couplings use $g$ as boundary condition,~while the tau Yukawa emerges from a higher-dimensional operator~and has significantly wider freedom. We use the standard tau~lepton mass \cite{Zyla:2020zbs} as an~input. 

We consider~uncertainties on the two important boundaries we consider, namely~$M_{GUT}$ and $M_{TeV}$, because of threshold corrections (for~a more comprehensive discussion see \cite{Kubo:1995cg}).~For simplicity we have considered degeneracy between~all supersymmetric particles that acquire masses at the $TeV$~scale.
The uncertainty of~the top and bottom Yukawa couplings on the GUT boundary is taken~to be $6\%$, while on the $TeV$ boundary is taken to be $2\%$.~For $\lambda_{1,2}$ the uncertainty is $8\%$ on both boundaries~and for $\lambda_{3,4}$ is $7\%$ at GUT and $5\%$ at $TeV$. 

Both top and bottom quark~masses are predicted within $2\sigma$ of their~experimental values (\refeq{mtmbexp}):
\begin{equation}
m_b(M_Z)=3.00~GeV~,~~~~~~~~~~\hat{m}_t=171.6~GeV~,
\end{equation}
while the light Higgs~boson mass is predicted within $1\sigma$ of \refeq{higgsexp}:
\begin{equation}
m_h=125.18~GeV~.
\end{equation}
The model features~a large $tan\beta\sim48$. This is necessary, since~the Yukawas begin from the same value at the GUT boundary, so~a large difference between the two vevs is needed to reproduce~the known fermion hierarchy. The pseudoscalar Higgs~boson is considered to have mass between $2000-3000~GeV$, in agreement with its non-detection \cite{Aad:2020zxo} at the LHC.

\chapter{Conclusions}\label{conclusions}

This thesis focuses on supersymmetric theories that have only a  few free parameters. In the first part, after a brief discussion on the ideas concerning the~reduction of couplings of renormalizable theories and~the theoretical tools which
have been developed to~confront the problem, as well as a review of the concept of finiteness in supersymmetric theories, 
updates and new results~were given for four specific  models, in which the reduction of~parameters has been theoretically explored and tested against the~experimental data. Important~updates w.r.t.\ previous analyses~are the~improved Higgs-boson mass predictions as provided by the~latest version~of {\tt FeynHiggs} (version 2.16.0), including in particular the~improved uncertainty evaluation.~Furthermore, the CDM
predictions of each model have been evaluated with \MO (version 5.0).~From a phenomenological point of view,~the reduction of couplings method~described in the thesis~provides~selection rules that single out~realistic models, whether applied to GUTs or directly to the MSSM. In each case the number~of free parameters is~decreased substantially and the model~becomes more predictive.

The analysis is focused on four models, namely the Minimal $N=1$~$SU(5)$, the Finite $N=1$ $SU(5)$,~the Two-Loop Finite  $N=1$~$SU(3)\otimes SU(3)\otimes SU(3)$ and the Reduced MSSM, which all share~similar features.~The Minimal $N=1$ $SU(5)$ model predicts the top quark~mass and the light~Higgs boson mass in agreement with LHC measurements,~as well as the full SUSY spectrum of the MSSM.~However, concerning the bottom-quark mass predictions, relatively small~values are obtained, and agreement with the experimental data can be~found at the $3\,\sigma$~level only if~additionally a $\sim 6 \mev$
theory uncertainty is~included, favoring an extremely heavy supersymmetric~spectrum. The Finite~$N=1$ $SU(5)$ model and the Finite $N=1$ $SU(3)\otimes SU(3)\otimes SU(3)$ model are in natural~agreement with all LHC measurements and searches and feature heavy spectra as well, while the Reduced MSSM - which is the model with the lightest supersymmetric spectrum among the four -  has a low pseudoscalar Higgs boson mass $M_A$ that is ruled out by the recent ATLAS results. The three GUTs evade detection at HL-LHC, while their discovery potential at FCC-hh is determined as well. Concerning~the DM~predictions, the three former~models have an excess of
CDM w.r.t.\ the experimental~measurements, while the latter has a lower~relic density than~required by experimental searches. In the case of the Reduced MSSM the situation can be ameliorated if additional sources of CDM are allowed, in order for the experimental value to be applied only as an upper limit. If not, bilinear R-parity violating terms (that preserve finiteness) should be introduced in each model.

In the second part of the thesis an $N=1$, $10D$ $E_8$ is considered on a compactified spacetime~$M_4 \times B_0/ \mathbf{Z}_3 $, where $B_0$ is the~nearly-K\"ahler~manifold
$SU(3)/U(1) \times U(1)$~and $\mathbf{Z}_3$ is a~freely acting discrete group on~$B_0$. Then $E_8$ is dimensionally reduced on this manifold and the Wilson flux breaking mechanism is employed, leading in four dimensions to an $N=1~SU(3)^3$ gauge theory with two extra global $U(1)$s. 

In the case considered the radii of the coset space are small, in order for the compactification scale to~match the unification~scale. This choice results~in a split-like supersymmetric~scenario, where gauginos,~Higgsinos (of the third generation)~and sleptons all acquire~masses at the TeV scale, and~the rest supesymmetric spectrum~is superheavy~($\sim M_{GUT}$). The global $U(1)_A$ conserves~Baryon number, a fact which~allows for the predicted unification~scale to be $\sim10^{15}\gev$. The 2 Higgs doublets model~employed below GUT~breaking predicts a light Higgs boson mass within the~experimental boundaries, while the top and bottom quark~masses are also in ($2\sigma$) agreement with~experimental measurements. The pseudoscalar Higgs boson is considered to have mass between $2000-3000~GeV$ and the neutralino, which is expected to be the LSP, acquires mass at $\sim1500~GeV$. 

The next step in this direction includes the prediction of the full (light)~supersymmetric spectrum, a 2-loop analysis~of the model, the application of more~experimental constraints (i.e. B-physics observables), the calculation of the CDM relic density and the model's discovery potential from present and/or future colliders. This is already planned for future~work \cite{future}. Another direction in the same framework would be the examination of the high energy potential of the theory, in order to test the model's agreement with the observed values of the cosmological constant. The torsion of the model is heavily involved in this endeavour, although its value is already constrained by the gaugino masses. A success in this direction could establish
this theory as the low energy limit of the heterotic string, thus achieving the long-term challenge of the development of~a framework in which~the above successes~of the field~theory models are combined~with gravity.


\appendix

\addtocontents{toc}{\protect\setcounter{tocdepth}{0}}

\chapter{Reducing the 10-dimensional 32-spinor to 8-spinor by Majorana-Weyl Condition}\label{appendixA}

Here the reduction of the Dirac spinor with $2^{D/2}=32$ components to a Weyl-Majorana spinor with 8 components in the case of a $N=1$ theory in $10D$ is demonstrated, in order to have the same degrees of freedom as the gauge fields. We choose the following representation for the $\Gamma$ matrices:

\begin{equation}
\Gamma^{\mu}=\gamma^{\mu}\otimes I_8\pan,\pan \mu=0,1,2,3~.
\end{equation}
The Dirac spinor can be written as
\begin{equation}
\psi=(\psi_1...\psi_4\chi_1...\chi_4)^T~,
\end{equation}
where all $\psi_i$, $\chi_i$ ($i=1,...,4$) transform as $SO(1,3)$ Dirac spinors. The rest $\Gamma$ matrices are presented:

\begin{align}
\Gamma^4 &=\gamma^5\otimes\sigma^1\otimes\sigma^2\otimes\sigma^2 &,\pan \Gamma^5 &=\gamma^5\otimes\sigma^2\otimes\sigma^2\otimes\sigma^2\pan,\nonumber\\
\Gamma^6 &=\gamma^5\otimes I_2\otimes\sigma^3\otimes\sigma^2 &,\pan \Gamma^7 &=\gamma^5\otimes I_4\otimes\sigma^1\pan,\\
\Gamma^8 &=\gamma^5\otimes\sigma^3\otimes\sigma^2\otimes\sigma^2 &,\pan \Gamma^9 &=\gamma^5\otimes I_2\otimes\sigma^1\otimes\sigma^2\pan\nonumber
\end{align}
and~hence 
\begin{equation}
\Gamma^{11}=\Gamma^0...\Gamma^9=-\gamma^5\otimes I_2\otimes I_2\otimes\sigma^3=\gamma^5\otimes
\begin{pmatrix*}-I_4 & 0 \\ 0 & I_4 \\
\end{pmatrix*}~.
\end{equation}
The spinor $\psi$ is reducible, $\Gamma^{11}\psi_{\pm}=\pm\psi_{\pm}$, where $\psi_{\pm}=\frac{1}{2}(1\pm\Gamma^{11})\psi$. Then the Weyl condition $\Gamma^{11}\psi=\psi$ selects $\psi_+$, where 
\begin{equation}
\psi_+=(L\psi_1...L\psi_4R\chi_1...R\chi_4)^T~,
\end{equation}
where $L=\frac{1}{2}(1-\gamma^5)$ (left-handed) and $R=\frac{1}{2}(1+\gamma^5)$ (right-handed). The $\psi_i$ form the $4$ and the $\chi_i$ the $\overline{4}$ representations of $SO(6)$. Imposing further~the Mayorana condition on the $10$-dimensional spinor
\begin{equation}
\psi=C_{10}\Gamma^0\psi^*~,
\end{equation}
where $C_{10}=C_4\otimes \sigma_2\otimes\sigma_2\otimes I_2$, we are led to the relations $ \chi_{1,3}=C\gamma_0\psi^*_{2,4}$ and $\chi_{2,4}=-C\gamma_0\psi^*_{1,3}$. Therefore, by~imposing the Weyl and~Majorana condition in $10D$, we obtain a Weyl spinor in 4 dimensions transforming as the $4$ representation of $SO(6)$:
\begin{equation}
\psi=(L\psi_1\pan L\psi_2\pan L\psi_3\pan L\psi_4\pan R\tilde{\psi}_1\pan R\tilde{\psi}_2\pan R\tilde{\psi}_3\pan R\tilde{\psi}_4\pan )^T\pan,\pan\tilde{\psi}_i=(-1)^iC\gamma_0\psi^*_{i}~.
\end{equation}
In addition, we~need the $\gamma$ matrices in the~coset space $SU(3)/U(1)\times U(1)$. The metric is given as $g_{ab}=diag(R_1^2,R_1^2,R_2^2,R_2^2,R_3^2,R_3^2)$, $g^{ab}=diag(\frac{1}{R_1^2}\frac{1}{R_1^2}\frac{1}{R_2^2}\frac{1}{R_2^2}\frac{1}{R_3^2}\frac{1}{R_3^2},)$ and therefore the $\Gamma$ matrices are~given by
\begin{align}
\Gamma^4 &=\frac{1}{R_1}\gamma^5\otimes\sigma^1\otimes\sigma^2\otimes\sigma^2 &,\pan \Gamma^5 &=\frac{1}{R_1}\gamma^5\otimes\sigma^2\otimes\sigma^2\otimes\sigma^2\pan,\nonumber\\
\Gamma^6 &=\frac{1}{R_2}\gamma^5\otimes I_2\otimes\sigma^3\otimes\sigma^2 &,\pan \Gamma^7 &=\frac{1}{R_2}\gamma^5\otimes I_4\otimes\sigma^1\pan,\\
\Gamma^8 &=\frac{1}{R_3}\gamma^5\otimes\sigma^3\otimes\sigma^2\otimes\sigma^2 &,\pan \Gamma^9 &=\frac{1}{R_3}\gamma^5\otimes I_2\otimes\sigma^1\otimes\sigma^2\pan.\nonumber
\end{align}

\chapter{Useful relations for the gaugino mass calculation}\label{appendixB}

We use the real metric of the coset,
 $g_{ab}=diag(R_1^2,R_1^2,R_2^2,R_2^2,R_3^2,R_3^2)$. Using the structure~constants of~$SU(3)$, $f_{12}^3=2$, $f_{45}^8=f_{67}^8=\sqrt{3}$, $f_{24}^6=f_{14}^7=-f_{36}^7=-f_{15}^6=f_{34}^5=1$ (where the indices $3$ and $8$ correspond to $U(1)\times U(1)$ and~the rest are~the coset~indices), we calculate $D_{abc}$:\small
\begin{align*}
D_{523}=D_{613}=D_{624}=D_{541}=-D_{514}=-D_{532}=-D_{631}=-D_{624}=\frac{1}{2}(c-a-b)\\
D_{235}=D_{136}=D_{264}=D_{154}=-D_{145}=-D_{253}=-D_{163}=-D_{264}=\frac{1}{2}(a-b-c)\\
D_{352}=D_{361}=D_{462}=D_{415}=-D_{451}=-D_{325}=-D_{316}=-D_{426}=\frac{1}{2}(b-c-a)
\end{align*}\normalsize
From the $D$'s we calculate the contorsion tensor
\begin{equation*}
\Sigma_{abc}=2\tau(D_{abc}+D_{bca}+D_{cba})~,
\end{equation*}
and then the~tensor
\begin{equation*}
G_{abc}=D_{abc}+\frac{1}{2}\Sigma_{abc}~,
\end{equation*}
which is\scriptsize
\begin{align*}
&G_{523}=G_{613}=G_{642}=G_{541}=-G_{514}=-G_{532}=-G_{631}=-G_{642}=\frac{1}{2}[(1-\tau)c-(1+\tau)a-(1+\tau)b]\\
&G_{235}=G_{136}=G_{246}=G_{154}=-G_{145}=-G_{253}=-G_{163}=-G_{264}=\frac{1}{2}[-(1-\tau)a+(1+\tau)b+(1+\tau)c]\\
&G_{352}=G_{361}=G_{462}=G_{415}=-G_{451}=-G_{325}=-G_{316}=-G_{426}=\frac{1}{2}[-(1+\tau)a+(1-\tau)b-(1+\tau)c]
\end{align*}\normalsize

\chapter{Calculation of \texorpdfstring{$D^A$}{Lg} terms}\label{appendixC}

Here the explicit calculation of the $D^A$ terms of \refeq{DA} is given.\footnote{\cite{Kephart:1981gf}}

\subsection*{Decomposition of 27 and 78 of \texorpdfstring{$E_6$}{Lg} under \texorpdfstring{$SU(3)_c\times SU(3)_L\times SU(3)_R$}{Lg}}

The decomposition of 27 and 78 of $E_6$ under the trinification gauge group is given as:
\begin{align}
27&=(1,3,\bar{3})\oplus (3,\bar{3},1)\oplus (\bar{3},1,3)\\
78&=(1,8,1)\oplus (1,1,8)\oplus (8,1,1)\oplus (3,3,3)\oplus (\bar{3},\bar{3},\bar{3})
\end{align}
First, for bookkeeping, we study the $E_6$ group as seen under its $SU(3)^3$ maximal subgroup. A vector, $\psi_\mu, \mu=1,\ldots, 27$, of the fundamental representation of $E_6$, i.e. the $27$, can be expressed as a triple, $(\Psi_1,\Psi_2,\Psi_3)$, of the three complex $3\times 3$ matrices, $H_i$, which transform under the maximal subgroup as: 
\begin{itemize}
    \item $(1,3,\bar{3})$ is related to $\Psi_3\rightarrow L_a^{~p}$
    \item $(3,\bar{3},1)$ is related to $\Psi_2\rightarrow Q_\alpha^{~a}$
    \item $(\bar{3},1,3)$ is related to $\Psi_1\rightarrow (q^c)_p^{~\alpha}$
\end{itemize}
In addition, the corresponding $\bar{\psi}_\mu$ vector of $\overline{27}$ is expressed as $(\Psi_1^\dag,\Psi_2^\dag,\Psi_3^\dag)$.
In the $L, Q, q^c$ notation the indices are:
\begin{itemize}
    \item $\alpha, \beta, \gamma=1,\ldots 3$: $SU(3)_c$ indices
    \item $a, b, c=1,\ldots,3$: $SU(3)_L$ indices
    \item $p, q, r=1,\ldots 3$: $SU(3)_R$ indices 
\end{itemize}
The $78$ generators of $E_6$ written in the $SU(3)_c\times SU(3)_L\times SU(3)_R$ meaningful notation are:
\begin{itemize}
    \item $t_A^c\sim (8,1,1)$
    \item $t_A^L\sim (1,8,1)$
    \item $t_A^R\sim (1,1,8)$
    \item $t_{\alpha ap}\sim (3,3,3)$
    \item $\bar{t}^{\alpha ap}\sim (\bar{3},\bar{3},\bar{3})$,
\end{itemize}
where the index $A$ is referring to the octet, $A=1,\ldots 8$, of each $SU(3)$. The commutation relations of the generators written in the above form are:
\begin{align}
    [t_A^c,t_B^c]&=if_{ABC}t^c_C\,,\quad [t_A^L,t_B^L]=if_{ABC}t_C^L\,,\quad [t_A^R,t_B^R]=if_{ABC}t_C^R\,,\\
    [t_A^c,t_B^L]&\,=\,[t_A^c,t_B^R]\,=\,[t_A^L,t_B^R]\,=\,0\,,\\
    [t_A^c,t_{\alpha ap}]&=-\tfrac{1}{2}(\lambda_A)_\alpha^{~\beta}t_{\beta ap}\,,\quad [t_A^L,t_{\alpha ap}]=-\tfrac{1}{2}(\lambda_A)_a^{~b}t_{\alpha bp}\,,\quad [t_A^R,t_{\alpha ap}]=-\tfrac{1}{2}(\lambda_A)_p^{~q} t_{\alpha aq}\\
    [t_A^c,\bar{t}^{\alpha ap}]&=\tfrac{1}{2}(\lambda_A)_\beta^{~\alpha}\bar{t}^{\beta ap}\,,\quad [t_A^L,\bar{t}^{\alpha ap}]=\tfrac{1}{2}(\lambda_A)_b^{~a}\bar{t}^{\alpha bp}\,,\quad [t_A^R,\bar{t}^{\alpha ap}]=\tfrac{1}{2}(\lambda_A)_q^{~p} \bar{t}^{\alpha aq}\,,\\
    [t_{\alpha ap},{t}_{\beta bq}]&=\epsilon_{\alpha\beta\gamma}\epsilon_{abc}\epsilon_{pqr}\bar{t}^{\gamma cr}\,,\quad 
    [\bar{t}^{\alpha ap},\bar{t}_{\beta bq}]=\epsilon^{\alpha\beta\gamma}\epsilon^{abc}\epsilon^{pqr}{t}_{\gamma cr}\,,\\
    [\bar{t}^{\alpha ap},t_{\beta bq}]&=(\lambda_A)_\beta^{~\alpha}\delta_b^{~a}\delta_q^{~p}t_A^c+\delta_\beta^{~\alpha}(\lambda_A)_b^{~a}\delta_q^{~p}t_A^L+\delta_\beta^{~\alpha}\delta_b^{~a}(\lambda_A)_q^{~p}t_A^R\,.
\end{align}
where $f_{ABC}$ are the standard $SU(3)$ structure constants:
\begin{align}
    f_{12}^{~~3}&=2\,,\quad f_{45}^{~~8}\,=\, f_{67}^{~~8}=\sqrt{3}\,,\\
    f_{24}^{~~6}\,&=\,f_{14}^{~~7}\,=\, f_{25}^{~~7}\,=\, -f_{36}^{~~7}\,=\,-f_{15}^{~~6}\,=\,-f_{34}^{~~5}=1
\end{align}
and $\lambda_A$ are the eight Gell-Mann matrices. Also, the action of the various tensors (generators) on the multiplets is given as follows:
\begin{align}
    t_A^c Q&=\tfrac{1}{2}\lambda_A Q\,,\quad t_A^cL=0\,,\quad t_A^c q^c=-\tfrac{1}{2}q^c\lambda_A\,,\\
    t_A^LQ&=-\tfrac{1}{2}Q\lambda_A\,,\quad t_A^LL=\tfrac{1}{2}\lambda_AL\,,\quad t_A^Lq^c=0\,,\\
    t_A^RQ&=0\,,\quad t_A^RL=-\tfrac{1}{2}L\lambda_A\,,\quad t_A^Rq^c=\tfrac{1}{2}\lambda_Aq^c\,,\\
    t_{\alpha ap}Q_b^{~b}&=\epsilon_{\alpha\beta\gamma}\delta_a^{~b}(q^c)_p^{~\gamma}\,,\quad \bar{t}^{\alpha ap}Q_\beta^{~b}=-\epsilon^{abc}\delta_\beta^{~\alpha}L_c^{~p}\,,\\
    t_{\alpha ap}L_b^{~q}&=\epsilon_{abc}\delta_p^{~q}Q_\alpha^{~c}\,,\quad\bar{t}^{\alpha ap}L_b^{~q}=-\epsilon^{pqr}\delta_b^{~a}(q^c)_r^{~\alpha}\,,\\
    t_{\alpha ap}(q^c)_q^{~\beta}&=\epsilon_{pqr}\delta_\alpha^{~\beta}L_\alpha^{~r}\,,\quad \bar{t}^{\alpha ap}(q^c)_q^{~\beta}=-\epsilon^{\alpha\beta\gamma}\delta_q^{~p}Q_\gamma^{~a}\,.
\end{align}
So, as mentioned above, the $78$ generators are written in the following set:
\begin{equation}
    G_S=\{t_A^c,\,t_A^L,\,t_A^R,\,t_{\alpha ap},\,\bar{t}^{\alpha ap}\}\,.
\end{equation}

\newpage

\subsection*{Study of the subgroup after the Wilson flux mechanism}

Performing the breaking with the Wilson flux, the resulting gauge group is $SU(3)_c\times SU(3)_L\times SU(3)_R$ and therefore its generators are given by the following set:
\begin{equation}
    t_A^s=\{t_A^c, t_A^L, t_A^R\}\,, \quad s=c,R,L\,,
\end{equation}
where $A=1,\ldots,8$. Therefore, we keep in mind the commutation relations and actions of the generators on the triplet only for the $24$ generators that remain unbroken, neglecting everything that involves the $t_{\alpha ap}$ and $\bar{t}^{\alpha ap}$.

\noindent At this point, everything that is necessary for the calculation of the $24$ $D$s is settled (where we lowered the $A$ index for more clarity). The starting point of the calculations is:
\begin{align}
    D_A^c&=\frac{1}{\sqrt{3}}\langle \Psi_i| t_A^c | \Psi_i\rangle\\
    D_A^L&=\frac{1}{\sqrt{3}}\langle \Psi_i| t_A^L | \Psi_i\rangle\\
    D_A^R&=\frac{1}{\sqrt{3}}\langle \Psi_i| t_A^R | \Psi_i\rangle\,.
\end{align}
From now on, we make use of the renamed $H_i$, that is the $Q, L, q^c$ and use the matrix notation instead of the Dirac one. In particular, the above relations become:
\begin{align}
    D_A^c&=\frac{1}{\sqrt{3}}\left[\bar{Q}_a^{~\alpha}(t_A^cQ)_\alpha^{~a}+(\bar{q}^c)_\alpha^{~p}(t_A^cq^c)_p^{~\alpha}+\bar{L}_p^{~a}(t_A^cL)_a^{~p}\right]\\
    D_A^L&=\frac{1}{\sqrt{3}}\left[\bar{Q}_a^{~\alpha}(t_A^LQ)_\alpha^{~a}+(\bar{q}^c)_\alpha^{~p}(t_A^Lq^c)_p^{~\alpha}+\bar{L}_p^{~a}(t_A^LL)_a^{~p}\right]\\
    D_A^R&=\frac{1}{\sqrt{3}}\left[\bar{Q}_a^{~\alpha}(t_A^RQ)_\alpha^{~a}+(\bar{q}^c)_\alpha^{~p}(t_A^Rq^c)_p^{~\alpha}+\bar{L}_p^{~a}(t_A^RL)_a^{~p}\right]
\end{align}
Acting with the generators on the multiplets, according to the corresponding rules given above, the above equations turn into:
\begin{align}
    D_A^c&=\frac{1}{2\sqrt{3}}\left[\overline{Q}_a^{~\alpha}(\lambda_A)_\alpha^{~\beta}Q_\beta^{~a}-(\overline{q}^c)_\alpha^{~p}(q^c)_p^{~\beta}(\lambda_A)_\beta^{~\alpha}\right]\\
    D_A^L&=\frac{1}{2\sqrt{3}}\left[-\overline{Q}_a^{~\alpha}Q_\alpha^{~b}(\lambda_A)_b^{~a}+\overline{L}_p^{~a}(\lambda_A)_a^{~b}L_b^{~p} \right]\\
    D_A^R&=\frac{1}{2\sqrt{3}}\left[(\overline{q}^c)_\alpha^{~p}(\lambda_A)_p^{~q}(q^c)_q^{~\alpha}-\overline{L}_p^{~a}L_a^{~q}(\lambda_A)_q^{~p} \right]\,.
\end{align}
Let us now demonstrate the way the above terms are obtained through the example of $D_3^L$, taking into consideration that $\lambda_3=\mathrm{diag}(1,-1,0)$:
\begin{align}
    D_3^L&=\frac{1}{2\sqrt{3}}\left[-\overline{Q}_a^{~\alpha}Q_{\alpha}^{~b}(\lambda_3)_b^{~a}+\overline{L}_p^{~a}(\lambda_3)_a^{~b}L_b^{~p}\right]\\
    &=\frac{1}{2\sqrt{3}}\left[-\overline{Q}_1^{~\alpha}Q_\alpha^{~1}(\lambda_3)_1^{~1}-\overline{Q}_2^{~\alpha}Q_\alpha^{~2}(\lambda_3)_2^{~2}+\overline{L}_p^{~1}(\lambda_3)_1^{~1}L_1^{~p}+\overline{L}_p^{~2}(\lambda_3)_2^{~2}L_2^{~p}\right]\\
    &=\frac{1}{2\sqrt{3}}\left[-\overline{Q}_1^{~\alpha}Q_\alpha^{~1}+\overline{Q}_2^{~\alpha}Q_\alpha^{~2}+\overline{L}_p^{~1}L_1^{~p}-\overline{L}_p^{~2}L_2^{~p}\right]\\
    &=\frac{1}{2\sqrt{3}}\left[-\overline{Q}_1^{~1}Q_1^{~1}-\overline{Q}_1^{~2}Q_2^{~1}-\overline{Q}_1^{~3}Q_3^{~1}+\overline{Q}_2^{~1}Q_1^{~2}+\overline{Q}_2^{~2}Q_2^{~2}+\overline{Q}_2^{~3}Q_3^{~2} \right.\\
    &~~~~~~~~~~~~+\overline{L}_1^{~1}L_1^{~1}+\overline{L}_2^{~1}L_1^{~2}+\overline{L}_3^{~1}L_1^{~3}-\overline{L}_1^{~2}L_2^{~1}-\overline{L}_2^{~2}L_2^{~2}-\overline{L}_3^{~2}L_2^{~3}\left. \right]\\
    &=\frac{1}{2\sqrt{3}}\left[-\bar{d}_L^{1}d_L^1-\bar{d}_L^2u_L^1-\bar{d}_L^3D_L^1+\bar{u}_L^1d_L^2+\bar{u}_L^2u_L^2+\bar{u}_L^3D_L^2 \right.\\
    &~~~~~~~~~~~~\left.+\bar{H}_d^0H_d^0+\bar{H}_d^-H_u^++\bar{v}_Rv_L-\bar{H}_u^+H_d^--\bar{H}_u^0H_u^0-\bar{e}_Re_L \right]
\end{align}
Accordingly, the rest of the $D$s are calculated and their results are listed as follows:

\subsection*{The $D^c_A$s}

\begin{align}
    D_1^c&=\frac{1}{2\sqrt{3}}[\bar{d}_L^1u_L^1+\bar{u}_L^1u_L^2+\bar{D}_L^1d_L^3+\bar{d}_L^2d_L^1+\bar{u}_L^2u_L^1+\bar{D}_L^2d_L^3\nonumber\\
    &~~~~~~~-\bar{d}_R^2d_R^1-\bar{u}_R^2d_R^2-\bar{D}_R^2d_R^3-\bar{d}_R^1u_R^1-\bar{u}_R^1u_R^2-\bar{D}_R^1u_R^3]\\
    D_2^c&=\frac{i}{2\sqrt{3}}[-\bar{d}_L^1u_L^1-\bar{u}_L^1u_L^2-\bar{D}_L^1d_L^3+\bar{d}_L^2d_L^1+\bar{u}_L^2u_L^1+\bar{D}_L^2d_L^3\nonumber\\
    &~~~~~~~+\bar{d}_R^2d_R^1+\bar{u}_R^2d_R^2+\bar{D}_R^2d_R^3-\bar{d}_R^1u_R^1-\bar{u}_R^1u_R^2-\bar{D}_R^1u_R^3]\\
    D_3^c&=\frac{1}{2\sqrt{3}}[\bar{d}_L^1d_L^1+d_L^2u_L^1+\bar{D}_R^1d_L^3-\bar{d}_L^2u_L^1-\bar{u}_L^2u_L^2-\bar{D}_L^2u_L^3\nonumber\\
     &~~~~~~~-\bar{d}_R^1d_R^1-\bar{u}_R^1d_R^2-\bar{D}_R^1d_R^3+\bar{d}_R^2u_R^1+\bar{u}_R^2u_R^2+\bar{D}_R^2u_R^3]\\
     D_4^c&=\frac{1}{2\sqrt{3}}[\bar{d}_L^1d_R^3+\bar{u}_L^1D_L^2+\bar{D}_L^1D_L^3+\bar{d}_L^3d_L^1+\bar{u}_L^3d_L^2+\bar{D}_L^3d_L^3\nonumber\\
     &~~~~~~~-\bar{d}_R^3d_R^1-\bar{u}_R^3d_R^2-\bar{D}_R^3d_R^3-\bar{d}_R^1D_R^1-\bar{u}_R^1D_R^2-\bar{D}_R^1D_R^3]\\
          D_5^c&=\frac{i}{2\sqrt{3}}[-\bar{d}_L^1d_R^3-\bar{u}_L^1D_L^2-\bar{D}_L^1D_L^3+\bar{d}_L^3d_L^1+\bar{u}_L^3d_L^2+\bar{D}_L^3d_L^3\nonumber\\
     &~~~~~~~+\bar{d}_R^3d_R^1+\bar{u}_R^3d_R^2+\bar{D}_R^3d_R^3-\bar{d}_R^1D_R^1-\bar{u}_R^1D_R^2-\bar{D}_R^1D_R^3]\\
     D_6^c&=\frac{1}{2\sqrt{3}}[\bar{d}_L^2D_L^1+\bar{u}_L^2D_L^2+\bar{D}_L^2D_L^3+\bar{d}_L^3u_L^1+\bar{u}_L^3u_L^2+\bar{D}_L^3u_L^3\nonumber
\end{align}
\begin{align}
     &~~~~~~~-\bar{d}_R^3u_R^1-\bar{u}_R^3u_R^2-\bar{D}_R^3u_R^3-\bar{d}_R^2D_R^1-\bar{u}_R^2D_R^2-\bar{D}_R^2D_R^3]\\
     D_7^c&=\frac{i}{2\sqrt{3}}[-\bar{d}_L^2D_L^1-\bar{u}_L^2D_L^2-\bar{D}_L^2D_L^3+\bar{d}_L^3u_L^1+\bar{u}_L^3u_L^2+\bar{D}_L^3u_L^3\nonumber\\
     &~~~~~~~+\bar{d}_R^3u_R^1+\bar{u}_R^3u_R^2+\bar{D}_R^3u_R^3-\bar{d}_R^2D_R^1-\bar{u}_R^2D_R^2-\bar{D}_R^2D_R^3]\\
     D_8^c&=\frac{1}{6}[\bar{d}_L^1d_L^1+\bar{u}_L^1d_L^2+\bar{D}_L^1d_L^3+\bar{d}_L^2u_L^1+\bar{u}_L^2u_L^2+\bar{D}_L^2u_L^3-2\bar{d}_L^3D_L^1-2\bar{u}_L^3D_L^2-2\bar{D}_L^3D_L^3\nonumber\\
     &~~~~-\bar{d}_R^1d_R^1-\bar{u}_R^1d_R^2-\bar{D}_R^1d_R^3-\bar{d}_R^2u_R^1-\bar{u}_R^2u_R^2-\bar{D}_R^2u_R^3+2\bar{d}_R^3D_R^1+2\bar{u}_R^3D_R^2+2\bar{D}_R^3D_R^3]
\end{align}

\subsection*{The $D^L_A$s}

\begin{align}
    D_1^L&=\frac{1}{2\sqrt{3}}[-\bar{u}_L^1d_L^1-\bar{u}_L^2u_L^1-\bar{u}_L^3D_L^1-\bar{d}_L^1d_L^2-\bar{d}_L^2u_L^2-\bar{d}_L^3D_L^2\nonumber\\
    &~~~~~~~+\bar{H}_d^0H_d^-+\bar{H}_d^-H_u^0+\bar{v}_Re_L+\bar{H}_u^+H_d^0+\bar{H}_u^0H_u^++\bar{e}_Rv_R]\\
    D_2^L&=\frac{i}{2\sqrt{3}}[\bar{u}_L^1d_L^1+\bar{u}_L^2u_L^1+\bar{u}_L^3D_L^1-\bar{d}_L^1d_L^2-\bar{d}_L^2u_L^2-\bar{d}_L^3D_L^2\nonumber\\
    &~~~~~~~-\bar{H}_d^0H_d^--\bar{H}_d^-H_u^0-\bar{v}_Re_L+\bar{H}_u^+H_d^0+\bar{H}_u^0H_u^++\bar{e}_Rv_R]\\
    D_3^L&=\frac{1}{2\sqrt{3}}[-\bar{d}_L^1d_L^1-\bar{d}_L^2u_L^1-\bar{d}_L^3D_L^1+\bar{u}_L^1d_L^2+\bar{u}_L^2u_L^2+\bar{u}_L^3D_L^2\nonumber\\
    &~~~~~~~+\bar{H}_d^0H_d^0+\bar{H}_d^-H_u^++\bar{v}_Rv_L-\bar{H}_u^+H_d^--\bar{H}_u^0H_u^0-\bar{e}_Re_L]\\
    D_4^L&=\frac{1}{2\sqrt{3}}[-\bar{D}_L^1d_L^1-\bar{D}_L^2u_L^1-\bar{D}_L^3D_L^1-\bar{d}_L^1d_L^3-\bar{d}_L^2u_L^3-\bar{d}_L^3D_L^3\nonumber\\
    &~~~~~~~+\bar{H}_d^0v_R+\bar{H}_d^-e_R+\bar{v}_RS+\bar{v}_LH_d^0+\bar{e}_LH_u^++\bar{S}v_L]\\
    D_5^L&=\frac{i}{2\sqrt{3}}[\bar{D}_L^1d_L^1+\bar{D}_L^2u_L^1+\bar{D}_L^3D_L^1-\bar{d}_L^1d_L^3-\bar{d}_L^2u_L^3-\bar{d}_L^3D_L^3\nonumber\\
    &~~~~~~~-\bar{H}_d^0v_R-\bar{H}_d^-e_R-\bar{v}_RS+\bar{v}_LH_d^0+\bar{e}_LH_u^++\bar{S}v_L]\\
    D^L_6&=\frac{1}{2\sqrt{3}}[-\bar{D}_L^1d_L^2-\bar{D}_L^2u_L^2-\bar{D}_L^3D_L^2-\bar{u}_L^1d_L^3-\bar{u}_L^2u_L^3-\bar{u}_L^3D_L^3\nonumber\\
    &~~~~~~~+\bar{H}_u^+v_R+\bar{H}_u^0e_R+\bar{e}_RS+\bar{v}_LH_d^-+\bar{e}_LH_u^0+\bar{S}e_L]\\
    D^L_7&=\frac{i}{2\sqrt{3}}[\bar{D}_L^1d_L^2+\bar{D}_L^2u_L^2+\bar{D}_L^3D_L^2-\bar{u}_L^1d_L^3-\bar{u}_L^2u_L^3-\bar{u}_L^3D_L^3\nonumber\\
    &~~~~~~~-\bar{H}_u^+v_R-\bar{H}_u^0e_R-\bar{e}_RS+\bar{v}_LH_d^-+\bar{e}_LH_u^0+\bar{S}e_L]\\
    D^L_8&=\frac{1}{6}[-\bar{d}_L^1d_L^1-\bar{d}_L^2u_L^1-\bar{d}_L^3D_L^1-\bar{u}_L^1d_L^2-\bar{u}_L^2u_L^2-\bar{u}_L^3D_L^2+2\bar{D}_L^1d_L^3+2\bar{D}_L^2u_L^3+2\bar{D}_L^3D_L^3\nonumber\\
    &~~~~+\bar{H}_d^0H_d^0+\bar{H}_d^-H_u^++\bar{v}_Rv_L+\bar{H}_u^+H_d^-+\bar{H}_u^0H_0+\bar{e}_Re_L-2\bar{v}_Lv_R-2\bar{e}_Le_R-2\bar{S}S]
\end{align}

\subsection*{The $D^R_A$s}

\begin{align}
    D_1^R&=\frac{1}{2\sqrt{3}}[\bar{d}_R^1d_R^2+\bar{d}_R^2u_R^2+\bar{d}_R^3D_R^2+\bar{u}_R^1d_R^1+\bar{u}_R^2u_R^1+\bar{u}_R^3D_R^1\nonumber\\
     &~~~~~~~-\bar{H}_d^-H_d^0-\bar{H}_u^0H_d^--\bar{e}_Lv_R-\bar{H}_d^0H_u^+-\bar{H}_u^+H_u^0-\bar{v}_Le_R]\\
     D_2^R&=\frac{i}{2\sqrt{3}}[-\bar{d}_R^1d_R^2-\bar{d}_R^2u_R^2-\bar{d}_R^3D_R^2+\bar{u}_R^1d_R^1+\bar{u}_R^2u_R^1+\bar{u}_R^3D_R^1\nonumber\\
     &~~~~~~~+\bar{H}_d^-H_d^0+\bar{H}_u^0H_d^-+\bar{e}_Lv_R-\bar{H}_d^0H_u^+-\bar{H}_u^+H_u^0-\bar{v}_Le_R]\\
     D_3^R&=\frac{1}{2\sqrt{3}}[\bar{d}_R^1d_R^1+\bar{d}_R^2u_R^1+\bar{d}_R^3D_R^1-\bar{u}_R^1d_R^2-\bar{u}_R^2u_R^2-\bar{u}_R^3D_R^2\nonumber\\
     &~~~~~~~-\bar{H}_d^0H_d^0-\bar{H}_u^+H_d^--\bar{v}_Lv_R+\bar{H}_d^-H_u^++\bar{H}_u^0H_u^0+\bar{e}_Le_R]\\
     D_4^R&=\frac{1}{2\sqrt{3}}[\bar{d}_R^1d_R^3+\bar{d}_R^2u_R^3+\bar{d}_R^3D_R^3+\bar{D}_R^1d_R^1+\bar{D}_R^2u_R^1+\bar{D}_R^3D_R^1\nonumber\\
     &~~~~~~~-\bar{v}_RH_d^0-\bar{e}_RH_d^--\bar{S}v_R-\bar{H}_d^0v_L-\bar{H}_u^+e_L-\bar{v}_LS]\\
     D_5^R&=\frac{i}{2\sqrt{3}}[-\bar{d}_R^1d_R^3-\bar{d}_R^2u_R^3-\bar{d}_R^3D_R^3+\bar{D}_R^1d_R^1+\bar{D}_R^2u_R^1+\bar{D}_R^3D_R^1\nonumber\\
     &~~~~~~~+\bar{v}_RH_d^0+\bar{e}_RH_d^-+\bar{S}v_R-\bar{H}_d^0v_L-\bar{H}_u^+e_L-\bar{v}_LS]\\
     D_6^R&=\frac{1}{2\sqrt{3}}[\bar{u}_R^1d_R^3+\bar{u}_R^2u_R^3+\bar{u}_R^3D_R^3+\bar{D}_R^1d_R^2+\bar{D}_R^2u_R^2+\bar{D}_R^3D_R^2\nonumber\\
     &~~~~~~~-\bar{v}_RH_u^+-\bar{e}_RH_u^0-\bar{S}e_R-\bar{H}_d^-v_L-\bar{H}_u^0e_L-\bar{e}_LS]\\
     D_7^R&=\frac{i}{2\sqrt{3}}[-\bar{u}_R^1d_R^3-\bar{u}_R^2u_R^3-\bar{u}_R^3D_R^3+\bar{D}_R^1d_R^2+\bar{D}_R^2u_R^2+\bar{D}_R^3D_R^2\nonumber\\
     &~~~~~~~+\bar{v}_RH_u^++\bar{e}_RH_u^0+\bar{S}e_R-\bar{H}_d^-v_L-\bar{H}_u^0e_L-\bar{e}_LS]\\
     D_8^R&=\frac{1}{6}[\bar{d}_R^1d_R^1+\bar{d}_R^2u_R^1+\bar{d}_R^3D_R^1+\bar{u}_R^1d_R^2+\bar{u}_R^2u_R^2+\bar{u}_R^3D_R^2-2\bar{D}_R^1d_R^3-2\bar{D}_R^2u_R^3-2\bar{D}_R^3D_R^3\nonumber\\
     &~~~~-\bar{H}_d^0H_d^0-\bar{H}_u^+H_d^--\bar{v}_Lv_R-\bar{H}_d^-H_u^+-\bar{H}_u^0H_u^0-\bar{e}_Le_R+2\bar{v}_Rv_L+2\bar{e}_Re_L+2\bar{S}S]
\end{align}

\chapter{Calculation of scalar potential terms}\label{appendixD}

Here the expressions of the~$F$-, $D$- and soft terms are calculated, following the~results of \refse{fluxsu33}.

\subsection*{Potential due to F-terms $V_F$}

First, we need to calculate~the following derivative:
\begin{align}
    F_{\Psi_i}&=\frac{\partial W}{\partial \Psi_i},\quad F_{\theta}=\frac{\partial W}{\partial \theta}~,\quad W=\sqrt{40}d_{abc}\Psi_1^a\Psi_2^b\Psi_3^c~,\quad W^\dag= \sqrt{40}d^{abc}\Psi_{1a}\Psi_{2b}\Psi_{3c}\label{superpotential}
\end{align}
\begin{align}
    F_{\Psi_1}&=\frac{\partial W}{\partial \Psi_1^m}=\sqrt{40}d_{abc}\delta_m^a\Psi_2^b\Psi_3^c~,\quad F_{\Psi_1^\dag}^\dag=\frac{\partial W^\dag}{\partial \Psi_{1m}}=\sqrt{40}d^{abc}\delta_a^m\Psi_{2b}\Psi_{3c}\nonumber\\
    F_{\Psi_2}&=\frac{\partial W}{\partial \Psi_2^n}=\sqrt{40}d_{abc}\Psi_1^a\delta_n^b\Psi_3^c~,\quad F_{\Psi_2^\dag}^\dag=\frac{\partial W^\dag}{\partial \Psi_{2n}}=\sqrt{40}d^{abc}\Psi_{1a}\delta_b^n\Psi_{3c}\nonumber\\
    F_{\Psi_3}&=\frac{\partial W}{\partial \Psi_3^p}=\sqrt{40}d_{abc}\Psi_1^a\Psi_2^b\delta_p^c~,\quad F_{\Psi_2^\dag}^\dag=\frac{\partial W^\dag}{\partial H_{3p}}=\sqrt{40}d^{abc}\Psi_{1a}\Psi_{2b}\delta_c^p\nonumber
\end{align}
Therefore, since $F_{\theta}=0$,~the $V_F$ of \refeq{fterm} will be:
\begin{align}
    V_F&=\sum_{i=1}^3|F_{\Psi_i}|^2=\sum_{i=1}^3{F_{\Psi_i^\dag}}^\dag F_{\Psi_i}={F_{\Psi_1^\dag}}^\dag F_{\Psi_1}+{F_{\Psi_2^\dag}}^\dag F_{\Psi_2}+{F_{\Psi_3^\dag}}^\dag F_{\Psi_3}\nonumber\\
    &=\sqrt{40}d^{def}\delta_d^m\Psi_{2e}\Psi_{3f}\sqrt{40}d_{abc}\delta_m^a\Psi_{2}^b\Psi_{3}^c\nonumber\\
    &+\sqrt{40}d^{def}\Psi_{1d}\delta_e^n\Psi_{3f}\sqrt{40}d_{abc}\Psi_1^a\delta_n^b\Psi_{3}^c\nonumber\\
    &+\sqrt{40}d^{def}\Psi_{1d}\Psi_{2e}\delta_f^p\sqrt{40}d_{abc}\Psi_1^a\Psi_2^b\delta_p^c\nonumber\\
    &=40\left(d^{aef}\Psi_{2e}\Psi_{3f}d_{abc}\Psi_2^b\Psi_3^c+d^{dbf}\Psi_{1d}\Psi_{3f}d_{abc}\Psi_1^a\Psi_3^c+d^{dec}\Psi_{1d}\Psi_{2e}d_{abc}\Psi_1^a\Psi_2^b\right)\label{vfsum}
\end{align}
For convenience, the three terms of~the above equation will be studied separately. Interchanging the~various indices produces:
\begin{align}
    d^{aef}\Psi_{2e}\Psi_{3f}d_{abc}\Psi_2^b\Psi_3^c&=d_{abc}d^{ade}\Psi_2^b\Psi_3^c\Psi_{2d}\Psi_{3c}=d_{cba}d^{cde}\Psi_2^b\Psi_3^a\Psi_{2d}\Psi_{3e}=d_{cab}d^{dce}\Psi_2^a\Psi_3^b\Psi_{2d}\Psi_{3e}\nonumber\\
    &=d_{abc}d^{cde}\Psi_2^a\Psi_3^b\Psi_{2d}\Psi_{3e}\nonumber\\
    d^{dbf}d_{abc}\Psi_{1d}\Psi_{3f}\Psi_1^a\Psi_3^c&=d_{acb}d^{dcf}\Psi_{1d}\Psi_{3f}\Psi_1^a\Psi_3^b=d_{abc}d^{cde}\Psi_2^a\Psi_3^b\Psi_{1d}\Psi_{3e}\nonumber\\
    d^{dec}d_{abc}\Psi_{1d}\Psi_{3f}\Psi_1^a\Psi_3^c&=d_{acb}d^{dcf}\Psi_1^a\Psi_2^b\Psi_{1d}\Psi_{2f}=d_{abc}d^{cde}\Psi_1^a\Psi_2^b\Psi_{1d}\Psi_{2e}\nonumber
\end{align}
Taking into consideration~the above relations and putting them back together into \refeq{vfsum}, we~obtain the potential of the F-terms \refeq{orf423}:
\begin{equation}
    V_F=40d_{abc}d^{cde}\left(\Psi_1^a\Psi_2^b\Psi_{1d}\Psi_{2e}+\Psi_2^a\Psi_3^b\Psi_{2d}\Psi_{3e}+\Psi_1^a\Psi_3^b\Psi_{1d}\Psi_{3e}\right)\label{vfresultintermsofH}
\end{equation}
Now we want to express the~above potential in terms of $3\times 3$ matrices, $(q^c, Q, L)$,~which compose an alternative description of the vectors~of $E_6$, i.e. the various $H_i$s that live in representation~27. First, we consider the following cubic invariant~along with its hermitian conjugate:
\begin{equation}
    I_3=d_{abc}\Psi_1^a\Psi_2^b\Psi_3^c~,\quad \bar{I}_3=d^{abc}\Psi_{1a}\Psi_{2b}\Psi_{3c}\label{cubicinvariants}
\end{equation}
Taking into~account \cite{Kephart:1981gf}, the above expressions of the cubic~invariants convert to the following ones:
\begin{equation}
    I_3=\mathrm{det}q^c+\mathrm{detq}Q+\mathrm{detq}L-\mathrm{Tr}(q^cQL)~,\quad \bar{I}_3=\mathrm{det}q^{c\dag}+\mathrm{detq}Q^\dag+\mathrm{detq}L^\dag-\mathrm{Tr}(q^{c\dag}Q^\dag L^\dag)
\end{equation}
The invariants of~\refeq{cubicinvariants} are actually the superpotential $W$ (and~$\bar{W}$) of \refeq{superpotential}. In order to obtain the potential~due to  the F-terms, $V_F$, in terms of $q^c,Q,L$, one needs to~calculate the derivatives $F_{\Psi_i}$ of \eqref{superpotential} in~terms of them:
\begin{align}
    F_{\Psi_1}&\rightarrow F_{q^c}=\sqrt{40}\frac{\partial I_3}{\partial q^c}\equiv \sqrt{40}3\hat{q}^c~,\quad F^\dag_{\Psi_1^\dag}\rightarrow F^\dag_{q^{c\dag}}=\sqrt{40}\frac{\partial \bar{I}_3}{\partial q^{c\dag}}\equiv \sqrt{40}3\hat{q}^{c\dag}\nonumber\\
    F_{\Psi_2}&\rightarrow F_{Q}=\sqrt{40}\frac{\partial I_3}{\partial Q}\equiv \sqrt{40}3\hat{Q}~,\quad F^\dag_{\Psi_2^\dag}\rightarrow F^\dag_{Q^\dag}=\sqrt{40}\frac{\partial \bar{I}_3}{\partial Q^\dag}\equiv \sqrt{40}3\hat{Q}^\dag\nonumber\\
    F_{\Psi_3}&\rightarrow F_{L}=\sqrt{40}\frac{\partial I_3}{\partial L}\equiv \sqrt{40}3\hat{L}~,\quad F^\dag_{\Psi_3^\dag}\rightarrow F^\dag_{L^\dag}=\sqrt{40}\frac{\partial \bar{I}_3}{\partial L^\dag}\equiv \sqrt{40}3\hat{L}^\dag~,\nonumber
\end{align}
where the definitions \refeq{orf420} were used.~Therefore, in accordance with \eqref{vfsum}, the potential~$V_F$ will be written as:
\begin{align}
    V_F=|F_{\Psi_i}|^2~\rightarrow~ V_F&=\mathrm{Tr}\left(F_{q^{c\dag}}^\dag F_{q^c}+F_{Q^\dag}^\dag F_Q+F_{L\dag}^\dag F_L^\dag\right)=40\mathrm{Tr}\left(9\hat{q}^{c\dag}\hat{q}^c+9\hat{Q}^\dag\hat{Q}+9\hat{L}^\dag\hat{L}\right)\nonumber\\
    &=360\mathrm{Tr}\left(\hat{q}^{c\dag}\hat{q}^c+\hat{Q}^\dag\hat{Q}+\hat{L}^\dag\hat{L}\right)~,\label{vfintermsofhattedmatrices}
\end{align}
which is that of \refeq{ftermspotentialhattedquantities}. Moreover, the explicit expressions of the hatted quantities, which will be of importance after the vev is considered, are (\cite{Kephart:1981gf}):
\begin{align}
    3(\hat{q}^{c})_a^{~\alpha}&=\frac{\partial I_3}{\partial ({q}^{c})_{~\alpha}^{a}}=\frac{1}{2}\epsilon_{abc}\epsilon^{\alpha\beta\gamma}({q}^c)_\beta^{~b}({q}^c)_\gamma^{~c}-Q_a^{~p}L_p^{~\alpha}\nonumber\\
    3\hat{Q}_p^{~a}&=\frac{\partial I_3}{\partial {Q}_a^{~p}}=\frac{1}{2}\epsilon_{pqr}\epsilon^{abc}{Q}_b^{~q}{Q}_c^{~r}-L_p^{~\alpha}({q}^c)_{\alpha}^{~a}\nonumber\\
    3\hat{L}_\alpha^{~p}&=\frac{\partial I_3}{\partial {L}_p^{~\alpha}}=\frac{1}{2}\epsilon_{\alpha\beta\gamma}\epsilon^{pqr}{L}_q^{~\beta}L_r^{~\gamma}-(q^c)_\alpha^{~a}Q_{a}^{~p}~.\label{expressionsofhattedmatrices}
\end{align}
The above expressions~are the ones from which we will obtain the matrix elements~and then use their~explicit expressions.
Since we will only calculate these terms at the vacuum, only $\hat{L}$ will be non-zero. Thus, we can re-write  \eqref{vfintermsofhattedmatrices} as:
\begin{equation}
    V_F=360\left(\mathrm{Tr}(\hat{q}^{c\dag}\hat{q}^c)+\mathrm{Tr}(\hat{Q}^{\dag}\hat{Q}^c)+\mathrm{Tr}(\hat{L}^{\dag}\hat{L}^c)\right)=360\mathrm{Tr}(\hat{L}^{\dag}\hat{L})~,\label{vfequalstraceofL}
\end{equation}
 In order to calculate the trace of \eqref{vfequalstraceofL}, we need to calculate the matrix elements of the $\hat{L}$, in this case of the vacuum state. The calculations begins with the expression mentioned in \eqref{expressionsofhattedmatrices}, which in this case becomes:
\begin{equation}
    3(\hat{L})_\alpha^{~p}=\frac{\partial I_3}{\partial (L)_p^{~\alpha}}=\frac{1}{2}\epsilon_{\alpha\beta\gamma}\epsilon^{pqr}L_q^{~\beta}L_r^{~\gamma}\,.
\end{equation}
We calculate straightforwardly the matrix elements:
\begin{align}
    \hat{L}_1^{~1}&=\frac{1}{2}\epsilon_{1\beta\gamma}\epsilon^{1qr}L_q^{~\beta}L_r^{~\gamma}=\frac{1}{2}\left(\epsilon_{12\gamma}\epsilon^{1qr}L_q^{~2}L_r^{~\gamma}+\epsilon_{13\gamma}\epsilon^{1qr}L_q^{~3}L_r^{~\gamma}\right)=\frac{1}{2}\left(\epsilon_{123}\epsilon^{1qr}L_q^{~2}L_r^{3}+\epsilon_{132}\epsilon^{1qr}L_q^{~3}L_r^{~3}\right)\nonumber\\
    &=\frac{1}{2}\left(\epsilon_{123}\epsilon^{12r}L_2^{~2}L_r^{3}+\epsilon_{123}\epsilon^{13r}L_3^{~2}L_r^{3}+
    \epsilon_{132}\epsilon^{12r}L_2^{~3}L_r^{~2}+\epsilon_{132}\epsilon^{13r}L_3^{~3}L_r^{2}\right)\nonumber\\
    &=\frac{1}{2}\left(\epsilon_{123}\epsilon^{123}L_2^{~2}L_3^{3}+\epsilon_{123}\epsilon^{132}L_3^{~2}L_2^{3}+
    \epsilon_{132}\epsilon^{123}L_2^{~3}L_3^{~2}+\epsilon_{132}\epsilon^{132}L_3^{~3}L_2^{2}\right)\,
\end{align}
and similarly
\begin{align}
    \hat{L}_1^{~2}&=\frac{1}{2}\left(\epsilon_{123}\epsilon^{213}L_1^{~2}L_3^{3}+\epsilon_{123}\epsilon^{231}L_3^{~2}L_1^{3}+
    \epsilon_{132}\epsilon^{213}L_1^{~3}L_3^{~2}+\epsilon_{132}\epsilon^{231}L_3^{~3}L_1^{2}\right)\\    
 \hat{L}_1^{~3}&=\frac{1}{2}\left(\epsilon_{123}\epsilon^{312}L_1^{~2}L_2^{3}+\epsilon_{123}\epsilon^{321}L_2^{~2}L_1^{3}+
    \epsilon_{132}\epsilon^{312}L_1^{~3}L_2^{~2}+\epsilon_{132}\epsilon^{321}L_2^{~3}L_1^{2}\right)\\    
 \hat{L}_2^{~1}&=\frac{1}{2}\left(\epsilon_{231}\epsilon^{123}L_2^{~3}L_3^{1}+\epsilon_{231}\epsilon^{132}L_3^{~3}L_1^{2}+
    \epsilon_{213}\epsilon^{123}L_2^{~1}L_3^{~3}+\epsilon_{213}\epsilon^{132}L_3^{~1}L_2^{3}\right)\\
 \hat{L}_2^{~2}&=\frac{1}{2}\left(\epsilon_{213}\epsilon^{213}L_1^{~1}L_3^{3}+\epsilon_{213}\epsilon^{231}L_3^{~1}L_1^{3}+
    \epsilon_{231}\epsilon^{213}L_1^{~3}L_3^{~1}+\epsilon_{231}\epsilon^{231}L_3^{~3}L_1^{1}\right)
\end{align}
\begin{align}
 \hat{L}_2^{~3}&=\frac{1}{2}\left(\epsilon_{213}\epsilon^{312}L_1^{~1}L_3^{2}+\epsilon_{213}\epsilon^{321}L_2^{~1}L_1^{3}+
    \epsilon_{231}\epsilon^{312}L_1^{~3}L_2^{~1}+\epsilon_{231}\epsilon^{321}L_2^{~3}L_1^{1}\right)\\ 
 \hat{L}_3^{~1}&=\frac{1}{2}\left(\epsilon_{312}\epsilon^{123}L_2^{~1}L_3^{2}+\epsilon_{132}\epsilon^{321}L_3^{~1}L_2^{2}+
    \epsilon_{321}\epsilon^{123}L_2^{~2}L_3^{~1}+\epsilon_{321}\epsilon^{132}L_3^{~2}L_2^{1}\right)\\  
 \hat{L}_3^{~2}&=\frac{1}{2}\left(\epsilon_{312}\epsilon^{213}L_1^{~1}L_3^{2}+\epsilon_{312}\epsilon^{231}L_3^{~1}L_1^{2}+
    \epsilon_{321}\epsilon^{213}L_1^{~2}L_3^{~1}+\epsilon_{321}\epsilon^{231}L_3^{~2}L_1^{1}\right)\\      
 \hat{L}_3^{~3}&=\frac{1}{2}\left(\epsilon_{312}\epsilon^{312}L_1^{~1}L_2^{2}+\epsilon_{312}\epsilon^{321}L_2^{~1}L_1^{2}+
    \epsilon_{321}\epsilon^{312}L_1^{~2}L_2^{~1}+\epsilon_{321}\epsilon^{321}L_2^{~2}L_1^{1}\right)  
\end{align}

\subsection*{Potential due to D-terms $V_D$}

The potential due to the~D-terms is given in \refeq{dterm}. Each term of this potential~is given in \refeqs{DA}-(\ref{D2}).
The first equation is set to zero when at a vacuum, due~to the fact that the coefficients $G^A$ are antisymmetric~on their gauge~indices. Therefore, we will not~deal with it any more. The second and third expressions~of the above are written in terms of the redefined quantities~as:
\begin{align}
    D_1&=3\sqrt{\frac{10}{3}}\left(\mathrm{Tr}(q^{c\dag}q^c)-\mathrm{Tr}(Q^\dag Q)\right)\nonumber\\
    D_2&=\sqrt{\frac{10}{3}}\left(\mathrm{Tr}(q^{c\dag}q^c)+\mathrm{Tr}(Q^\dag Q)-2\mathrm{Tr}(L^\dag L)-2|\theta|^2\right)~,\label{dtermintermsofmatrices}
\end{align}
where we have made use of the fact~that $\langle \Psi_1|\Psi_1\rangle=\mathrm{Tr}(q^{c\dag}q^c)$, $\langle \Psi_2|\Psi_2\rangle=\mathrm{Tr}(Q^\dag Q)$~and $\langle \Psi_3|\Psi_3\rangle=\mathrm{Tr}(L^\dag L)$.

\subsection*{Potential due to soft terms $V_{soft}$}

The potential due to the soft terms, $V_{soft}$, is given in \refeq{softterms}. It can be  re-expressed in terms of the three matrices as:
\begin{align}
    V_{soft}&=\left(\frac{4R_1^2}{R_2^2R_3^2}-\frac{8}{R_1^2}\right)\mathrm{Tr}(q^{c\dag}q^c)+\left(\frac{4R_2^2}{R_1^2R_3^2}-\frac{8}{R_2^2}\right)\mathrm{Tr}(Q^\dag Q)\nonumber\\
    &+\left(\frac{4R_3^2}{R_1^2R_2^2}-\frac{8}{R_3^2}\right)\left(\mathrm{Tr}(L^\dag L)+|\theta|^2\right)+80\sqrt{2}\left(\frac{R_1}{R_2R_3}+\frac{R_2}{R_1R_3}+\frac{R_3}{R_1R_2}\right)(I_3+\bar{I}_3)~.\label{softtermintermsofmatrices}
\end{align}

\chapter{Trinification supersymmetry breaking terms at the vacuum}\label{appendixE}

The GUT breaking of \refse{breakings} is performed by considering the following vevs:
\begin{equation}
    L_0^{(3)}=\left(\begin{array}{ccc}
       0 & 0 & 0 \\
        0 & 0 & 0\\
        0 & 0 & V 
    \end{array}\right)~, \quad L_0^{(2)}=\left(\begin{array}{ccc}
       0 & 0 & 0 \\
        0 & 0 & 0\\
        V & 0 & 0 
    \end{array}\right)\label{gutvevs}
\end{equation}
We calculate the useful quantities:
\begin{align}
\mathrm{Tr}\left( (L_0^{(3)})^\dag L_0^{(3)}\right)=
\mathrm{Tr}\left( (L_0^{(2)})^\dag L_0^{(2)}\right)=&V^2\label{traces}\\
    \mathrm{det}\left(L_0^{(3)}\right)= \mathrm{det}\left(L_0^{(3)\dag}\right)=&0~,\\ \mathrm{det}\left(L_0^{(2)}\right)= \mathrm{det}\left(L_0^{(2)\dag}\right)=&0\,.\label{determinantsbothgenerations}
\end{align}
It is also obvious that $I_3=\bar{I}_3=0$ for each generation. 

Because of the way the vevs are placed in the two multiplets in \refeq{gutvevs}, no element of \refeq{vfequalstraceofL} survives. This means that every $F$-term vanishes, but so do the $D_1$-terms, since they only contain fields that do not acquire any vev:
\begin{equation}
    V_{D_1}^{vev}=\frac{1}{2}D_1^2=3\sqrt{\frac{10}{3}}\frac{1}{2}\left(\mathrm{Tr}(q^{c\dag}q^c)-\mathrm{Tr}(Q^\dag Q)\right)^2=0~.
\end{equation}
However, from $D_A^{L,R}$ of \refapp{appendixC} and from \refeq{dtermintermsofmatrices} of \refapp{appendixD}:
\begin{align}
    V_{D^A}^{vev}=&\frac{1}{2}D_A^2=...=\frac{V^4}{9}\label{contributionofdaterm}\\
    V_{D_2}^{vev}=&\frac{1}{2}D_2^2=\nonumber\\
    =&\frac{1}{2}\sqrt{\frac{10}{3}}\Big[\left(-2\mathrm{Tr}(L^{(3)\dag} L^{(3)})-2|\theta_0^{(3)}|^2\right)^2+
    \left(-2\mathrm{Tr}(L^{(2)\dag} L^{(2)})-2|\theta_0^{(2)}|^2\right)^2-
    \left(2|\theta_0^{(1)}|^2\right)^2\Big]=\nonumber\\
    =&\frac{20}{3}\Big[(V^2+|\theta_0^{(3)}|^2)^2+(V^2+|\theta_0^{(2)}|^2)^2+|\theta_0^{(1)}|^4\Big]\label{contributionofd2term}
\end{align}
Accordingly, beginning from \eqref{softtermintermsofmatrices}, we obtain the corresponding result for the potential due to the soft terms, again making use of the $L^\dag L$, \eqref{traces}, along with $\mathrm{Tr}(q^{c\dag}q^c)=\mathrm{Tr}(Q^\dag Q)=0$:
\begin{align}
    V_{soft}^{vev}=&\left(\frac{4R_3^2}{R_1^2R_2^2}-\frac{8}{R_3^2}\right)\Big(\mathrm{Tr}(L^{(3)\dag} L^{(3)})+|\theta_0^{(3)}|^2+\mathrm{Tr}(L^{(2)\dag} L^{(3)})+|\theta_0^{(2)}|^2+|\theta_0^{(1)}|^2\Big)\nonumber\\
    &+3\cdot80\sqrt{2}\left(\frac{R_1}{R_2R_3}+\frac{R_2}{R_1R_3}+\frac{R_3}{R_1R_2}\right)\Big(I_3+\bar{I}_3\Big)\nonumber\\
    =&\left(\frac{4R_3^2}{R_1^2R_2^2}-\frac{8}{R_3^2}\right)\Big(2V^2+|\theta_0^{(3)}|^2+|\theta_0^{(2)}|^2+|\theta_0^{(1)}|^2\Big)\label{contributionofsoftterm}
\end{align}
Therefore, restoring the coupling constant, the scalar potential for all generations at the vacuum just after the GUT breaking will be:
\begin{align}
\frac{2}{g^2}V_{scalar}^{GUT}=&\phantom{+}\frac{6}{5}\left(\frac{1}{R_1^4}+\frac{1}{R_2^4}+\frac{1}{R_3^4}\right)+\frac{V^4}{9}\nonumber\\
&+\frac{20}{3}\Big[(V^2+(\theta_0^{(3)})^2)^2+(V^2+(\theta_0^{(2)})^2)^2+(\theta_0^{(1)})^4\Big]\nonumber\\
&+\left(\frac{4R_3^2}{R_1^2R_2^2}-\frac{8}{R_3^2}\right)\Big(2V^2+|\theta_0^{(3)}|^2+|\theta_0^{(2)}|^2+|\theta_0^{(1)}|^2\Big)\label{scalarpotgut}
\end{align}

\chapter{Analytical expressions for \texorpdfstring{$\beta$}{Lg} functions}\label{appendixF}

Here the calculation of the 1-loop $\beta$ functions used in \refpa{part2} is carried out. The general gauge $\beta$ function expression can be found in \cite{Jones:1981we}, while the general $\lambda_i$ $\beta$ functions can be found in \cite{Haber:1993an}.

\subsection*{Gauge \texorpdfstring{$\beta$}{Lg} functions}

For the gauge structure $G_1\otimes G_2$ with couplings $g_1$ and $g_2$ respectively, the 1-loop $\beta_{g_1}$ is given by \cite{Jones:1981we}:
\begin{align*}
\beta_{g_1}=&\Big[(16\pi^2)^{-1}g_1^3[\frac{2}{3}T(R_1)d(R_2)+\frac{1}{3}T(S_1)d(S_2)-\frac{11}{3}C_2(G_1)]\Big]=(16\pi^2)^{-1}g_1^3b_1~.
\end{align*}
The fermion multiplets transform according to the $R_1$ representation for $G_1$ and according to the $R_2$ representation for $G_2$, while the bosonic multiplets transform according to the $S_1$ representation for $G_1$ and according to the $S_2$ representation for $G_2$.

$C_2(R)$ is the quadratic Casimir operator, while  $G$ is the adjoint representation. We have:
\begin{align}\label{beta1}
&R^{\alpha}R^{\alpha}=C_2(R)I~,\nonumber\\
 &Tr[R^{\alpha},R^{\beta}]=T(R)\delta^{\alpha,\beta}~,\\
 &C_2(R)d(R)=T(R)r~,\nonumber
\end{align}
where $R^{\alpha}$ is a matrix representation of the generators of the group, $r$ the number of generators of the group and $d(R)$ the dimension of the representation. For non-abelian groups we have
\begin{equation}\label{beta2}
C_2(G)=N~,
\end{equation}
where the convention is selected, 
\begin{equation}\label{beta3}
T(fundamental)=\frac{1}{2}~,
\end{equation}
while for the abelian $U(1)$:
\begin{align}\label{beta4}
 &C_2(G)=0~,\nonumber\\
 &C_2(R)=T(R)=Y^2~,
\end{align}
where $Y$ is the hypercharge (in the appropriate formalism).
Moreover, it is worth noting that we have taken the scalar representations~to be complex and the fermion representations complex \textit{and} chiral, as in the theories of interest left- and right- handed~projections of a field often~transform differently.

\noindent For the Standard Model gauge group $SU(3)\times SU(2)\times U(1)$ we have:
\begin{align}
b_1=&\frac{2}{3}T(R_1)d(R_2)d(R_3)+\frac{1}{3}T(S_1)d(S_2)d(S_3)-\frac{11}{3}C_s(G_1)~,\nonumber\\
b_2=&\frac{2}{3}T(R_2)d(R_1)d(R_3)+\frac{1}{3}T(S_2)d(S_1)d(S_3)-\frac{11}{3}C_s(G_2)~,\\
b_3=&\frac{2}{3}T(R_3)d(R_2)d(R_1)+\frac{1}{3}T(S_3)d(S_2)d(S_1)-\frac{11}{3}C_s(G_3)~,\nonumber
\end{align}
which, after the insertion of gauginos (and the rest of the supersymmetric spectrum) and the substitution of each field, become:
\begin{align}
b_1=&\frac{4}{9}n_Q+\frac{32}{9}n_u+\frac{8}{9}n_d+\frac{4}{3}n_L+\frac{8}{3}n_e+\frac{2}{9}n_{\tilde{Q}}+\frac{16}{9}n_{\tilde{u}}+\frac{4}{9}n_{\tilde{d}}+\frac{2}{3}n_{\tilde{L}}+\frac{4}{3}n_{\tilde{e}}+\frac{2}{3}n_H+\frac{4}{3}n_{\tilde{H}}~,\nonumber\\
b_2=&n_Q+\frac{1}{3}n_L+\frac{1}{2}n_{\tilde{Q}}+\frac{1}{6}n_{\tilde{L}}+\frac{1}{6}n_H+\frac{1}{3}n_{\tilde{H}}-2(\frac{11}{3}-\frac{2}{3}n_{\tilde{W}})~,\nonumber\\
b_3=&\frac{2}{3}n_Q+\frac{1}{3}n_u+\frac{1}{3}n_d+\frac{1}{3}n_{\tilde{Q}}+\frac{1}{6}n_{\tilde{u}}+\frac{1}{6}n_{\tilde{d}}-3(\frac{11}{3}-\frac{2}{3}n_{\tilde{g}})~,
\end{align}
where $n_i$ is the number of the fields that belong to the same multiplet, $Q$ being the quark doublet, $u,d$ the quark singlets (possibly also counting any exotic particles of the theory, such as the superheavy $D$ in the trinification model), $L$ and $e$ the lepton doublet and singlet respecitively, $H$ the Higgs doublet and their "tilded" versions their supersymmetric counterparts. $\tilde{W}$ and $\tilde{g}$ are the Winos and the gluinos, respectively. Therefore, one can now construct the 1-loop gauge $\beta$ functions of any supersymmetric version of the Standard Model, even a split-like one. It should be noted that in the $U(1)$ coefficient we use the SM formalism. The GUT formalism can be easily achieved by multiplying with a $3/20$ factor.

\subsection*{Yukawa \texorpdfstring{$\beta$}{Lg} functions}

The easiest of the $\beta$ functions needed in this thesis are the (third generation) Yukawa $\beta$ functions, since they are not very sensitive in changes at the supersymmetric content. Thus we only have to differentiate between their supersymmetric and non-supersymmetric versions:
\begin{align*}
16\pi^2\beta_{t}^{SM}=&y_t\Big[\frac{9}{2}y_t^2+\frac{3}{2}y_b^2+y_{\tau}^2-8g_3^2-\frac{9}{4}g_2^2-\frac{17}{20}g_1^2\Big]~,\\
16\pi^2\beta_{b}^{SM}=&y_b\Big[\frac{9}{2}y_b^2+\frac{3}{2}y_t^2+y_{\tau}^2-8g_3^2-\frac{9}{4}g_2^2-\frac{1}{4}g_1^2\Big]~,\\
16\pi^2\beta_{\tau}^{SM}=&y_{\tau}\Big[\frac{5}{2}y_{\tau}^2+3y_t^2+3y_b^2-\frac{9}{4}g_2^2-\frac{9}{4}g_1^2\Big]
\end{align*}
\begin{align*}
16\pi^2\beta_{t}^{SSM}=&y_t\Big[6y_t^2+y_b^2+-\frac{16}{3}g_3^2-3g_2^2-\frac{13}{15}g_1^2\Big]~,\\
16\pi^2\beta_{b}^{SSM}=&y_b\Big[6y_b^2+y_{\tau}^2-\frac{16}{3}g_3^2-3g_2^2-\frac{7}{15}g_1^2\Big]~,\\
16\pi^2\beta_{\tau}^{SSM}=&y_{\tau}\Big[4y_{\tau}^2+3y_b^2-3g_2^2-\frac{9}{5}g_1^2\Big]
\end{align*}
The above expressions are given in GUT formalism.

\subsection*{Higgs couplings \texorpdfstring{$\beta$}{Lg} functions}

The general expressions for these $\beta$ functions can be found in the Appendix A of \cite{Haber:1993an}. Here their expressions for the three energy regions (and with the appropriate field content) explained in \refcha{pheno-su33} are listed. Since $\lambda_{5,6,7}$ vanish in the model examined, only $\lambda_{1,2,3,4}$ are listed. The below expressions are in Standard Model formalism.

\vspace{1cm}

\subsubsection*{\texorpdfstring{$M_{GUT}-M_{int}$}{Lg}}

\begin{align*}
16\pi^2\beta_{\lambda_1}^{high}=&~~ 6\lambda_1^2+\lambda_3^2+(\lambda_3+\lambda_4)^2+\frac{3}{8}\Big[g_2^4+(g_2^2+g_1^2)^2\Big]\\
&-2y_{\tau}^4+(y_{\tau}^2-\frac{1}{2}g_1^2)^2+y_{\tau}^4-\frac{1}{2}y_{\tau}^2(-g_1^2+g_2^2)+\frac{1}{8}(g_2^4+g_1^4)\\
&+3\Big[-2y_b^4+(y_b^2-\frac{1}{6}g_1^2)^2+\frac{1}{9}g_1^4+y_b^4-\frac{1}{2}y_b^2(\frac{1}{3}g_1^2+g_2^2)+\frac{1}{8}(g_2^4+\frac{1}{9}g_1^4)\Big]\\
&-\frac{5}{2}g_2^4-g_1^2g_2^2-\frac{1}{2}g_1^4-\frac{1}{2}\lambda_1\Big[9g_2^2+3g_1^2-4y_{\tau}^2-12y_b^2\Big]\\
16\pi^2\beta_{\lambda_2}^{high}=&~~
6\lambda_2^2+\lambda_3^2+(\lambda_3+\lambda_4)^2+\frac{3}{8}\Big[g_2^4+(g_2^2+g_1^2)^2\Big]+(\frac{1}{2}g_1^2)^2+\frac{1}{8}(g_2^4+g_1^4)\\
&+3\Big[-2y_t^4+(\frac{1}{6}g_1^2)^2+(y_t^2-\frac{1}{3}g_1^2)^2+y_t^4+\frac{1}{2}y_t^2(\frac{1}{3}g_1^2-g_2^2)+\frac{1}{8}(g_2^4+\frac{1}{9}g_1^4)\Big]\\
&-\frac{5}{2}g_2^4-g_1^2g_2^2-\frac{1}{2}g_1^4-\frac{1}{2}\lambda_2\Big[9g_2^2+3g_1^2-12y_t^2\Big]
\end{align*}
\begin{align*}
16\pi^2\beta_{\lambda_3}^{high}=&~~
(\lambda_1+\lambda_2)(3\lambda_3+\lambda_4)+2\lambda_3^2+\lambda_4^2+\frac{3}{8}\Big[2g_2^4+(g_2^2-g_2^2)^2\Big])\\
&+\frac{1}{2}g_1^2(y_{\tau}^2-\frac{1}{2}g_1^2)-\frac{1}{4}y_{\tau}^2(g_1^2+g_2^2)+\frac{1}{8}(g_2^4-g_1^4)\\
&+3\Big[-2y_t^2y_b^2+\frac{1}{6}g_1^2(y_b^2-\frac{1}{6}g_1^2)+\frac{1}{3}g_1^2(y_t^2-\frac{1}{3}g_1^2)+y_t^2y_b^2+y_t^2y_b^2\\
&-\frac{1}{4}y_t^2(\frac{1}{3}g_1^2+g_2^2)+\frac{1}{4}y_b^2(\frac{1}{3}g_1^2-g_2^2)+\frac{1}{8}(g_2^4-\frac{1}{9}g_1^4)\Big]\\
&-\frac{5}{2}g_2^4+g_1^2g_2^2-\frac{1}{2}g_1^4-\frac{1}{4}\lambda_3\Big[2(9g_2^2+3g_1^2-6g_2^2-2g_1^2)-4y_{\tau}^2-12y_b^2-12y_t^2\Big]\\
16\pi^2\beta_{\lambda_4}^{high}=&~~
\lambda_4(\lambda_1+\lambda_2+4\lambda_3+2\lambda_4)+\frac{3}{2}g_2^2g_1^2+\frac{1}{2}g_2^2(y_{\tau}^2-\frac{1}{2}g_2^2)\\
&+3\Big[2y_t^2y_b^2-y_t^2y_b^2-(y_t^2-\frac{1}{2}g_2^2)(y_b^2-\frac{1}{2}g_2^2)\Big]\\
&+2g_2^4-2g_1^2g_2^2-\frac{1}{4}\lambda_3\Big[2(9g_2^2+3g_1^2-6g_2^2-2g_1^2)-4y_{\tau}^2-12y_b^2-12y_t^2\Big]
\end{align*}

\vspace{0.7cm}

\subsubsection*{\texorpdfstring{$M_{int}-M_{TeV}$}{Lg}}

\begin{align*}
16\pi^2\beta_{\lambda_1}^{inter.}=&~~ 6\lambda_1^2+\lambda_3^2+(\lambda_3+\lambda_4)^2+\frac{3}{8}\Big[g_2^4+(g_2^2+g_1^2)^2\Big]\\
&-2y_{\tau}^4+(y_{\tau}^2-\frac{1}{2}g_1^2)^2+y_{\tau}^4-\frac{1}{2}y_{\tau}^2(-g_1^2+g_2^2)+\frac{1}{8}(g_2^4+g_1^4)\\
&-6y_b^4-\frac{5}{2}g_2^4-g_1^2g_2^2-\frac{1}{2}g_1^4-\frac{1}{2}\lambda_1\Big[9g_2^2+3g_1^2-4y_{\tau}^2-12y_b^2\Big]\\
16\pi^2\beta_{\lambda_2}^{inter.}=&~~
6\lambda_2^2+\lambda_3^2+(\lambda_3+\lambda_4)^2+\frac{3}{8}\Big[g_2^4+(g_2^2+g_1^2)^2\Big]+(\frac{1}{2}g_1^2)^2+\frac{1}{8}(g_2^4+g_1^4)\\
&-6y_t^4-\frac{5}{2}g_2^4-g_1^2g_2^2-\frac{1}{2}g_1^4-\frac{1}{2}\lambda_2\Big[9g_2^2+3g_1^2-12y_t^2\Big]\\
16\pi^2\beta_{\lambda_3}^{inter.}=&~~
(\lambda_1+\lambda_2)(3\lambda_3+\lambda_4)+2\lambda_3^2+\lambda_4^2+\frac{3}{8}\Big[2g_2^4+(g_2^2-g_2^2)^2\Big])\\
&+\frac{1}{2}g_1^2(y_{\tau}^2-\frac{1}{2}g_1^2)-\frac{1}{4}y_{\tau}^2(g_1^2+g_2^2)+\frac{1}{8}(g_2^4-g_1^4)\\
&-6y_t^2y_b^2-\frac{5}{2}g_2^4+g_1^2g_2^2-\frac{1}{2}g_1^4-\frac{1}{4}\lambda_3\Big[2(9g_2^2+3g_1^2-6g_2^2-2g_1^2)-4y_{\tau}^2-12y_b^2-12y_t^2\Big]\\
16\pi^2\beta_{\lambda_4}^{inter.}=&~~
\lambda_4(\lambda_1+\lambda_2+4\lambda_3+2\lambda_4)+\frac{3}{2}g_2^2g_1^2+\frac{1}{2}g_2^2(y_{\tau}^2-\frac{1}{2}g_2^2)\\
&+6y_t^2y_b^2+2g_2^4-2g_1^2g_2^2-\frac{1}{4}\lambda_3\Big[2(9g_2^2+3g_1^2-6g_2^2-2g_1^2)-4y_{\tau}^2-12y_b^2-12y_t^2\Big]
\end{align*}

\subsubsection*{\texorpdfstring{$M_{TeV}-M_{EW}$}{Lg}}

\begin{align*}
16\pi^2\beta_{\lambda_1}^{low}=&~~ 6\lambda_1^2+\lambda_3^2+(\lambda_3+\lambda_4)^2+\frac{3}{8}\Big[g_2^4+(g_2^2+g_1^2)^2\Big]\\
&-2y_{\tau}^4-6y_b^4-\frac{1}{2}\lambda_1\Big[9g_2^2+3g_1^2-4y_{\tau}^2-12y_b^2\Big]\\
16\pi^2\beta_{\lambda_2}^{low}=&~~
6\lambda_2^2+\lambda_3^2+(\lambda_3+\lambda_4)^2+\frac{3}{8}\Big[g_2^4+(g_2^2+g_1^2)^2\Big]\\
&-6y_t^4-\frac{1}{2}\lambda_2\Big[9g_2^2+3g_1^2-12y_t^2\Big]\\
16\pi^2\beta_{\lambda_3}^{low}=&~~
(\lambda_1+\lambda_2)(3\lambda_3+\lambda_4)+2\lambda_3^2+\lambda_4^2+\frac{3}{8}\Big[2g_2^4+(g_2^2-g_2^2)^2\Big]\\
&-6y_t^2y_b^2-\frac{1}{4}\lambda_3\Big[2(9g_2^2+3g_1^2)-4y_{\tau}^2-12y_b^2-12y_t^2\Big]\\
16\pi^2\beta_{\lambda_4}^{low}=&~~
\lambda_4(\lambda_1+\lambda_2+4\lambda_3+2\lambda_4)+\frac{3}{2}g_2^2g_1^2\\
&+6y_t^2y_b^2-\frac{1}{4}\lambda_3\Big[2(9g_2^2+3g_1^2)-4y_{\tau}^2-12y_b^2-12y_t^2\Big]
\end{align*}

\newpage
\thispagestyle{empty}
\mbox{} 
\newpage

\thispagestyle{empty}
\mbox{}

\end{document}